\documentclass[a4paper,fleqn,usenatbib]{mnras}

\usepackage{amsmath}	
\usepackage{amssymb}	
\usepackage{txfonts}

\usepackage[T1]{fontenc}
\usepackage{ae,aecompl}

\usepackage{graphicx}	
\usepackage{placeins}
\usepackage{pdflscape}
\usepackage[caption=false]{subfig}
\usepackage{xcolor}
\usepackage{url}

\newcommand{\AV}{\hbox{$A_{\rm V}$}}

\newcommand{\Bl}{\hbox{$B_{\rm \ell}$}}
\newcommand{\BmV}{\hbox{${\rm B-V}$}}
\newcommand{\chisqr}{\hbox{$\chi^2_{\rm r}$}}
\newcommand{\dom}{\hbox{$d\Omega$}}
\newcommand{\kms}{\hbox{km\,s$^{-1}$}}
\newcommand{\lkca}{\hbox{LkCa{~}4}}
\newcommand{\logg}{\hbox{$\log g$}}

\newcommand{\lstar}{\hbox{$L_{\star}$}}
\newcommand{\lsun}{\hbox{${\rm L}_{\odot}$}}
\newcommand{\mjup}{\hbox{${\rm M}_{\rm Jup}$}}
\newcommand{\mps}{\hbox{m\,s$^{-1}$}}
\newcommand{\mrd}{\hbox{mrad\,d$^{-1}$}}
\newcommand{\msini}{\hbox{$M\sin i$}}
\newcommand{\mstar}{\hbox{$M_{\star}$}}
\newcommand{\msun}{\hbox{${\rm M}_{\odot}$}}
\newcommand{\omeq}{\hbox{$\Omega_{\rm eq}$}}

\newcommand{\Prot}{\hbox{$P_{\rm rot}$}}
\newcommand{\rpd}{\hbox{rad\,d$^{-1}$}}
\newcommand{\rstar}{\hbox{$R_{\star}$}}
\newcommand{\rsun}{\hbox{${\rm R}_{\odot}$}}
\newcommand{\rvfil}{\hbox{RV$_{\rm filt}$}}
\newcommand{\rvraw}{\hbox{RV$_{\rm raw}$}}
\newcommand{\sn}{\hbox{S/N}}
\newcommand{\sti}{\hbox{Stokes $I$}}
\newcommand{\stv}{\hbox{Stokes $V$}}
\newcommand{\tap}{\hbox{TAP{~}26}}
\newcommand{\teff}{\hbox{$T_{\rm eff}$}}

\newcommand{\vsini}{\hbox{$v \sin i$}}
\newcommand{\vt}{\hbox{V410{~}Tau}}
\newcommand{\vtt}{\hbox{V830{~}Tau}}

\newcommand{\caii}{\hbox{Ca$\;${\sc ii}}}

\newcommand{\hei}{\hbox{He$\;${\sc i}}}
\newcommand{\hal}{\hbox{H${\alpha}$}}

\title[Mag. field, activity \& companions of \vt]{Magnetic field, activity and companions of \vt}

\author[L.{~}Yu et al.]{L.{~}Yu$^1$\thanks{E-mail: louise.yu@irap.omp.eu}, J.-F.{~}Donati$^1$, K.{~}Grankin$^2$, A.{~}Collier Cameron$^3$, C.{~}Moutou$^4$, G.{~}Hussain$^{5,1}$,
	\newauthor C.{~}Baruteau$^1$, L. Jouve$^1$ and the MaTYSSE collaboration\\\\
	All affiliations are listed at the end of the paper
}

\date{Accepted 2019 September 3. Received 2019 August 28; in original form 2019 July 11}

\pubyear{2018}

\begin{document}
	\label{firstpage}
	\pagerange{\pageref{firstpage}--\pageref{lastpage}}
	\maketitle

	\begin{abstract}
		We report the analysis, conducted as part of the MaTYSSE programme, of a spectropolarimetric monitoring of the $\sim$0.8{~}Myr, $\sim$1.4{~}\msun\ disc-less weak-line T Tauri star \vt\ with the ESPaDOnS instrument at the Canada-France-Hawaii Telescope and NARVAL at the T\'{e}lescope Bernard Lyot, between 2008 and 2016. With Zeeman-Doppler Imaging, we reconstruct the surface brightness and magnetic field of \vt, and show that the star is heavily spotted and possesses a $\sim$550{~}G relatively toroidal magnetic field.

		We find that \vt\ features a weak level of surface differential rotation between the equator and pole $\sim$5 times weaker than the solar differential rotation. The spectropolarimetric data exhibit intrinsic variability, beyond differential rotation, which points towards a dynamo-generated field rather than a fossil field. Long-term variations in the photometric data suggest that spots appear at increasing latitudes over the span of our dataset, implying that, if \vt\ has a magnetic cycle, it would have a period of more than 8 years.

		Having derived raw radial velocities (RVs) from our spectra, we filter out the stellar activity jitter, modeled either from our Doppler maps or using Gaussian Process Regression. Thus filtered, our RVs exclude the presence of a hot Jupiter-mass companion below $\sim$0.1{~}au, which is suggestive that hot Jupiter formation may be inhibited by the early depletion of the circumstellar disc, which for \vt\ may have been caused by the close (few tens of au) M dwarf stellar companion.
	\end{abstract}




	\section{Introduction}
	\label{sec:int}

	Investigating the birth and youth of low-mass stars ($<3${~}\msun) and of their planetary systems heavily contributes to unveiling the origin and history of the Sun and of its planets, in particular the life-hosting Earth. We know that stars and their planets form from the collapse of parsec-sized molecular clouds which progressively flatten into massive accretion discs, until finally settling as pre-main-sequence (PMS) stars surrounded by protoplanetary discs. T Tauri stars (TTSs) are PMS stars that have emerged from their dust cocoons and are gravitationally contracting towards the main sequence (MS); typically aged 1-15{~}Myr, they are classical TTSs (cTTSs) when they are still surrounded by a massive accretion disc (where planets are potentially forming), and weak-line TTSs (wTTSs) when their accretion has stopped and their inner disc has dissipated. Large-scale magnetic fields are known to play a crucial role in the early life of low-mass stars, as they can open a magnetospheric gap at the center of the disc, funnel accreting disc material onto the star, induce stellar winds and prominences, and thus impact the angular momentum evolution of TTSs \citep[][]{Donati09}. Observing and understanding the magnetic topologies of TTSs is therefore a necessary endeavour to complete our understanding of stellar and planetary formation \citep[e.g.][]{Bouvier07}.

	Since the first detection of a magnetic field around a cTTS nearly 20 years ago \citep[][]{Johns99b}, the large-scale topologies of a dozen cTTSs were mapped \citep[e.g.][]{Donati07,Hussain09,Donati10,Donati13} thanks to the MaPP (Magnetic Protostars and Planets) Large Observing Programme allocated on the 3.6{~}m Canada-France-Hawaii Telescope (CFHT) with the ESPaDOnS (Echelle SpectroPolarimetric Device for the Observation of Stars) high-resolution spectropolarimeter, using Zeeman-Doppler Imaging (ZDI), a tomography technique designed for imaging the brightness features and magnetic topologies at the surfaces of active stars \citep[eg][]{Brown91, Donati97c}. This first exploration showed that the topologies of cTTSs are either quite simple or rather complex depending on whether the stars are fully convective or largely radiative respectively \citep{Gregory12,Donati13}. Moreover, these fields are reported to vary with time \citep[e.g.][]{Donati11,Donati12,Donati13} and resemble those of mature stars with similar internal structure \citep[e.g.][]{Morin08b}, suggesting that they are produced through dynamo processes within the bulk of the convective zone.

	The MaTYSSE (Magnetic Topologies of Young Stars and the Survival of close-in giant Exoplanets) Large Programme aims at mapping the large-scale magnetic topologies of $\sim$35 wTTSs, comparing them to those of cTTSs and MS stars, and probing the potential presence of massive close-in exoplanets (hot Jupiters/hJs) around its targets. It was allocated at CFHT over semesters 2013a to 2016b (510{~}h) with complementary observations from the ESPaDOnS twin NARVAL on the T\'{e}lescope Bernard Lyot (TBL) at Pic du Midi in France and from the HARPS spectropolarimeter at the ESO Telescope at La Silla in Chile. Up to now, about a dozen wTTSs were studied with MaTYSSE for their magnetic topologies and activity, for example \vt\ \citep[][]{Skelly10}, \lkca\ \citep[][]{Donati14} and \vtt\ \citep[][]{Donati17}. These studies showed that the fields of wTTSs are much more diverse than those of cTTSs, with for example \vt\ and LkCa{~}4 displaying strong toroidal components despite being fully convective, as opposed to the results obtained on cTTSs \citep[see discussion in][]{Donati14}. MaTYSSE fostered the detection of two hJs around wTTSs, the 2{~}Myr-old \vtt{~}b \citep[][]{Donati16, Donati17} and the 17{~}Myr-old TAP{~}26{~}b \citep[][]{Yu17}.

	This new study focuses on \vt, a very young \citep[$\sim$1{~}Myr in][]{Skelly10} disc-less wTTS \citep[][]{Luhman10} with a well-constrained rotation period of 1.872{~}d \citep[][]{Stelzer03}. One of the most observed wTTSs, \vt\ has been the target of both photometric and spectropolarimetric observation campaigns. High variability detected in its light curve \citep[][]{Bouvier89, Sokoloff08, Grankin08} indicates a high level of activity, confirmed with Doppler maps \citep[][]{Skelly10,Rice11,Carroll12} showing that the photosphere features large polar and equatorial cool spots, responsible for these modulations. Magnetic maps made by \cite{Skelly10} and \cite{Carroll12} have shown a highly toroidal and non-axisymmetric large-scale field despite the mostly convective structure of \vt.

	We first describe our data, comprising new NARVAL data from MaTYSSE added to previous spectropolarimetric data taken with ESPaDOnS and NARVAL in 2008-2011, and contemporaneous photometric observations taken at the Crimean Astronomical Observatory (CrAO) and from the Super Wide Angle Search for Planets (SuperWASP) campaign (Section{~}\ref{sec:obs}). We then derive the general properties of \vt\ (Section{~}\ref{sec:phy}), after which we pursue the investigation of both its photosphere and its magnetic field, using ZDI with a model including both bright plages and cool spots (Section{~}\ref{sec:mod}). Then, we disentangle the activity jitter from the actual radial velocities (RVs) in the RV curve, using models from our ZDI maps and Gaussian Process Regression (GPR), in order to look for a potential planet signature (Section{~}\ref{sec:rv}), before finally discussing our results and concluding (Section{~}\ref{sec:ccl}).

	\section{Observations}
	\label{sec:obs}

	Our spectropolarimetric data set spans from 2008 Oct to 2016 Jan, totalling 144 high-resolution optical spectra, both unpolarized (\sti) and circularly polarized (\stv). It is composed of 8 runs, most of which cover around 15 days, taken during 4 different seasons: 2008b-2009a, 2011a, 2013b and 2015b-2016a. The full journal of observations is available in Table{~}\ref{tab:sob}. The 2008b data set and 4 points in the 2009a data set were taken with the ESPaDOnS echelle spectropolarimeter at CFHT, while the rest were taken with the ESPaDOnS twin NARVAL installed at TBL.

	The raw frames are processed with the nominal reduction package {\sc Libre Esprit} as described in e.g.\ \cite{Donati97b, Donati11}, yielding a typical root-mean-square (rms) RV precision of 20-30{~}\mps \citep[][]{Moutou07,Donati08}. The peak signal-to-noise ratios (\sn, per 2.6{~}\kms\ velocity bin) reached on the spectra range between 82 and 238 for the majority (3 spectra have a \sn\ lower than 70 and were rejected for ZDI and the RV analysis), with a median of 140.

	Time is counted in units of stellar rotation, using the same reference date and rotation period as in \cite{Skelly10}, namely ${\mbox{BJD}_0=2,454,832.58033}$ and ${\Prot=1.871970\pm0.000010}${~}d \citep[][]{Stelzer03} respectively:
	\begin{equation}
	c = ( \mbox{BJD} - \mbox{BJD}_0 ) / \Prot .
	\label{eq:eph}
	\end{equation}
	The stellar phase is defined as the decimal part of the cycle $c$.

	The emission core of the \caii\ infrared triplet (IRT) presents an average equivalent width (EW) of $\simeq$13{~}\kms\ (0.37{~}\AA). The \hei\ $D_3$ line is relatively weak with an average EW of 13{~}\kms\ as well (0.25{~}\AA), in agreement with the non-accreting status of \vt. The \hal\ line has an average EW of 14{~}\kms\ (0.33{~}\AA) and a rms EW of 27{~}\kms\ and exhibits a periodicity of period $1.8720\pm 0.0009${~}d (see Appendix{~}\ref{anx:act}). From the \hei\ $D_3$ line, we detected small flares on 2008 Dec 10 (rotational cycle -15+3.514, as per Table~\ref{tab:sob}), on the night of 2013 Dec 08 to 2013 Dec 09 (rotational cycles 959+4.090 and 959+4.151), and on the night of 2016 Jan 20 (rotational cycles 1376+0.021 and 1376+0.040). One big flare, on 2008 Dec 15 (rotational cycle -15+6.181), was visible not only in \hei\ $D_3$ (EW $\simeq 30${~}\kms) but also in \hal\ (EW $\simeq 230${~}\kms) and the \caii\ IRT (core emission EW $\simeq 40${~}\kms). We removed the 6 flare-subjected observations from our data sets in order to proceed with the mapping of the photosphere and surface magnetic field, as well as the RV analysis.

	Least-squares deconvolution \citep[LSD, see][]{Donati97b} was applied to all our spectra in order to add up information from all spectral lines and boost the resulting S/N of both Stokes \textit{I} and \textit{V} LSD profiles. The spectral mask we employed for LSD was computed from an {\sc Atlas9} LTE model atmosphere \citep{Kurucz93} featuring \teff=4,500{~}K and \logg=3.5, and involves about 7{~}800 spectral features \citep[with about 40{~}\% from Fe{~}{\sc i}, see e.g.][for more details]{Donati10b}. \sti\ and \stv\ LSD profiles shown in Section{~}\ref{sec:mod} display distorsions that betray the stellar activity with a periodicity corresponding to the rotation of the star. Moonlight pollution, which affects 15 of our \sti\ LSD profiles, was filtered out using a two-step tomographic imaging process described in \cite{Donati16}. The \sn\ in the \sti\ LSD profiles, ranging from 1633 to 2930 (per 1.8{~}\kms\ velocity bin) with a median of 2410, is measured from continuum intervals, including not only the noise from photon statistics, but also the (often dominant) noise introduced by LSD (see Table{~}\ref{tab:sob}). The \sn\ in \stv\ LSD profiles, dominated by photon statistics, range from 1817 to 6970 with a median value of 3584.

	Phase coverage is of varying quality depending on the observation epoch. The 2008b data set, with only 6 points, covers only half the surface of the star (phases -0.20 to 0.30). The 2009a data set, although the densest with 48 points in 16 days and including data from both instruments, lacks observations between phases 0.05 and 0.20. The 2011a data set presents a large gap between phases -0.05 and 0.15, and a smaller one between phases 0.65 and 0.80. The 2013b and 2015b data sets are well sampled, and the 2016a data set, with only 9 points, lacks observations between phases 0.25 to 0.45 and -0.15 to 0.05.

	Contemporaneous BVR$_{\rm J}$I$_{\rm J}$ photometric measurements, documented in Table{~}\ref{tab:pob}, were taken from the Crimean Astrophysical Observatory 1.25{~}m and 0.60{~}m telescopes between August 2008 and March 2017, counting 420 observations distributed over 9 runs at a rate of one run per year, each run covering 3 to 7 months. In each run, the visible magnitude presents modulations of a period $\sim$1.87{~}d and amplitude varying from 0.04 to 0.24{~}mag (see Appendix{~}\ref{anx:pha}). The visible magnitude reaches a global minimum of 10.563 during the 2014b run. We also used 2703 data points of visible magnitude from the Wide Angle Search for Planets \citep[WASP][]{Pollacco06} photometric campaign covering semesters 2010b-2011a. Plots of the photometric data contemporaneous to our spectropolarimetric runs (i.e.\ 2008b+2009a, 2010b+2011a, 2013b+2014a and 2015b+2016a) are visible in Section{~}\ref{sec:mod}.

	\section{Evolutionary status of \vt}
	\label{sec:phy}

	V410{~}Tau is a very well-observed three-star system located in the Taurus constellation at ${d = 129.0\pm 0.5}${~}pc from Earth \citep[][ we chose this value over the Gaia result, ${130.4\pm 0.9}${~}pc, because it is both in agreement with it and more precise]{Galli18}. \vt{~}B was estimated to have a mass ${0.2\pm 0.1}$ times that of \vt{~}A, and \vt{~}C to have a mass $0.08^{+0.10}_{-0.08}$ times that of \vt{~}AB \citep{Kraus11}. The sky-projected separation between \vt{~}A and \vt{~}B was measured at ${0.13\pm 0.01}${~}arcsec, i.e.\ ${16.8\pm 1.4}${~}au, and that between \vt{~}AB and \vt{~}C was measured at ${0.28\pm 0.01}${~}arcsec, i.e.\ ${36\pm 3}${~}au. Given that \vt{~}A is much brighter than \vt{~}B and \vt{~}C in the optical bandwidth \citep{Ghez97}, we consider that the spectra analysed in this study characterize the light of \vt{~}A predominantly. Applying the automatic spectral classification tool developped within the frame of the MaPP and MaTYSSE projects \citep[][]{Donati12}, we constrain the temperature and logarithmic gravity of \vt{~}A to, respectively, ${\teff = 4500\pm 100}${~}K and ${\logg = 3.8\pm 0.2}$.

	Its rotation period was previously estimated to ${\Prot=1.871970\pm 0.000010}${~}d \citep{Stelzer03}, a value which we use throughout this paper to phase our data (see Eq.{~}\ref{eq:eph}). Comparing both our contemporary measurements (Table{~}\ref{tab:pob}) and those found in \cite{Grankin08}, we find that the minimum magnitude measured on \vt\ is ${10.52\pm 0.02}$, value that we use as a reference to compute the unspotted magnitude.

	Our photometric measurements yield a mean \BmV\ index of ${1.17\pm 0.02}$, and since the theoretical \BmV\ at 4500{~}K is ${1.04\pm 0.02}$ \citep[][Table{~}6]{Pecaut13}, the amount of visual extinction is ${\AV = 3.1\cdot (1.17 - 1.04) = 0.40\pm 0.10}$. The bolometric correction at \teff\ being equal to ${-0.64\pm 0.05}$ \citep[][Table{~}6]{Pecaut13}, and the distance modulus to ${-5 \cdot \log_{10} (d / 10) = -5.55\pm 0.01}$, we find an absolute magnitude of ${3.93\pm 0.11}$.

	The value of \vsini\footnote{line-of-sight-projected equatorial rotation velocity} found from the spectra, ${73.2\pm 0.2}${~}\kms\ (see Section{~}\ref{sec:mod}), indicates that the minimum radius of the star \rstar${\sin i}$ is equal to ${2.708\pm 0.007}${~}\rsun, which implies a maximum absolute unspotted magnitude of ${3.67\pm 0.10}$ given the photospheric temperature. The discrepancy with the value found in the previous paragraph indicates the presence of dark spots on the photosphere even when the star is the brightest. If we assume a spot coverage at maximum brightness of $\sim$25{~}\%, typical of active stars, \citep[like it was done in ][]{Donati14,Donati15,Yu17}, then the unspotted absolute magnitude would be ${3.61\pm 0.32}$, which corresponds to an inclination\footnote{angle between the stellar rotation axis and the line of sight} of ${77\pm 22}$\degr. However, the models best fitting our spectra have an inclination of ${50\pm 10}$\degr\ (see Sec 4), which would require the spot coverage at maximum brightness to actually be $\sim$50{~}\%. Such a high permanent spot coverage is unusual but not unconceivable, since another wTTS, LkCa4, was observed to have as much as 80{~}\% of its surface covered with spots \citep[][]{Gully-Santiago17}. Assuming a spot coverage at maximum brightness of ${50\pm 15}${~}\% for \vt, we derive an absolute unspotted magnitude of ${3.17\pm 0.33}$, a logarithmic luminosity ${\log (\lstar / \lsun) = 0.63\pm 0.13}$, and a stellar radius ${\rstar = \sqrt{\lstar / \lsun} \cdot (T_{\odot} / T_{\star})^2 = 3.4\pm 0.5}${~}\rsun. This value for the radius, combined with the \vsini\ derived from the spectra, yields an inclination of ${53\pm 11}$\degr.

\begin{figure}
		\centering
		\includegraphics[totalheight=0.2\textheight]{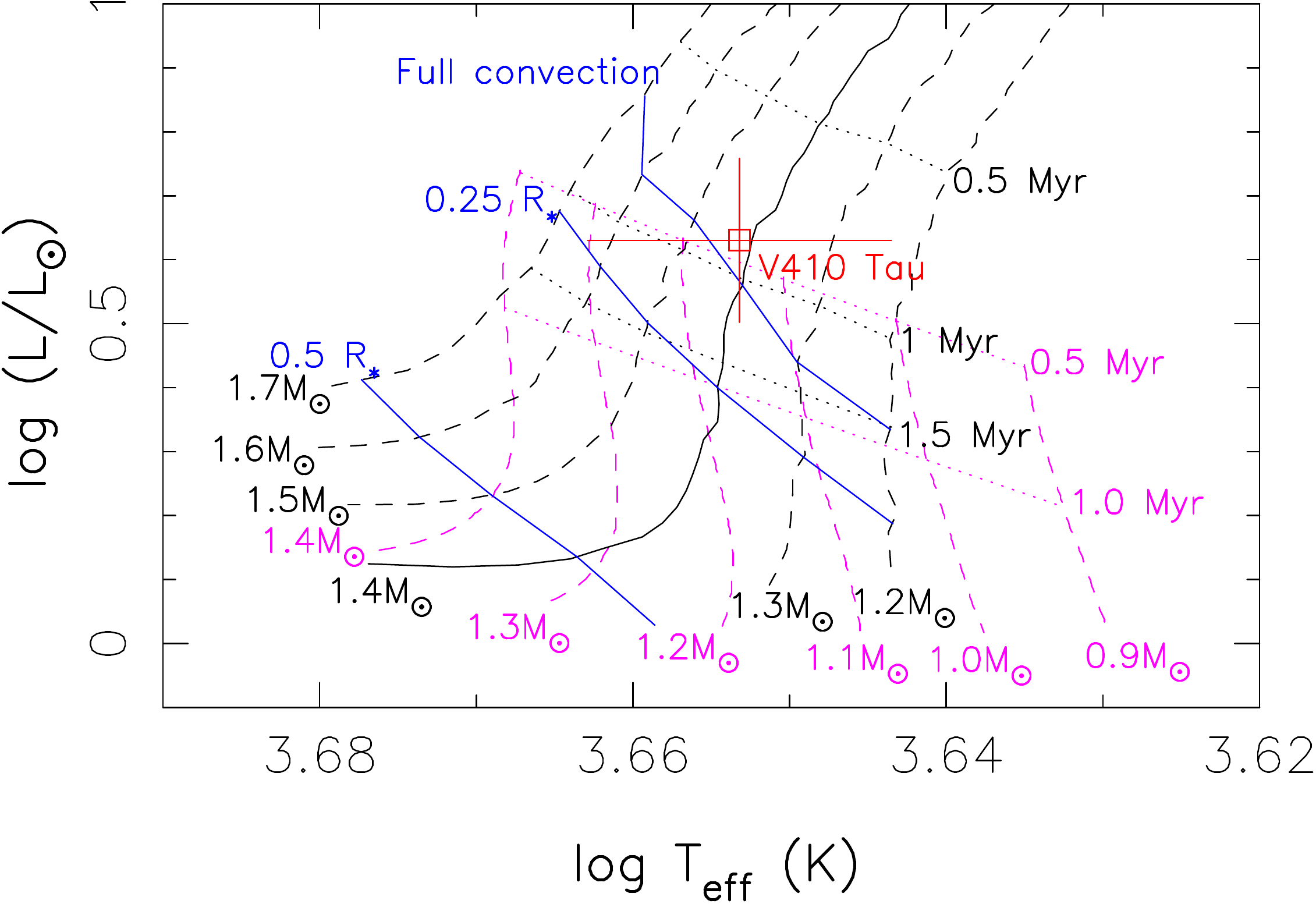}
		\caption{Position of \vt\ (red) in the Hertzsprung-Russell diagram. The curves yielded by the Siess models (with solar metallicity and overshooting) are represented in black and those yielded by the Baraffe models are represented in magenta. In both cases, evolution tracks are displayed in dashed lines, except the Siess 1.4{~}\msun\ track, the one we chose to model the evolution of \vt, which is shown as a full line. Isochrones are displayed in dotted lines. The thresholds where the radiative core starts developping ("Full convection") and where it reaches 25\% of the stellar radius, according to the Siess models, are marked in blue.}
		\label{fig:hrd}
	\end{figure}

	The position of \vt\ on the Hertzsprung-Russell diagram is displayed in Figure{~}\ref{fig:hrd}. According to \cite{Siess00} stellar evolution models for pre-main sequence stars, with solar metallicity and overshooting, \vt\ is a ${1.42\pm 0.15}${~}\msun\ star, aged ${0.84\pm 0.20}${~}Myr and fully convective. Baraffe models \citep{Baraffe15} disagree with the Siess models for stars as young as \vt\ and yield an age of <0.5{~}Myr with a mass of ${1.14\pm 0.10}${~}\msun. However, for the sake of consistency with the other MaPP and MaTYSSE studies, we will consider the values yielded by the Siess models in this paper. Our values are in good agreement with \cite{Welty95} and \cite{Skelly10}, who had previously derived masses of $\sim$1.5{~}\msun\ and ${1.4\pm 0.2}${~}\msun\ respectively, radii of $\sim$2.64{~}\rsun\ and $\sim$3.0{~}\rsun\ respectively, and ages of ${1-2}${~}Myr ${1.2\pm 0.3}${~}Myr respectively. Moreover, \cite{Skelly10} had deduced that \vt\ could have a radiative core of radius between 0.0{~}\rstar\ and 0.28{~}\rstar. Table{~}\ref{tab:php} sums up the stellar parameters of \vt\ found in this study.

	\begin{table}
		\centering
		\caption{Physical parameters of wTTS \vt. From top to bottom: distance from Earth, effective temperature, rotation period, luminosity, minimum stellar radius, stellar radius, line-of-sight-projected equatorial velocity, inclination, mass and age.}
		\begin{tabular}{ccc}
			\hline
			Parameter & Value & Reference \\
			\hline
			$d$ & 129.0{~}$\pm${~}0.5{~}pc & \cite{Galli18} \\
			\teff & 4500{~}$\pm${~}100{~}K & \\
			\Prot & 1.87197{~}$\pm${~}0.00010{~}d & \cite{Stelzer03} \\
			$\log (\lstar /\lsun)$ & 0.63{~}$\pm${~}0.13 & \\
			\rstar$\sin i$ & 2.708{~}$\pm${~}0.007{~}\rsun & \\
			\rstar & 3.4{~}$\pm${~}0.5{~}\rsun & \\
			\vsini & 73.2{~}$\pm${~}0.2{~}\kms & ZDI (Section{~}\ref{sec:mod}) \\
			$i$ & 50{~}$\pm${~}10\degr & ZDI (Section{~}\ref{sec:mod}) \\
			\mstar & 1.42{~}$\pm${~}0.15{~}\msun & \\
			Age & 0.84{~}$\pm${~}0.20{~}Myr & \\
			\hline
		\end{tabular}
		\label{tab:php}
	\end{table}

	\section{Stellar tomography}
	\label{sec:mod}

	To map the surface brightness and magnetic topology of \vt, we use the tomographic technique ZDI \citep[][]{Brown91, Donati97c}, which inverts simultaneous time-series of \sti\ and \stv\ LSD profiles into brightness and magnetic field surface maps. At each observation date, \sti\ and \stv\ profiles are synthesized from model maps by integrating the spectral contribution of each map cell over the visible half of the stellar surface, Doppler-shifted according to the local RV (i.e.\ line-of-sight-projected velocity) and weighted according to the local brightness, cell sky-projected area and limb darkening. The main modifier of local RV at the surface of the star is, in ZDI, the assumed rotation profile at the stellar surface, e.g.\ the solid-body rotation of the star or a square-cosine-type latitudinal differential rotation. Local \sti\ and \stv\ line profiles are computed from the Unno-Rachkovsky analytical solution to the polarized radiative transfer equations in a Milne-Eddington model atmosphere \citep[this is where the local magnetic field and the Zeeman effect intervene, see][]{Landi04}. To fit the LSD profiles of \vt\ in this study, we chose a spectral line of mean wavelength, Doppler width, Land\'{e} factor and equivalent width of respective values 640{~}nm, 1.8{~}\kms, 1.2 and 3.8{~}\kms.

	ZDI uses a conjugate gradient algorithm to iteratively reconstruct maps whose synthetic profiles can fit the LSD profiles down to a user-provided reduced chi-square (\chisqr) level. To lift degeneracy among the multiple solutions compatible with the data at the given reduced chi square, ZDI looks for the maximal-entropy solution, considering that the minimized information from the resulting maps is the most reliable. While the brightness value can vary freely from cell to cell, the surface magnetic field is modelled as a combination of poloidal and toroidal fields, both represented as weighted sums of spherical harmonics and projected onto the spherical coordinate space \citep[][for the equations]{Donati06b}. In this study, the magnetic field was fitted with spherical harmonics of orders $l=1$ to $l=15$.

	Because ZDI does not reconstruct intrinsic temporal variability except for differential rotation, there is a limit to the duration a fittable data set can span. At the same time, ZDI needs a good phase coverage from the data to build a complete map. For those reasons, ZDI was not applied to runs 2008 Oct and 2013 Nov; moreover, we reconstructed a different set of brightness and magnetic images for each of the runs on which ZDI was applied.

	Using ZDI on our data yielded values for \vsini\ and $i$ of ${73.2\pm 0.5}${~}\kms\ and ${50\pm 10}$\degr\ respectively. We also adjusted the systemic RV of \vt\ with ZDI, and noticed a drift in the optimal value with time (see Table{~}\ref{tab:zdi}).

	\begin{table*}
		\caption{Characteristics of the ZDI models for \vt\ at each observation epoch. \textit{Column 1}: observation epoch. \textit{Column 2}: number of spectropolarimetric observations used for ZDI. \textit{Column 3}: contribution of cool ("spots") and hot ("plages") areas on the brightness map. \textit{Column 4}: average magnetic strength, defined as the square root of the average squared magnetic field over the surface of the star. \textit{Columns 5 to 7}: normalized contribution of the poloidal field, part of the poloidal field that is dipolar and part of the poloidal field that is symmetric. \textit{Columns 8-9}: part of the toroidal field that is dipolar and part of the toroidal field that is symmetric. \textit{Column 10}: dipole characteristics: field strength, tilt with respect to the rotation axis and phase of the pole. \textit{Column 11}: systemic RV of the star as measured with ZDI, the error bar on those values is 0.20{~}\kms. Error bars on the magnetic field ratios are typically of 0.1.}
		\begin{tabular}{ccccccccccc}
			Date & N$_{\rm obs}$ & Spot+plage & B & $r_{\rm pol}$ & $r_{\rm dip/pol}$ & $r_{\rm sym/pol}$ & $r_{\rm dip/tor}$ & $r_{\rm sym/tor}$ & Dipole strength (G), & RV$_{\rm bulk}$ \\
			& & coverage (\%) & (G) & & & & & & tilt \& phase & (\kms) \\
			\hline
			2008 Dec & 6 & 5.8+4.4 & 486 & 0.32 & 0.13 & 0.37 & 0.89 & 0.96 & 129, 23\degr\ \& 0.71 & 16.30 \\
			2009 Jan & 48 & 9.6+7.1 & 556 & 0.55 & 0.26 & 0.09 & 0.54 & 0.79 & 165, 54\degr\ \& 0.54 & 16.30 \\
			2011 Jan & 20 & 8.1+6.6 & 560 & 0.40 & 0.24 & 0.23 & 0.72 & 0.85 & 239, 44\degr\ \& 0.62 & 16.40 \\
			2013 Dec & 25 & 11.0+7.5 & 568 & 0.49 & 0.23 & 0.34 & 0.66 & 0.81 & 254, 18\degr\ \& 0.56 & 16.50 \\
			2015 Dec & 21 & 8.9+6.7 & 600 & 0.68 & 0.37 & 0.45 & 0.62 & 0.78 & 458, 30\degr\ \& 0.54 & 16.65 \\
			2016 Jan & 9 & 7.9+6.5 & 480 & 0.77 & 0.38 & 0.30 & 0.68 & 0.87 & 400, 44\degr\ \& 0.51 & 16.65 \\
		\end{tabular}
		\label{tab:zdi}
	\end{table*}

		\subsection{Brightness and magnetic imaging}

\begin{figure*}
			\subfloat[2008 Dec]{\includegraphics[totalheight=0.25\textheight]{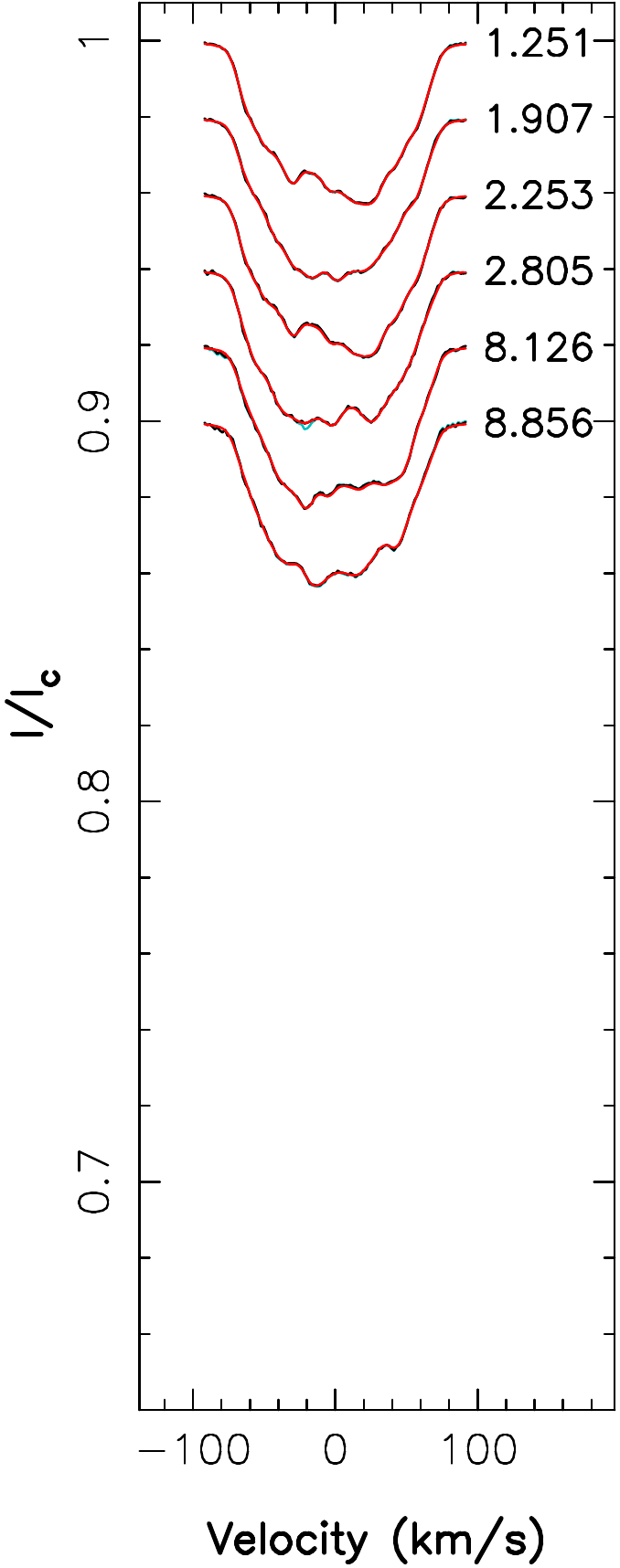}
				\includegraphics[totalheight=0.25\textheight]{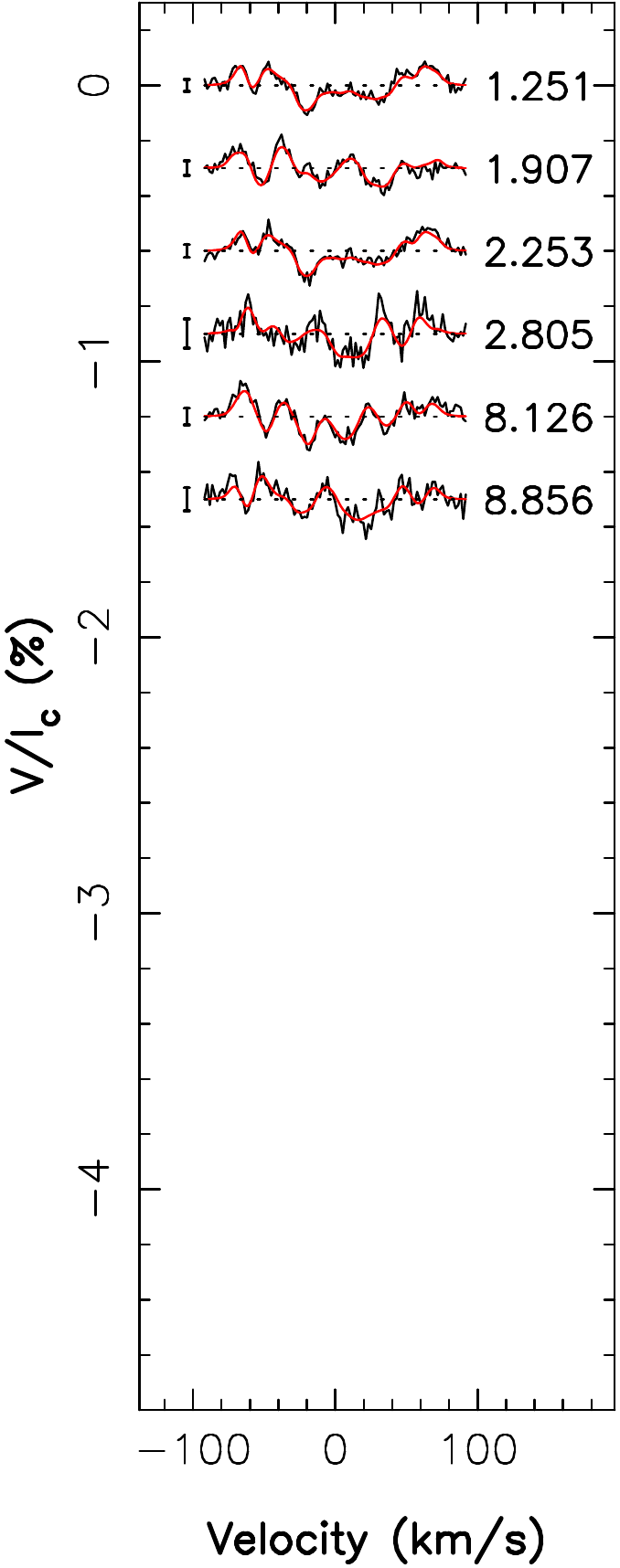}}
			\hspace{5mm}
			\subfloat[2009 Jan]{\includegraphics[totalheight=0.25\textheight]{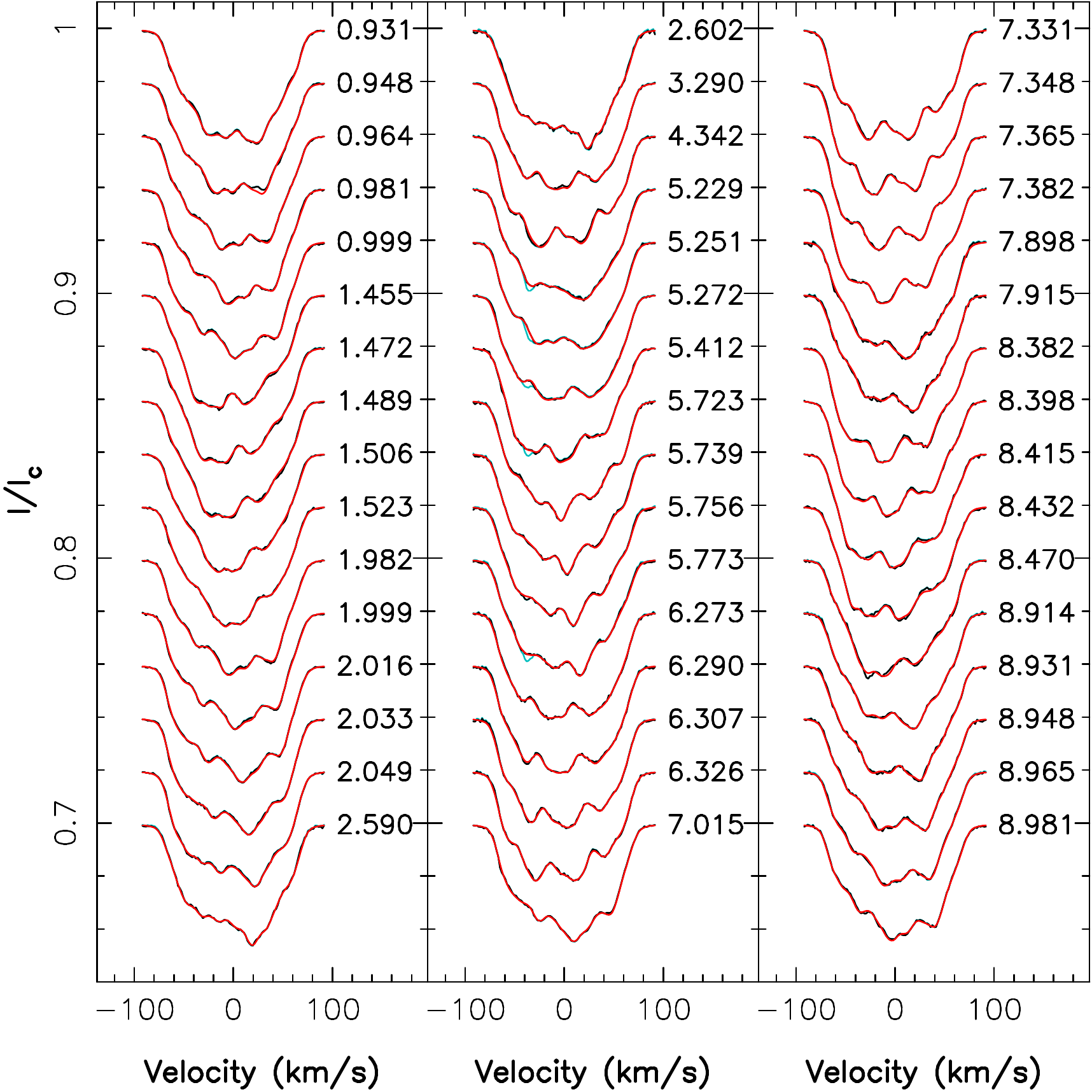}
				\includegraphics[totalheight=0.25\textheight]{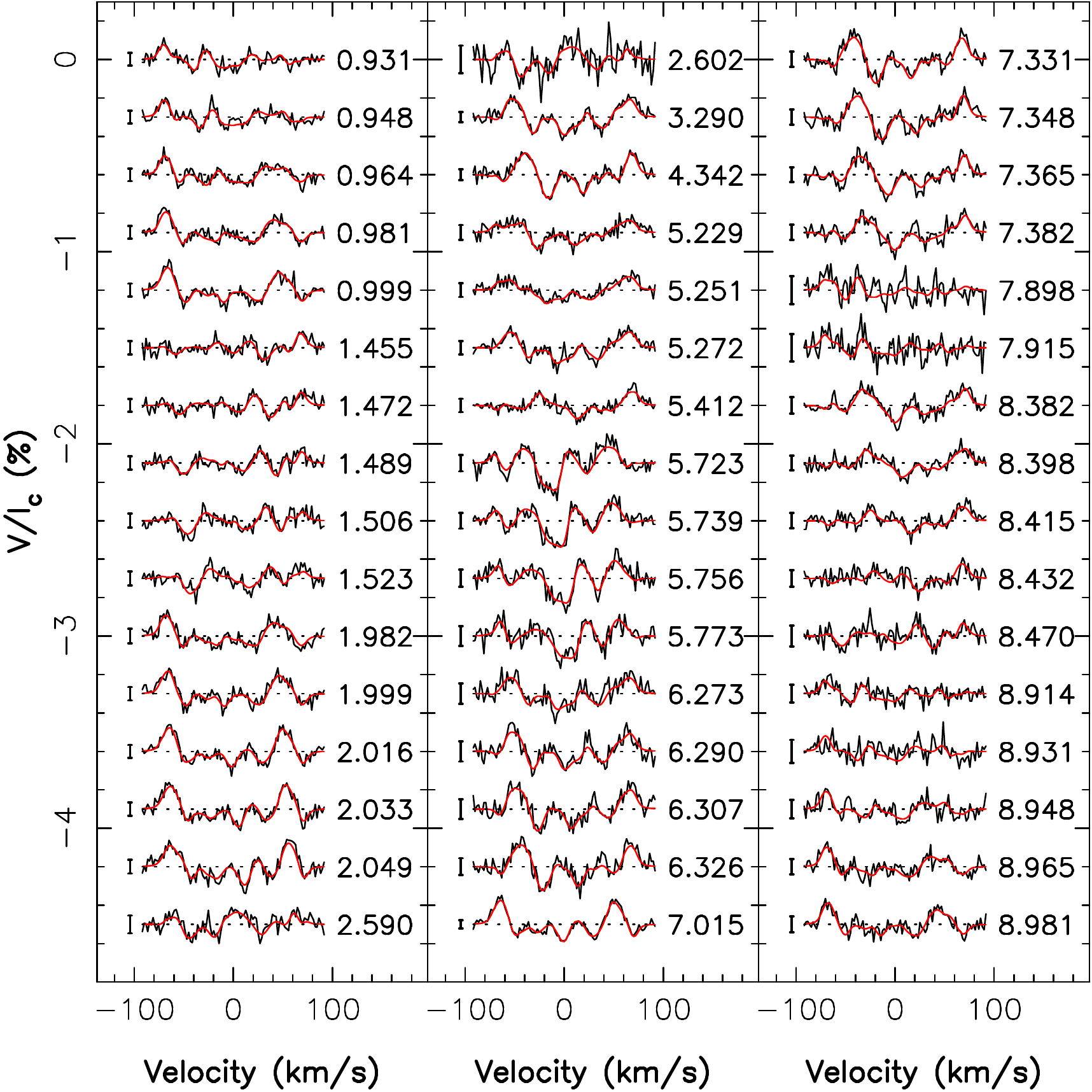}}

			\subfloat[2011 Jan]{\includegraphics[totalheight=0.25\textheight]{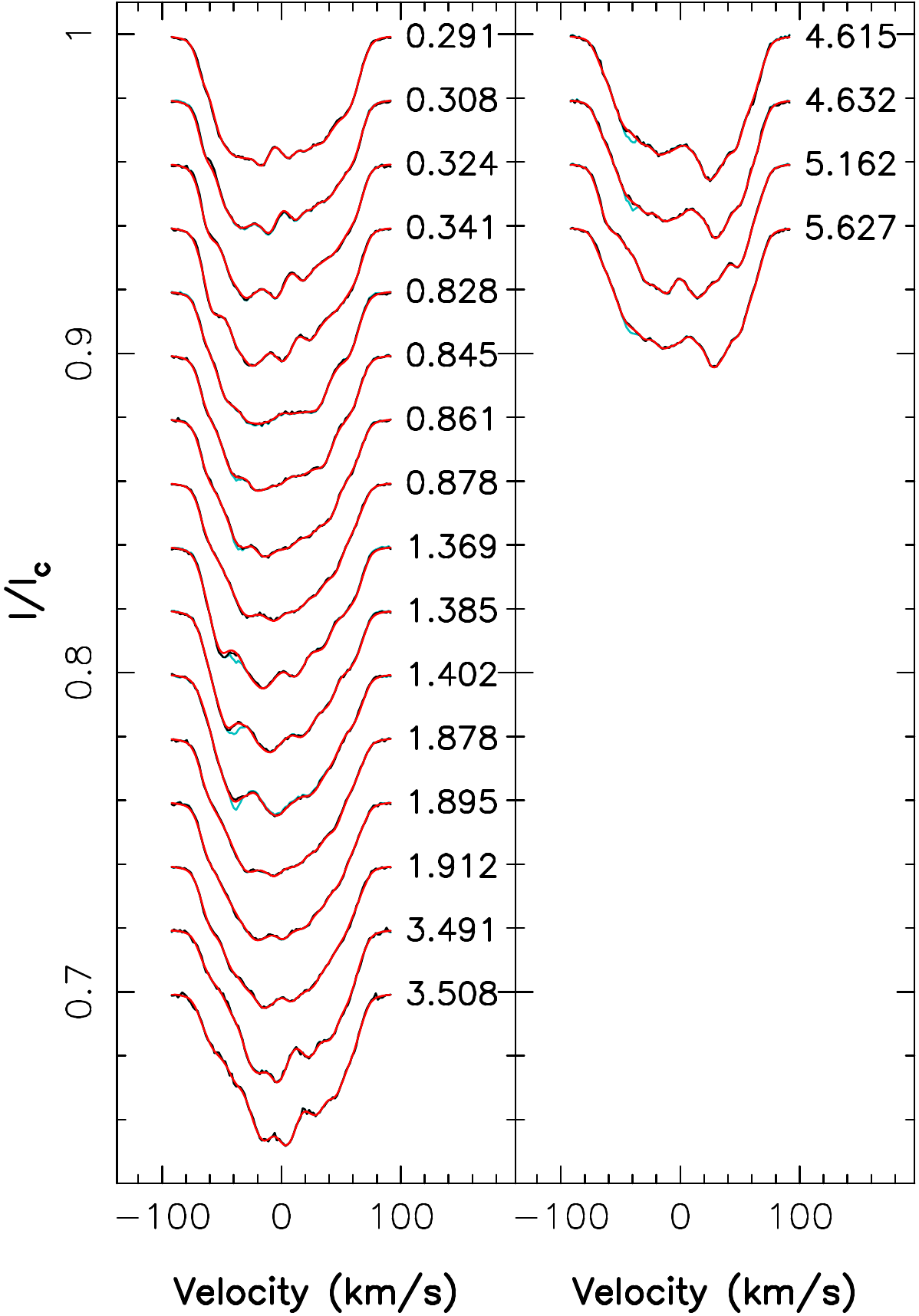}
				\includegraphics[totalheight=0.25\textheight]{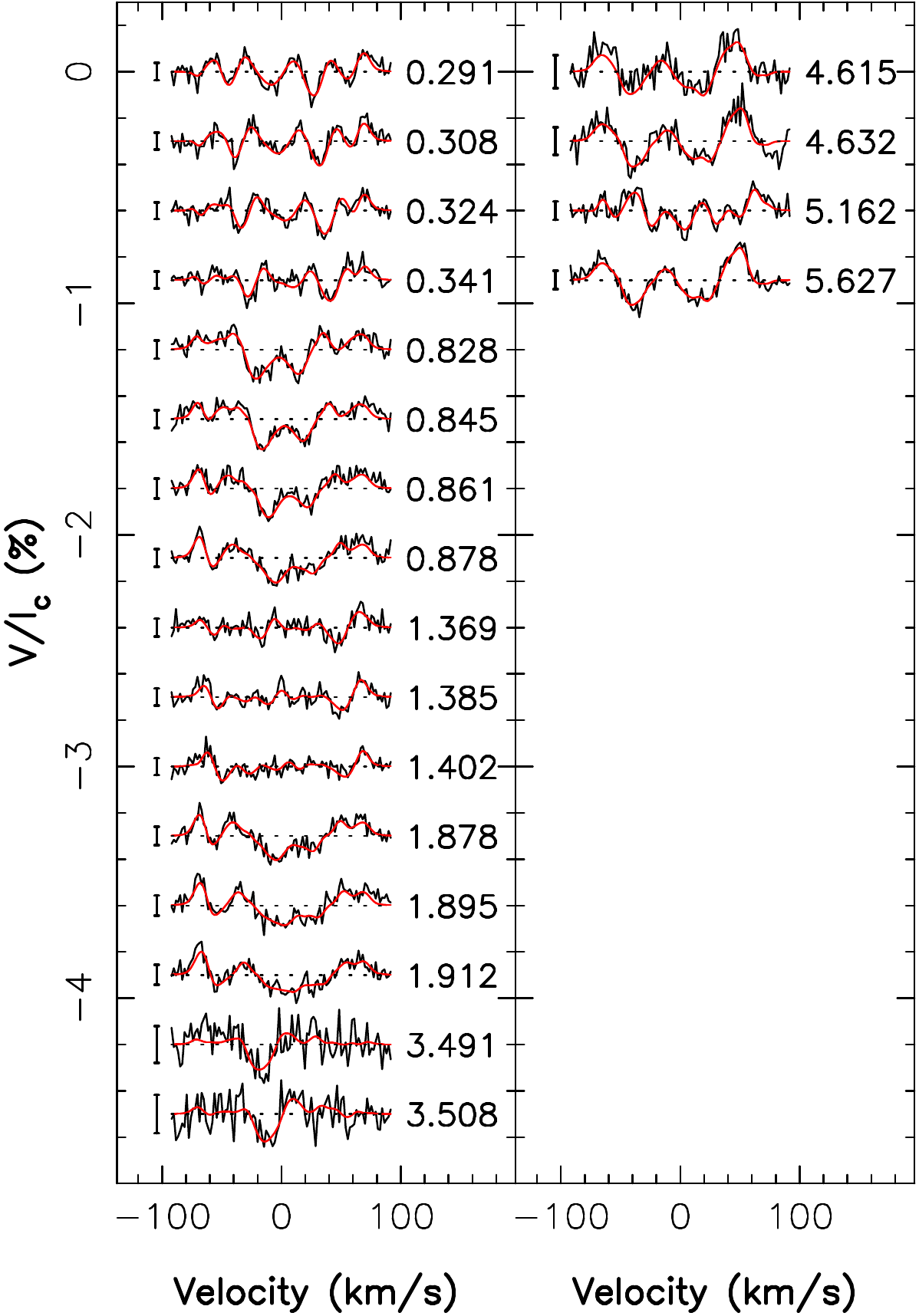}}
			\hspace{5mm}
			\subfloat[2013 Dec]{\includegraphics[totalheight=0.25\textheight]{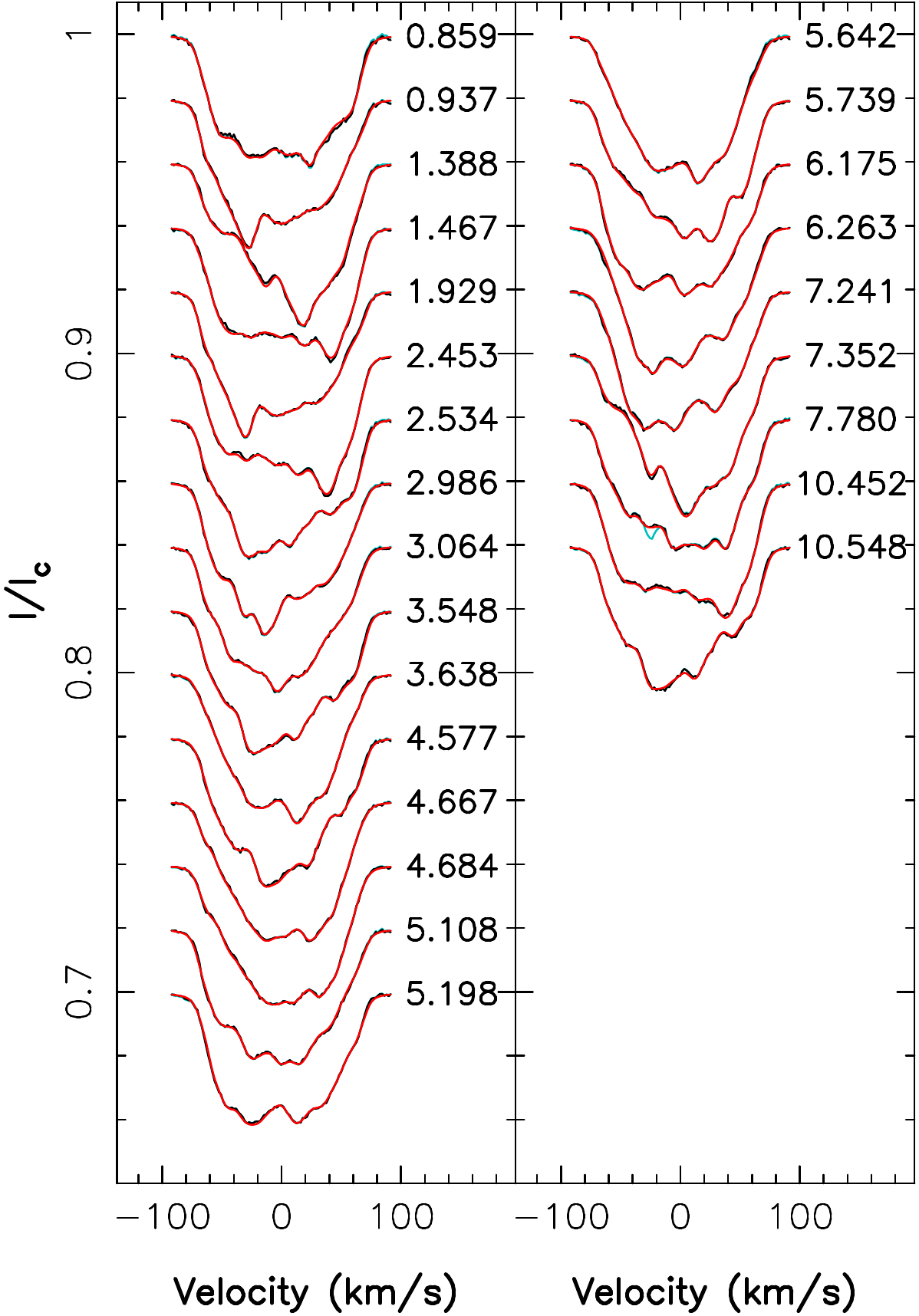}
				\includegraphics[totalheight=0.25\textheight]{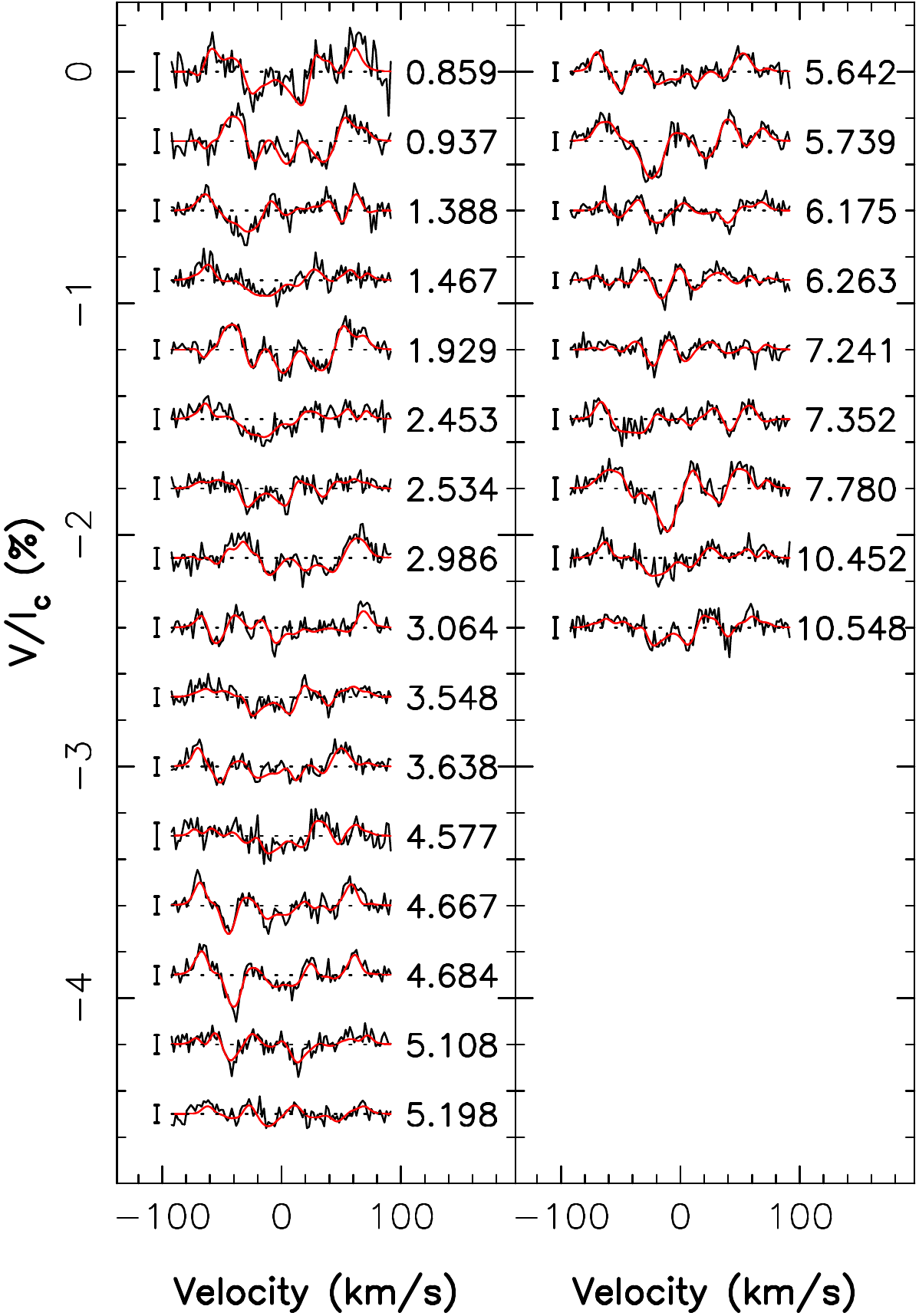}}

			\subfloat[2015 Dec]{\includegraphics[totalheight=0.25\textheight]{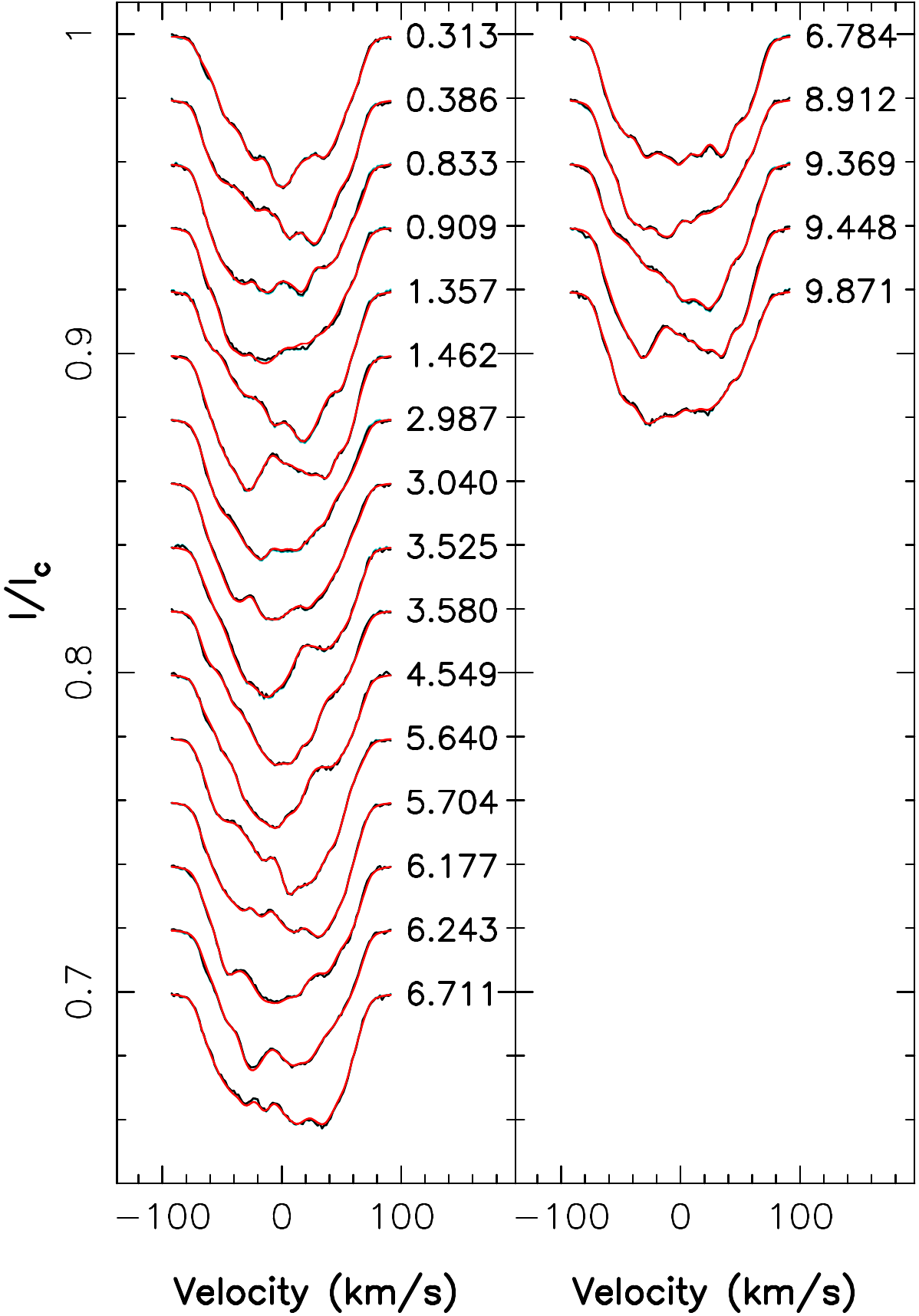}
				\includegraphics[totalheight=0.25\textheight]{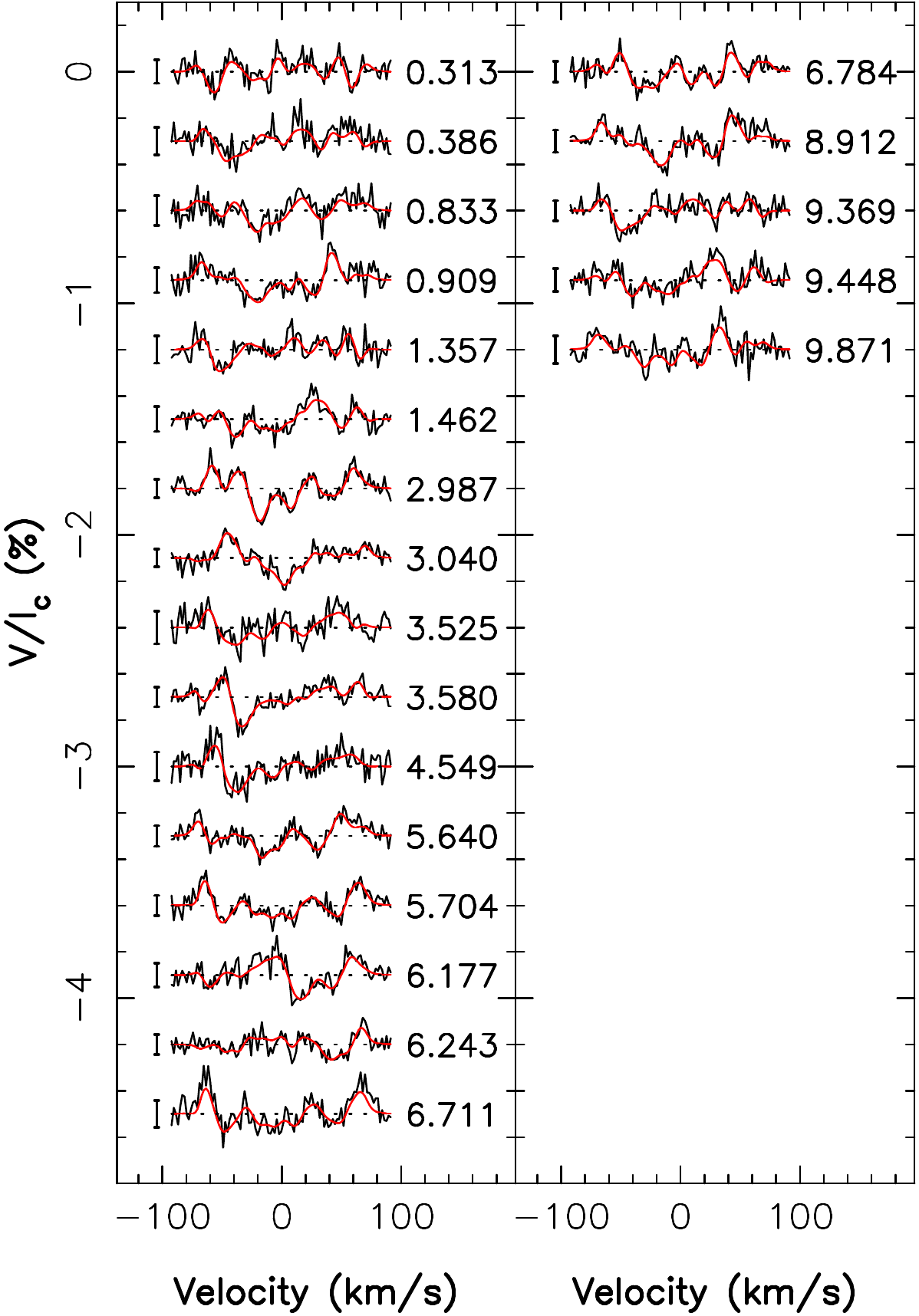}}
			\hspace{5mm}
			\subfloat[2016 Jan]{\includegraphics[totalheight=0.25\textheight]{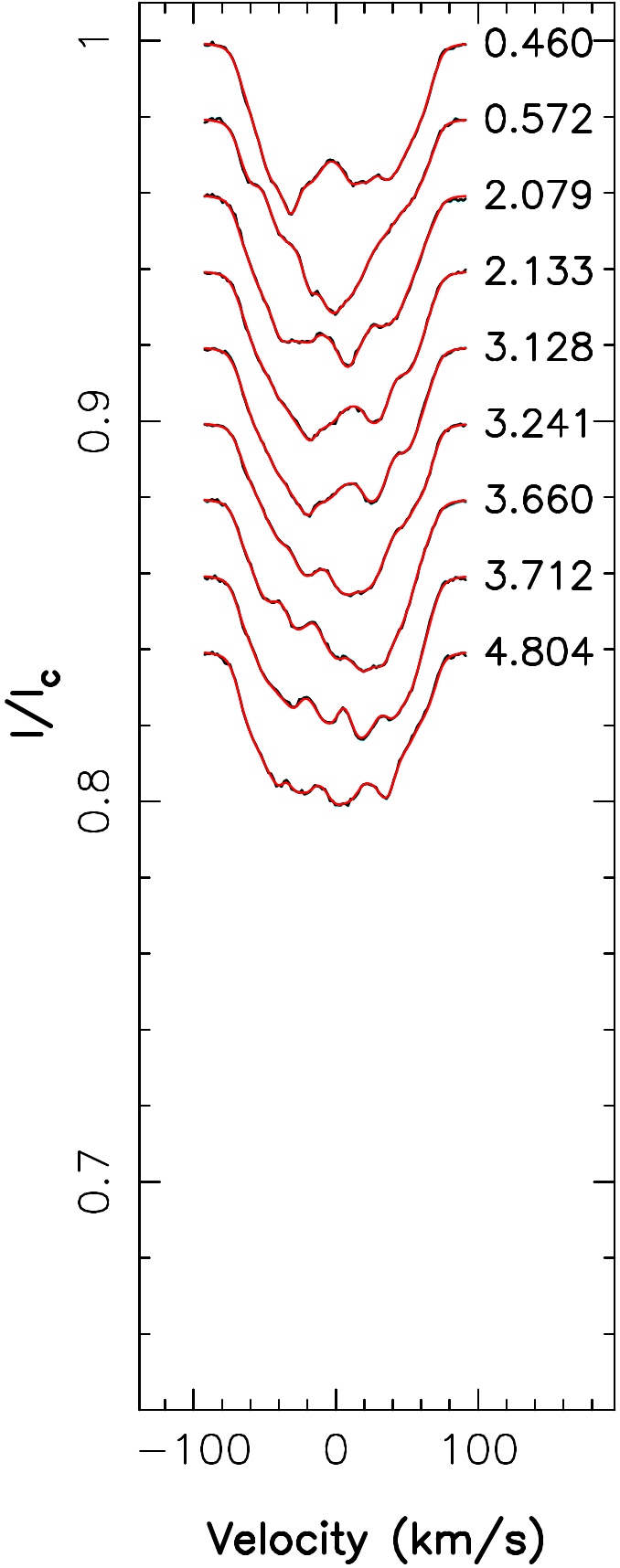}
				\includegraphics[totalheight=0.25\textheight]{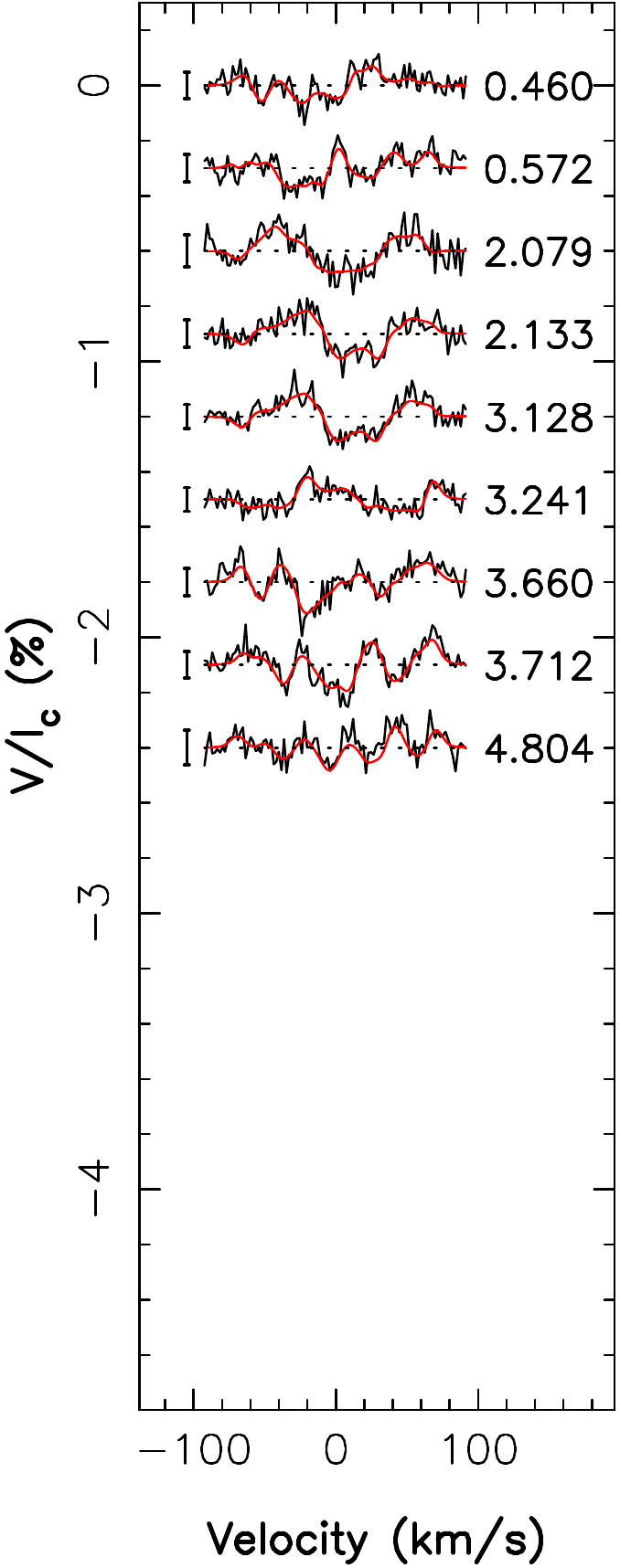}}
			\caption{LSD profiles for observation epochs 2008 Dec (a), 2009 Jan (b), 2011 Jan (c), 2013 Dec (d), 2015 Dec (e) and 2016 Jan (f). On the right of each profile is written the corresponding rotation cycle as indicated in Table{~}\ref{tab:sob}. The cyan, black and red lines represent respectively the profiles before removal of Moon pollution, after removal of Moon pollution and the fit obtained with Zeeman Doppler Imaging. For each epoch, \sti\ profiles are on the left and \stv\ profiles on the right. 3$\sigma$-error bars are displayed beside each \stv\ profile.}
			\label{fig:sto}
		\end{figure*}

\begin{figure*}
			\includegraphics[totalheight=0.8\textheight]{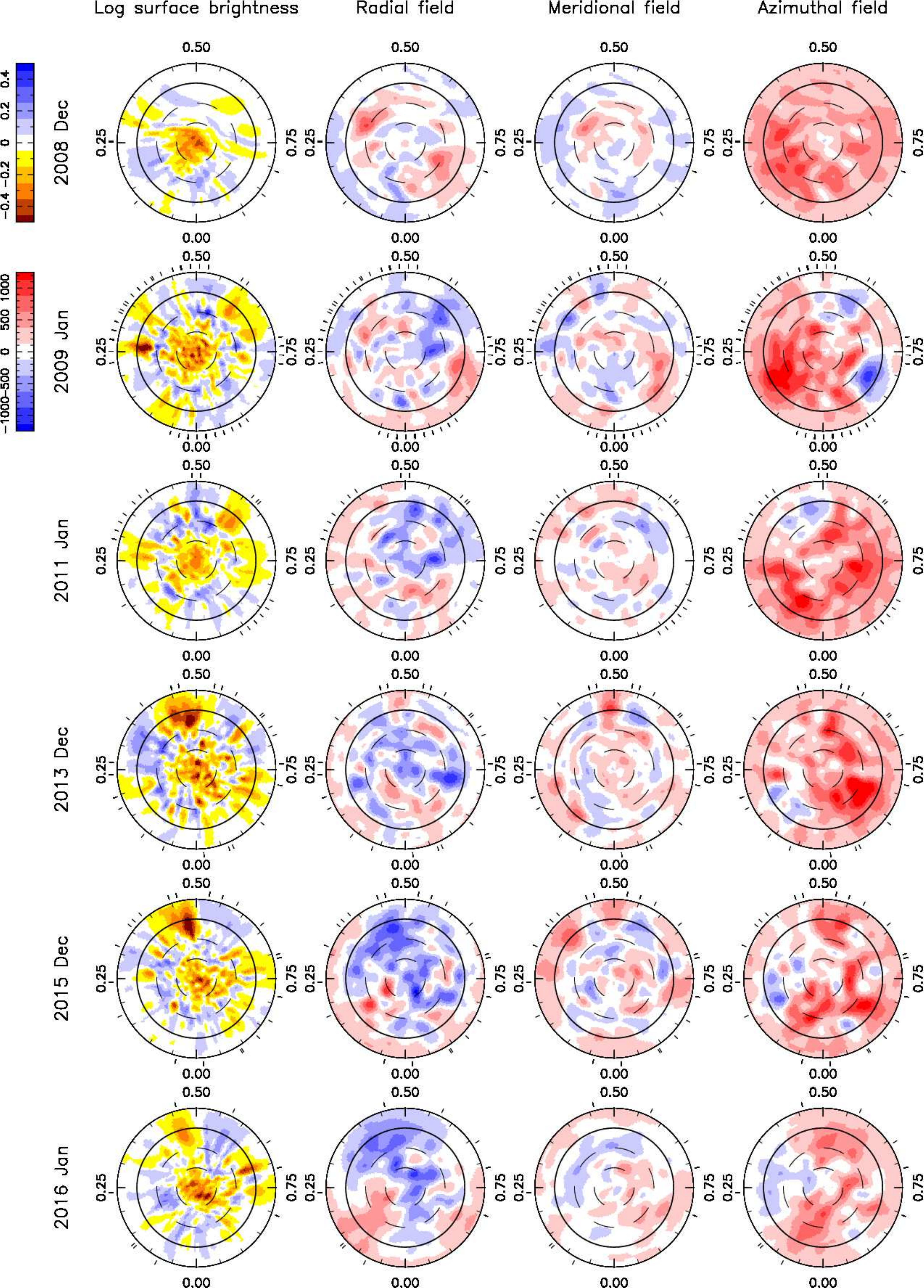}
			\caption{ZDI maps of the logarithmic relative surface brightness (first column), and the radial, meridional and azimuthal magnetic field (second to fourth columns) of \vt, reconstructed from data collected in 2008 Dec, 2009 Jan, 2011 Jan, 2013 Dec, 2015 Dec and 2016 Jan (top to bottom rows). Each map is shown as a flattened polar view, with the equator being represented as a full line, and 60\degr, 30\degr, and -30\degr\ latitude parallels as dashed lines, and ticks around the star mark the spectropolarimetric observations. For the brightness maps, cool spots are colored in brown and bright plages in blue. For the magnetic maps, red represents outwards and anti-clockwise field on the radial and azimuthal field maps respectively, and the direction of the visible pole on the meridional field maps.}
			\label{fig:qbm}
		\end{figure*}

		Time-series of \sti\ and \stv\ LSD profiles are shown in Figure{~}\ref{fig:sto}, both before and after removal of lunar pollution, as well as synthetic profiles generated from the reconstructed ZDI maps. The corresponding maps are shown in Figure{~}\ref{fig:qbm}, with brightness maps in the first column and radial, meridional and azimuthal components of the surface magnetic field in the second to fourth columns. Properties of these reconstructed maps are listed in Table{~}\ref{tab:zdi}. Since the 2008 Dec data set has a phase coverage of only half the star, the derived parameters characterizing the global field topology at this epoch are no more than weakly meaningful and were not used for the following analysis and discussion. Our data have been fitted down to ${\chisqr = 1}$ with a feature coverage between 15{~}\% and 18{~}\% depending on the epochs, and a large-scale field strength of 0.5-0.6{~}kG. Since ZDI is only sensitive to mid- to large-scale surface features, and returns the maximum-entropy solution, this amount of spot coverage is not discrepant with the assumption made in Section{~}\ref{sec:phy}; it further suggests that {~}30\% of the star is more or less evenly covered with small-scale dark features.

		Brightness maps display a complex structure with many relatively small-scale features, and a high contrast. At all epochs, a large concentration of dark spots is observed at the pole. In 2009 Jan, 2013 Dec and 2015 Dec, the brightness map exhibits a strong equatorial spot, respectively at phases 0.27, 0.48 and 0.48. The presence of a strong polar spot is consistent with the maps published in \cite{Skelly10}, \cite{Rice11} and \cite{Carroll12} for data set 2009 Jan. At that particular epoch, the equatorial spot at phase 0.27, and another equatorial spot at phase 0.60, are also visible in both \cite{Skelly10} and \cite{Rice11} (figure 8), albeit less contrasted compared to other features than they are on our map. A remnant of the 2015 Dec equatorial spot is observed on the 2016 Jan map, where its intensity seems to have decreased, but this has to be taken with caution since ZDI maps are somewhat dependent on phase coverage. Dark spots and bright plages contribute to the feature coverage at about 9{~}\%{~}/{~}7{~}\% respectively.

\begin{figure*}
			\subfloat[08b \& 09a]{\includegraphics[width=0.45\linewidth]{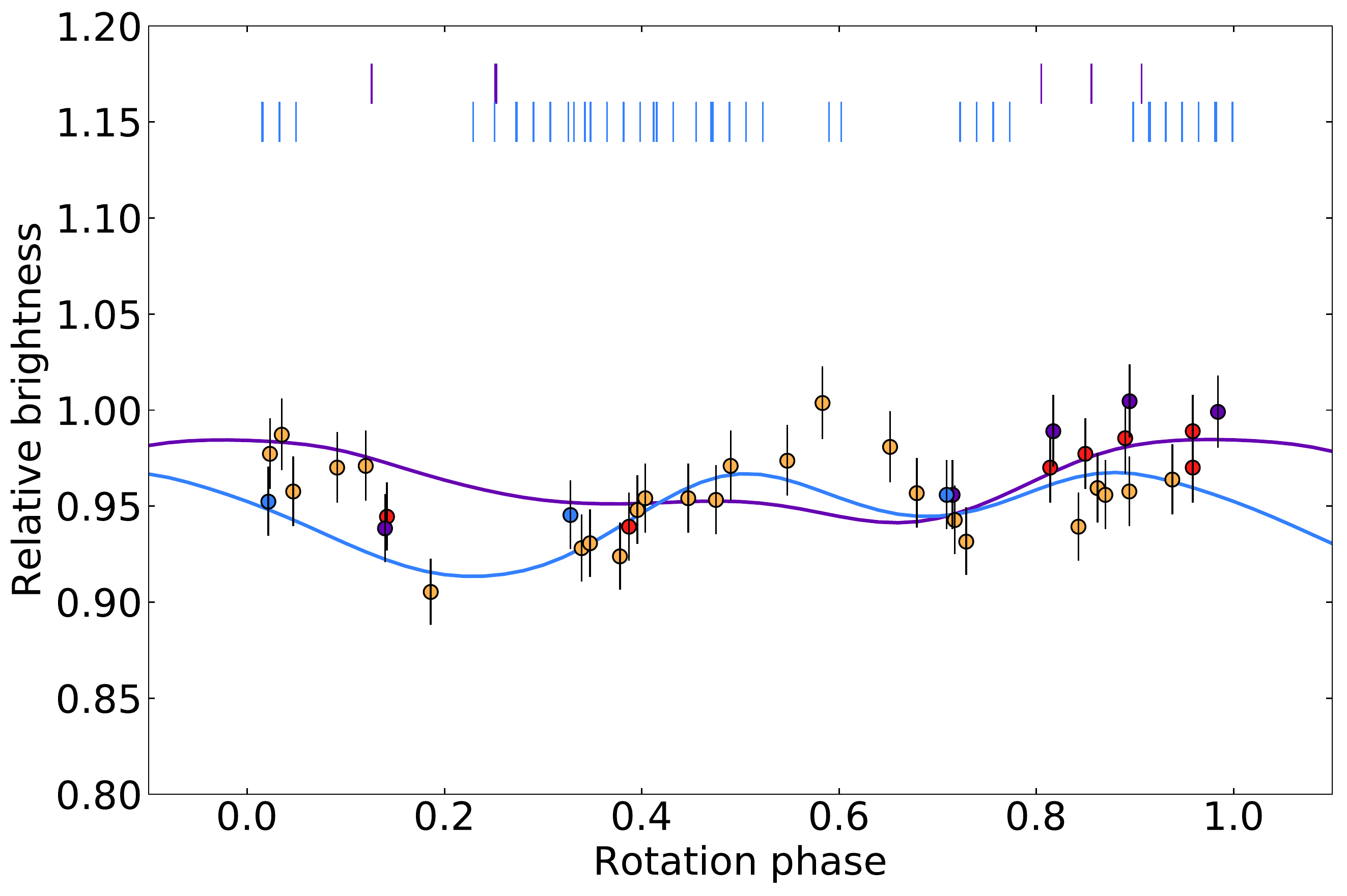}}
			\hfill
			\subfloat[10b+11a]{\includegraphics[width=0.45\linewidth]{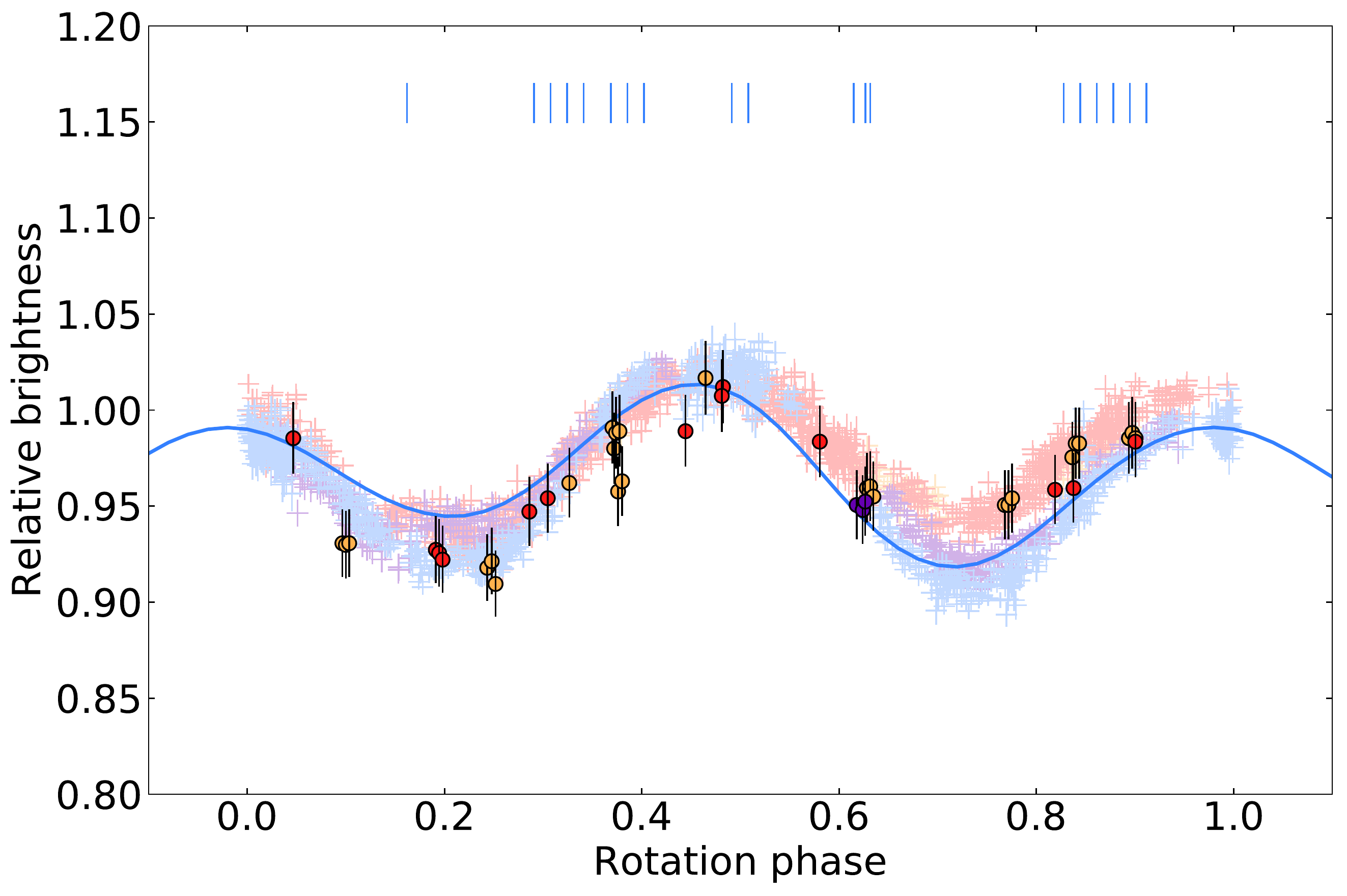}}

			\subfloat[13b+14a]{\includegraphics[width=0.45\linewidth]{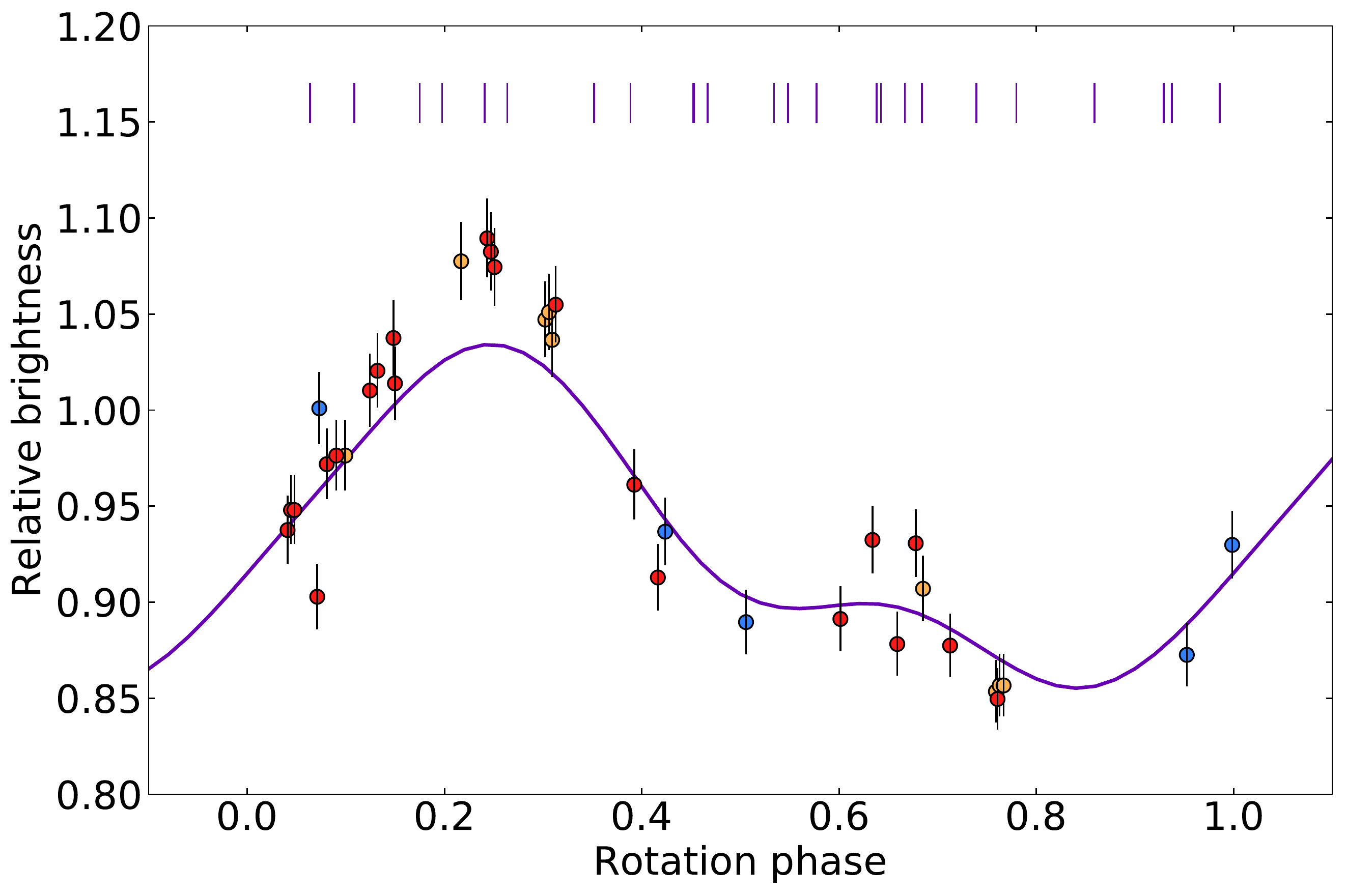}}
			\hfill
			\subfloat[15b \& 16a]{\includegraphics[width=0.45\linewidth]{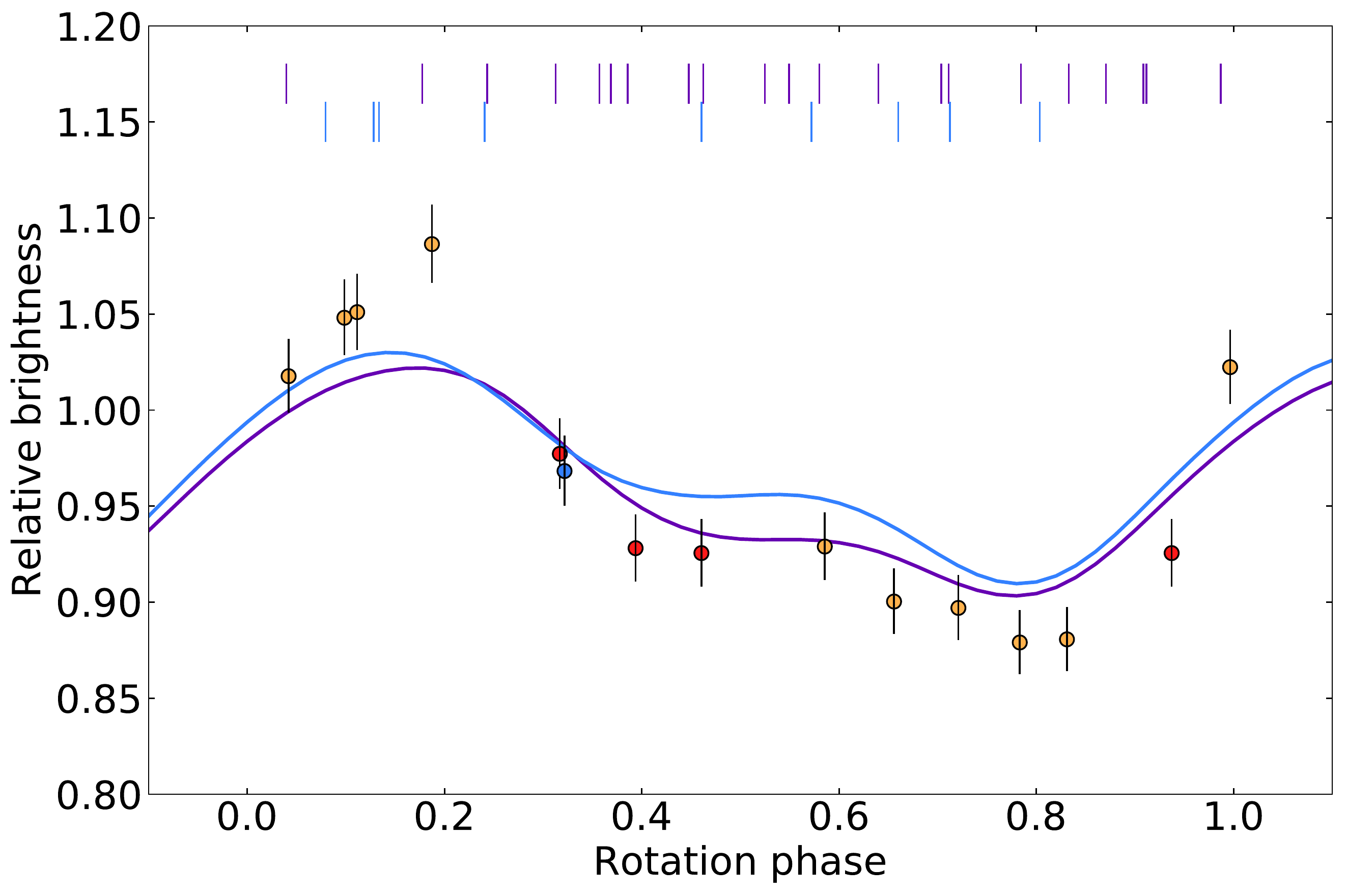}}
			\caption{Phase-folded photometry data (dots with 1$\sigma$ error bars) and ZDI models (lines) for observation epochs 08b \& 09a (a), 10b+11a (b), 13b+14a (c) and 15b \& 16a (d). In the case of 08b \& 09a and 15b \& 16a, two ZDI curves are plotted for the two ZDI maps reconstructed within each epoch. Orange, red, purple and blue colors each indicate a quarter of the total time span of the observations (photometric and spectropolarimetric together), in chronological order. Spectropolarimetric observations are marked by ticks above the light curves. In Figure b, WASP data were added as desaturated crosses, with the size of the cross branches indicating their 1$\sigma$ error bars.}
			\label{fig:pho}
		\end{figure*}

		Photometry curves from the ZDI brightness maps were synthesized and a comparison to contemporary CrAO data, and WASP data in the case of 2011 Jan, is shown in Figure{~}\ref{fig:pho}. Despite a slightly underestimated amplitude at phase 0.60 in 2008b-2009a, at phase 0.20 in 2011a, at phase 0.20 in 2013b and at phases 0.20 and 0.80 in 2015b-2016a, ZDI manages to retrieve the measured photometric variations of \vt\ rather satisfyingly. We notice a small temporal evolution of the light curve in the WASP data during season 2010b-2011a, where the regions around phases 0.20 and 0.70 globally darken by 0.02-0.03{~}mag ($\simeq 4\sigma$) over the 4 months that the data set spans.

\begin{figure}
			\centering
			\subfloat[2009 Jan]{\includegraphics[angle=-90,width=0.5\linewidth]{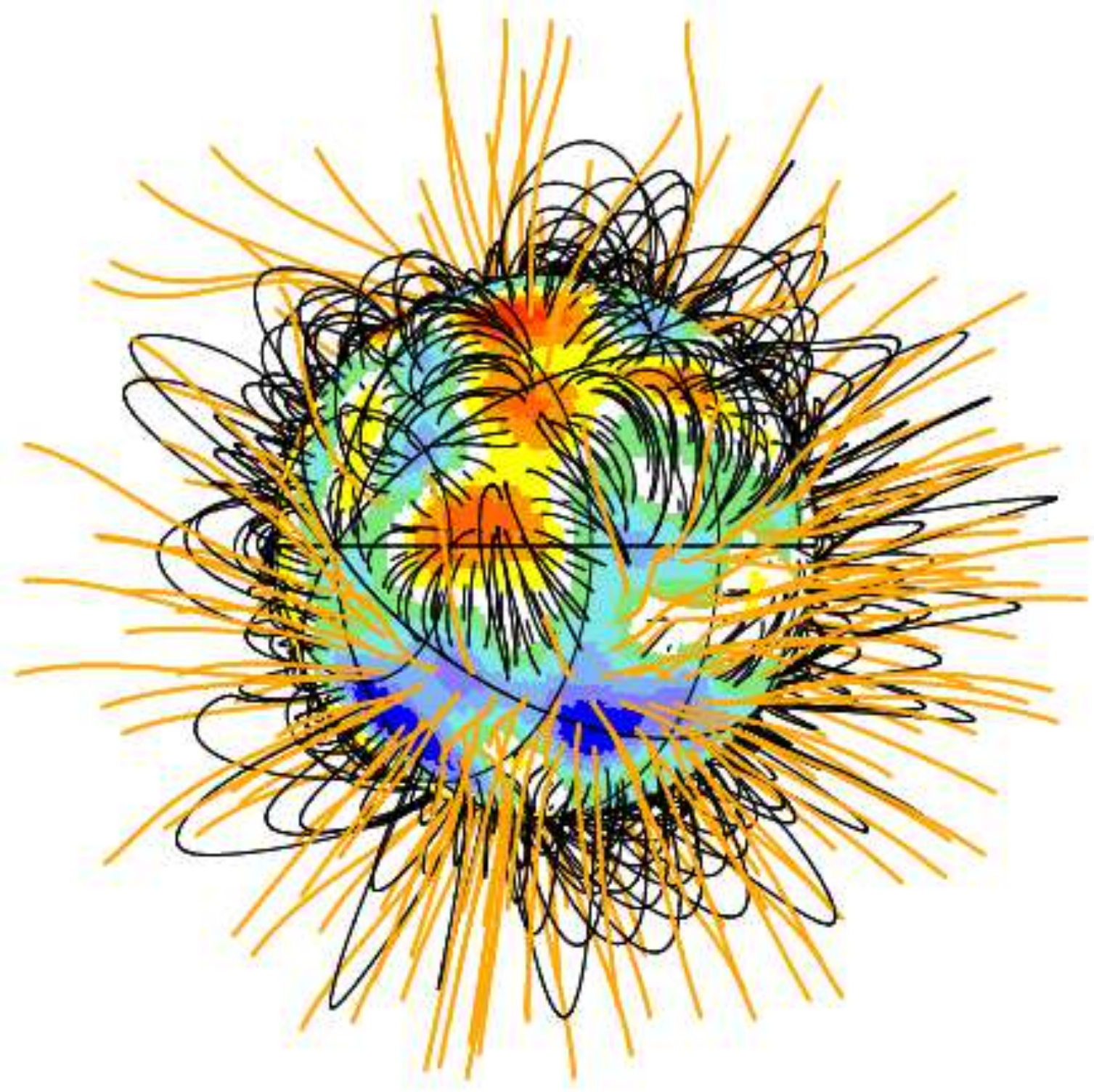}}
			\subfloat[2011 Jan]{\includegraphics[angle=-90,width=0.5\linewidth]{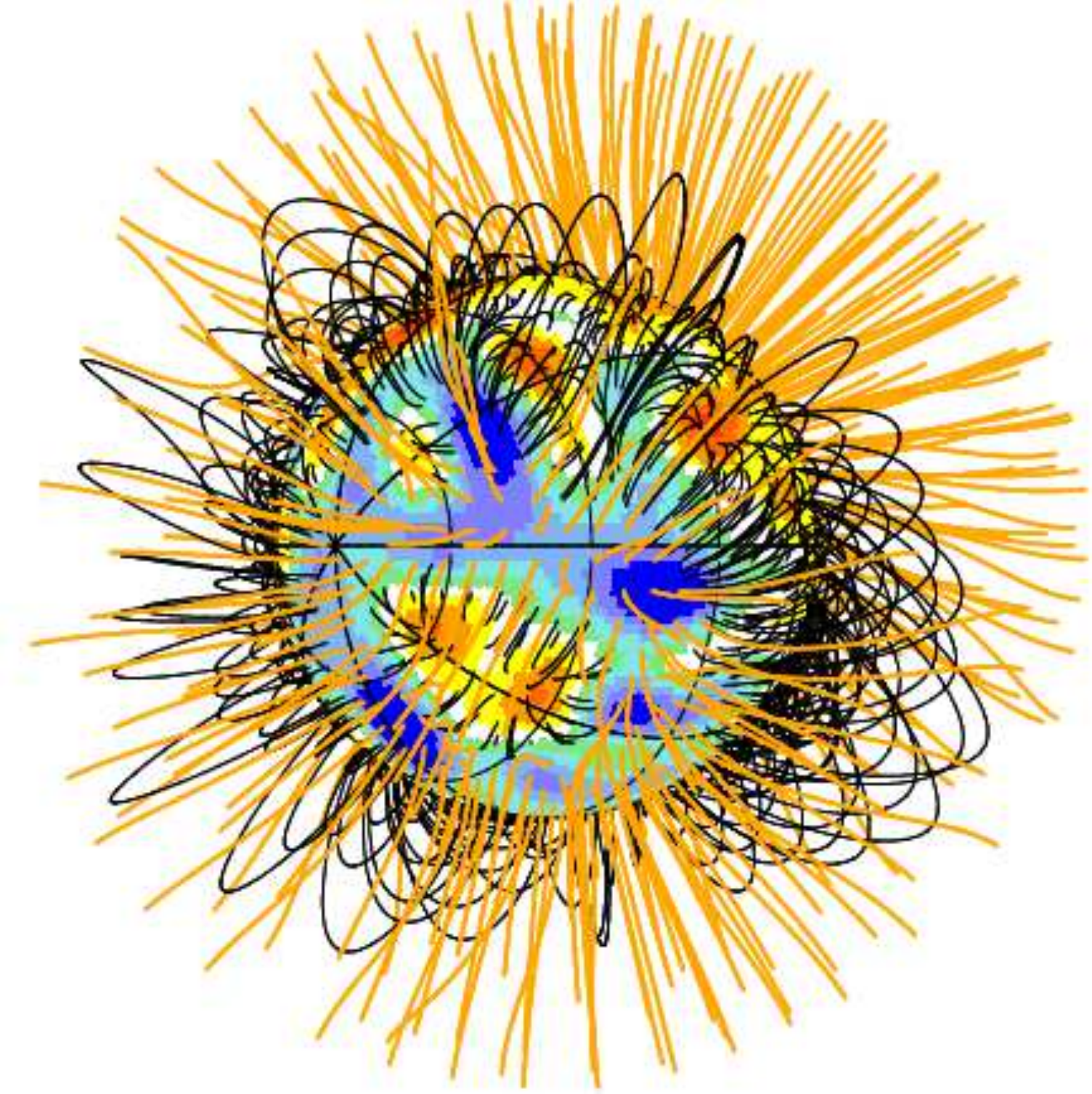}}

			\subfloat[2013 Dec]{\includegraphics[angle=-90,width=0.5\linewidth]{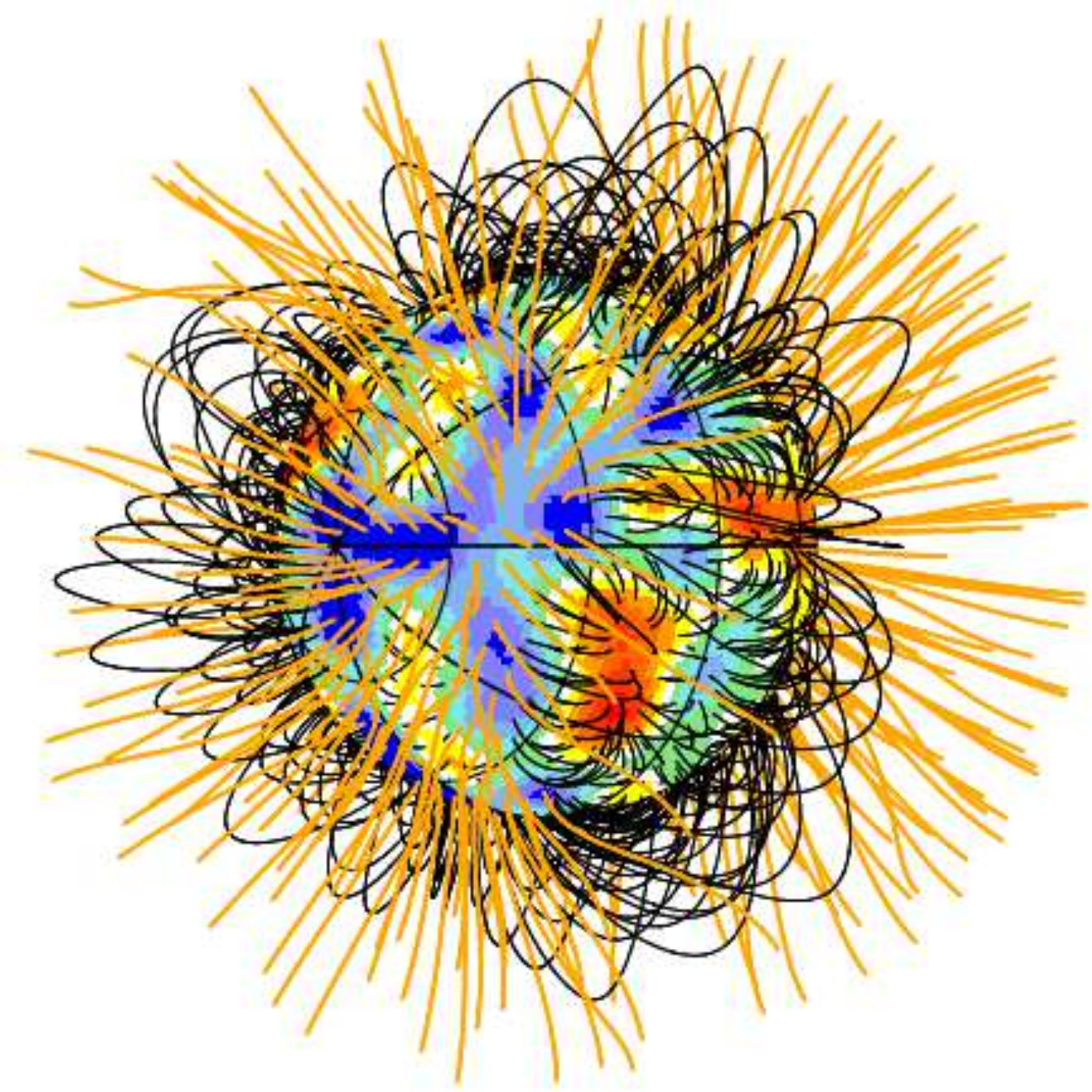}}
			\subfloat[2015 Dec]{\includegraphics[angle=-90,width=0.5\linewidth]{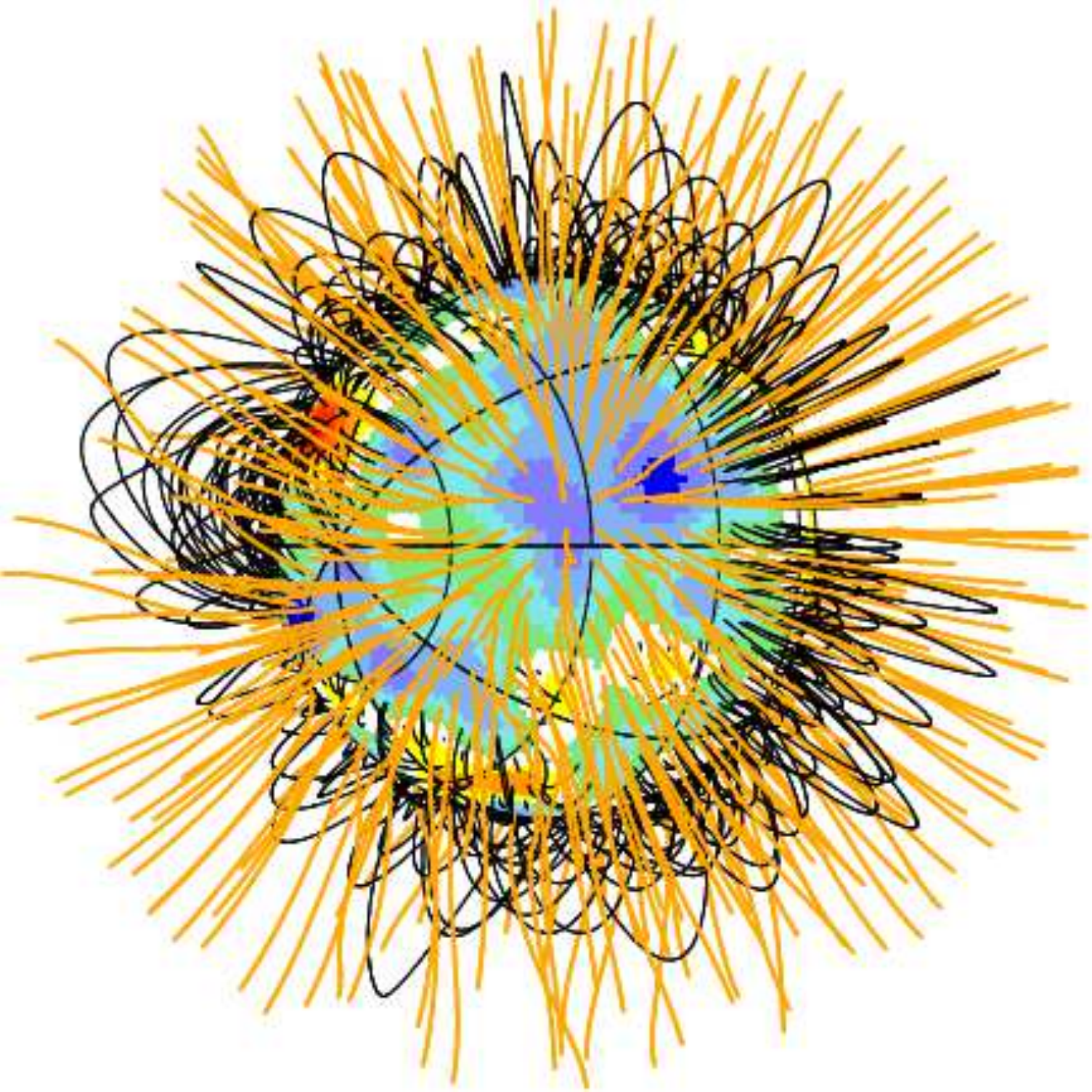}}

			\subfloat[2016 Jan]{\includegraphics[angle=-90,width=0.5\linewidth]{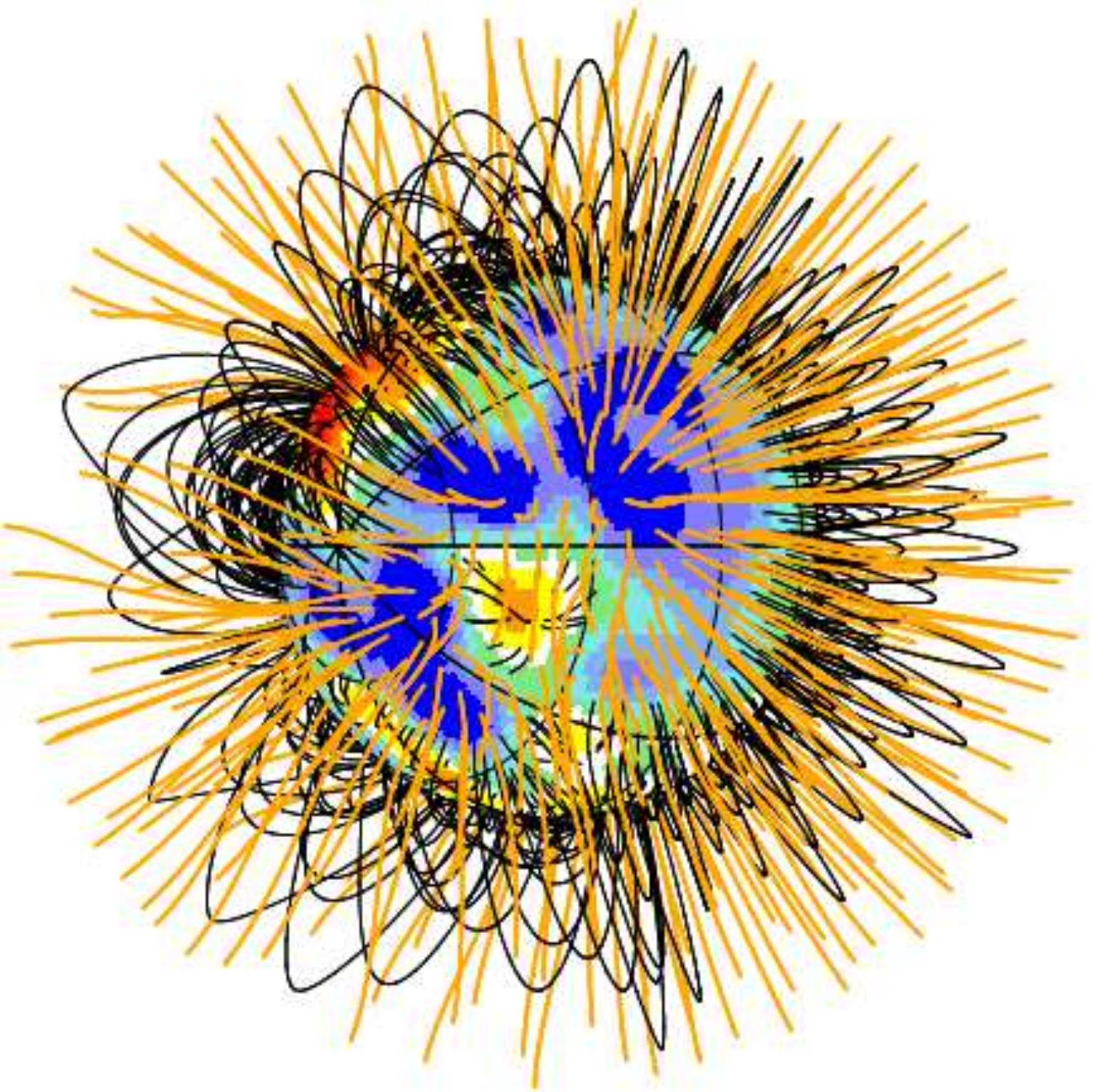}}
			\caption{Potential field extrapolations of the ZDI-reconstructed surface radial field, as seen by an Earth-based observer, for observation epochs 2009 Jan (a), 2011 Jan (b), 2013 Dec (c), 2015 Dec (d) and 2016 Jan (e) at phase 0.50. Open/closed field lines are shown in orange/black respectively, and colours on the stellar surface depict the local value of the radial field (in G, as shown in the left-hand panels of Fig.{~}\ref{fig:qbm}). The source surface at which field lines open is set to 2.1{~}\rstar, corresponding to the corotation radius and beyond which field lines are expected to quickly open under centrifugal forces given the high rotation rate of V410 Tau.}
			\label{fig:b3d}
		\end{figure}

		The magnetic field maps also show a high complexity, with a poloidal component that has a weak dipolar contribution and that is rather non-axisymmetric, and a toroidal component contributing to $\sim$50\% of the overall magnetic energy in 2009, 2011 and 2013, and decreasing towards ${\simeq 30}${~}\% in 2015-2016, that is both strongly dipolar and highly axisymmetric. The dipole pole is tilted at various angles depending on the epoch, with a tilt as high as 54\degr\ in 2009 Jan, down to 18\degr\ in 2013 Dec. The phase of the pole is always around 0.50-0.60, and the intensity of the poloidal dipole increases over time, from 165{~}G in 2009 Jan to ${\simeq 400}${~}G in 2015-2016. We note that the maximum emission of \hal\ corresponds to the phase at which the dipole is tilted (Fig.{~}\ref{fig:hal}). For visualisation purposes, 3-dimensional potential fields were extrapolated from the radial components of the magnetic maps, and displayed in Figure{~}\ref{fig:b3d}, with phase 0.50 facing the reader.

		We do not observe a particular correlation between our brightness and our magnetic maps, meaning the areas with strong magnetic field are not necessarily crowded with dark spots, according to the ZDI reconstruction.

		\subsection{Differential rotation}
		\label{sec:dr}
		Without differential rotation, ZDI cannot fit an extended data set, such as 2008 Dec + 2009 Jan, 2013 Nov + 2013 Dec or 2015 Dec + 2016 Jan (shortened in this subsection to 08b+09a, 13b and 15b+16a respectively), down to \chisqr=1, it only manages to reach values of 1.66, 1.20 and 2.64 respectively. This implies that some level of variability exists and impacts the data on time scales of a few months, which could come from the presence of differential rotation at the surface of \vt. We model differential rotation with the following law:
		\[\Omega(\theta) = \omeq - (\cos \theta)^2 \dom\]
		where $\theta$ is the colatitude, \omeq\ the equatorial rotation rate and \dom\ the pole-to-equator rotation rate difference. We constrain \omeq\ and \dom\ by pre-setting the amount of information ZDI is allowed to reconstruct, and having ZDI minimize the \chisqr\ in these conditions.

\begin{figure}
			\centering
			\includegraphics[totalheight=0.6\textheight]{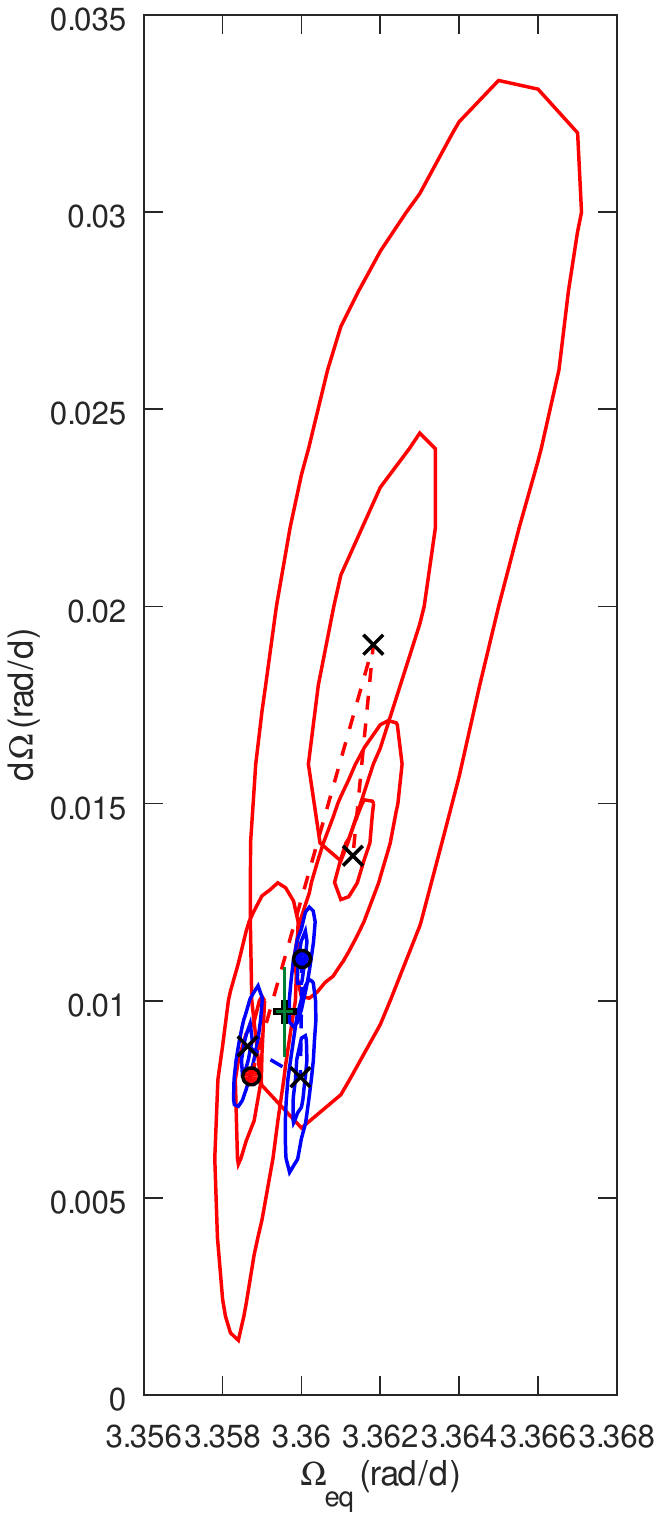}
			\caption{Evolution of the differential rotation of \vt\ as measured from \sti\ (blue) and \stv\ (red) profiles. The points corresponding to observation epoch 2008b-2009a are marked with o symbols, then the dashed lines link the epochs in chronological order (2013b-2014a and 2015b-2016a are marked with x symbols). 68.3\% and 99.7\% contours of confidence are displayed for each observation epoch. The weighted average of the six measurements, chosen to produce the maps shown in Section{~}\ref{sec:mod}, is represented as a black +, with overlayed error bars in green.}
			\label{fig:dr3}
		\end{figure}

		We performed this analysis on the three afore-mentioned extended data sets, and on \sti\ and \stv\ time-series separately, reconstructing only brightness or only magnetic field respectively. From the resulting \chisqr\ maps over the \{\omeq,\dom\} space, one can plot the contours of the 1$\sigma$- (68.3\%) and 3$\sigma$- (99.7\%) areas of confidence for each observation epoch. Figure{~}\ref{fig:dr3}, which shows such contours, highlights clear minima surrounded by almost elliptic areas of confidence at each epoch, and shows that each 3$\sigma$-confidence area overlaps at least two other 3$\sigma$-confidence areas. Numerical results for each epoch are given in Table{~}\ref{tab:dir}. We chose to use a unique set of parameters to reconstruct all images shown in Section{~}\ref{sec:mod}: the weighted means of the six seasonal minima, ${\omeq = 3.35957\pm 0.00006}${~}\rpd\ and ${\dom = 0.0097\pm 0.0003}${~}\rpd.

		Following the method described in \cite{Donati03b}, we computed, for each epoch, the colatitude at which the rotation rate is constant along the confidence ellipse major axis. This value corresponds to the colatitude where the barycenter of the brightness/magnetic features imposing a correlation between \omeq\ and \dom\ are located. For both \sti\ and \stv, we note a slight increase with time of the cosine of this colatitude (Table{~}\ref{tab:dir}), i.e.\ an increase in the barycentric latitude of the dominant features of ${5\pm 2}$\degr\ and ${15\pm 5}$\degr\ respectively.

		\begin{table*}
			\caption{Summary of differential rotation parameters obtained for \vt\ on each season. All rotation rates are given in \mrd. Column 2 gives the total number of data points used in the imaging process, then columns 3 to 7 correspond to \sti\ data while column 8 to 12 correspond to \stv\ data. Columns 3 and 8 list the derived equatorial rotation rate \omeq, with its 68\% (i.e.\ 1$\sigma$) confidence interval, columns 4 and 9 the difference in rotation rate \dom\ between the equator and pole, with its 68\% confidence interval, columns 5 and 10 give the reduced chi square of the ZDI model compared to the data, columns 6 and 11 give the inverse slope of the ellipsoid in the \omeq-\dom\ plane (also equal to $\cos^2\,\theta_s$, where $\theta_s$ denotes the colatitude of the gravity centre of the spot distribution, see \citealt{Donati00}), and columns 7 and 12 give the rotation rate $\Omega_s$ at colatitude $\theta_s$.}
			\tiny
			\begin{tabular}{cccccccccccccc}
				& & | & \multicolumn{5}{c}{\sti\ data / brightness reconstruction} & | & \multicolumn{5}{c}{\stv\ data / magnetic field reconstruction} \\
				Epoch & $n$ & | & \omeq & \dom & \chisqr & $\cos^2\,\theta_s$ & $\Omega_s$ & | & \omeq & \dom & \chisqr & $\cos^2\,\theta_s$ & $\Omega_s$ \\
				\hline
				08b+09a & 5562 & | & 3360.0{~}$\pm${~}0.1 & 11.1{~}$\pm${~}0.6 & 1.276 & 0.12{~}$\pm${~}0.03 & 3358.7{~}$\pm${~}0.4 & | & 3358.7{~}$\pm${~}0.3 & 8.1{~}$\pm${~}1.8 & 1.127 & 0.11{~}$\pm${~}0.03 & 3357.9{~}$\pm${~}0.5 \\
				13b & 2781 & | & 3360.0{~}$\pm${~}0.1 & 8.1{~}$\pm${~}0.7 & 1.341 & 0.11{~}$\pm${~}0.03 & 3359.1{~}$\pm${~}0.3 & | & 3361.8{~}$\pm${~}1.3 & 19.0{~}$\pm${~}4.3 & 1.038 & 0.23{~}$\pm${~}0.03 & 3354.6{~}$\pm${~}2.1 \\
				15b+16a & 3090 & | & 3358.6{~}$\pm${~}0.1 & 8.8{~}$\pm${~}0.5 & 2.583 & 0.18{~}$\pm${~}0.03 & 3357.0{~}$\pm${~}0.4 & | & 3361.3{~}$\pm${~}0.4 & 13.7{~}$\pm${~}1.0 & 1.046 & 0.32{~}$\pm${~}0.03 & 3352.7{~}$\pm${~}0.8
			\end{tabular}
			\label{tab:dir}
		\end{table*}

		These models exclude solid-body rotation at a level of 3.6 to 22$\sigma$ depending on the epoch. We note that, even with differential rotation, ZDI cannot fit the data of 08b+09a and of 15b+16a down to \chisqr=1, no matter the amount of information allowed. This indicates that surface features are also altered by a significant level of intrinsic variability within the 2-month span of our data set. This issue is further discussed in section{~}\ref{sec:evo}.

	\section{Radial velocities}
	\label{sec:rv}

\begin{figure*}
			\subfloat[2009 Jan. Respective rms: 1.20, 0.13, 0.08{~}\kms]{\includegraphics[totalheight=0.19\textheight]{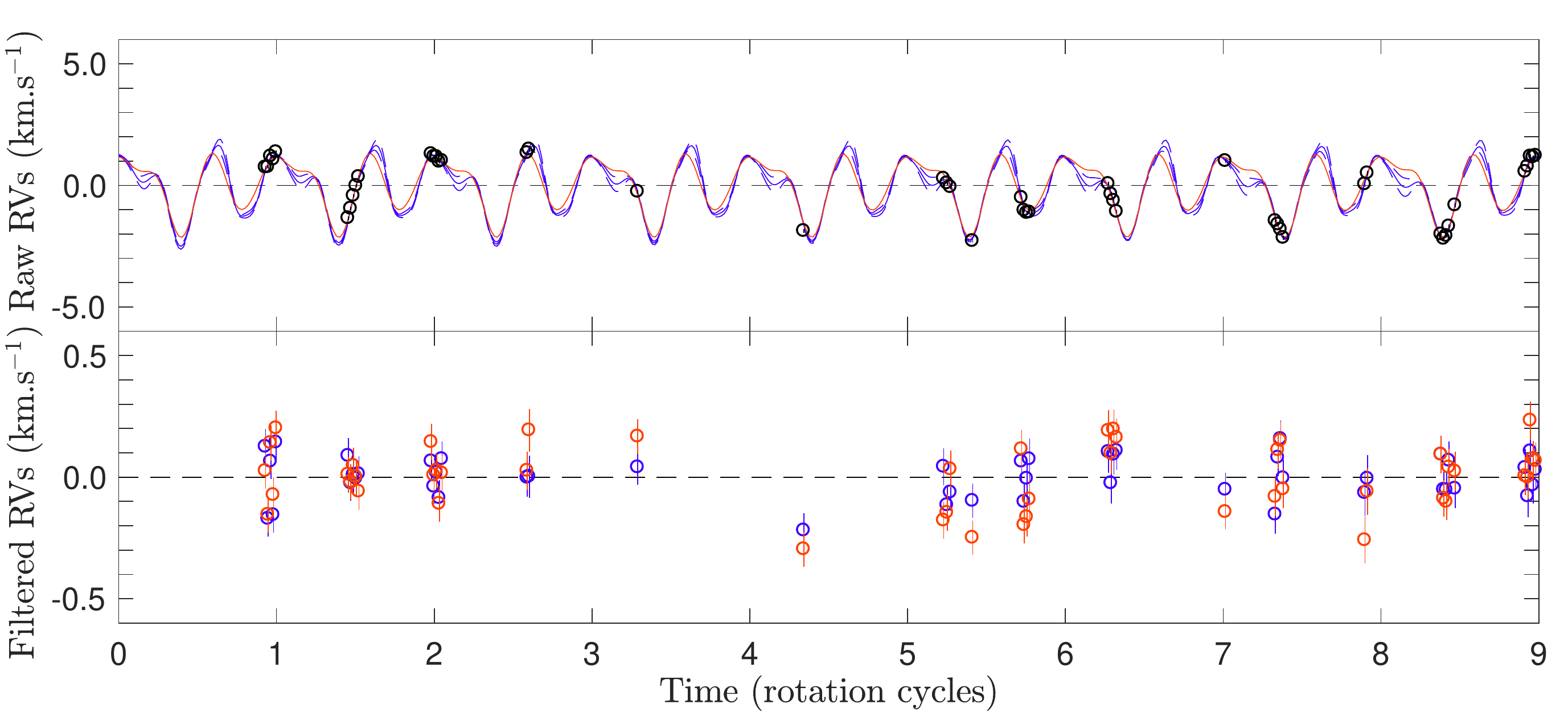}}
			\hfill
			\subfloat[2011 Jan. Respective rms: 2.40, 0.14, 0.06{~}\kms]{\includegraphics[totalheight=0.19\textheight]{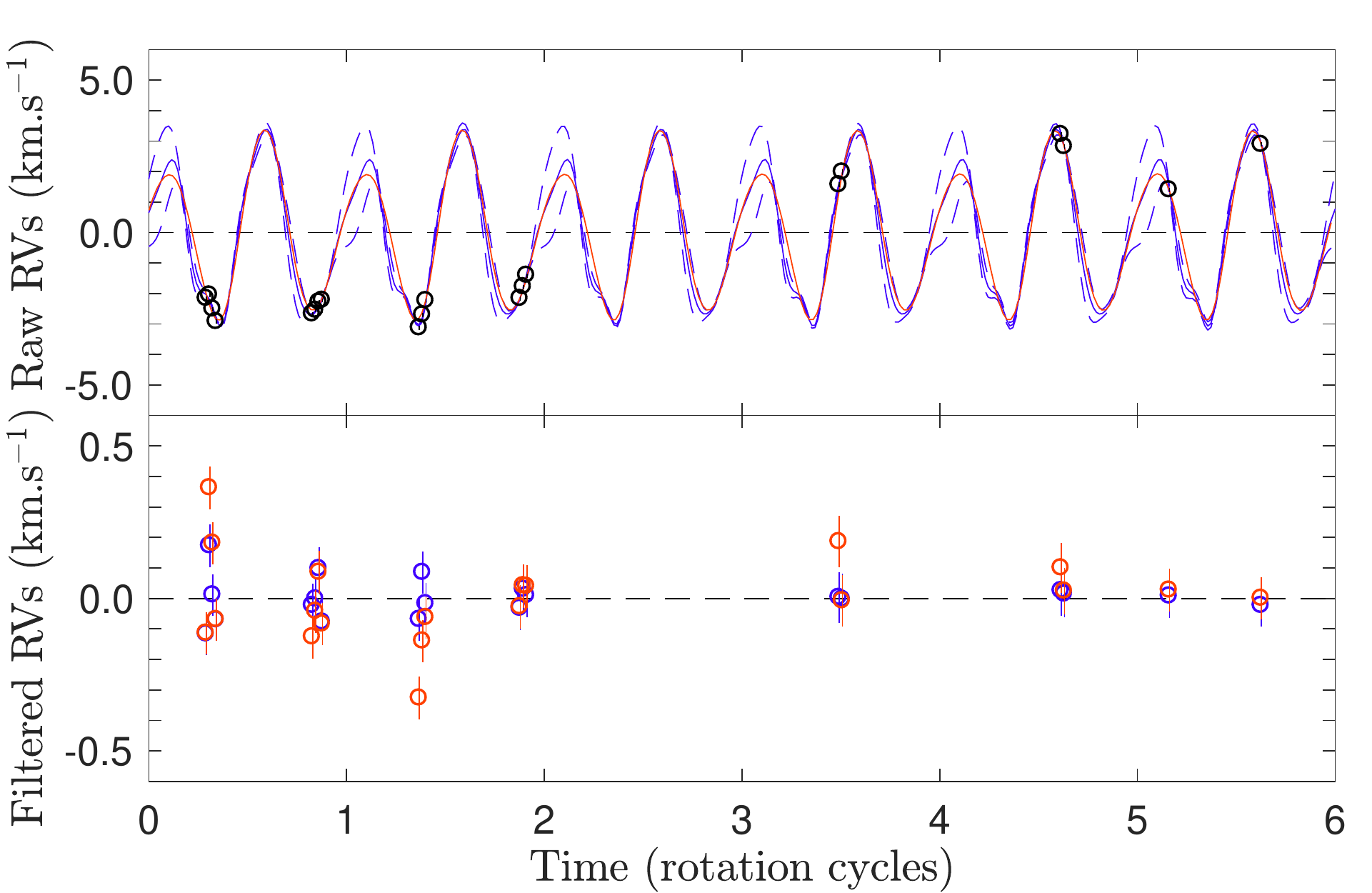}}

			\subfloat[2013 Dec. Respective rms: 2.43, 0.22, 0.09{~}\kms]{\includegraphics[totalheight=0.19\textheight]{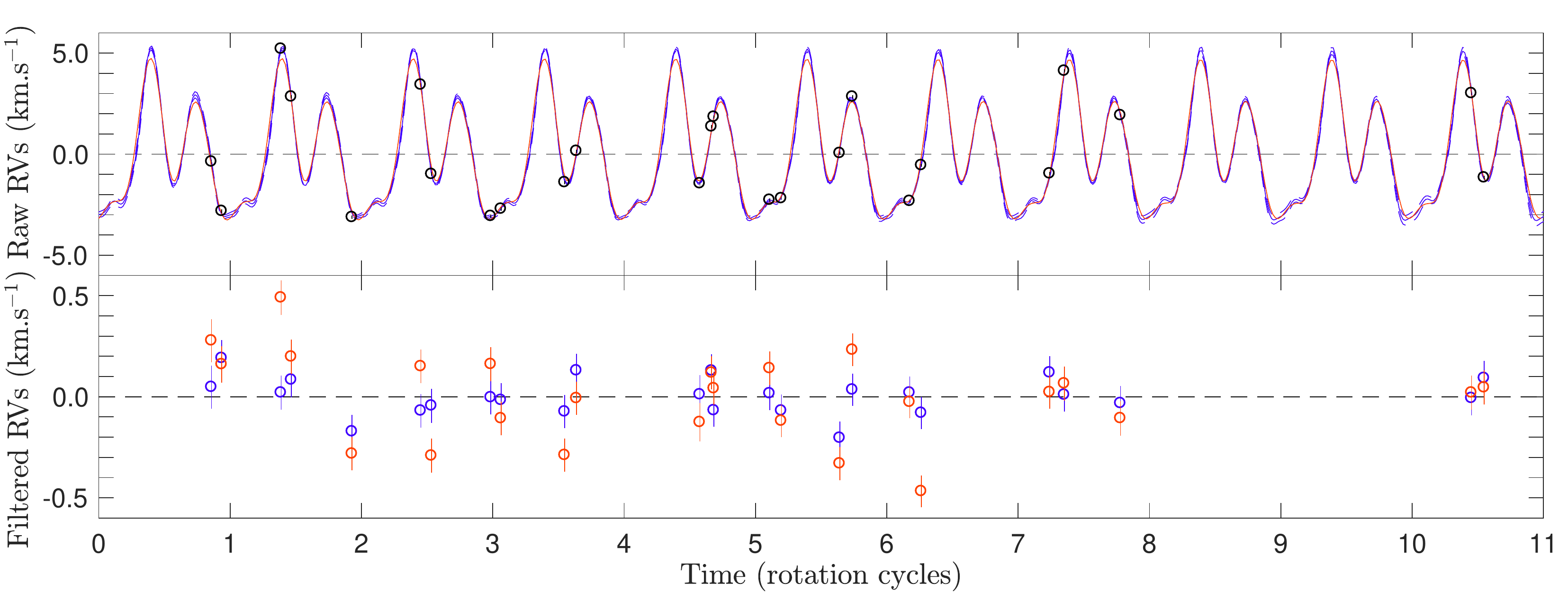}}

			\subfloat[2015 Dec. Respective rms: 1.93, 0.22, 0.08{~}\kms]{\includegraphics[totalheight=0.19\textheight]{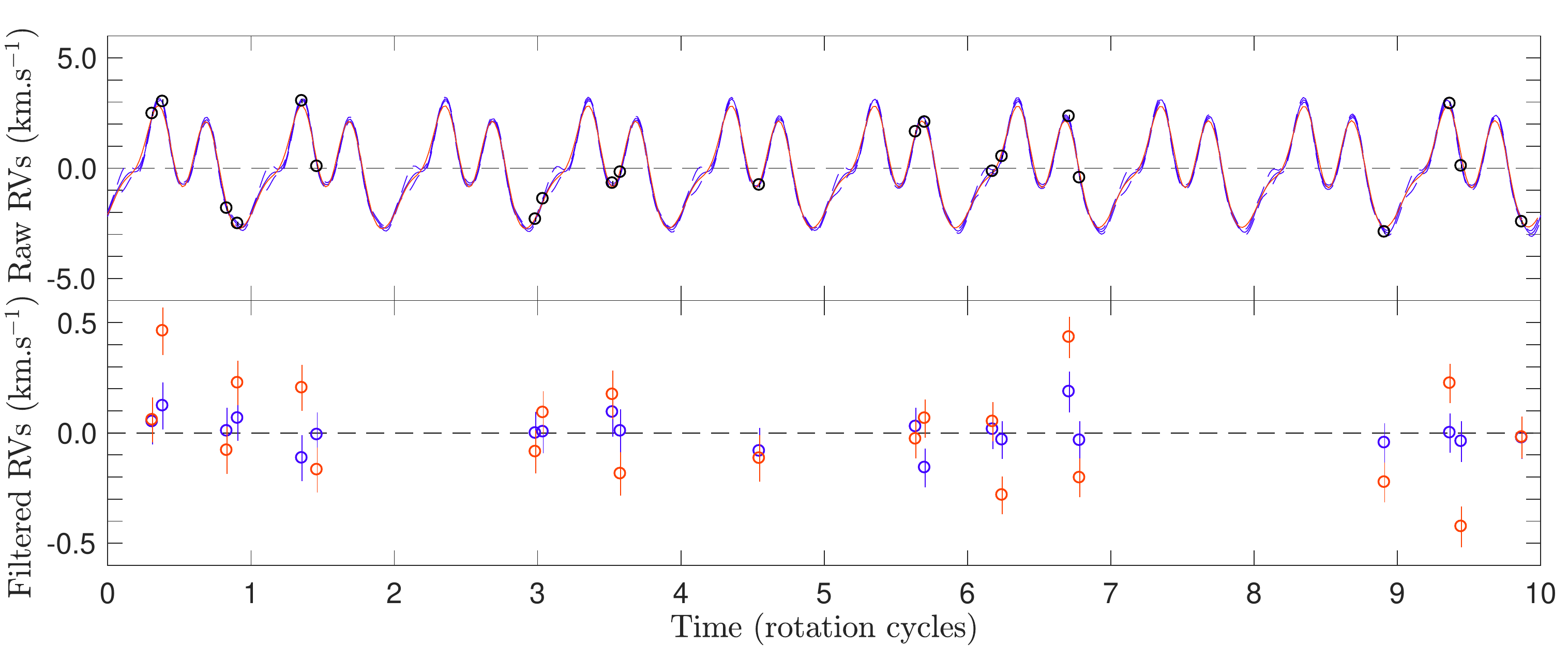}}
			\hfill
			\subfloat[2016 Jan. Respective rms: 1.41, 0.09, 0.01{~}\kms]{\includegraphics[totalheight=0.19\textheight]{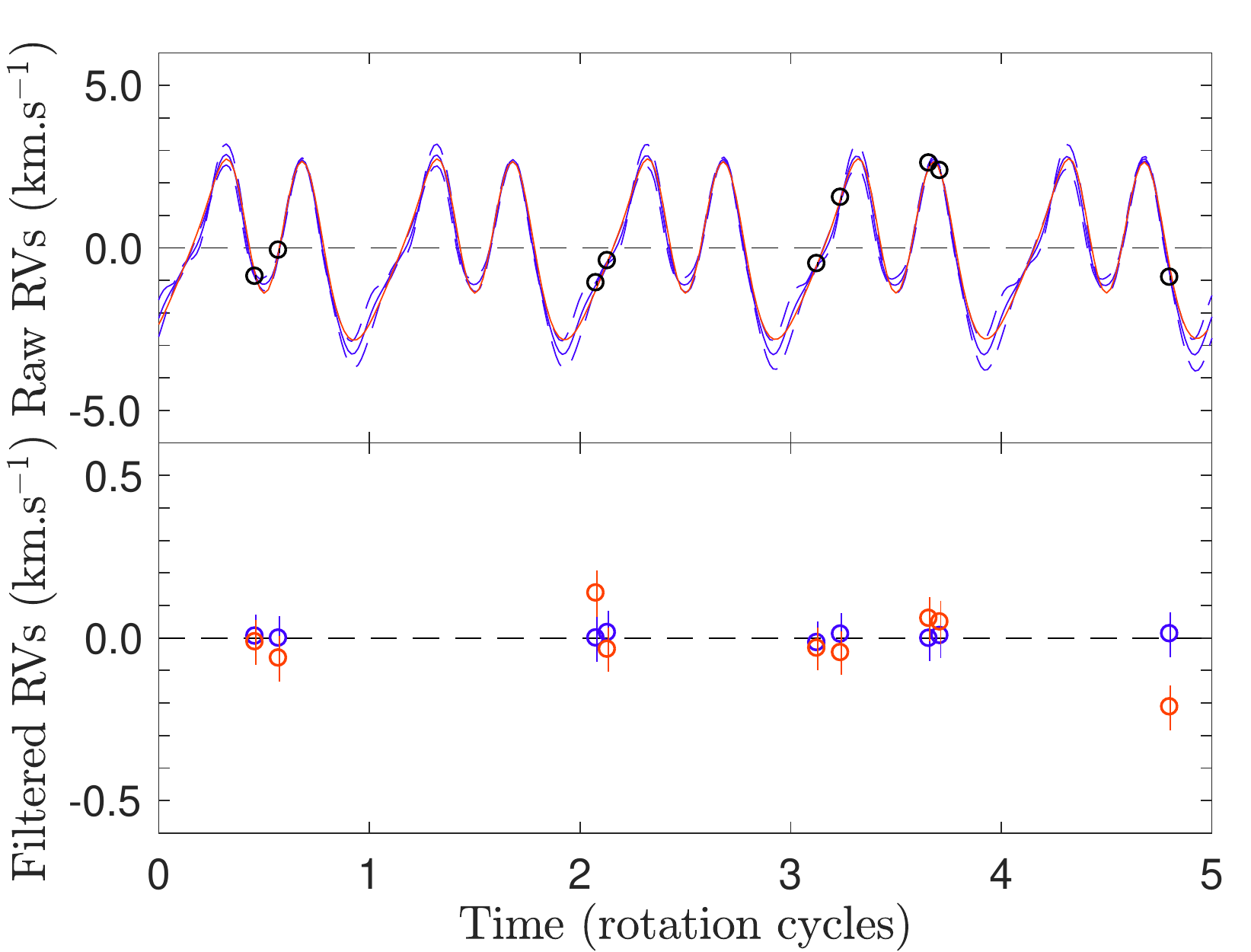}}
			\caption{Raw and filtered RVs of \vt\ for each observation epoch. On each figure, the top plot depicts the raw RVs (red dots), the ZDI reconstruction (red full line) and the GP fit (blue full line with 1-$\sigma$ area of confidence marked as blue dashed lines, see Section{~}\ref{sec:gpr}). The bottom plot depicts the RVs filtered from the ZDI-modelled activity (red dots) and the RVs filtered from the GP-modelled activity (blue dots). The subcaptions indicate the rms of the raw RVs, the ZDI-filtered RVs and the GPR-filtered RVs respectively. All rotational cycles are displayed as in Table{~}\ref{tab:sob}.}
			\label{fig:rvs}
		\end{figure*}

	Radial velocity values were derived as the first-order moment of the continuum-subtracted \sti\ LSD profiles, for all spectra except the 3 with low \sn\ and the 6 in which we identified flares (see Table{~}\ref{tab:sob}). The raw RVs we obtain contain a contribution from the inhomogeneities on the photosphere, called activity jitter, which we aim to filter out in order to access the actual RV of the star, and look for a potential planet signature. The activity jitter is modelled with two different techniques, ZDI and Gaussian Process Regression. Raw RVs and jitter models are plotted in Figure{~}\ref{fig:rvs} and listed in Table{~}\ref{tab:sob}. For the 2015-2016 points, a new version of ZDI, with the logarithmic brightness of surface features allowed to lineary vary with time, was tested (section{~}\ref{sec:evo}). The raw RVs present modulations whose amplitude vary between 4 and 8.5{~}\kms, with a global rms of 1.8{~}\kms. Like with the photometric data, the RV variations are the lowest in 2009 Jan and the strongest in 2013 Dec.

		\subsection{Activity jitter}
		\label{sec:gpr}
		The first method consists in deriving the activity jitter from the ZDI models (see Fig.{~}\ref{fig:sto}), computed as the first-order moment of the continuum-subtracted synthetic \sti\ profiles. Indeed, when computing the raw RV from the observed \sti\ LSD profiles, this activity jitter is added on top of the radial motion of the star as a whole. We model the activity jitter separately for epochs 2009 Jan, 2011 Jan, 2013 Dec, 2015 Dec and 2016 Jan (excluding 2008 Dec because of the poor phase coverage).

\begin{figure}
			\centering
			\includegraphics[angle=-90,width=\linewidth]{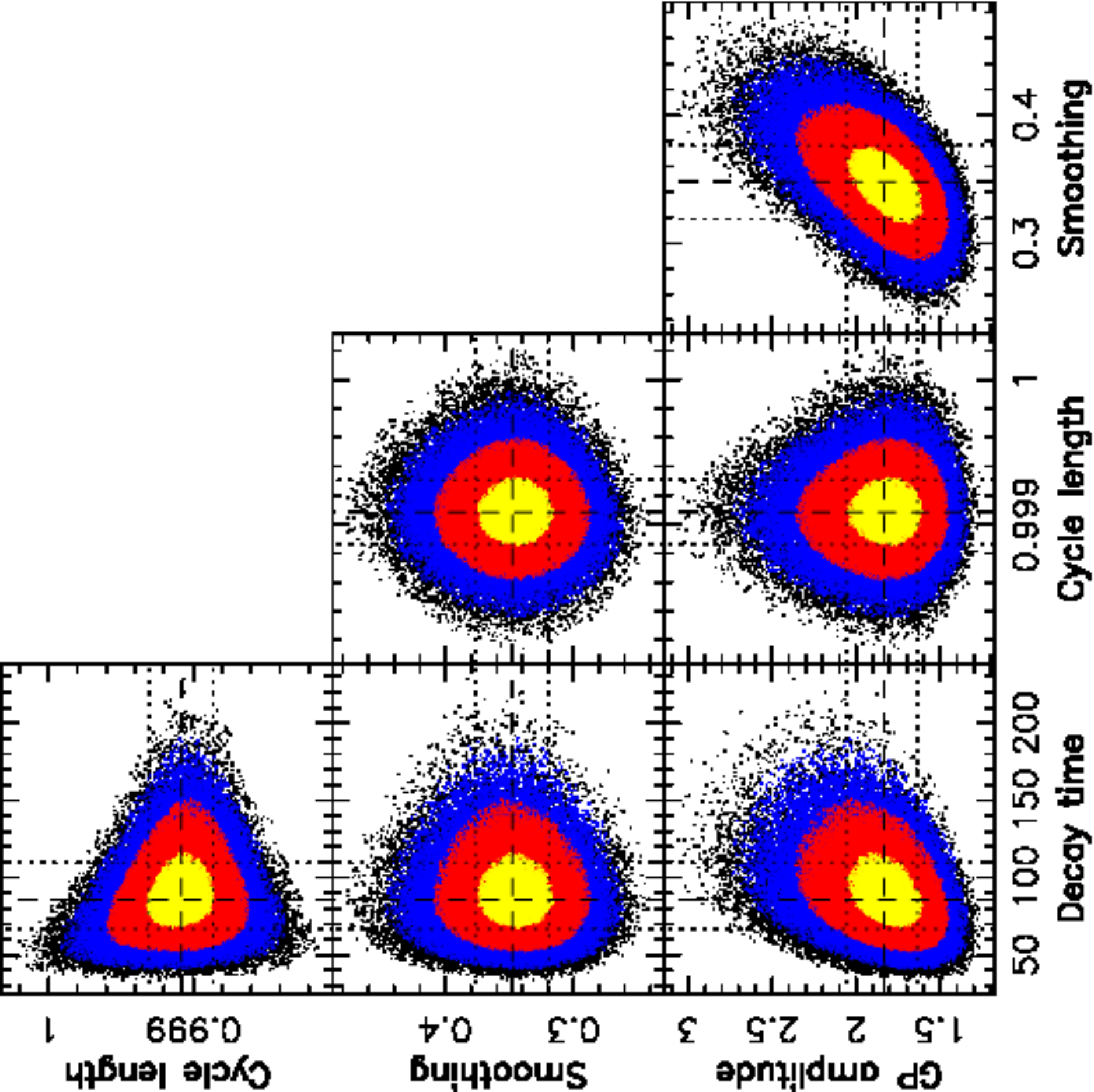}
			\caption{Phase plot of the MCMC-GPR run on the raw RVs, model without planet. The yellow, red and blue colors indicate respectively the 1$\sigma$-, 2$\sigma$- and 3$\sigma$-areas of confidence, and the optimal values for the hyperparameters are marked with black dashed lines, with 1$\sigma$-intervals marked with black dotted lines. GP amplitude ($\theta_1$): $1.8^{+0.2}_{-0.2}${~}\kms, Cycle length ($\theta_2$): $0.9991\pm 0.0002${~}\Prot, Decay time ($\theta_3$): $86^{+24}_{-19}${~}\Prot, Smoothing ($\theta_4$): $0.35\pm 0.03$.}
			\label{fig:gp1}
		\end{figure}

		The second method uses Gaussian Process Regression \citep{Haywood14,Donati17}, a numerical method focusing on the statistical properties of the model. In short, GPR extrapolates a continuous curve described by a given covariance function from some given data points. To describe the activity jitter here, we use a pseudo-periodic covariance function:
		\begin{equation}
		K(t,t')=\theta_1^2.\exp\left[-\frac{(t-t')^2}{\theta_3^2}-\frac{\sin^2\left(\frac{\pi(t-t')}{\theta_2}\right)}{\theta_4^2}\right]
		\end{equation}
		where $t$ and $t'$ are the dates of the two RV points between which the covariance is computed, $\theta_1$ is the amplitude of the GP, $\theta_2$ the recurrence time scale (expected to be close to \Prot), $\theta_3$ the decay time scale (i.e., the typical spot lifetime in the present case) and $\theta_4$ a smoothing parameter (within [0, 1]) setting the amount of high frequency structures that we allow the fit to include. The modelling process therefore consists in optimizing the 4 parameters $\theta_1$, $\theta_2$, $\theta_3$ and $\theta_4$, called hyperparameters. To do so, we use a Markov Chain Monte-Carlo algorithm, and allocate to each point of the hyperparameter space a likelihood value, which takes into account both the quality of the fit and some penalizations on the hyperparameters (for example we penalize high amplitudes, low decay times and low smoothings). The priors are listed in Table{~}\ref{tab:gpr}. The phase plot of the MCMC is displayed in Figure{~}\ref{fig:gp1} and the best fit is shown in Figure{~}\ref{fig:rvs}, together with the ZDI fits. We note that, contrary to ZDI, GPR, being capable of describing intrinsic variability in a consistent way, is able to fit our whole 8-year-long data set with one model. We obtain ${\theta_1 = 1.8 ^{+0.2}_{-0.2}}${~}\kms, ${\theta_2 = 0.9991\pm 0.0002}${~}\Prot, ${\theta_3 = 86^{+24}_{-19}}${~}\Prot\ and ${\theta_4 = 0.35\pm 0.03}${~}\Prot.
		\begin{table}
			\centering
			\caption{Priors for our GP-MCMC run on our raw RVs. For the modified Jeffreys prior, the knee value is given, for the Gaussian prior we give the mean and standard deviation, and for the Jeffreys and the uniform priors we give the lower and upper boundaries.}
			\begin{tabular}{cc}
				\hline
				Hyperparameter & Prior \\
				\hline
				$\theta_1$ (\kms) & Modified Jeffreys ($\sigma_{\rm RV}$) \\
				$\theta_2$ (\Prot) & Gaussian (1.0000, 0.1000) \\
				$\theta_3$ (\Prot) & Jeffreys(0.1, 500.0) \\
				$\theta_4$ & Uniform (0, 1) \\
				\hline
			\end{tabular}
			\label{tab:gpr}
		\end{table}

		The rms of the filtered RVs for each epoch and each method are summarized in Table{~}\ref{tab:rms}. The RV curve filtered from the ZDI model presents a global rms of 0.167{~}\kms, i.e.\ ${\sim 2 <\sigma_{\rm RV}>}$ (see Table{~}\ref{tab:sob}). The epoch where the filtering is most efficient is 2009 Jan, although the rms of the filtered RVs is only at 1.5${<\sigma_{\rm RV}>}$, and it goes up to 3${<\sigma_{\rm RV}>}$ in 2011 Jan and 2013 Dec. On the other hand, the GPR model filters the RV out down to 0.076{~}\kms{~}={~}0.94${<\sigma_{\rm RV}>}$.
		\begin{table}
			\centering
			\caption{Rms of RVs. All rms RVs are given in \kms.}
			\label{tab:rms}
			\begin{tabular}{ccccccc}
				\hline
				Epoch & 2009 & 2011 & 2013 & 2015 & 2016 & All \\
				\hline
				Raw & 1.200 & 2.392 & 2.429 & 1.932 & 1.411 & 1.8 \\
				ZDI filt. & 0.131 & 0.141 & 0.215 & 0.222 & 0.094 & 0.167 \\
				GP filt. & 0.084 & 0.064 & 0.087 & 0.075 & 0.009 & 0.076 \\
				\hline
			\end{tabular}
		\end{table}

		\subsection{Periodograms}
		Lomb-Scargle periodograms for both raw and filtered RVs, for both methods (Fig.{~}\ref{fig:pe5} for each individual epoch, \ref{fig:pea} for the whole data set), show that the stellar rotation period or its first harmonic are clearly present in 2009 Jan and 2011 Jan, but not well retrieved in 2013 Dec, 2015 Dec and 2016 Jan. However the periodogram for the whole ${\rm RV}_{\rm raw}$ data set presents neat peaks at \Prot\ and its first two harmonics. \Prot\ and its first harmonic are well filtered out by both modelling methods, and the second harmonic is well filtered out in the GP residuals. A weak signal remains at \Prot/3 in the ZDI residuals but looking at a phase-folded plot does not reveal any particularly obvious tendency, leading us to suspect that it mostly reflects the contribution of a few stray points. No other period stands out with a false-alarm-probability lower than 5\%, which allows us to conclude that no planet signature is found in this data set with our filtering methods.

		\begin{figure}
			\subfloat[2009 Jan]{\includegraphics[width=\linewidth]{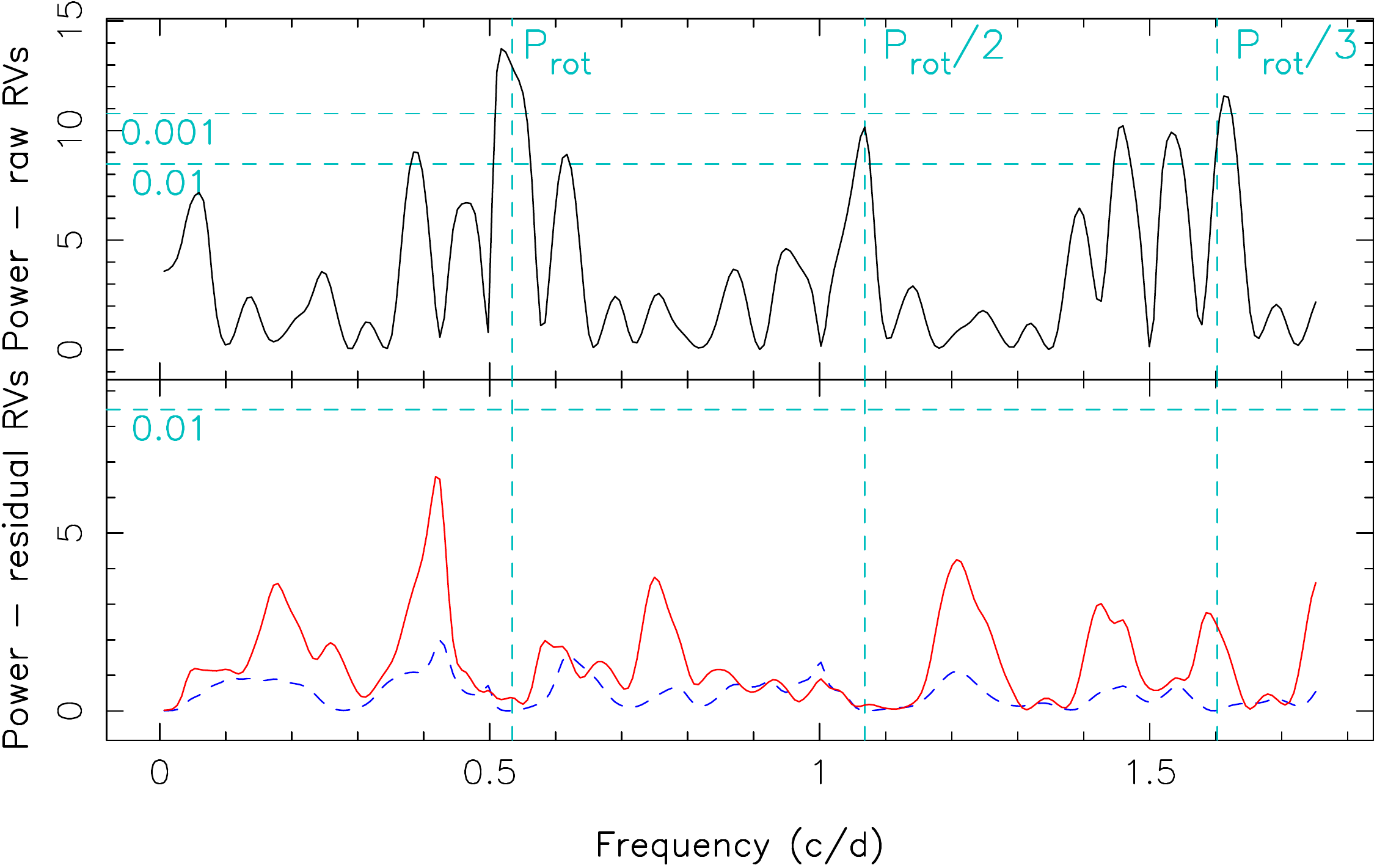}}
			\hfill
			\subfloat[2011 Jan]{\includegraphics[width=\linewidth]{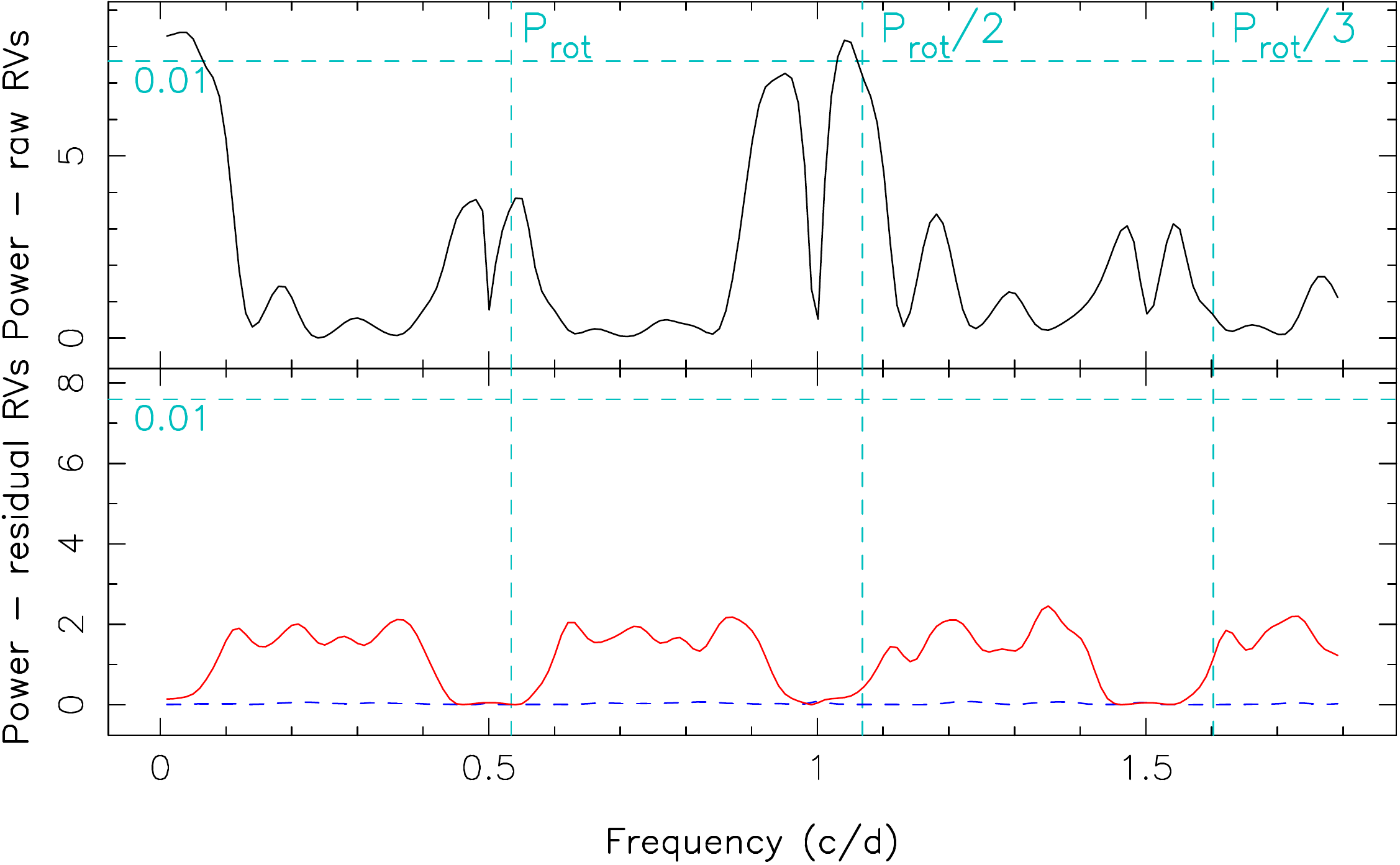}}

			\subfloat[2013 Dec]{\includegraphics[width=\linewidth]{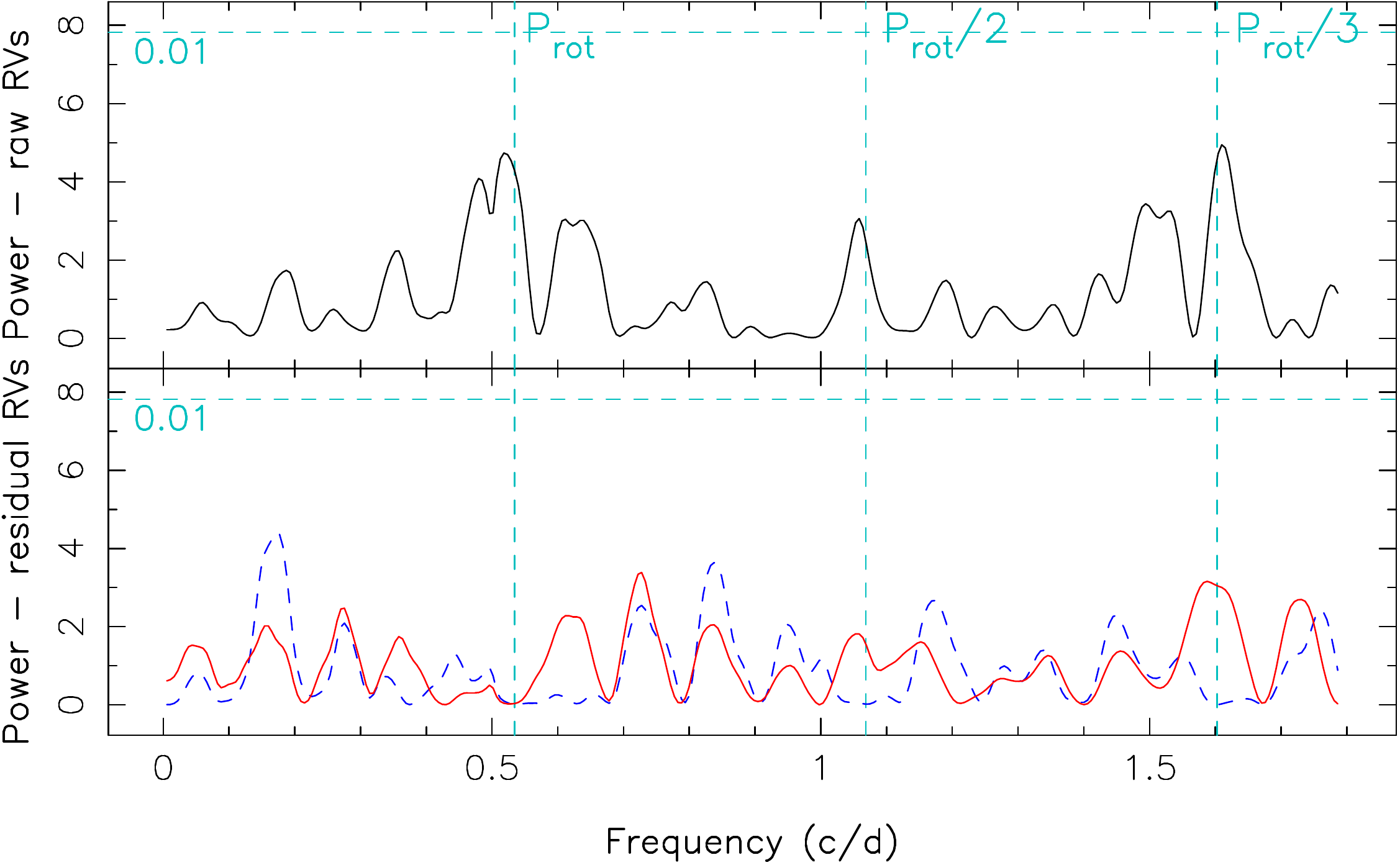}}
			\caption{Periodograms of the raw RVs (top), of the ZDI-filtered RVs (bottom, red full line) and of the GP-filtered RVs (bottom, blue dashed line), for observation epochs 2009 Jan (a), 2011 Jan (b), 2013 Dec (c), 2015 Dec (d) and 2016 Jan (e). False-alarm probability levels of 1\% and 0.1\% are represented as horizontal cyan dashed lines, and \Prot\ and its first two harmonics as vertical cyan dashed lines (continuing next page).}
			\label{fig:pe5}
		\end{figure}
		\begin{figure}
			\ContinuedFloat
			\subfloat[2015 Dec]{\includegraphics[width=\linewidth]{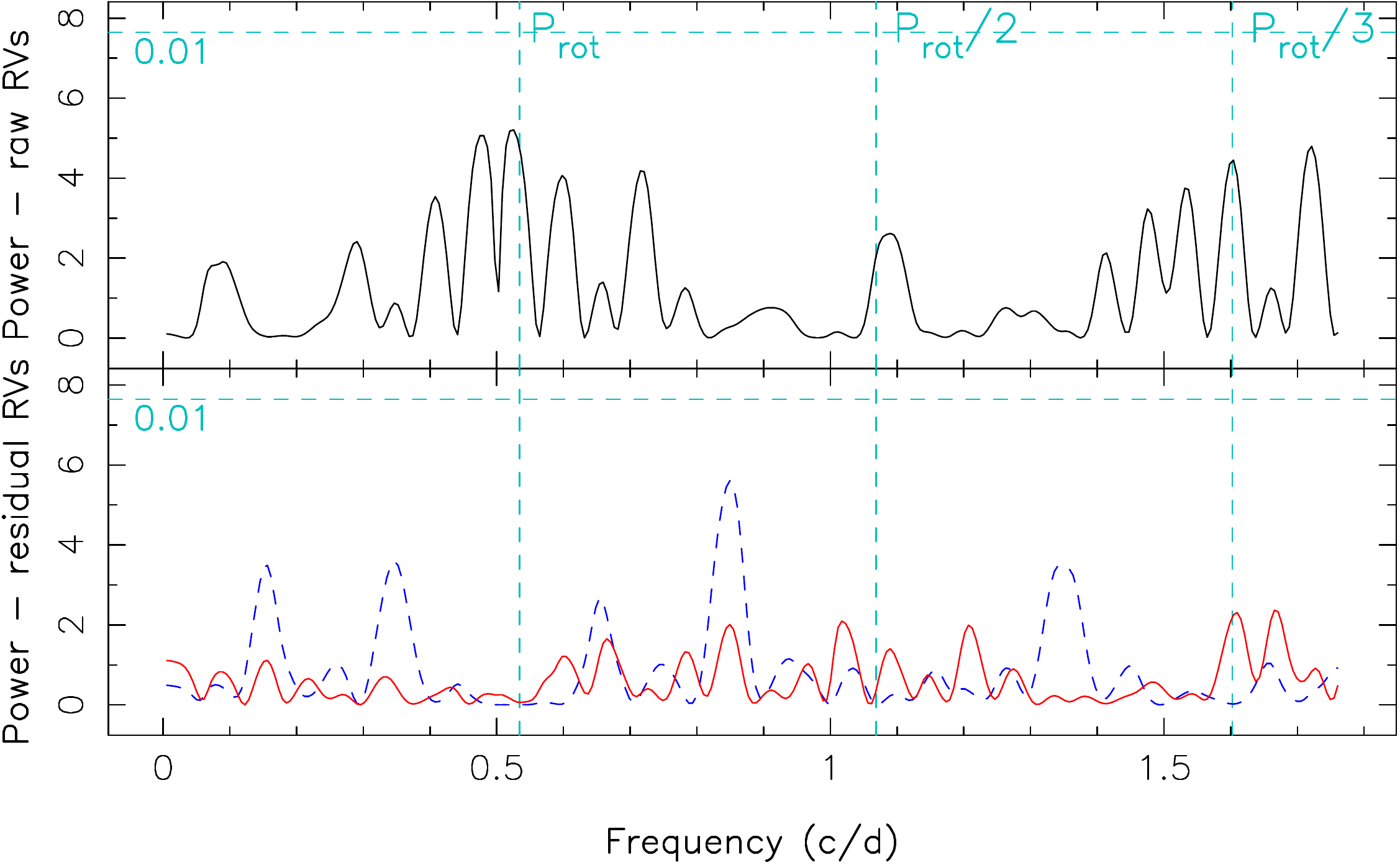}}

			\subfloat[2016 Jan]{\includegraphics[width=\linewidth]{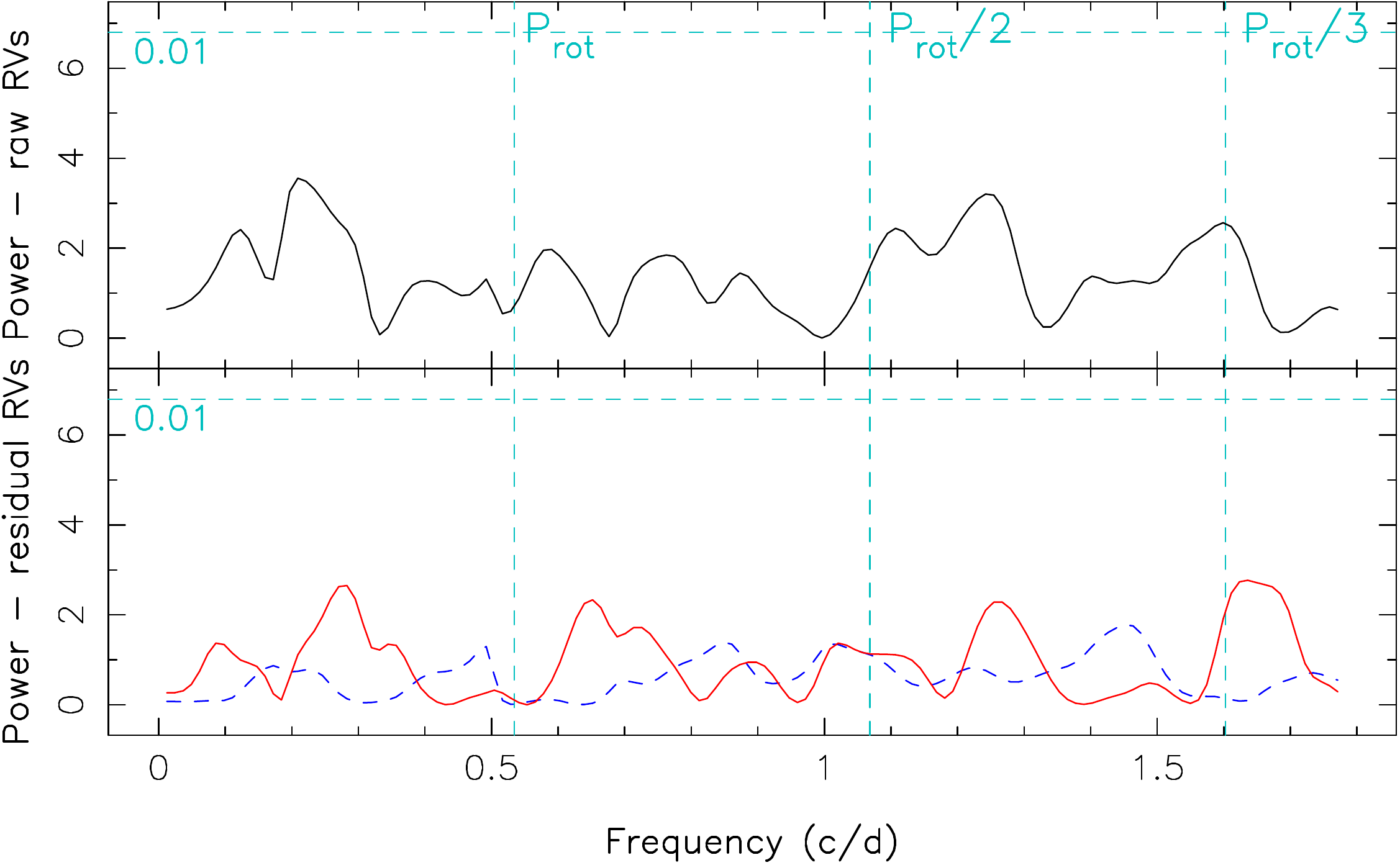}}
			\caption{(Continued from previous page).}
		\end{figure}
		\begin{figure*}
			\includegraphics[width=\linewidth]{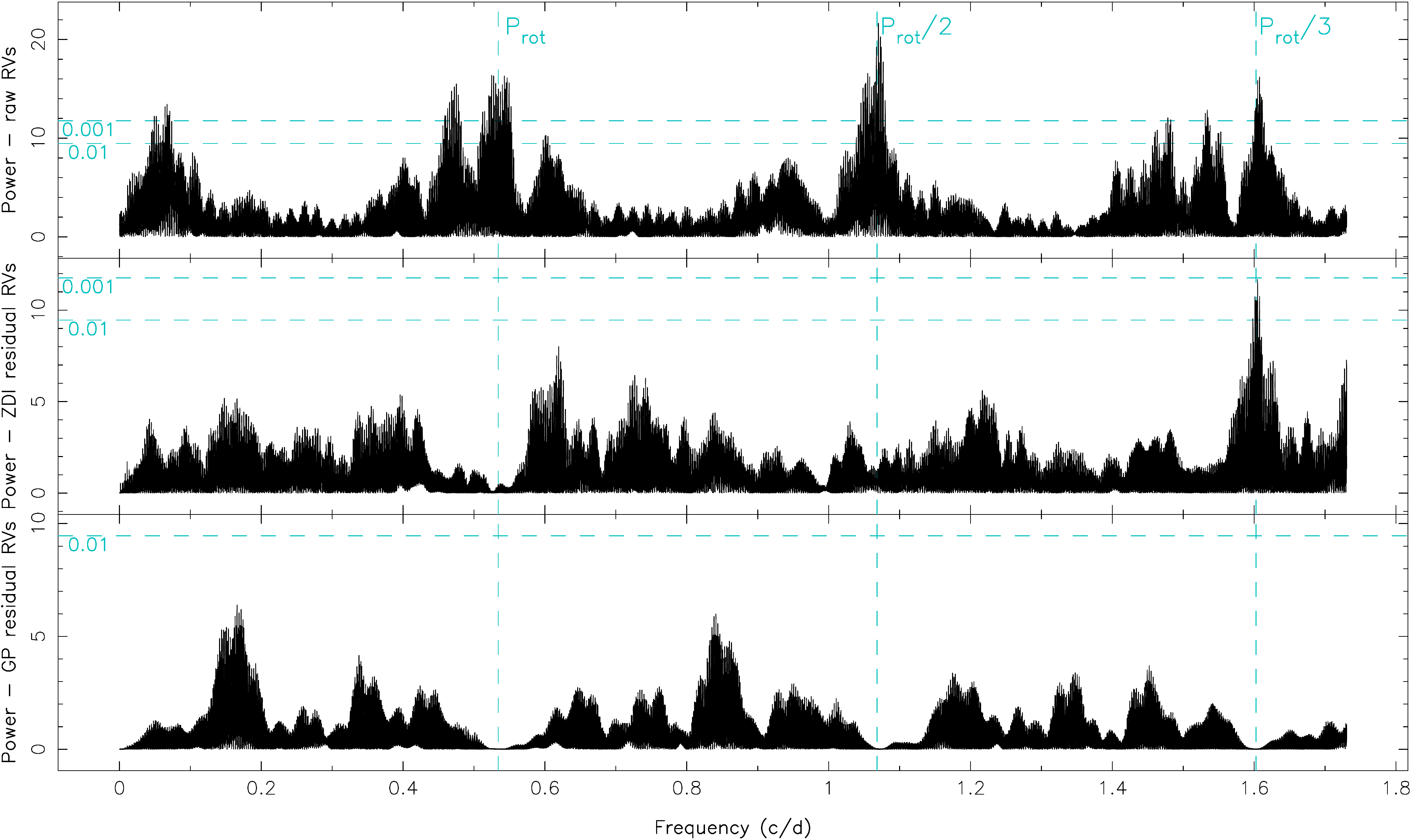}
			\caption{Periodograms of the raw RVs (top), of the RVs filtered from ZDI-modelled activity (middle) and of the RVs filtered from GP-modelled activity (bottom, blue dashed line), for observation epochs 2009  Jan (a), 2011 Jan (b), 2013 Dec (c), 2015 Dec (d) and 2016 Jan (e). Periodograms of the whole data set raw RVs (top), RVs filtered from ZDI-modelled activity (middle) and RVs filtered from GP-modelled activity (bottom). False-alarm probability levels of 1\% and 0.1\% are represented as horizontal cyan dashed lines, and \Prot\ and its first two harmonics as vertical cyan dashed lines.}
			\label{fig:pea}
		\end{figure*}

		\subsection{New ZDI: with short-time intrinsic evolution}
		\label{sec:evo}
\begin{figure}
			\includegraphics[width=\linewidth]{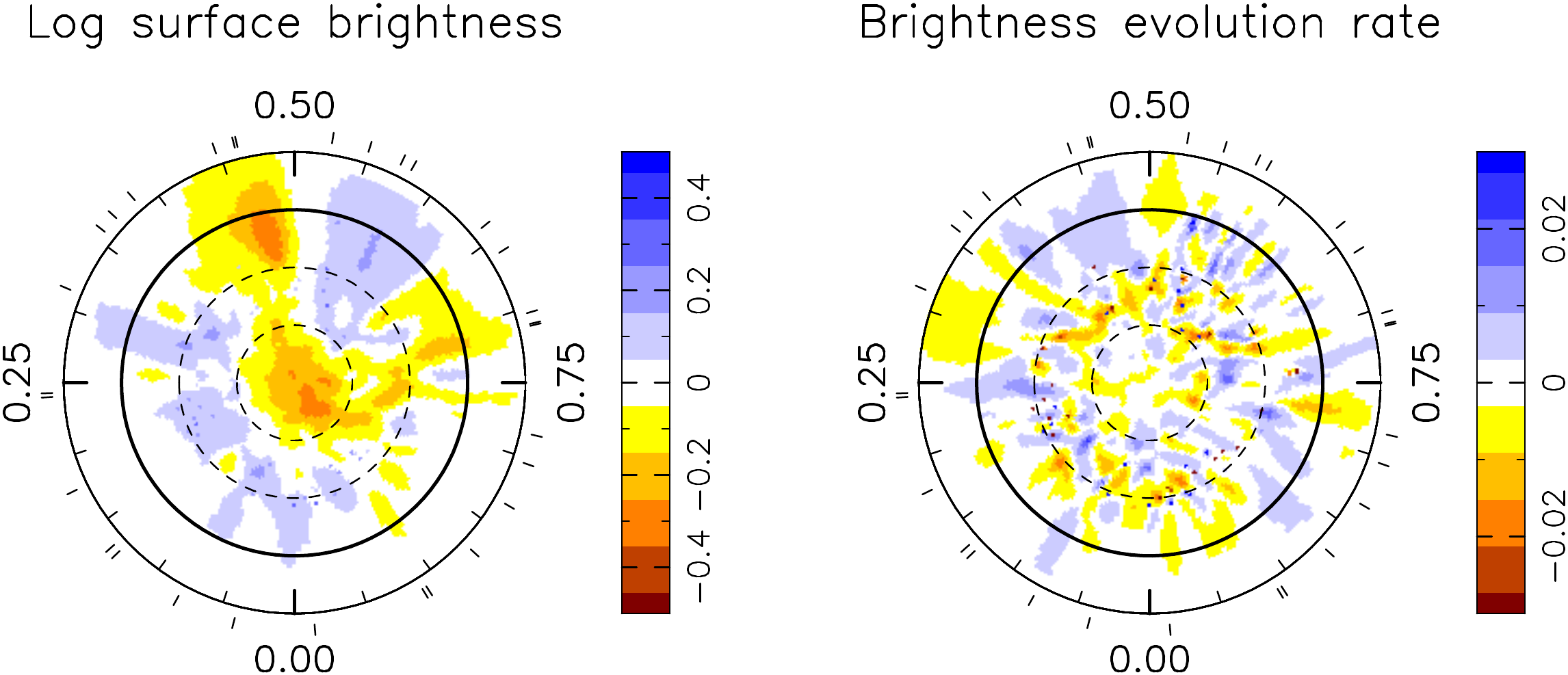}
			\caption{Brightness map and evolution rate reconstructed by ZDI on data set Dec 2015-Jan 2016. Pole-on view with the equator being represented as a full line, and 60\degr, 30\degr, and -30\degr\ latitude parallels as dashed lines. Cool spots are colored in brown and bright plages in blue, and ticks around the star mark the spectropolarimetric observations.}
			\label{fig:qem}
		\end{figure}

\begin{figure*}
			\includegraphics[totalheight=0.19\textheight]{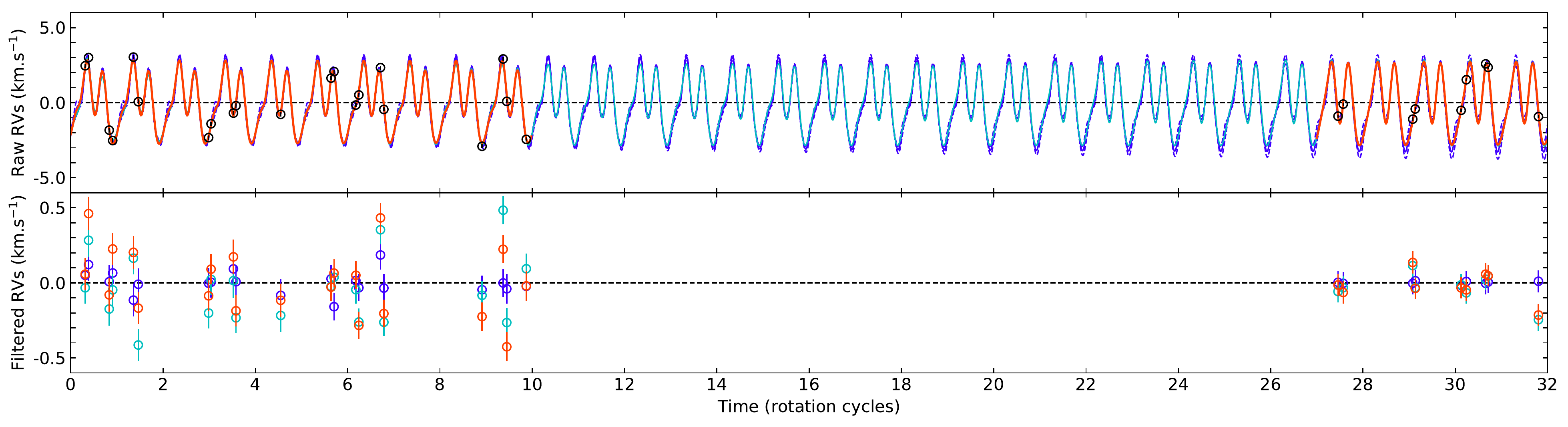}
			\caption{Comparison between the GP model, the new ZDI model and the classical ZDI models for \vt\ RVs in season 2015b-2016a. Rotation cycles are offset to concur with Table{~}\ref{tab:sob}. Top: raw RVs (black dots) with $1\sigma$-error bars, GP model (purple full line), new ZDI model (cyan full line) and classical ZDI models for both observation epochs 2015 Dec and 2016 Jan (red full lines). Bottom: RVs filtered from the GP model (purple dots), from the new ZDI model (cyan dots) or from the classical ZDI models (red dots). The rms of the filtered RVs with GP, new ZDI and classical ZDI are respectively 0.065, 0.194 and 0.193{~}\kms.}
			\label{fig:qer}
		\end{figure*}

\begin{figure}
			\includegraphics[width=\linewidth]{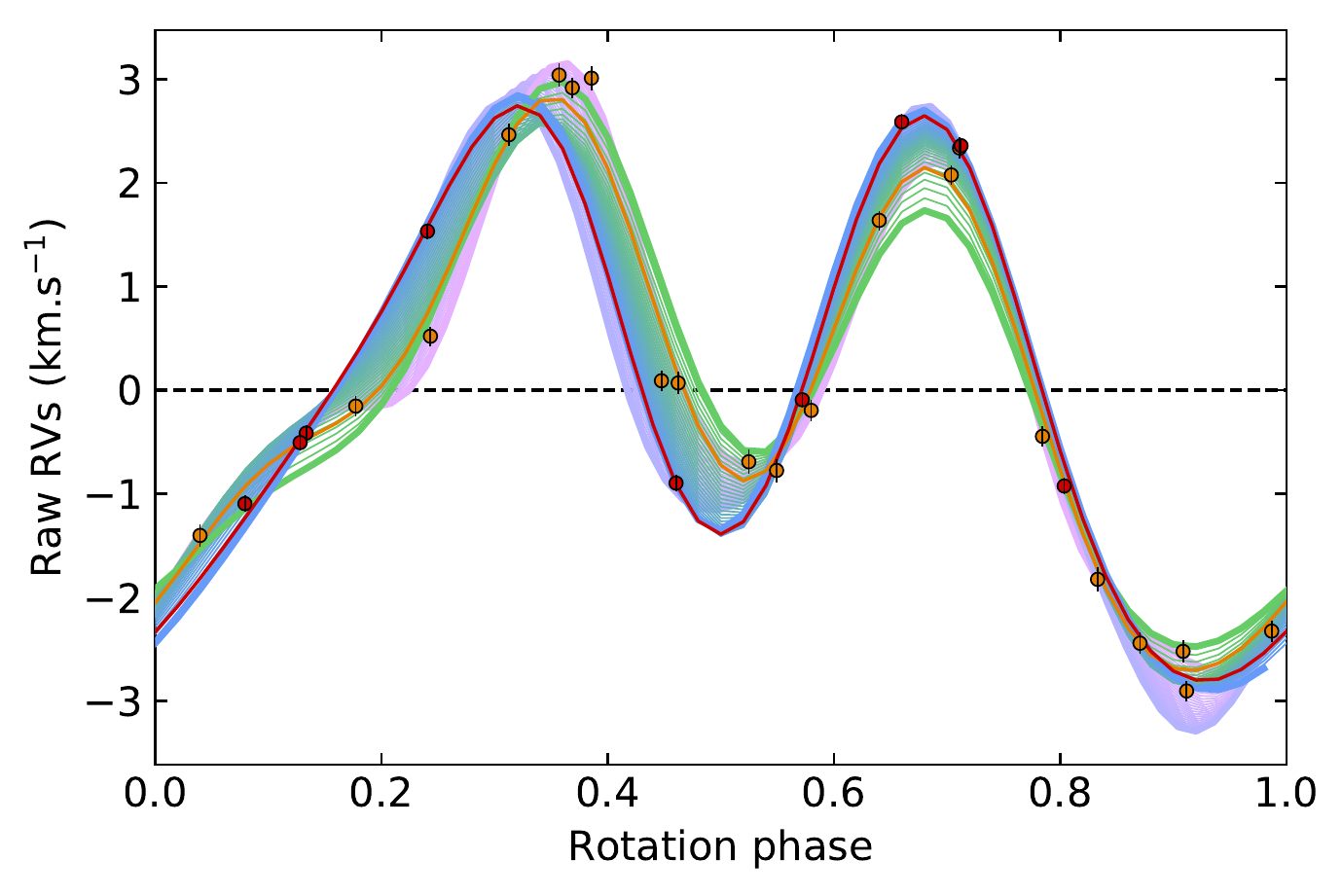}
			\caption{Raw RVs of \vt\ in the 2015b-2016a season, between cycles 1349 and 1381 as referenced in Table{~}\ref{tab:sob}, plotted against stellar rotation phase. The GPR and new ZDI models are represented by full lines colored in gradients, from earliest to latest cycle, respectively pink to purple and green to blue, while the classical ZDI models for 2015 Dec and 2016 Jan are plotted in orange and red respectively. Observations are plotted as dots with 1$\sigma$-error bars, orange for 2015 Dec and 2016 Jan.}
			\label{fig:fol}
		\end{figure}

		Seeing that the filtered RVs when using GPR have a rms twice lower than when using ZDI (Table{~}\ref{tab:rms}), we try to improve our ZDI filtering process by implementing a new feature: instead of only having one brightness value in each cell, we give it a brightness value and an evolution parameter, so that ZDI brightness maps are allowed to evolve with time to better fit time-series of LSD profiles with variability. Thus we reconstruct two maps for the brightness: the brightness at time 0 and the map of the evolution parameter. We choose, for now, a simple model where the logarithmic relative brightness of each cell $k$ is allowed to evolve linearily with time:
		\begin{equation}
			\log Q_k(t)=\log Q_k(0)+m_kt,
		\end{equation}
		where $Q_k(t)$ is the local surface brightness and $m_k$ is the evolution parameter. Applying this new method to the 2015-2016 extended data set, we manage to fit the whole data set down to a \chisqr\ of 1 where classical ZDI, even with differential rotation, could not reach lower than \chisqr=2.5 (see Section{~}\ref{sec:dr}). Maps associated to this reconstruction are shown in Fig.{~}\ref{fig:qem}, and derived RVs are plotted in Fig.{~}\ref{fig:qer} and \ref{fig:fol}, to be compared with RVs derived from classical ZDI maps. The rms of the filtered RVs here, 0.194{~}\kms, does not decrease compared to when using classical ZDI, which means our model is still too simple and cannot fully account for the observed variability. However, Fig.{~}\ref{fig:fol} shows that global trends in the temporal evolution of the RV curve are well-reproduced by this new ZDI model, such as the jitter maximum moving from phase 0.37 to 0.32, or the local minimum at phase 0.54 in 2015 Dec moving to 0.50 in 2016 Jan.
	\section{Summary and discussion}
		\label{sec:ccl}

		\begin{figure}
			\includegraphics[width=\linewidth]{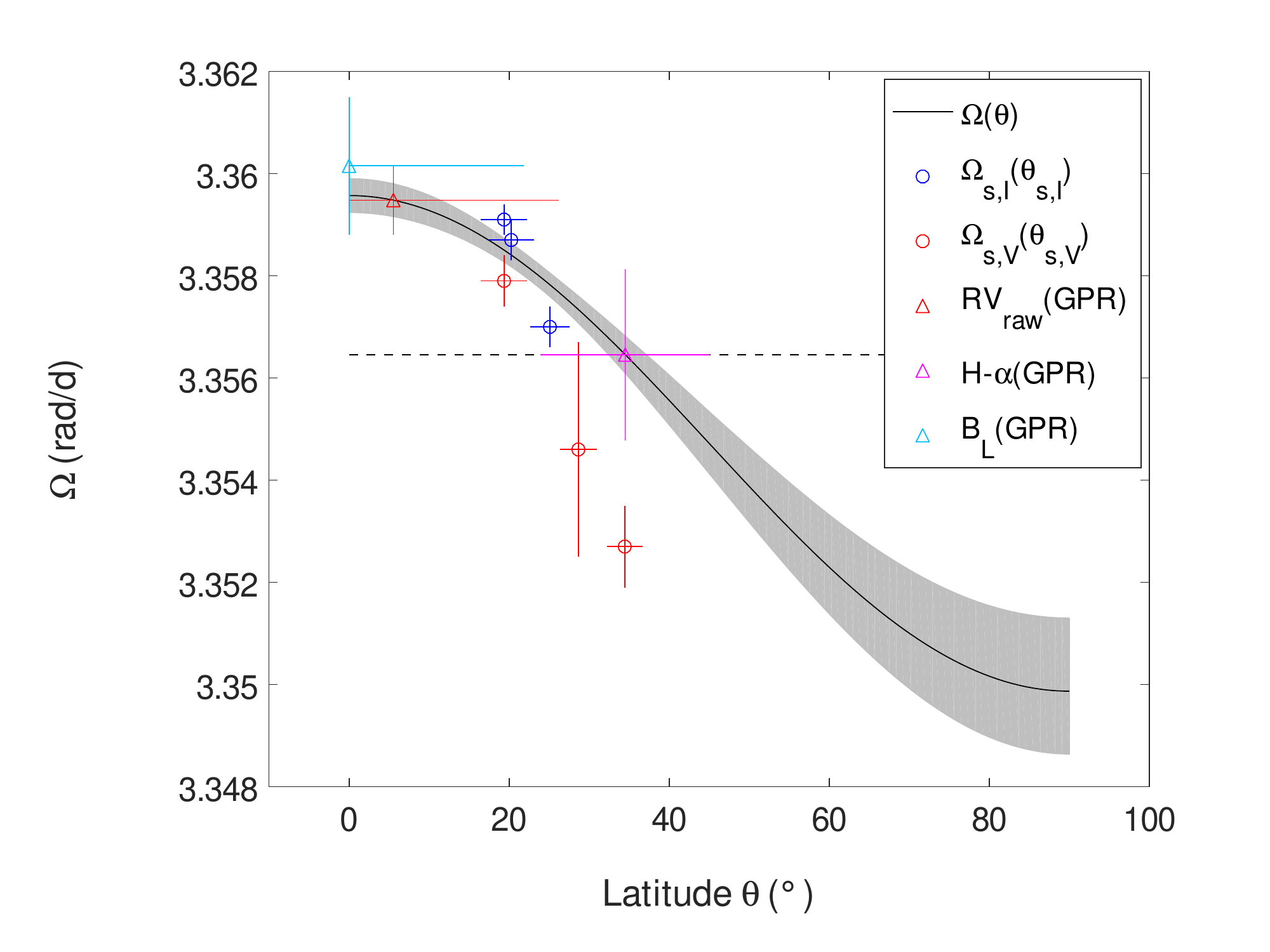}
			\caption{Differential rotation curve of \vt\ (black full line) with 1$\sigma$ uncertainty in gray, with ${\omeq = 3.35957\pm 0.00006}${~}\rpd\ and ${\dom = 0.0097\pm 0.0003}${~}\rpd. The stellar rotation rate chosen to phase our data is represented as a dashed horizontal line. The rotation rates derived from the RVs (red), from the \hal\ equivalent widths (magenta) and from the longitudinal field measurements (cyan) are positioned on the differential rotation curve as triangles with 1$\sigma$ error bars, thus yielding the barycentric latitude of the features determining the period. The dots represent couples ${\lbrace 90 - \theta_s, \Omega_s \rbrace }$ derived in our epoch-wise differential rotation measurements, those coming from \sti\ / \stv\ data being plotted in blue / red respectively.}
			\label{fig:dhr}
		\end{figure}
		This paper reports the analysis of an extended spectropolarimetric data set on the $\sim$0.8{~}Myr wTTS \vt, taken with the instruments ESPaDOnS at CFHT and NARVAL at TBL, spanning eight years and split between six observation epochs (2008b, 2009a, 2011a, 2013b, 2015b and 2016a), the last three of which were observed as part of the MaTYSSE observation programme. Contemporaneous photometric observations from the CrAO and from the WASP programme complemented the study. ESPaDOnS, NARVAL and CrAO observations are documented in Appendix{~}\ref{anx:obs}.

		V410{~}Tau is composed of an inner close binary (\vt\ A-B) around which orbits a third component \citep[C][]{Ghez97}, with \vt\ A being much brighter than the other two in the optical domain, and thus the star that our data inform. The stellar parameters derived in this work are summed up in Table{~}\ref{tab:php}: at $\sim$0.8{~}Myr, \vt\ is a ${1.42\pm 0.15}${~}\msun\ and ${3.4\pm 0.5}${~}\rsun\ wTTS.

		\subsection{Activity and magnetic field of \vt}
		Applying LSD then ZDI on our data set, we estimated the \vsini\ and inclination of \vt\ at ${73.2\pm 0.5}${~}\kms\ and ${50\pm 10}$\degr\ respectively. Considering the well-determined rotation period of ${1.871970\pm 0.000010}$ \citep{Stelzer03} and the minimal observed visible magnitude of 10.52 \citep{Grankin08}, this implies a relatively high level ($\sim$50\%) of spot coverage. We reconstructed brightness and magnetic surface maps at each observation epoch, constrained the differential rotation and found a drift in the bulk radial velocity. Our ZDI brightness maps display a relatively highly spotted surface: the spot coverage reaches 6.5 to 11.5 percent depending on the epoch (not counting 2008 Dec where only half the star was imaged) and the plage coverage is found around 7 percent at all epochs. Since ZDI mostly recovers large non-axisymmetric features and misses small ones evenly distributed over the star, the spot and plage coverage is underestimated, which makes this result compatible with the spot coverage obtained from the aforementioned V magnitude measurements. We note that \vt\ being heavily spotted makes it difficult to pinpoint its age. We fit a 2-temperature model (photosphere at 4500{~}K and fixed-temperature spots with a varying filling factor) into our B-V and V magnitude data, and found an optimal spot temperature of around 3750{~}K, which implies a contrast of $\sim$750{~}K between dark spots and the photosphere (see Fig.{~}\ref{fig:vbk}). This contrast is slightly lower than the one retrieved for the 2{~}Myr wTTS \lkca\ in \cite{Gully-Santiago17}. \vt\ always presents a high concentration of dark spots around the pole, and several big patches of dark spots on the equator.

		V410{~}Tau has a relatively strong large-scale magnetic field, with an average surface intensity that is roughly constant over the years at ${550\pm 50}${~}G. Its radial field reaches local values beyond -1{~}kG and +1{~}kG in several epochs. The brightness and magnetic surface maps both present some variability from epoch to epoch (Fig.{~}\ref{fig:qbm}, Table{~}\ref{tab:zdi}), which points to a dynamo-generated magnetic field rather than a fossil one. The magnetic energy is, at all epochs, equally distributed between the poloidal and toroidal components of the field, with the poloidal component being rather non-dipolar and non-axisymmetric, whereas the toroidal component is mostly dipolar and axisymmetric. The poloidal dipole, tilted towards a phase that stays within ${0.6\pm 0.1}$ during the whole survey, but at an angle varying between 20\degr\ and 55\degr\ depending on the epoch, sees its intensity increase almost monotonously from 165{~}G to 458{~}G over 8 years, and the dipolar contribution to the poloidal field also increases from $\sim$25\% to $\sim$40\% (see Table{~}\ref{tab:zdi}).

		The toroidal component, which displays a constant orientation throughout our data set, is unusually strong compared to other fully convective rapidly-rotating stars \citep[e.g.\ \vtt\ is 90 percent poloidal, see][]{Donati17}. A similarly strong toroidal field was observed on one other MaTYSSE target, \lkca\ \citep{Donati14}. The origin of this strong toroidal field is still unclear: could it be maintained by an $\alpha^2$ dynamo, like in the simulations of low-Rossby fully convective stars by \cite{Yadav15}? The remnants of a subsurfacic radial shear between internal layers accelerating due to contraction, and disc-braked outer layers? 
		Or would the even earlier toroidal energy, from right after the collapse of the second Larson core \citep[as found in the simulations of][]{Vaytet18}, somehow not have entirely subsided yet? Would the early dissipation of the disc, a common factor between \lkca\ and \vt, have something to do with this?

		At $\sim$0.8{~}Myr, \vt\ is one of the youngest observed wTTSs \citep[][Fig.{~}3]{Kraus12}. Assuming that, when the disc was present, \vt\ was magnetically locked to it at a rotation period of $\sim$8{~}d with a cavity of $\sim$0.085{~}au \citep[similarly to cTTSs BP Tau, AA Tau and GQ Lup, see][resp.]{Donati08,Donati10b, Donati12}, then \vt\ should have had a radius of $\sim$7{~}\rsun\ when the disc dissipated, to match the angular momentum that we measure today \citep[][]{Bouvier07c}. According to the Siess models \citep{Siess00}, this corresponds to an age of $\sim$0.2{~}Myr. With a radius of $\sim$7{~}\rsun, \vt\ would have needed a magnetic dipole barely above $100${~}G to maintain the assumed magnetospheric cavity, even with an accretion rate of ${\sim 10^{-8}}${~}\msun\,${\rm yr}^{-1}$ just before disc dissipation. That value is compatible with the 200-400{~}G dipole we measure on the $\sim$3.5{~}\rsun\ star today. \citealt{Kraus12} (Fig.{~}1) shows a correlation between the presence of a close companion and the early depletion of the accretion disc, which indicates that \vt{~}B, observed at a projected separation of ${16.8\pm 1.4}${~}au \citep[][]{Ghez95}, could have been responsible for the early depletion of the disc.

\begin{figure*}
			\subfloat[Phase 0.95]{\includegraphics[width=0.2\linewidth]{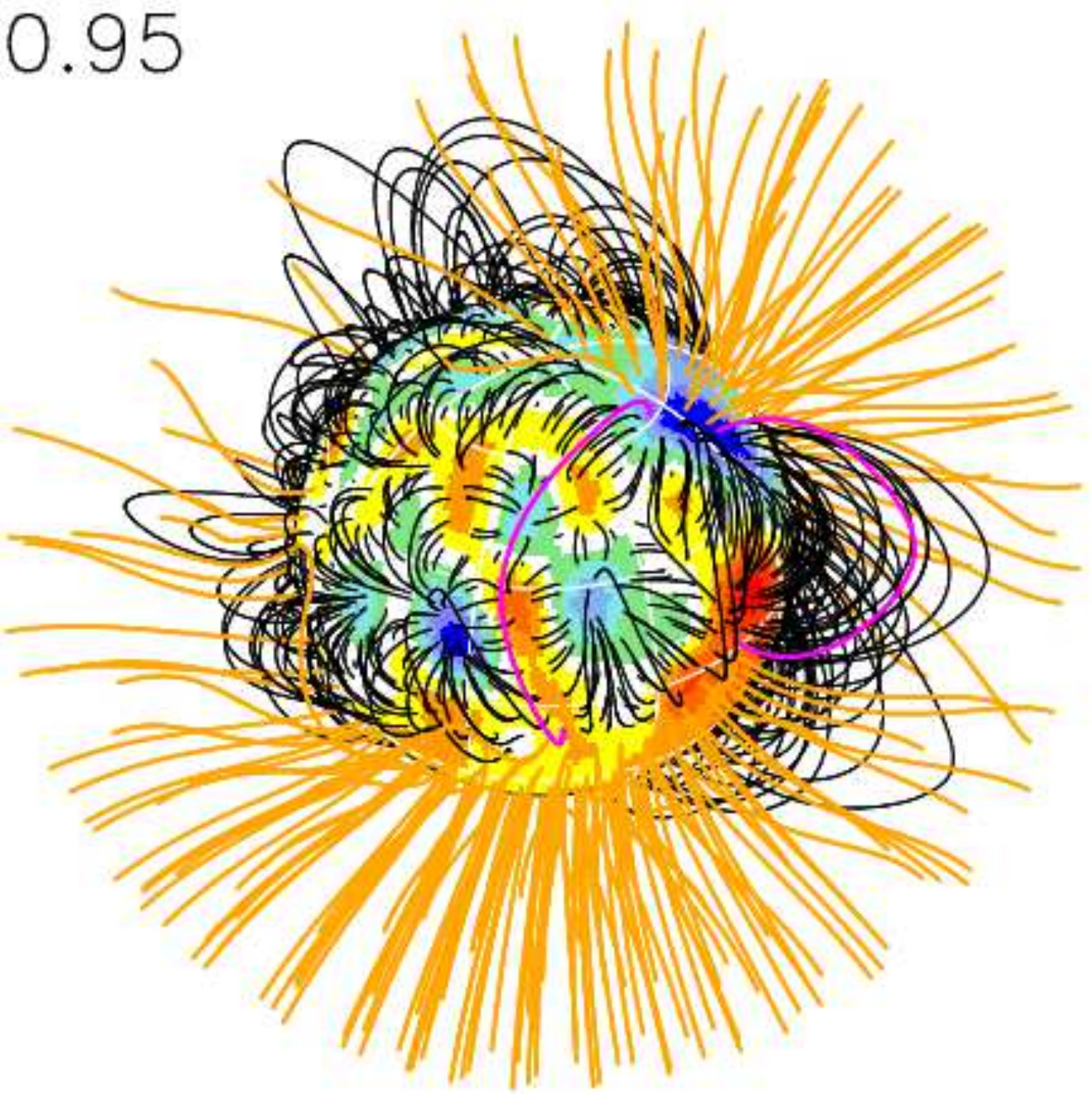}}
			\subfloat[Phase 0.15]{\includegraphics[width=0.2\linewidth]{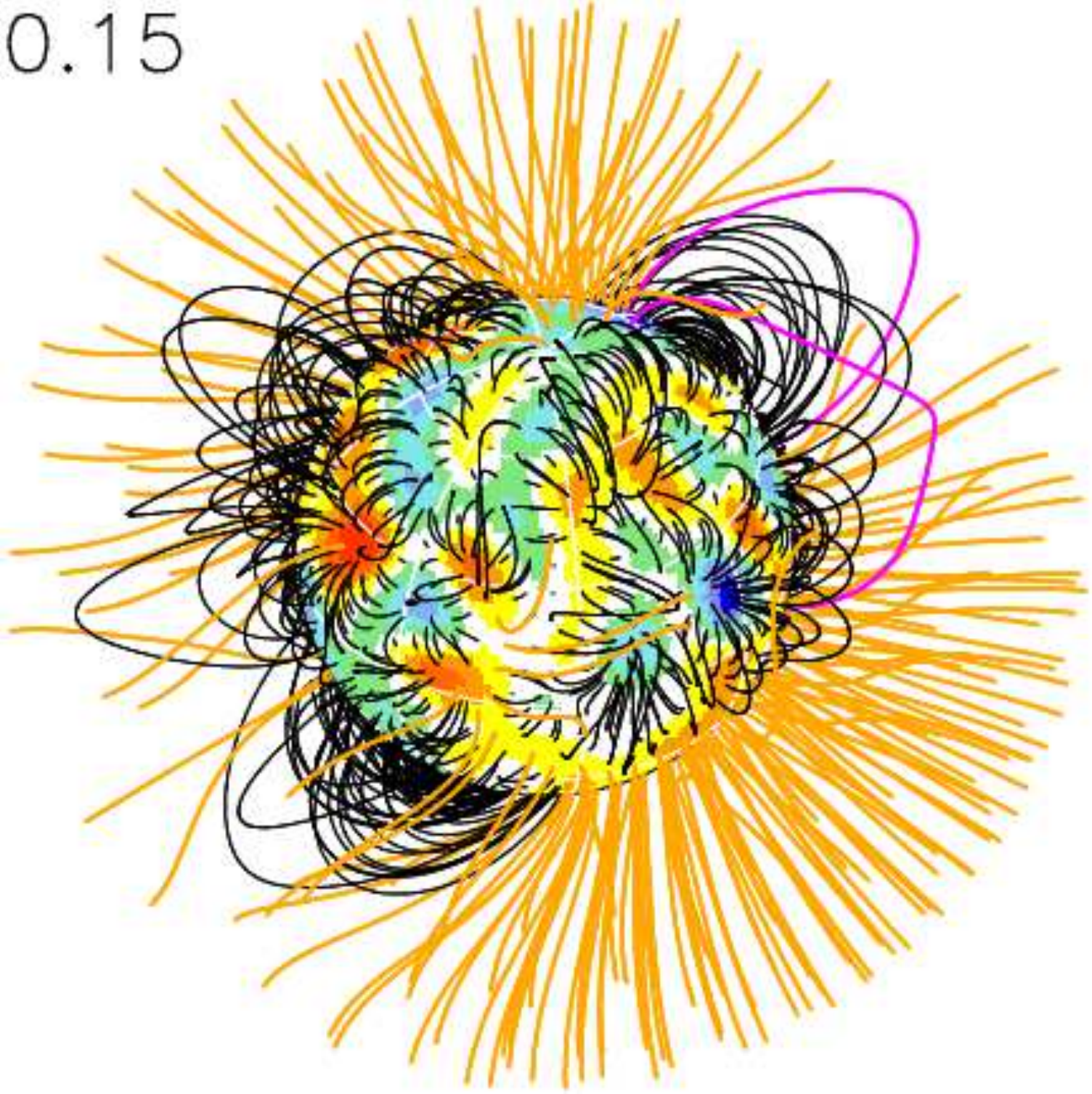}}
			\subfloat[Phase 0.35]{\includegraphics[width=0.2\linewidth]{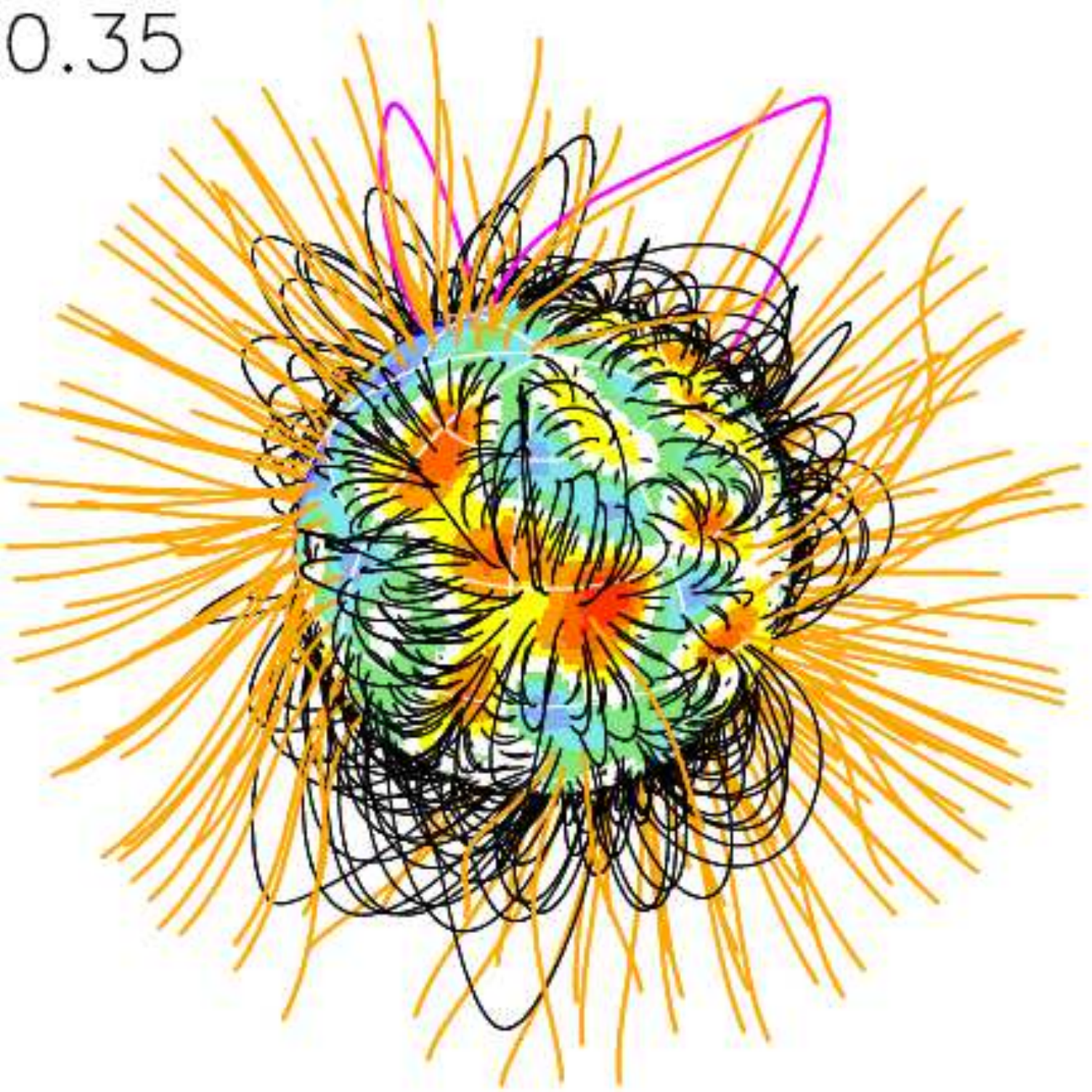}}
			\subfloat[Phase 0.55]{\includegraphics[width=0.2\linewidth]{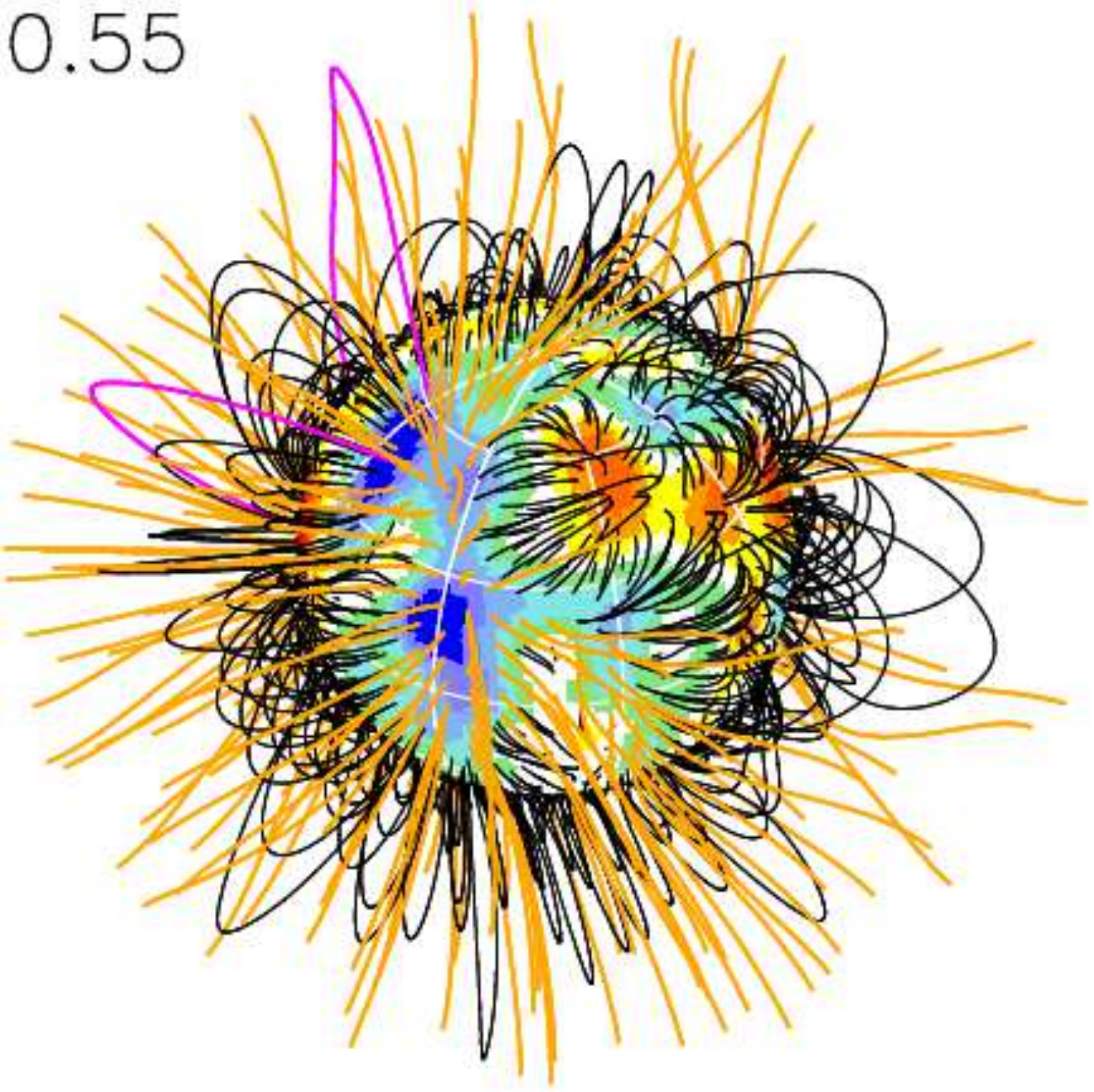}}
			\subfloat[Phase 0.75]{\includegraphics[width=0.2\linewidth]{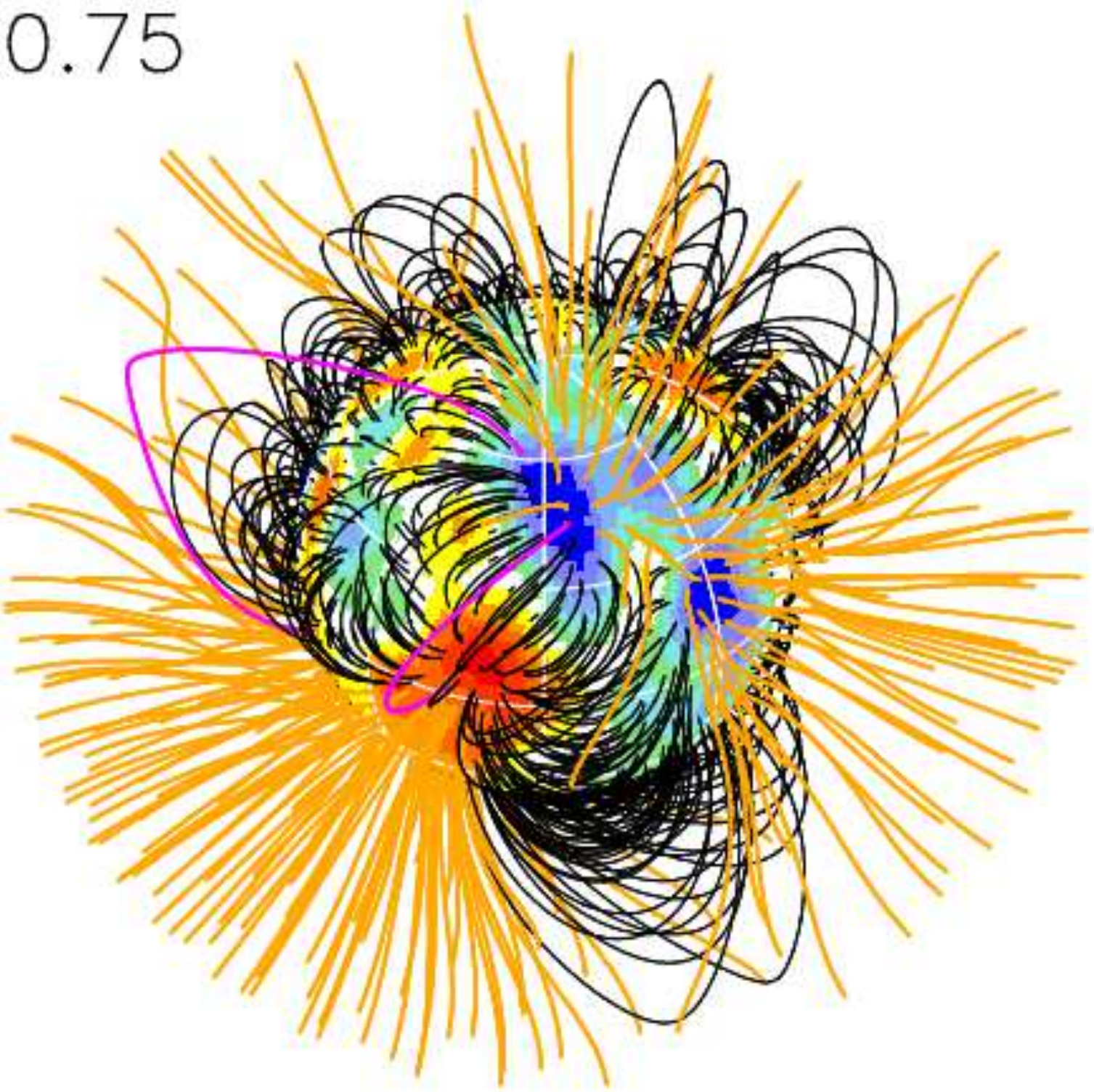}}

			\subfloat[Phase 0.35]{\includegraphics[width=0.2\linewidth]{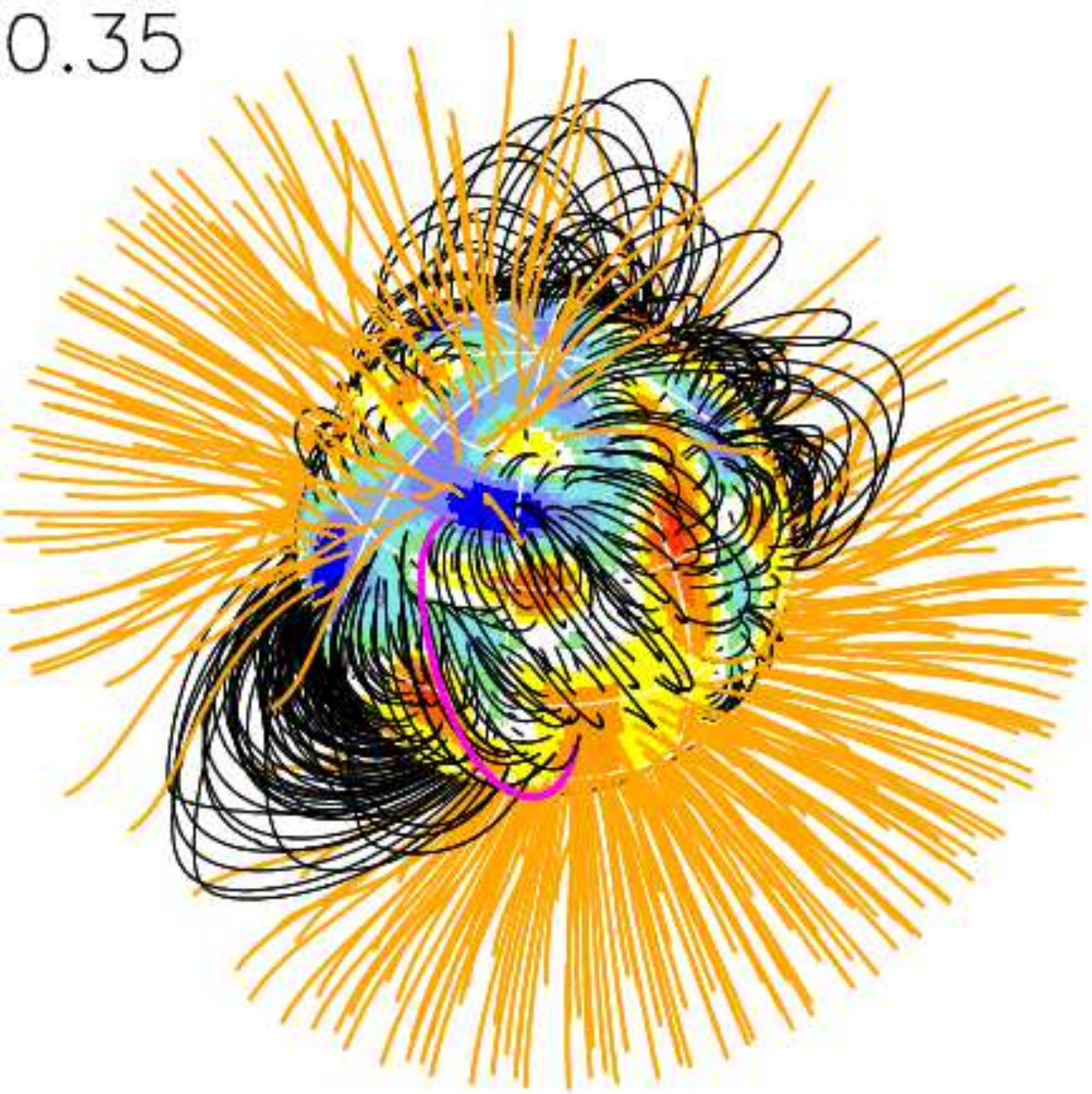}}
			\subfloat[Phase 0.55]{\includegraphics[width=0.2\linewidth]{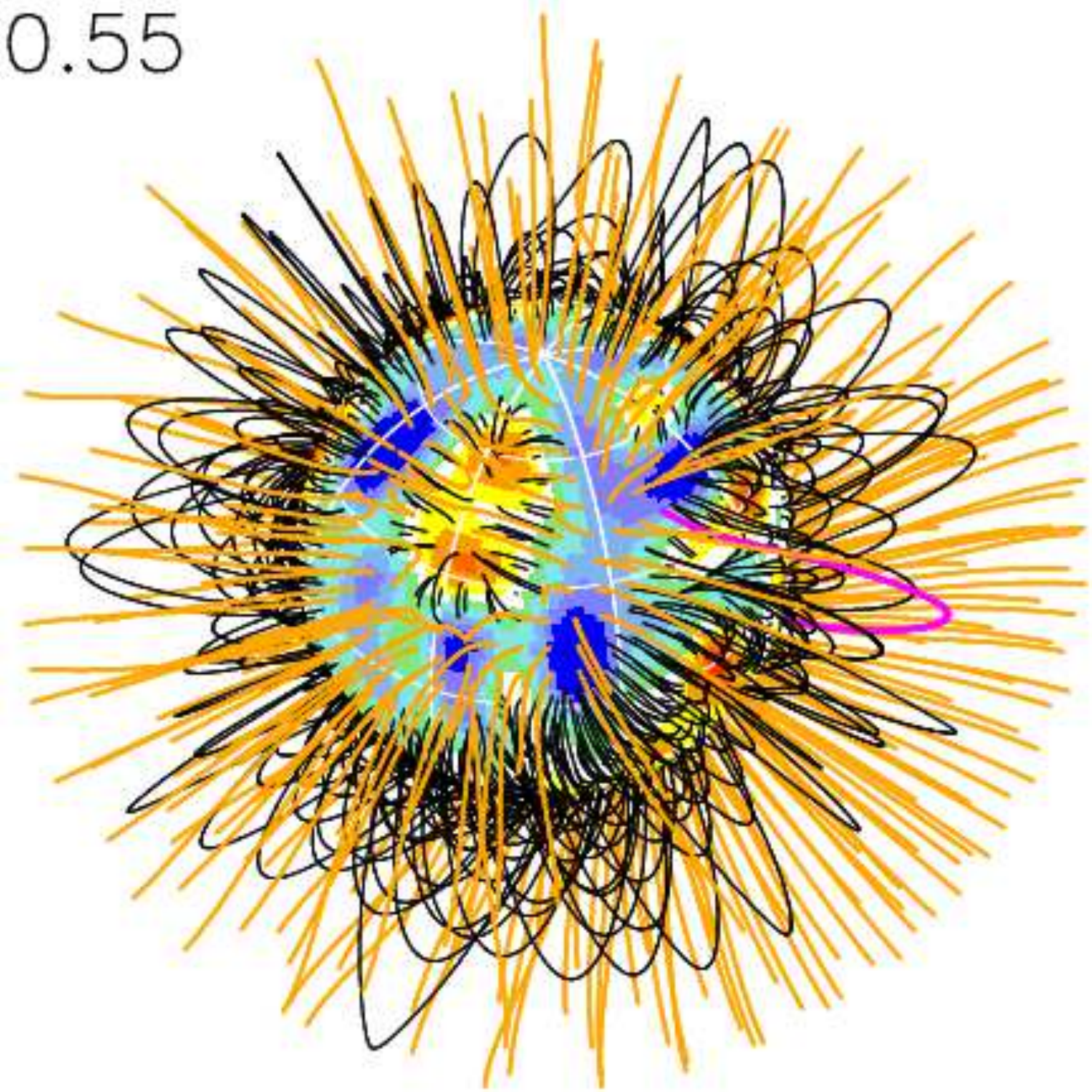}}
			\subfloat[Phase 0.75]{\includegraphics[width=0.2\linewidth]{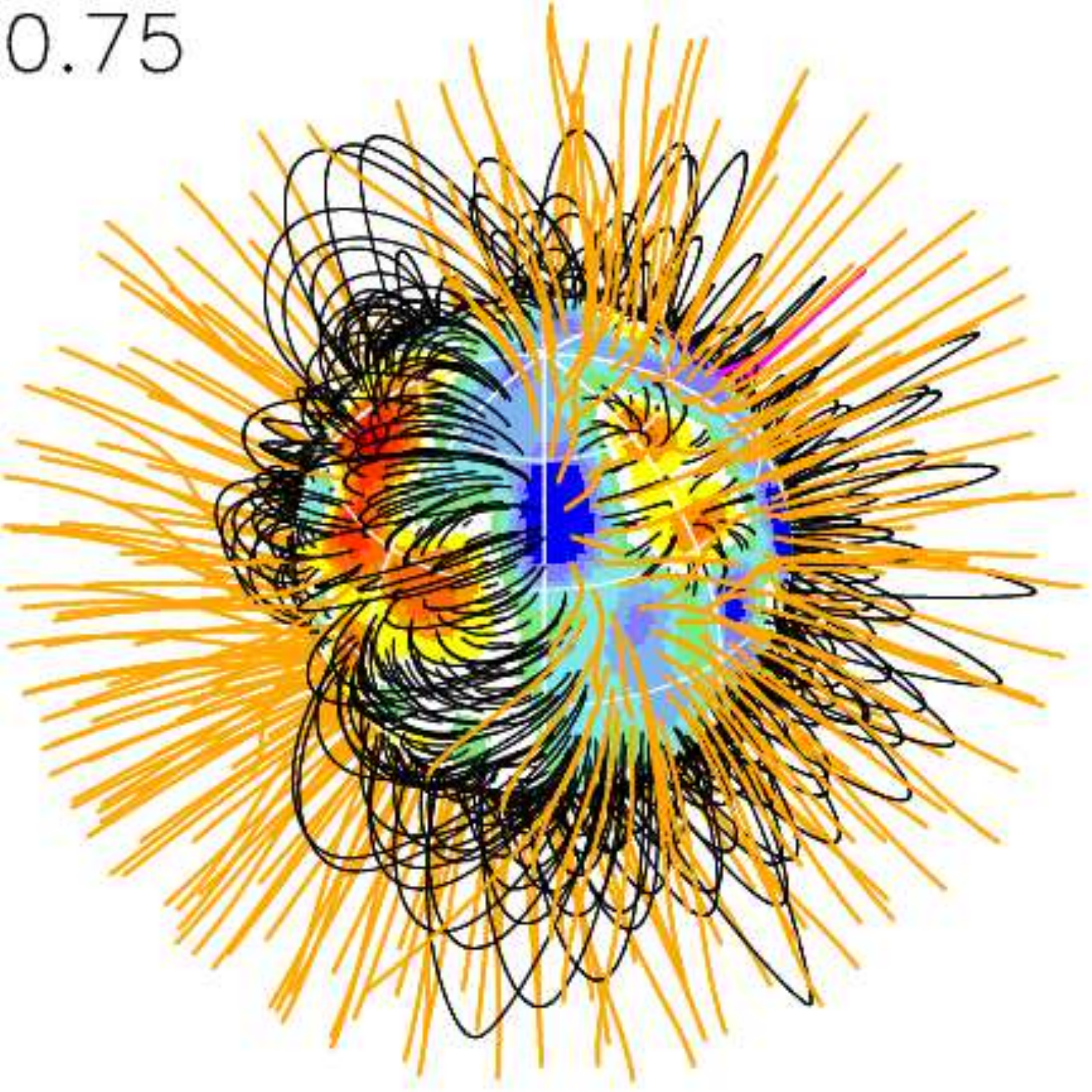}}
			\subfloat[Phase 0.95]{\includegraphics[width=0.2\linewidth]{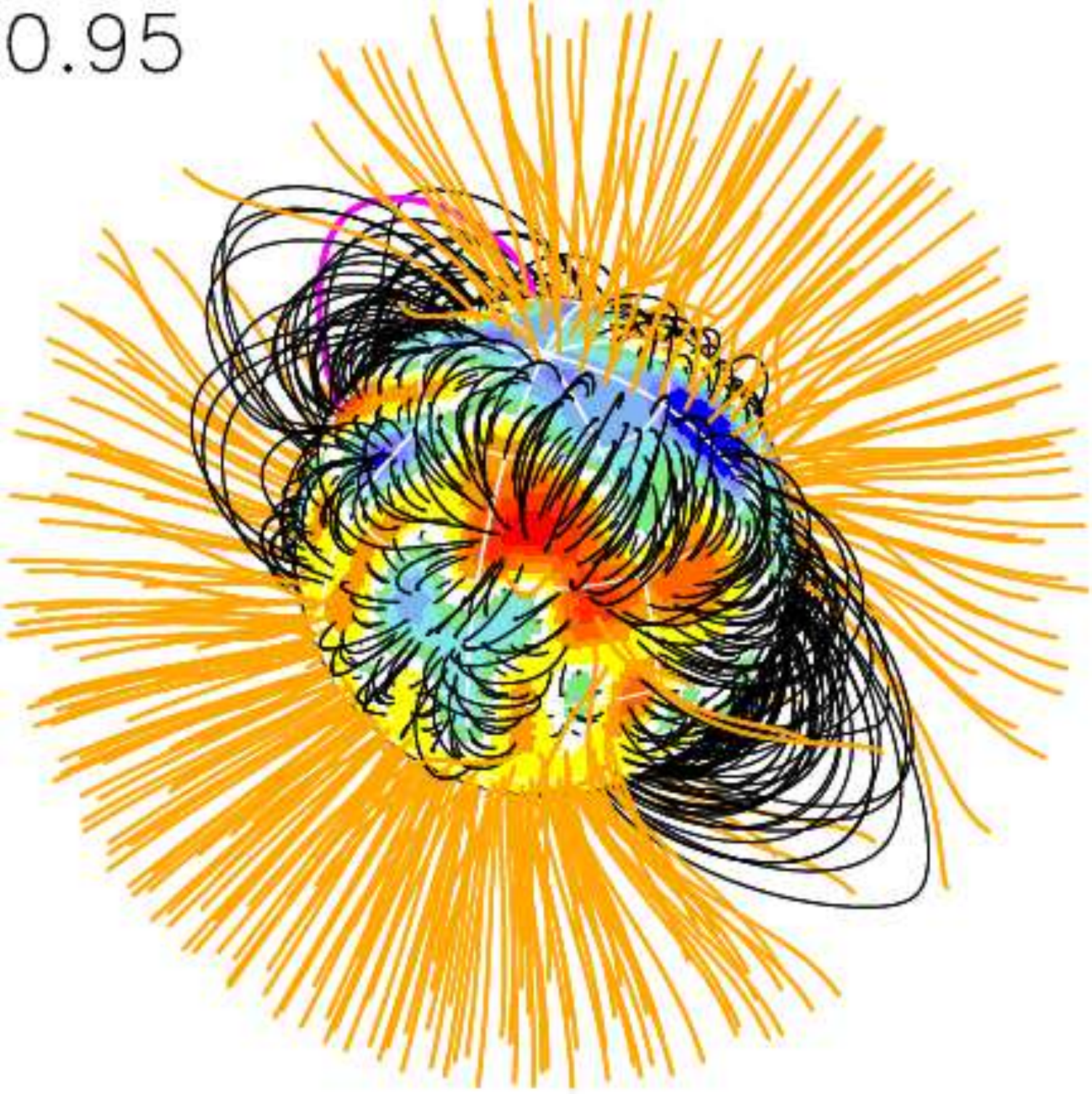}}
			\subfloat[Phase 0.15]{\includegraphics[width=0.2\linewidth]{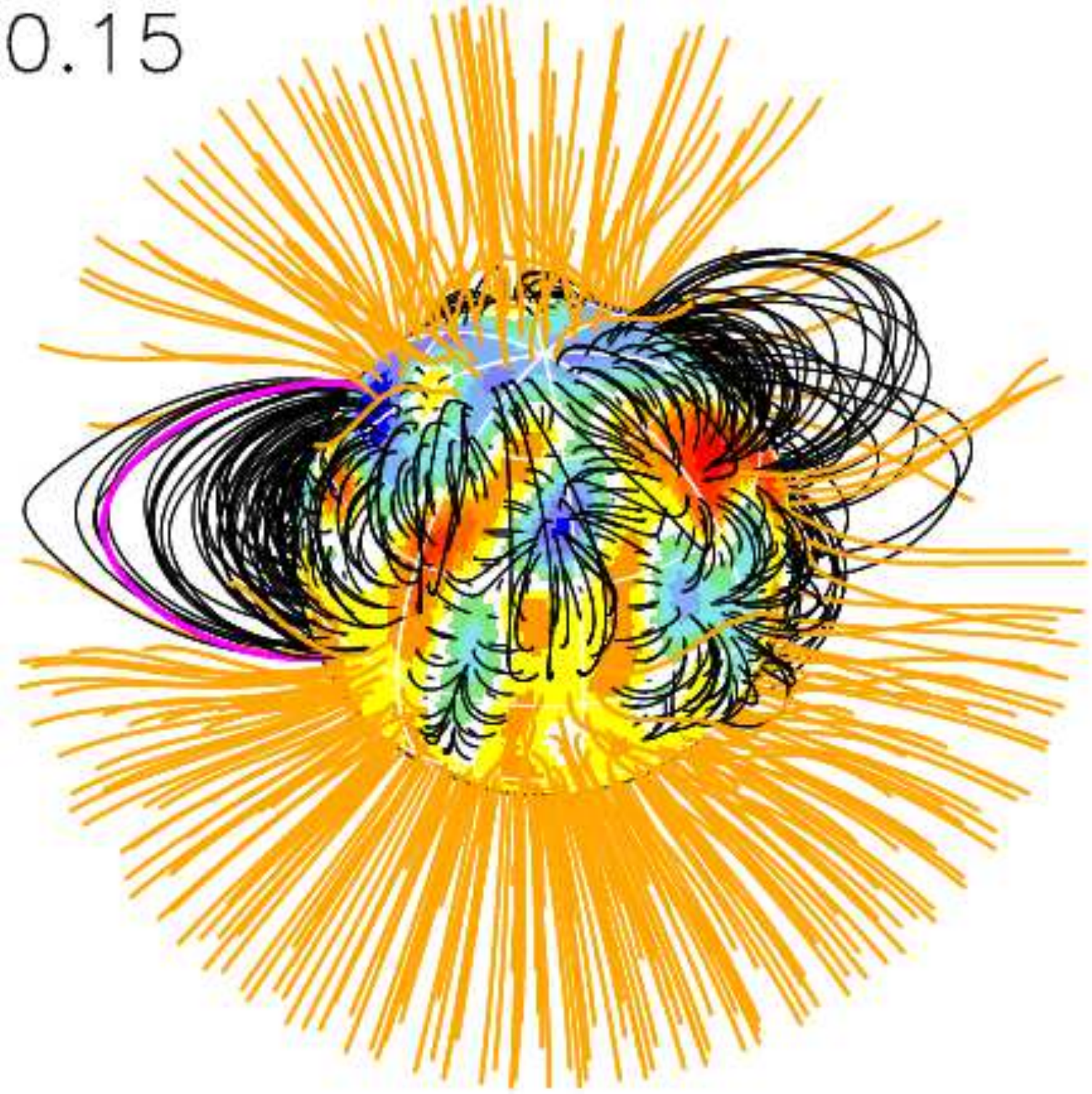}}
			\caption{Potential field extrapolations of the ZDI-reconstructed surface radial field, as seen by an Earth-based observer, for observation epochs 2009 Jan (top) and 2011 Jan (bottom) at different phases. Open/closed field lines are shown in orange/black respectively, and colours at the stellar surface depict the local value of the radial field (in G, as shown in the left-hand panels of Fig.{~}\ref{fig:qbm}). The source surface at which field lines open is set to 2.18{~}\rstar. The field lines that would carry the potential observed prominences (phase 0.95 and 0.8 in 2009, phase 0.35 in 2011) are colored in magenta. Animated versions with the star rotating are available at \url{http://userpages.irap.omp.eu/~lyu/jan09a.gif} and \url{http://userpages.irap.omp.eu/~lyu/jan11n.gif}.}
			\label{fig:pro}
		\end{figure*}

		In our \hal\ dynamic spectra, we observe a conspicuous absorption feature in the second part of the 2009 Jan run around phase 0.95 (Fig.{~}\ref{fig:hal}), that could be the signature of a prominence \citep[see e.g.][]{Cameron92}. Fitting a sine curve in the absorption feature yields an amplitude of $\sim$2{~}\vsini, corresponding to a prominence $\sim$2{~}\rstar\ away from the center of \vt, confirming that the prominence is located close to the corotation radius. Plotting the 3D potential field extrapolation of the reconstructed surface radial field for 2009 Jan, at phases 0.95, 0.20, 0.45 and 0.70, we observe the presence of closed field lines reaching $\sim$2{~}\rstar\ at phase 0.95 (Fig.{~}\ref{fig:pro}), which may be able to support the observed prominence. We also observe similar absorption features in 2009 Jan around phase 0.8 and in 2011 Jan around phase 0.35, but they are less well-covered by our observations. We however found corresponding field lines at the right phase for each (see Fig.{~}\ref{fig:pro} for 2009 Jan).

		We also constrained the differential rotation of \vt\ with ZDI: we obtained six values for the equatorial rotation rate \omeq\ and for the pole-to-equator rotation rate difference \dom, by using separately our \sti\ and \stv\ LSD profiles from each of the three data sets 2008b+2009a, 2013b and 2015b+2016a. Overall mean values are \omeq=${3.35957\pm 0.00022}${~}\rpd\ and \dom=${0.0097\pm 0.0011}${~}\rpd. The differential rotation of \vt\ is thus relatively weak, with a pole-to-equator rotation rate difference 5.6 times smaller than that of the Sun, and a lap time of ${648\pm 73}${~}d. Compared to other wTTSs previously analyzed within the MaTYSSE programme, the differential rotation of \vt\ is similar to that of \vtt\ \citep{Donati17} but much smaller than that of \tap, which is almost of solar level, consistent with the fact that TAP 26 is no longer fully convective and has developped a radiative core \citep[of size 0.6 \rstar, ][]{Yu17}.

		\subsection{Mid-term variability of \vt}
		Even with differential rotation, it is impossible for our current version of ZDI to model data sets spanning a few months down to noise level, which shows that the surface of \vt\ undergoes significant instrinsic variability, corroborating the hypothesis of a dynamo-generated field. The variations of the photosphere and of the surface magnetic field over the years might be the manifestation of a magnetic cycle, whose existence has been suggested by previous studies \citep[][]{Stelzer03, Hambalek19}. No clear change in \dom\ is observed while the dipole grows in intensity (Table{~}\ref{tab:dir}), which could indicate a time lag in the dynamo interaction between the magnetic field and the rotation profile.

\begin{figure}
			\centering
			\includegraphics[width=0.8\linewidth]{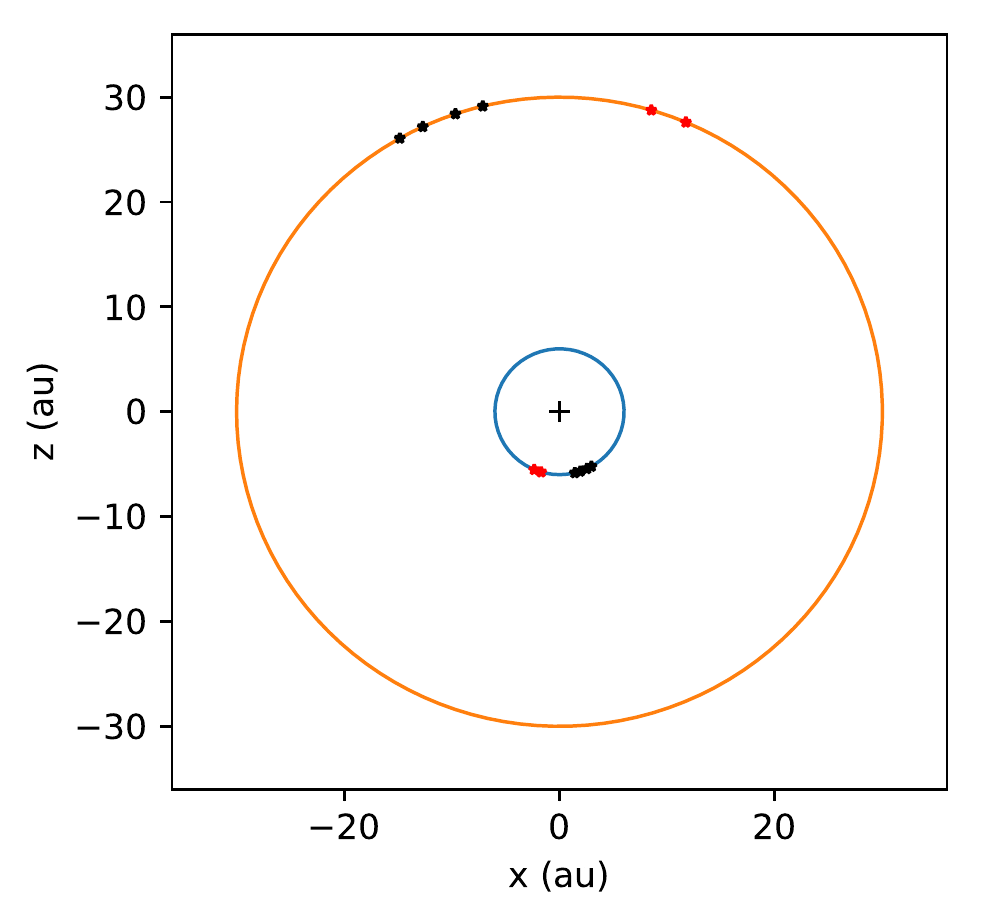}
			\includegraphics[width=\linewidth]{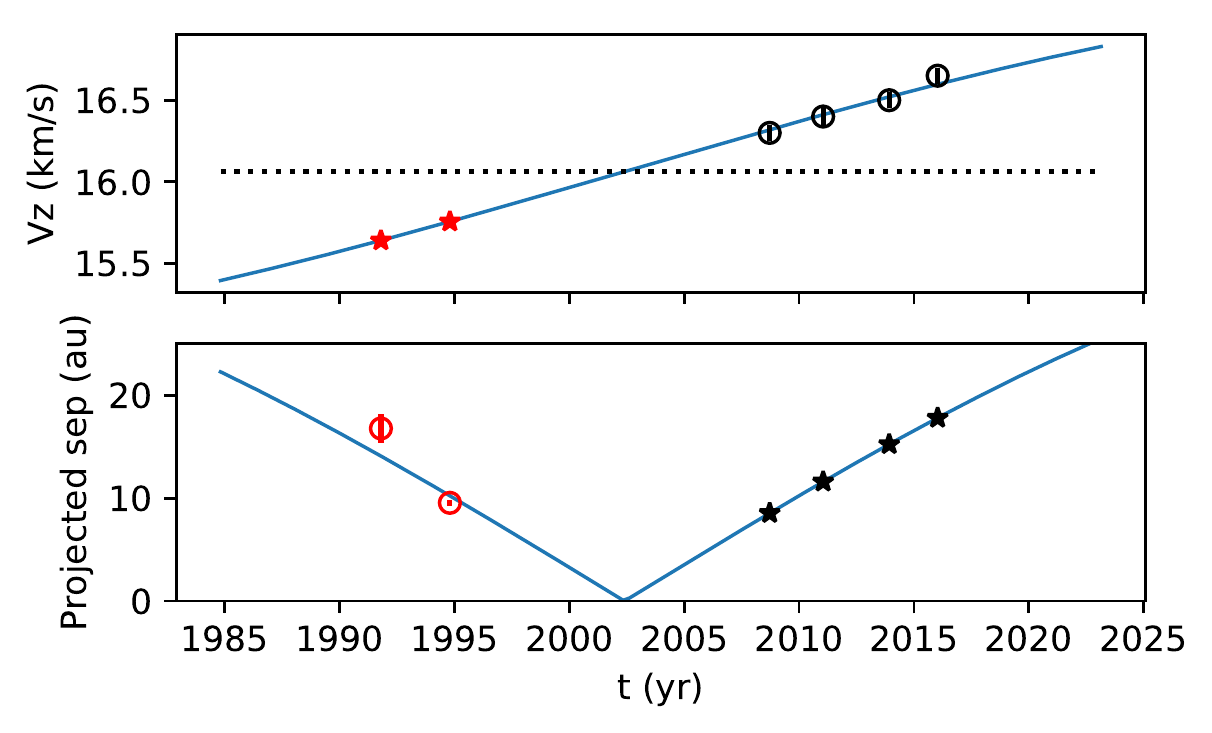}
			\caption{Circular model for the binary motion of \vt{~}A and \vt{~}B: edge-on orbit, separation 36.0{~}au, period 166{~}a and systemic radial velocity 16.06{~}\kms. \textit{Top}: top-view of the model orbit, with the z-axis parallel to the line-of-sight, where the positions of \vt{~}A and B according to the model are marked by red and black stars at the times of the separation measurements and of our spectropolarimetric seasons (2008b-2009a, 2011a, 2013b and 2015b-2016a) respectively. \textit{Middle}: RV$_{\rm bulk}$ of \vt{~}A with time, as measured by us in black dots with $1\sigma$ error bars and as derived from the model orbit in blue. The predicted RV$_{\rm bulk}$ at the times of the separation measurements are represented by red stars. \textit{Bottom}: Sky-projected binary separation as a function of time, as measured by \citealt{Ghez95} in red dots with $1\sigma$ error bars, and as derived from the model orbit in blue. The predicted sky-projected separations at the dates of our observing seasons are marked in black stars.}
			\label{fig:rvb}
		\end{figure}

		The bulk RV of \vt\ exhibits a drift throughout our 8-year campaign, from $16.30\pm 0.05${~}\kms\ in 2008b-2009a to $16.65\pm 0.05${~}\kms\ in 2015b-2016a. One explanation could be a variation in the suppression of convective blueshift in regions of strong magnetic field \citep{Haywood16,Meunier10}, which could further support a secular evolution of the magnetic topology. It could also be a manifestation of the binary motion of \vt~A-B. The central binary of \vt\ was observed twice, with a sky-projected separation of $16.8\pm 1.4${~}au in 1991 Oct and $9.5\pm 0.3${~}au in 1994 Oct \citep[$0.13\pm 0.01${~}arcsec and $0.074\pm 0.002${~}arcsec resp. in][]{Ghez95}, and a mass ratio of $0.20\pm 0.10$ \citep[][]{Kraus11}. Assuming a mass ratio of 0.2 and an edge-on circular orbit, we find that an orbit of the primary star of radius 6.0{~}au, i.e.\ binary separation 36.0{~}au and period 166{~}a, fits our bulk RVs and the sky-projected separations at a level of $2\sigma$ (see Fig.{~}\ref{fig:rvb}). No binary motion was detected in the 2013 to 2017 astrometry measurements of \cite{Galli18}, which is consistent with our model where the sky-projected velocity varies by only 0.13{~}${{\rm mas}\, a^{-1}}$ over these 3.5 years (roughly a 50th of the orbital period). More measurements would enable to estimate the eccentricity and potentially fit the sky-projected separations to a better level, as well as to decide whether the binary motion can explain the RV drift observed in this study.

		The rotation period derived from our V magnitude measurements, in each observing season, also displays long-term variations. Placing the periods found from the photometric data on a period-latitude diagram representing the modeled differential rotation (Fig.{~}\ref{fig:drp}), we observe that the latitudes corresponding to the successive periods tend to increase from 0 in 2008 to $\sim$50\degr\ in 2016. We note that this trend is observed with both the periods derived from sine fits to the photometric data and those derived from GPR (see \ref{anx:pha}). This implies that the largest features, ie those with the biggest impact on the photometric curve, underwent a poleward \textbf{migration}, reminiscent of the Solar butterfly diagram (albeit reversed). This would suggest that the dynamo wave, if cyclic, has a period of at least 8{~}a and likely much longer (16{~}a if our data covers only one half of a full cycle). Previous studies using different data have suggested the existence of an activity cycle on \vt, with periods of 5.4{~}a and 15{~}a respectively \citep[][]{Stelzer03, Hambalek19}. We further note that our differential rotation measurements confirm that the barycenter of surface features migrates to higher latitudes over time (see Fig{~}\ref{fig:dr3}).

		Applying GPR with MCMC parameter exploration to our \hal\ equivalent widths and longitudinal magnitude field measurements \citep[\Bl, first-order moment of the \stv\ LSD profiles, ][]{Donati97c}, we also found rotation periods from which we derive mean barycentric latitudes of features constraining the modeling of each quantity (see Fig.{~}\ref{fig:dhr}). The period found from \hal\ is equal within error bars to the one derived in \citealt{Stelzer03} from photometry, whereas the period found from \Bl\ seems tied to equatorial features. It is worth mentioning that we also find long decay times for these two activity proxies: ${589 ^{+774}_{-335}}${~}d and ${604 ^{+553}_{-289}}${~}d respectively, which suggests, with the caution needed with such high error bars, that the \hal\ and \Bl\ modulations are particularly sensitive to large, long-lasting features. The phase plots are displayed in Appendix \ref{anx:act}.

		\begin{table}
			\centering
			\caption{Various evolution time scales.}
			\label{tab:tsc}
			\begin{tabular}{l|c}
				\hline
				Quantity & Time scale (d) \\
				\hline
				RV decay time & $160 ^{+45}_{-35}$ \\
				& \\
				V mag decay time & $314 ^{+31}_{-29}$ \\
				& \\
				\hal\ decay time & $589 ^{+774}_{-335}$ \\
				& \\
				\Bl\ decay time & $604 ^{+553}_{-289}$ \\
				& \\
				Differential rotation lap time & $648\pm 73$ \\
				\hline
			\end{tabular}
		\end{table}

		\subsection{Radial velocity modulations}
		We modeled the activity RV jitter from line profiles synthetized from our ZDI maps, and filtered it out from the RV curve of \vt. From a rms of 1.802{~}\kms\ in the raw RVs, we get residuals with a rms of 0.167{~}\kms. We also applied GPR to our raw RVs and found a jitter of periodicity ${1.87029\pm 0.00037}${~}d and decay time ${160^{+45}_{-35}}${~}d, with residuals of rms 0.076{~}\kms. The period derived from the GPR on our raw RVs is shorter than the period we used to phase our data, and corresponds to a latitude of 5.5\degr. This period is much closer to the period derived with GPR from \Bl\ than to the period derived from \hal, showing that in this case, \Bl\ is a better activity proxy than \hal\ \citep[for a more systematic study of the correlation of \Bl\ with stellar activity, see][]{Hebrard16}. The decay time associated to RVs is much shorter than the differential rotation lap time and the decay times of the V magnitude, \hal\ and \Bl\ (see Table{~}\ref{tab:tsc}), which suggests that RVs are more sensitive to small-scale short-lived features while the photometry, \hal\ and \Bl\ are more sensitive to large-scale long-lasting features.

\begin{figure}
			\includegraphics[width=\linewidth]{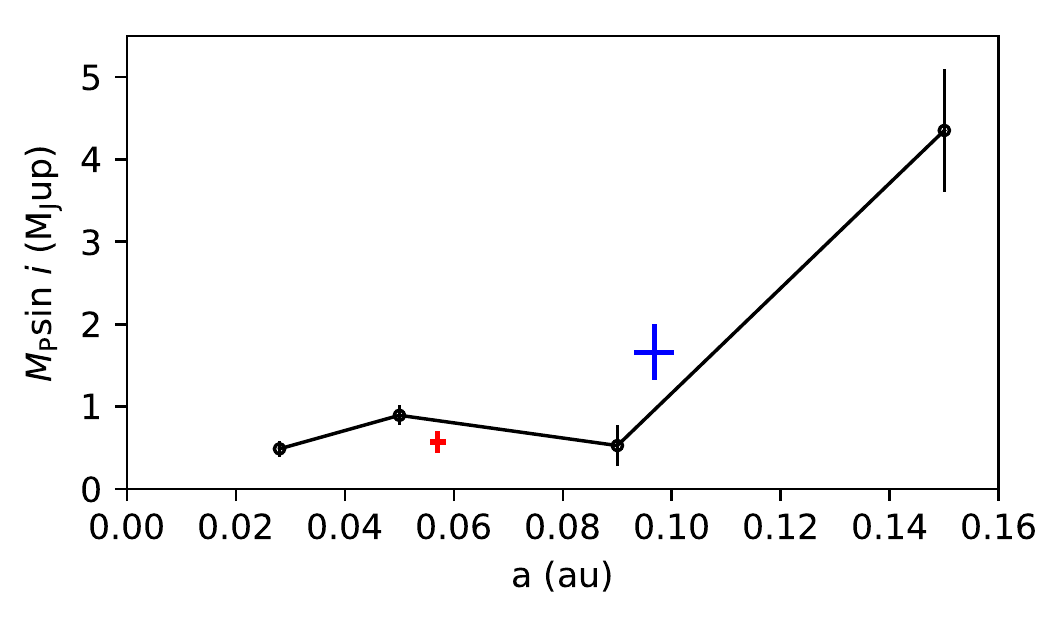}
			\caption{Detectability threshold in terms of \msini\ for planets at various $a$, with the RV filtering technique involving GPR. \vtt{~}b is plotted in red \citep[parameters from][]{Donati17} and TAP{~}26{~}b in blue \citep[parameters from][]{Yu17}.}
			\label{fig:det}
		\end{figure}

		Through both processes, the residual RVs present no significant periodicity which would betray the presence of a potential planet. To estimate the planet mass detection threshold, GPR-MCMC was run on simulated data sets, composed of a base activity jitter (our GP model from Section{~}\ref{sec:rv}), and a circular planet signature, plus a white noise of level 0.081{~}\kms. Various planet separations and masses were tested, and for each case, GPR-MCMC was run several times with different randomization seeds, to mitigate statistical bias. For every randomization seed, GPR-MCMC was run with a model including a planet and a model including no planet, and the difference of logarithmic marginal likelihood between them (hereafter $\Delta\mathcal{L}$) was computed. Finally, the detection threshold was set at $\Delta\mathcal{L}=10$ and the minimum detectable mass at each separation was interpolated from the mass/$\Delta\mathcal{L}$ curve. Fig.{~}\ref{fig:det} shows the planet mass detection threshold as a function of planet-star separation: we thus obtained a detectability threshold of $\sim$1{~}\mjup\ for ${a < 0.09}${~}au and $\sim$4.6{~}\mjup\ for ${a = 0.15}${~}au. The figure also shows the parameters of \vtt{~}b and TAP{~}26{~}b, showing that we would likely have detected a planet like TAP{~}26{~}b but not one like \vtt{~}b. Planets beyond $a=0.15${~}au are difficult to detect due to the temporal coverage of our data, that never exceeds 19{~}d at any given epoch. The early depletion of the disc may have prevented the formation and/or the migration of giant exoplanets. \citealt{Kraus16} outlines a correlation between the presence of a close companion and a lack of planets, in a sample of binary stars with mass ratios $q > 0.4$, which could support the hypothesis that \vt{~}B, although having a slightly lower mass ratio \citep[$q=0.2\pm 0.1$, ][]{Kraus11}, played a role in the early disc dissipation, which in turn prevented the formation of a hot Jupiter.

		In terms of methodology, GPR fits the data down to a significantly lower \chisqr\ than ZDI because it is capable of accounting for most of the mid-term variability, contrarily to ZDI, which for now only integrates differential rotation and a simplistic description of intrinsic variability. Small strutures evolve on time scales of $\sim$few weeks, so we need to be able to model their temporal evolution in a more elaborate way to be able to match the capability of GPR to fit time-variable RV curves. Self-consistent methods that combine the physical faithfulness of ZDI and the flexibility of GPR will be developped in the near future and applied to more MaTYSSE data, as well as to data from the SPIRou (Spectropolarimetre InfraRouge) Legacy Survey (SLS). Finally, observing \vt\ and other wTTSs with SPIRou will yield spectra in the near infrared, where we expect a smaller jitter than in the optical bandwidth, and will offer an opportunity to benchmark our activity jitter filtering technique performances.

	\section*{Acknowledgements}

	This paper is based on observations obtained at the CFHT, operated by the National Research Council of Canada (CNRC), the Institut National des Sciences de l'Univers (INSU) of the Centre National de la Recherche Scientifique (CNRS) of France and the University of Hawaii, and at the TBL, operated by Observatoire Midi-Pyr\'en\'ees and by INSU / CNRS. We thank the QSO teams of CFHT and TBL for their great work and efforts at collecting the high-quality MaTYSSE data presented here, without which this study would not have been possible. MaTYSSE is an international collaborative research programme involving experts from more than 10 different countries (France, Canada, Brazil, Taiwan, UK, Russia, Chile, USA, Ireland, Switzerland, Portugal, China and Italy).

	JFD also warmly thanks the IDEX initiative at Universit\'e F\'ed\'erale Toulouse Midi-Pyr\'en\'ees (UFTMiP) for funding the STEPS collaboration program between IRAP/OMP and ESO. JFD acknowledges funding from the European Research Council (ERC) under the H2020 research \& innovation programme (grant agreements \#740651 NewWorlds). We acknowledge funding from the LabEx OSUG@2020 that allowed purchasing the ProLine PL230 CCD imaging system installed on the 1.25-m telescope at CrAO.

	Finally, we warmly thank the referee for taking the time to review this research.

	This research has made use of the SIMBAD database, operated at CDS, Strasbourg, France, and of the {\sc Matplotlib} python module \citep[][]{Hunter07}.




	\bibliographystyle{mnras}
	\bibliography{v410tau} 

\begin{thebibliography}{}
\makeatletter
\relax
\def\mn@urlcharsother{\let\do\@makeother \do\$\do\&\do\#\do\^\do\_\do\%\do\~}
\def\mn@doi{\begingroup\mn@urlcharsother \@ifnextchar [ {\mn@doi@}
  {\mn@doi@[]}}
\def\mn@doi@[#1]#2{\def\@tempa{#1}\ifx\@tempa\@empty \href
  {http://dx.doi.org/#2} {doi:#2}\else \href {http://dx.doi.org/#2} {#1}\fi
  \endgroup}
\def\mn@eprint#1#2{\mn@eprint@#1:#2::\@nil}
\def\mn@eprint@arXiv#1{\href {http://arxiv.org/abs/#1} {{\tt arXiv:#1}}}
\def\mn@eprint@dblp#1{\href {http://dblp.uni-trier.de/rec/bibtex/#1.xml}
  {dblp:#1}}
\def\mn@eprint@#1:#2:#3:#4\@nil{\def\@tempa {#1}\def\@tempb {#2}\def\@tempc
  {#3}\ifx \@tempc \@empty \let \@tempc \@tempb \let \@tempb \@tempa \fi \ifx
  \@tempb \@empty \def\@tempb {arXiv}\fi \@ifundefined
  {mn@eprint@\@tempb}{\@tempb:\@tempc}{\expandafter \expandafter \csname
  mn@eprint@\@tempb\endcsname \expandafter{\@tempc}}}

\bibitem[\protect\citeauthoryear{{Baraffe}, {Homeier}, {Allard}  \&
  {Chabrier}}{{Baraffe} et~al.}{2015}]{Baraffe15}
{Baraffe} I.,  {Homeier} D.,  {Allard} F.,   {Chabrier} G.,  2015, \mn@doi
  [\aap] {10.1051/0004-6361/201425481}, \href
  {http://adsabs.harvard.edu/abs/2015A%26A...577A..42B} {577, A42}

\bibitem[\protect\citeauthoryear{{Bouvier}}{{Bouvier}}{2007}]{Bouvier07c}
{Bouvier} J.,  2007, in {Bouvier} J.,  {Appenzeller} I.,  eds,  IAU Symposium
  Vol. 243, IAU Symposium. pp 231--240, \mn@doi{10.1017/S1743921307009593}

\bibitem[\protect\citeauthoryear{{Bouvier} \& {Bertout}}{{Bouvier} \&
  {Bertout}}{1989}]{Bouvier89}
{Bouvier} J.,  {Bertout} C.,  1989, \aap, \href
  {http://adsabs.harvard.edu/abs/1989A%26A...211...99B} {211, 99}

\bibitem[\protect\citeauthoryear{{Bouvier}, {Alencar}, {Harries}, {Johns-Krull}
   \& {Romanova}}{{Bouvier} et~al.}{2007}]{Bouvier07}
{Bouvier} J.,  {Alencar} S.~H.~P.,  {Harries} T.~J.,  {Johns-Krull} C.~M.,
  {Romanova} M.~M.,  2007, in {Reipurth} B.,  {Jewitt} D.,   {Keil} K.,  eds,
  Protostars and Planets V. pp 479--494

\bibitem[\protect\citeauthoryear{{Brown}, {Donati}, {Rees}  \& {Semel}}{{Brown}
  et~al.}{1991}]{Brown91}
{Brown} S.~F.,  {Donati} J.-F.,  {Rees} D.~E.,   {Semel} M.,  1991, \aap, \href
  {http://adsabs.harvard.edu/abs/1991A%26A...250..463B} {250, 463}

\bibitem[\protect\citeauthoryear{{Carroll}, {Strassmeier}, {Rice}  \&
  {K{\"u}nstler}}{{Carroll} et~al.}{2012}]{Carroll12}
{Carroll} T.~A.,  {Strassmeier} K.~G.,  {Rice} J.~B.,   {K{\"u}nstler} A.,
  2012, \mn@doi [\aap] {10.1051/0004-6361/201220215}, \href
  {http://adsabs.harvard.edu/abs/2012A%26A...548A..95C} {548, A95}

\bibitem[\protect\citeauthoryear{{Collier Cameron} \& {Woods}}{{Collier
  Cameron} \& {Woods}}{1992}]{Cameron92}
{Collier Cameron} A.,  {Woods} J.~A.,  1992, \mn@doi [\mnras]
  {10.1093/mnras/258.2.360}, \href
  {http://adsabs.harvard.edu/abs/1992MNRAS.258..360C} {258, 360}

\bibitem[\protect\citeauthoryear{{Donati} \& {Brown}}{{Donati} \&
  {Brown}}{1997}]{Donati97c}
{Donati} J.-F.,  {Brown} S.~F.,  1997, \aap, \href
  {http://adsabs.harvard.edu/abs/1997A%26A...326.1135D} {326, 1135}

\bibitem[\protect\citeauthoryear{{Donati} \& {Landstreet}}{{Donati} \&
  {Landstreet}}{2009}]{Donati09}
{Donati} J.,  {Landstreet} J.~D.,  2009, \mn@doi [\araa]
  {10.1146/annurev-astro-082708-101833}, \href
  {http://adsabs.harvard.edu/abs/2009ARA%26A..47..333D} {47, 333}

\bibitem[\protect\citeauthoryear{{Donati}, {Semel}, {Carter}, {Rees}  \&
  {Collier Cameron}}{{Donati} et~al.}{1997}]{Donati97b}
{Donati} J.-F.,  {Semel} M.,  {Carter} B.~D.,  {Rees} D.~E.,   {Collier
  Cameron} A.,  1997, \mnras, \href
  {http://adsabs.harvard.edu/abs/1997MNRAS.291..658D} {291, 658}

\bibitem[\protect\citeauthoryear{{Donati}, {Mengel}, {Carter}, {Marsden},
  {Collier Cameron}  \& {Wichmann}}{{Donati} et~al.}{2000}]{Donati00}
{Donati} J.-F.,  {Mengel} M.,  {Carter} B.~D.,  {Marsden} S.,  {Collier
  Cameron} A.,   {Wichmann} R.,  2000, \mnras, \href
  {http://adsabs.harvard.edu/abs/2000MNRAS.316..699D} {316, 699}

\bibitem[\protect\citeauthoryear{{Donati}, {Collier Cameron}  \&
  {Petit}}{{Donati} et~al.}{2003}]{Donati03b}
{Donati} J.-F.,  {Collier Cameron} A.,   {Petit} P.,  2003, \mnras, 345, 1187

\bibitem[\protect\citeauthoryear{{Donati} et~al.,}{{Donati}
  et~al.}{2006}]{Donati06b}
{Donati} J.-F.,  et~al., 2006, \mn@doi [\mnras]
  {10.1111/j.1365-2966.2006.10558.x}, \href
  {http://adsabs.harvard.edu/abs/2006MNRAS.370..629D} {370, 629}

\bibitem[\protect\citeauthoryear{{Donati} et~al.,}{{Donati}
  et~al.}{2007}]{Donati07}
{Donati} J.-F.,  et~al., 2007, \mn@doi [\mnras]
  {10.1111/j.1365-2966.2007.12194.x}, \href
  {http://adsabs.harvard.edu/abs/2007MNRAS.380.1297D} {380, 1297}

\bibitem[\protect\citeauthoryear{{Donati} et~al.,}{{Donati}
  et~al.}{2008}]{Donati08}
{Donati} J.-F.,  et~al., 2008, \mn@doi [\mnras]
  {10.1111/j.1365-2966.2008.13111.x}, \href
  {http://adsabs.harvard.edu/abs/2008MNRAS.386.1234D} {386, 1234}

\bibitem[\protect\citeauthoryear{{Donati} et~al.,}{{Donati}
  et~al.}{2010a}]{Donati10}
{Donati} J.,  et~al., 2010a, \mn@doi [\mnras]
  {10.1111/j.1365-2966.2009.15998.x}, \href
  {http://adsabs.harvard.edu/abs/2010MNRAS.402.1426D} {402, 1426}

\bibitem[\protect\citeauthoryear{{Donati} et~al.,}{{Donati}
  et~al.}{2010b}]{Donati10b}
{Donati} J.,  et~al., 2010b, \mn@doi [\mnras]
  {10.1111/j.1365-2966.2010.17409.x}, \href
  {http://adsabs.harvard.edu/abs/2010MNRAS.409.1347D} {409, 1347}

\bibitem[\protect\citeauthoryear{{Donati} et~al.,}{{Donati}
  et~al.}{2011}]{Donati11}
{Donati} J.,  et~al., 2011, \mn@doi [\mnras]
  {10.1111/j.1365-2966.2010.18069.x}, \href
  {http://adsabs.harvard.edu/abs/2011MNRAS.412.2454D} {412, 2454}

\bibitem[\protect\citeauthoryear{{Donati} et~al.,}{{Donati}
  et~al.}{2012}]{Donati12}
{Donati} J.-F.,  et~al., 2012, \mn@doi [\mnras]
  {10.1111/j.1365-2966.2012.21482.x}, \href
  {http://adsabs.harvard.edu/abs/2012MNRAS.425.2948D} {425, 2948}

\bibitem[\protect\citeauthoryear{{Donati} et~al.,}{{Donati}
  et~al.}{2013}]{Donati13}
{Donati} J.-F.,  et~al., 2013, \mn@doi [\mnras] {10.1093/mnras/stt1622}, \href
  {http://adsabs.harvard.edu/abs/2013MNRAS.436..881D} {436, 881}

\bibitem[\protect\citeauthoryear{{Donati} et~al.,}{{Donati}
  et~al.}{2014}]{Donati14}
{Donati} J.-F.,  et~al., 2014, \mn@doi [\mnras] {10.1093/mnras/stu1679}, \href
  {http://adsabs.harvard.edu/abs/2014MNRAS.444.3220D} {444, 3220}

\bibitem[\protect\citeauthoryear{{Donati} et~al.,}{{Donati}
  et~al.}{2015}]{Donati15}
{Donati} J.-F.,  et~al., 2015, \mn@doi [\mnras] {10.1093/mnras/stv1837}, \href
  {http://adsabs.harvard.edu/abs/2015MNRAS.453.3706D} {453, 3706}

\bibitem[\protect\citeauthoryear{{Donati} et~al.,}{{Donati}
  et~al.}{2016}]{Donati16}
{Donati} J.~F.,  et~al., 2016, \mn@doi [\nat] {10.1038/nature18305}, \href
  {http://adsabs.harvard.edu/abs/2016Natur.534..662D} {534, 662}

\bibitem[\protect\citeauthoryear{{Donati} et~al.,}{{Donati}
  et~al.}{2017}]{Donati17}
{Donati} J.-F.,  et~al., 2017, \mn@doi [\mnras] {10.1093/mnras/stw2904}, \href
  {http://adsabs.harvard.edu/abs/2017MNRAS.465.3343D} {465, 3343}

\bibitem[\protect\citeauthoryear{{Galli} et~al.,}{{Galli}
  et~al.}{2018}]{Galli18}
{Galli} P.~A.~B.,  et~al., 2018, \mn@doi [\apj] {10.3847/1538-4357/aabf91},
  \href {http://adsabs.harvard.edu/abs/2018ApJ...859...33G} {859, 33}

\bibitem[\protect\citeauthoryear{{Ghez}, {Weinberger}, {Neugebauer}, {Matthews}
   \& {McCarthy}}{{Ghez} et~al.}{1995}]{Ghez95}
{Ghez} A.~M.,  {Weinberger} A.~J.,  {Neugebauer} G.,  {Matthews} K.,
  {McCarthy} Jr. D.~W.,  1995, \mn@doi [\aj] {10.1086/117560}, \href
  {http://adsabs.harvard.edu/abs/1995AJ....110..753G} {110, 753}

\bibitem[\protect\citeauthoryear{{Ghez}, {White}  \& {Simon}}{{Ghez}
  et~al.}{1997}]{Ghez97}
{Ghez} A.~M.,  {White} R.~J.,   {Simon} M.,  1997, \mn@doi [\apj]
  {10.1086/304856}, \href {http://adsabs.harvard.edu/abs/1997ApJ...490..353G}
  {490, 353}

\bibitem[\protect\citeauthoryear{{Grankin}, {Bouvier}, {Herbst}  \&
  {Melnikov}}{{Grankin} et~al.}{2008}]{Grankin08}
{Grankin} K.~N.,  {Bouvier} J.,  {Herbst} W.,   {Melnikov} S.~Y.,  2008,
  \mn@doi [\aap] {10.1051/0004-6361:20078476}, \href
  {http://adsabs.harvard.edu/abs/2008A%26A...479..827G} {479, 827}

\bibitem[\protect\citeauthoryear{{Gregory}, {Donati}, {Morin}, {Hussain},
  {Mayne}, {Hillenbrand}  \& {Jardine}}{{Gregory} et~al.}{2012}]{Gregory12}
{Gregory} S.~G.,  {Donati} J.-F.,  {Morin} J.,  {Hussain} G.~A.~J.,  {Mayne}
  N.~J.,  {Hillenbrand} L.~A.,   {Jardine} M.,  2012, \mn@doi [\apj]
  {10.1088/0004-637X/755/2/97}, \href
  {http://adsabs.harvard.edu/abs/2012ApJ...755...97G} {755, 97}

\bibitem[\protect\citeauthoryear{{Gully-Santiago} et~al.,}{{Gully-Santiago}
  et~al.}{2017}]{Gully-Santiago17}
{Gully-Santiago} M.~A.,  et~al., 2017, \mn@doi [\apj]
  {10.3847/1538-4357/836/2/200}, \href
  {http://adsabs.harvard.edu/abs/2017ApJ...836..200G} {836, 200}

\bibitem[\protect\citeauthoryear{{Hamb{\'a}lek}, {Va{\r{A}}ko}, {Paunzen}  \&
  {Smalley}}{{Hamb{\'a}lek} et~al.}{2019}]{Hambalek19}
{Hamb{\'a}lek} {\"A}.,  {Va{\r{A}}ko} M.,  {Paunzen} E.,   {Smalley} B.,  2019,
  \mn@doi [\mnras] {10.1093/mnras/sty3151}, \href
  {https://ui.adsabs.harvard.edu/abs/2019MNRAS.483.1642H} {483, 1642}

\bibitem[\protect\citeauthoryear{{Haywood} et~al.,}{{Haywood}
  et~al.}{2014}]{Haywood14}
{Haywood} R.~D.,  et~al., 2014, \mn@doi [\mnras] {10.1093/mnras/stu1320}, \href
  {http://adsabs.harvard.edu/abs/2014MNRAS.443.2517H} {443, 2517}

\bibitem[\protect\citeauthoryear{{Haywood} et~al.,}{{Haywood}
  et~al.}{2016}]{Haywood16}
{Haywood} R.~D.,  et~al., 2016, \mn@doi [\mnras] {10.1093/mnras/stw187}, \href
  {http://adsabs.harvard.edu/abs/2016MNRAS.457.3637H} {457, 3637}

\bibitem[\protect\citeauthoryear{{H{\'e}brard}, {Donati}, {Delfosse}, {Morin},
  {Moutou}  \& {Boisse}}{{H{\'e}brard} et~al.}{2016}]{Hebrard16}
{H{\'e}brard} {\'E}.~M.,  {Donati} J.-F.,  {Delfosse} X.,  {Morin} J.,
  {Moutou} C.,   {Boisse} I.,  2016, \mn@doi [\mnras] {10.1093/mnras/stw1346},
  \href {http://adsabs.harvard.edu/abs/2016MNRAS.461.1465H} {461, 1465}

\bibitem[\protect\citeauthoryear{{Hunter}}{{Hunter}}{2007}]{Hunter07}
{Hunter} J.~D.,  2007, \mn@doi [Computing in Science and Engineering]
  {10.1109/MCSE.2007.55}, \href
  {https://ui.adsabs.harvard.edu/abs/2007CSE.....9...90H} {9, 90}

\bibitem[\protect\citeauthoryear{{Hussain} et~al.,}{{Hussain}
  et~al.}{2009}]{Hussain09}
{Hussain} G.~A.~J.,  et~al., 2009, \mn@doi [\mnras]
  {10.1111/j.1365-2966.2009.14881.x}, \href
  {http://adsabs.harvard.edu/abs/2009MNRAS.tmp..997H} {pp 997--+}

\bibitem[\protect\citeauthoryear{{Johns-Krull}, {Valenti}  \&
  {Koresko}}{{Johns-Krull} et~al.}{1999}]{Johns99b}
{Johns-Krull} C.~M.,  {Valenti} J.~A.,   {Koresko} C.,  1999, \mn@doi [\apj]
  {10.1086/307128}, \href {http://adsabs.harvard.edu/abs/1999ApJ...516..900J}
  {516, 900}

\bibitem[\protect\citeauthoryear{{Kraus}, {Ireland}, {Martinache}  \&
  {Hillenbrand}}{{Kraus} et~al.}{2011}]{Kraus11}
{Kraus} A.~L.,  {Ireland} M.~J.,  {Martinache} F.,   {Hillenbrand} L.~A.,
  2011, \mn@doi [\apj] {10.1088/0004-637X/731/1/8}, \href
  {http://adsabs.harvard.edu/abs/2011ApJ...731....8K} {731, 8}

\bibitem[\protect\citeauthoryear{{Kraus}, {Ireland}, {Hillenbrand}  \&
  {Martinache}}{{Kraus} et~al.}{2012}]{Kraus12}
{Kraus} A.~L.,  {Ireland} M.~J.,  {Hillenbrand} L.~A.,   {Martinache} F.,
  2012, \mn@doi [\apj] {10.1088/0004-637X/745/1/19}, \href
  {http://adsabs.harvard.edu/abs/2012ApJ...745...19K} {745, 19}

\bibitem[\protect\citeauthoryear{{Kraus}, {Ireland}, {Huber}, {Mann}  \&
  {Dupuy}}{{Kraus} et~al.}{2016}]{Kraus16}
{Kraus} A.~L.,  {Ireland} M.~J.,  {Huber} D.,  {Mann} A.~W.,   {Dupuy} T.~J.,
  2016, \mn@doi [\aj] {10.3847/0004-6256/152/1/8}, \href
  {http://adsabs.harvard.edu/abs/2016AJ....152....8K} {152, 8}

\bibitem[\protect\citeauthoryear{{Kurucz}}{{Kurucz}}{1993}]{Kurucz93}
{Kurucz} R.,  1993, CDROM \#~13 (ATLAS9 atmospheric models) and \#~18 (ATLAS9
  and SYNTHE routines, spectral line database).
Smithsonian Astrophysical Observatory, Washington D.C.

\bibitem[\protect\citeauthoryear{{Landi degl'Innocenti} \& {Landolfi}}{{Landi
  degl'Innocenti} \& {Landolfi}}{2004}]{Landi04}
{Landi degl'Innocenti} E.,  {Landolfi} M.,  2004, {Polarisation in spectral
  lines}.
Dordrecht/Boston/London: Kluwer Academic Publishers

\bibitem[\protect\citeauthoryear{{Luhman}, {Allen}, {Espaillat}, {Hartmann}  \&
  {Calvet}}{{Luhman} et~al.}{2010}]{Luhman10}
{Luhman} K.~L.,  {Allen} P.~R.,  {Espaillat} C.,  {Hartmann} L.,   {Calvet} N.,
   2010, \mn@doi [\apjs] {10.1088/0067-0049/186/1/111}, \href
  {https://ui.adsabs.harvard.edu/abs/2010ApJS..186..111L} {186, 111}

\bibitem[\protect\citeauthoryear{{Meunier}, {Desort}  \& {Lagrange}}{{Meunier}
  et~al.}{2010}]{Meunier10}
{Meunier} N.,  {Desort} M.,   {Lagrange} A.-M.,  2010, \mn@doi [\aap]
  {10.1051/0004-6361/200913551}, \href
  {http://adsabs.harvard.edu/abs/2010A%26A...512A..39M} {512, A39}

\bibitem[\protect\citeauthoryear{{Morin} et~al.,}{{Morin}
  et~al.}{2008}]{Morin08b}
{Morin} J.,  et~al., 2008, \mn@doi [\mnras] {10.1111/j.1365-2966.2008.13809.x},
  \href {http://adsabs.harvard.edu/abs/2008MNRAS.390..567M} {390, 567}

\bibitem[\protect\citeauthoryear{{Moutou} et~al.,}{{Moutou}
  et~al.}{2007}]{Moutou07}
{Moutou} C.,  et~al., 2007, \mn@doi [\aap] {10.1051/0004-6361:20077795}, \href
  {http://adsabs.harvard.edu/abs/2007A%26A...473..651M} {473, 651}

\bibitem[\protect\citeauthoryear{{Pecaut} \& {Mamajek}}{{Pecaut} \&
  {Mamajek}}{2013}]{Pecaut13}
{Pecaut} M.~J.,  {Mamajek} E.~E.,  2013, \mn@doi [\apjs]
  {10.1088/0067-0049/208/1/9}, \href
  {http://adsabs.harvard.edu/abs/2013ApJS..208....9P} {208, 9}

\bibitem[\protect\citeauthoryear{{Pollacco} et~al.,}{{Pollacco}
  et~al.}{2006}]{Pollacco06}
{Pollacco} D.~L.,  et~al., 2006, \mn@doi [\pasp] {10.1086/508556}, \href
  {https://ui.adsabs.harvard.edu/abs/2006PASP..118.1407P} {118, 1407}

\bibitem[\protect\citeauthoryear{{Rice}, {Strassmeier}  \& {Kopf}}{{Rice}
  et~al.}{2011}]{Rice11}
{Rice} J.~B.,  {Strassmeier} K.~G.,   {Kopf} M.,  2011, \mn@doi [\apj]
  {10.1088/0004-637X/728/1/69}, \href
  {http://adsabs.harvard.edu/abs/2011ApJ...728...69R} {728, 69}

\bibitem[\protect\citeauthoryear{{Siess}, {Dufour}  \& {Forestini}}{{Siess}
  et~al.}{2000}]{Siess00}
{Siess} L.,  {Dufour} E.,   {Forestini} M.,  2000, \aap, 358, 593

\bibitem[\protect\citeauthoryear{{Skelly}, {Donati}, {Bouvier}, {Grankin},
  {Unruh}, {Artemenko}  \& {Petrov}}{{Skelly} et~al.}{2010}]{Skelly10}
{Skelly} M.~B.,  {Donati} J.-F.,  {Bouvier} J.,  {Grankin} K.~N.,  {Unruh}
  Y.~C.,  {Artemenko} S.~A.,   {Petrov} P.,  2010, \mn@doi [\mnras]
  {10.1111/j.1365-2966.2009.16132.x}, \href
  {http://adsabs.harvard.edu/abs/2010MNRAS.403..159S} {403, 159}

\bibitem[\protect\citeauthoryear{{Sokoloff}, {Nefedov}, {Ermash}  \&
  {Lamzin}}{{Sokoloff} et~al.}{2008}]{Sokoloff08}
{Sokoloff} D.~D.,  {Nefedov} S.~N.,  {Ermash} A.~A.,   {Lamzin} S.~A.,  2008,
  \mn@doi [Astronomy Letters] {10.1134/S1063773708110054}, \href
  {http://adsabs.harvard.edu/abs/2008AstL...34..761S} {34, 761}

\bibitem[\protect\citeauthoryear{{Stelzer} et~al.,}{{Stelzer}
  et~al.}{2003}]{Stelzer03}
{Stelzer} B.,  et~al., 2003, \mn@doi [\aap] {10.1051/0004-6361:20031414}, \href
  {http://adsabs.harvard.edu/abs/2003A%26A...411..517S} {411, 517}

\bibitem[\protect\citeauthoryear{{Vaytet}, {Commer{\c c}on}, {Masson},
  {Gonz{\'a}lez}  \& {Chabrier}}{{Vaytet} et~al.}{2018}]{Vaytet18}
{Vaytet} N.,  {Commer{\c c}on} B.,  {Masson} J.,  {Gonz{\'a}lez} M.,
  {Chabrier} G.,  2018, \mn@doi [\aap] {10.1051/0004-6361/201732075}, \href
  {http://adsabs.harvard.edu/abs/2018A%26A...615A...5V} {615, A5}

\bibitem[\protect\citeauthoryear{{Welty} \& {Ramsey}}{{Welty} \&
  {Ramsey}}{1995}]{Welty95}
{Welty} A.~D.,  {Ramsey} L.~W.,  1995, \mn@doi [\aj] {10.1086/117524}, \href
  {https://ui.adsabs.harvard.edu/abs/1995AJ....110..336W} {110, 336}

\bibitem[\protect\citeauthoryear{{Yadav}, {Christensen}, {Morin}, {Gastine},
  {Reiners}, {Poppenhaeger}  \& {Wolk}}{{Yadav} et~al.}{2015}]{Yadav15}
{Yadav} R.~K.,  {Christensen} U.~R.,  {Morin} J.,  {Gastine} T.,  {Reiners} A.,
   {Poppenhaeger} K.,   {Wolk} S.~J.,  2015, \mn@doi [\apjl]
  {10.1088/2041-8205/813/2/L31}, \href
  {https://ui.adsabs.harvard.edu/abs/2015ApJ...813L..31Y} {813, L31}

\bibitem[\protect\citeauthoryear{{Yu} et~al.,}{{Yu} et~al.}{2017}]{Yu17}
{Yu} L.,  et~al., 2017, \mnras, 467, 1342

\makeatother
\end{thebibliography}

	\vspace{7mm}
	{\small \it $^1$ Univ. de Toulouse, CNRS, IRAP, 14 avenue Edouard Belin, 31400 Toulouse, France\\
		$^2$ Crimean Astrophysical Observatory, Nauchny, Crimea 298409\\
		$^3$ SUPA, School of Physics \& Astronomy, Univ. of St Andrews, St Andrews, Scotland KY16 9SS, UK\\
		$^4$ CFHT Corporation, 65-1238 Mamalahoa Hwy, Kamuela, Hawaii 96743, USA\\
		$^5$ ESO, Karl-Schwarzschild-Str 2, D-85748 Garching, Germany}


\newpage
\FloatBarrier
	\appendix

	\section{Observations}
		\label{anx:obs}
		This appendix informs all the observations, both spectropolarimetric (Table{~}\ref{tab:sob}) and photometric (Table{~}\ref{tab:pob}), that we used in this study, excluding the WASP data. The spectropolarimetric data are spread over 8 runs (2008 Oct, 2008 Dec, 2009 Jan, 2011 Jan, 2013 Nov, 2013 Dec, 2015 Dec and 2016 Jan) and the photometric data are spread over 9 seasons: 08b+09a (short for 2008b + 2009a; all the other seasons follow the same naming convention), 09b+10a, 10b, 11b+12a, 12b+13a, 13b+14a, 14b, 15b+16a and 16b+17a.

		The instruments with which the spectropolarimetric data was taken, ESPaDOnS and NARVAL, are twin spectropolarimeters and cover a 370 to 1000{~}nm wavelength domain, with respective resolving powers of 65{~}000 (i.e.\ resolved velocity element of 4.6{~}\kms) and 60{~}000 (resolved velocity element of 5.0{~}\kms). Each polarization exposure sequence consists of four subexposures of 600{~}s each, taken in different polarimeter configurations to allow the removal of all spurious polarization signatures at first order \citep[][]{Donati97b}, except three observations comprised of only two subexposures of 600{~}s (2008 Dec 05 at phase 0.827, 2009 Jan 05 at phase 0.602, and 2013 Nov 07 at phase 0.541), and three observations comprised of four subexposures of 800{~}s (2009 Jan 10 at phases 0.229, 0.251 and 0.272).

		\begin{table*}
			\caption{Information on the \vt\ spectropolarimetric data. The first three columns contain the time at which the observations were taken: Coordinated Universal Time in the 1st column, Barycentric Julian Date in the 2nd and corresponding rotational cycle of \vt, $c$, in the 3rd, as is defined in equation \ref{eq:eph}. The 4th column indicates the instrument used for the observation (E: ESPaDOnS, N: NARVAL) and column 5 the spectrum \sn. Column 6 indicates rejected spectra or the presence of moon pollution ($^a$: not used in ZDI, $^b$: not used in GPR, $^c$: not used for period retrieval from \hal\ EW). Columns 7 and 8 contain the \sn\ in the \sti\ and \stv\ LSD profiles respectively. Columns 9 to 12 show the raw RVs, the RVs filtered with ZDI, the RVs filtered with GPR and the 1$\sigma$ RV error bar respectively, column 13 lists the equivalent width of the \hal\ line (with a typical error bar of 10{~}\kms) and column 14 informs the longitudinal projection of the magnetic field integrated over the visible surface (with a typical error bar of 50{~}G).}
			\label{tab:sob}
			\tiny
			\subfloat[Observations of the late 2008 and early 2009 runs.]{\begin{tabular}{cccccccccccccc}
				UTC & BJD & Cycle & Instr. & \sn & Comment & \sn$_I$ & \sn$_V$ & \rvraw\ & \rvfil$_{\rm /ZDI}$ & \rvfil$_{\rm /GP}$ & $\sigma_{\rm RV}$ & EW$_{\hal}$ & B$_{\rm long}$ \\
				2008 Oct & 2454700+ & -42+ & & & & & & (\kms) & (\kms) & (\kms) & (\kms) & (\kms) & (G) \\
				\hline
				15 11:42:34 & 54.992 & 0.552 & E & 228 & Isolated$^a$ & 2313 & 6167 & -0.119 & & -0.006 & 0.078 & 54.188 & -156 \\
				16 11:09:41 & 55.969 & 1.075 & E & 107 & Isolated$^a$ & 2184 & 2743 & 1.681 & & 0.004 & 0.084 & -0.857 & 102 \\
				19 09:09:16 & 58.886 & 2.633 & E & 227 & Isolated$^a$ & 2346 & 6305 & 1.069 & & -0.001 & 0.077 & 42.548 & -142 \\
				19 13:53:44 & 59.083 & 2.738 & E & 204 & Isolated$^a$ & 2295 & 5491 & 0.696 & & 0.006 & 0.079 & 12.192 & -167 \\
				\hline
				2008 Dec & 2454800+ & -15+ & & & & & & & & & & & \\
				\hline
				05 13:01:49 & 6.049 & 0.827 & E & 61 & Bad S/N$^{a,b}$ & & & & & & & 10.874 & \\
				06 08:06:19 & 6.843 & 1.251 & E & 238 & & 2669 & 6759 & 0.792 & 0.016 & -0.068 & 0.071 & 31.589 & -50 \\
				07 13:32:32 & 8.070 & 1.907 & E & 209 & & 2672 & 5784 & -0.516 & -0.043 & -0.000 & 0.071 & 17.226 & 240 \\
				08 05:05:16 & 8.718 & 2.253 & E & 230 & & 2634 & 6437 & 0.913 & 0.135 & 0.097 & 0.072 & 19.302 & -59 \\
				09 05:54:32 & 9.752 & 2.805 & E & 110 & Moon & 2514 & 2847 & -1.484 & -0.129 & -0.010 & 0.075 & -1.567 & -40 \\
				10 13:46:11 & 11.079 & 3.514 & E & 201 & He$\;${\sc i} $D_3$ flare$^{a,b}$ & & & & & & & 50.202 & \\
				15 13:36:10 & 16.072 & 6.181 & E & 212 & Big flare$^{a,b,c}$ & & & & & & & 231.256 & \\
				19 04:58:16 & 19.712 & 8.126 & E & 185 & & 2619 & 4876 & 0.425 & -0.234 & -0.010 & 0.072 & 18.761 & 46 \\
				20 13:45:25 & 21.078 & 8.856 & E & 142 & & 2537 & 3502 & -1.188 & -0.072 & -0.009 & 0.074 & 5.692 & 150 \\
				\hline
				2009 Jan & 2454800+ & 0+ & & & & & & & & & & & \\
				\hline
				02 19:38:17 & 34.323 & 0.931 & N & 167 & & 2610 & 4531 & 0.742 & 0.025 & 0.125 & 0.073 & -12.991 & 149 \\
				02 20:23:28 & 34.354 & 0.948 & N & 166 & & 2570 & 4408 & 0.754 & -0.155 & -0.171 & 0.073 & -9.125 & 177 \\
				02 21:08:39 & 34.386 & 0.964 & N & 166 & & 2622 & 4566 & 1.192 & 0.141 & 0.064 & 0.072 & -14.394 & 97 \\
				02 21:53:50 & 34.417 & 0.981 & N & 167 & & 2601 & 4546 & 1.065 & -0.074 & -0.156 & 0.073 & -18.321 & 126 \\
				02 22:41:08 & 34.450 & 0.999 & N & 159 & & 2609 & 4438 & 1.369 & 0.201 & 0.143 & 0.073 & -14.991 & 114 \\
				03 19:11:19 & 35.304 & 1.455 & N & 160 & & 2551 & 4235 & -1.347 & 0.010 & 0.088 & 0.074 & 29.299 & -124 \\
				03 19:56:30 & 35.336 & 1.472 & N & 157 & & 2586 & 4427 & -0.955 & -0.020 & -0.024 & 0.073 & 30.601 & -161 \\
				03 20:42:21 & 35.367 & 1.489 & N & 160 & & 2565 & 4321 & -0.430 & 0.047 & 0.009 & 0.074 & 32.998 & -151 \\
				03 21:28:13 & 35.399 & 1.506 & N & 151 & & 2524 & 4080 & -0.018 & 0.000 & -0.006 & 0.075 & 35.709 & -216 \\
				03 22:13:25 & 35.431 & 1.523 & N & 148 & & 2544 & 4053 & 0.346 & -0.059 & 0.013 & 0.074 & 43.994 & -158 \\
				04 18:52:52 & 36.291 & 1.982 & N & 149 & & 2552 & 3997 & 1.289 & 0.145 & 0.065 & 0.074 & -8.025 & 169 \\
				04 19:38:04 & 36.323 & 1.999 & N & 156 & & 2608 & 4291 & 1.176 & 0.007 & -0.039 & 0.073 & -7.124 & 81 \\
				04 20:23:16 & 36.354 & 2.016 & N & 159 & & 2610 & 4421 & 1.171 & 0.026 & 0.018 & 0.073 & -2.797 & 114 \\
				04 21:08:28 & 36.385 & 2.033 & N & 158 & & 2594 & 4337 & 0.973 & -0.110 & -0.086 & 0.073 & 1.774 & 117 \\
				04 21:53:40 & 36.417 & 2.049 & N & 158 & & 2583 & 4235 & 1.014 & 0.016 & 0.074 & 0.073 & 6.863 & 18 \\
				05 22:10:16 & 37.428 & 2.590 & N & 127 & & 2355 & 3240 & 1.332 & 0.026 & -0.002 & 0.079 & 43.598 & -141 \\
				05 22:44:08 & 37.452 & 2.602 & N & 86 & & 2118 & 2064 & 1.480 & 0.193 & 0.002 & 0.087 & 42.265 & -311 \\
				07 05:38:42 & 38.740 & 3.290 & E & 199 & & 2681 & 5764 & -0.259 & 0.167 & 0.040 & 0.071 & 31.322 & -82 \\
				09 04:54:60 & 40.709 & 4.342 & E & 193 & & 2650 & 5144 & -1.874 & -0.296 & -0.219 & 0.072 & 39.064 & -26 \\
				10 20:45:05 & 42.369 & 5.229 & N & 158 & Moon & 2453 & 3917 & 0.281 & -0.178 & 0.043 & 0.076 & 7.610 & -9 \\
				10 21:43:38 & 42.409 & 5.251 & N & 158 & Moon & 2495 & 4210 & 0.087 & -0.147 & -0.115 & 0.075 & 15.255 & -79 \\
				10 22:42:09 & 42.450 & 5.272 & N & 165 & Moon & 2463 & 4271 & -0.069 & 0.033 & -0.063 & 0.076 & 23.200 & -67 \\
				11 04:58:44 & 42.712 & 5.412 & E & 163 & Moon & 2686 & 4299 & -2.287 & -0.249 & -0.097 & 0.071 & 19.993 & -137 \\
				11 18:56:02 & 43.293 & 5.723 & N & 136 & & 2391 & 3495 & -0.506 & 0.115 & 0.064 & 0.078 & 14.943 & -200 \\
				11 19:41:16 & 43.324 & 5.739 & N & 142 & & 2467 & 3711 & -1.032 & -0.197 & -0.102 & 0.076 & 8.644 & -167 \\
				11 20:26:27 & 43.356 & 5.756 & N & 131 & & 2360 & 3369 & -1.132 & -0.165 & -0.006 & 0.079 & 7.370 & -57 \\
				11 21:11:39 & 43.387 & 5.773 & N & 116 & Moon & 2214 & 2911 & -1.109 & -0.091 & 0.074 & 0.084 & 4.870 & 37 \\
				11 21:56:50 & 43.419 & 5.790 & N & 17 & Bad S/N$^{a,b}$ & & & & & & & 2.131 & \\
				12 19:41:03 & 44.324 & 6.274 & N & 117 & & 2185 & 2788 & 0.062 & 0.191 & 0.104 & 0.084 & 27.575 & -123 \\
				12 20:26:17 & 44.356 & 6.290 & N & 114 & & 2212 & 2847 & -0.360 & 0.097 & -0.025 & 0.084 & 33.439 & 4 \\
				12 21:11:29 & 44.387 & 6.307 & N & 117 & & 2253 & 2927 & -0.630 & 0.196 & 0.092 & 0.082 & 37.490 & -116 \\
				12 22:01:04 & 44.421 & 6.325 & N & 132 & & 2415 & 3410 & -1.081 & 0.162 & 0.108 & 0.078 & 41.752 & -93 \\
				14 04:59:12 & 45.712 & 7.015 & E & 232 & & 2723 & 6970 & 1.004 & -0.143 & -0.052 & 0.070 & -37.801 & 69 \\
				14 19:12:20 & 46.304 & 7.331 & N & 144 & & 2325 & 3772 & -1.462 & -0.081 & -0.154 & 0.080 & 39.919 & -93 \\
				14 19:57:30 & 46.335 & 7.348 & N & 146 & & 2425 & 3848 & -1.603 & 0.111 & 0.080 & 0.077 & 30.220 & -111 \\
				14 20:42:42 & 46.367 & 7.365 & N & 147 & & 2470 & 3862 & -1.816 & 0.150 & 0.157 & 0.076 & 28.237 & -93 \\
				14 21:27:54 & 46.398 & 7.381 & N & 147 & & 2448 & 3821 & -2.156 & -0.050 & -0.004 & 0.077 & 28.349 & -82 \\
				15 20:41:08 & 47.366 & 7.898 & N & 90 & & 1893 & 2066 & 0.045 & -0.259 & -0.065 & 0.096 & 11.370 & 278 \\
				15 21:26:20 & 47.397 & 7.915 & N & 92 & & 1893 & 2089 & 0.499 & -0.060 & -0.006 & 0.096 & 7.425 & 218 \\
				16 18:24:01 & 48.270 & 8.382 & N & 137 & & 2429 & 3520 & -2.014 & 0.093 & 0.093 & 0.077 & 20.336 & -58 \\
				16 19:09:13 & 48.302 & 8.398 & N & 150 & & 2513 & 3951 & -2.196 & -0.088 & -0.052 & 0.075 & 18.041 & -97 \\
				16 19:54:26 & 48.333 & 8.415 & N & 145 & & 2484 & 3779 & -2.079 & -0.102 & -0.053 & 0.076 & 20.508 & -99 \\
				16 20:39:38 & 48.365 & 8.432 & N & 133 & & 2410 & 3442 & -1.687 & 0.040 & 0.068 & 0.078 & 21.507 & 41 \\
				16 21:24:49 & 48.396 & 8.449 & N & 16 & Bad S/N$^{a,b}$ & & & & & & & 21.522 & \\
				16 22:23:00 & 48.436 & 8.470 & N & 113 & & 2317 & 2783 & -0.820 & 0.024 & -0.047 & 0.080 & 22.599 & -126 \\
				17 18:20:13 & 49.268 & 8.914 & N & 127 & & 2374 & 3291 & 0.560 & 0.004 & 0.038 & 0.079 & -2.751 & 153 \\
				17 19:05:25 & 49.299 & 8.931 & N & 103 & & 2116 & 2484 & 0.778 & -0.002 & -0.079 & 0.087 & -6.869 & 113 \\
				17 19:50:37 & 49.330 & 8.948 & N & 121 & & 2394 & 3126 & 1.191 & 0.233 & 0.106 & 0.078 & -10.239 & 157 \\
				17 20:35:49 & 49.362 & 8.965 & N & 140 & & 2514 & 3740 & 1.158 & 0.075 & -0.033 & 0.075 & -15.775 & 106 \\
				17 21:21:01 & 49.393 & 8.981 & N & 137 & & 2483 & 3616 & 1.222 & 0.066 & 0.030 & 0.076 & -22.467 & 115 \\
			\end{tabular}}
		\end{table*}

		\begin{table*}
			\ContinuedFloat
			\caption{(Continued from previous page).}
			\tiny
			\subfloat[Observations of the early 2011 run.]{\begin{tabular}{cccccccccccccc}
				UTC & BJD & Cycle & Instr. & \sn & Comment & \sn$_I$ & \sn$_V$ & \rvraw\ & \rvfil$_{\rm /ZDI}$ & \rvfil$_{\rm /GP}$ & $\sigma_{\rm RV}$ & EW$_{\hal}$ & B$_{\rm long}$ \\
				2011 Jan & 2455500+ & 397+ & & & & & & (\kms) & (\kms) & (\kms) & (\kms) & (\kms) & (G) \\
				\hline
				14 19:01:45 & 76.297 & 0.291 & N & 145 & & 2789 & 4035 & -2.162 & -0.114 & -0.117 & 0.069 & -15.778 & -43 \\
				14 19:46:58 & 76.328 & 0.308 & N & 152 & & 2757 & 4175 & -2.052 & 0.363 & 0.172 & 0.070 & -24.729 & -147 \\
				14 20:32:12 & 76.360 & 0.324 & N & 155 & & 2789 & 4172 & -2.511 & 0.181 & 0.011 & 0.069 & -28.375 & -41 \\
				14 21:17:25 & 76.391 & 0.341 & N & 150 & & 2765 & 4113 & -2.923 & -0.070 & -0.071 & 0.069 & -25.398 & -103 \\
				15 19:09:27 & 77.302 & 0.828 & N & 138 & & 2710 & 3877 & -2.672 & -0.127 & -0.023 & 0.071 & 6.283 & -30 \\
				15 19:54:41 & 77.334 & 0.845 & N & 144 & Moon & 2739 & 3915 & -2.552 & -0.043 & -0.002 & 0.070 & 2.985 & -24 \\
				15 20:39:55 & 77.365 & 0.861 & N & 133 & Moon & 2693 & 3584 & -2.287 & 0.085 & 0.097 & 0.071 & 0.295 & -61 \\
				15 21:25:07 & 77.396 & 0.878 & N & 143 & & 2761 & 3909 & -2.226 & -0.084 & -0.078 & 0.069 & -3.177 & -48 \\
				16 19:27:12 & 78.314 & 1.369 & N & 146 & Moon & 2710 & 3867 & -3.130 & -0.327 & -0.069 & 0.071 & 0.040 & -29 \\
				16 20:12:25 & 78.346 & 1.385 & N & 148 & Moon & 2735 & 4044 & -2.706 & -0.140 & 0.085 & 0.070 & 3.591 & 30 \\
				16 20:57:37 & 78.377 & 1.402 & N & 151 & Moon & 2802 & 4122 & -2.236 & -0.063 & -0.018 & 0.069 & 10.411 & -39 \\
				17 18:20:45 & 79.268 & 1.878 & N & 135 & & 2706 & 3739 & -2.165 & -0.028 & -0.033 & 0.071 & -0.629 & -33 \\
				17 19:05:56 & 79.299 & 1.895 & N & 130 & & 2671 & 3441 & -1.782 & 0.041 & 0.030 & 0.071 & -1.757 & -25 \\
				17 19:51:09 & 79.331 & 1.912 & N & 132 & & 2698 & 3503 & -1.406 & 0.039 & 0.009 & 0.071 & -6.538 & 70 \\
				20 18:49:53 & 82.288 & 3.491 & N & 86 & & 2194 & 1903 & 1.561 & 0.186 & 0.003 & 0.084 & 19.874 & 126 \\
				20 19:35:05 & 82.320 & 3.508 & N & 82 & & 2173 & 1817 & 1.979 & -0.008 & -0.003 & 0.085 & 27.037 & 18 \\
				22 21:18:44 & 84.391 & 4.615 & N & 93 & Moon & 2249 & 2087 & 3.210 & 0.100 & 0.025 & 0.083 & 34.575 & 15 \\
				22 22:03:58 & 84.423 & 4.632 & N & 106 & Moon & 2484 & 2539 & 2.815 & 0.024 & 0.014 & 0.076 & 29.770 & -70 \\
				23 21:53:42 & 85.415 & 5.162 & N & 147 & & 2714 & 3773 & 1.397 & 0.027 & 0.008 & 0.070 & 6.158 & -40 \\
				24 18:46:11 & 86.285 & 5.627 & N & 140 & Moon & 2779 & 3766 & 2.885 & 0.000 & -0.023 & 0.069 & 37.130 & -104 \\
			\end{tabular}}

			\subfloat[Observations of the late 2013 run.]{\begin{tabular}{cccccccccccccc}
				UTC & BJD & Cycle & Instr. & \sn & Comment & \sn$_I$ & \sn$_V$ & \rvraw\ & \rvfil$_{\rm /ZDI}$ & \rvfil$_{\rm /GP}$ & $\sigma_{\rm RV}$ & EW$_{\hal}$ & B$_{\rm long}$ \\
				2013 Nov & 2456600+ & 946+ & & & & & & (\kms) & (\kms) & (\kms) & (\kms) & (\kms) & (G) \\
				\hline
				07 22:44:14 & 4.453 & 0.528 & N & 118 & Isolated$^a$ & 1792 & 2855 & -0.186 & & 0.059 & 0.101 & 42.094 & -87 \\
				07 23:18:58 & 4.477 & 0.541 & N & 88 & Isolated$^a$ & 1783 & 2191 & -0.922 & & -0.033 & 0.102 & 48.184 & -50 \\
				\hline
				2013 Dec & 2456600+ & 959+ & & & & & & & & & & & \\
				\hline
				02 21:38:55 & 29.408 & 0.859 & N & 86 & & 1700 & 1941 & -0.374 & 0.277 & 0.048 & 0.106 & -0.205 & -279 \\
				03 01:09:58 & 29.554 & 0.937 & N & 123 & & 2042 & 3010 & -2.818 & 0.160 & 0.191 & 0.090 & 21.666 & -60 \\
				03 21:25:55 & 30.399 & 1.388 & N & 134 & & 2193 & 3409 & 5.209 & 0.490 & 0.020 & 0.085 & 27.245 & -147 \\
				04 00:57:14 & 30.545 & 1.467 & N & 130 & & 2161 & 3175 & 2.841 & 0.198 & 0.084 & 0.086 & 52.594 & 19 \\
				04 21:43:24 & 31.411 & 1.929 & N & 150 & & 2282 & 3959 & -3.122 & -0.282 & -0.172 & 0.082 & 24.550 & -47 \\
				05 21:15:09 & 32.391 & 2.453 & N & 135 & & 2261 & 3536 & 3.431 & 0.150 & -0.070 & 0.082 & 53.441 & -61 \\
				06 00:54:07 & 32.543 & 2.534 & N & 131 & & 2192 & 3327 & -0.989 & -0.292 & -0.044 & 0.085 & 56.121 & -54 \\
				06 21:12:11 & 33.389 & 2.986 & N & 130 & & 2211 & 3431 & -3.067 & 0.161 & -0.004 & 0.084 & 18.241 & -76 \\
				07 00:41:21 & 33.534 & 3.063 & N & 147 & & 2222 & 3798 & -2.708 & -0.107 & -0.017 & 0.084 & -9.025 & -158 \\
				07 22:27:39 & 34.441 & 3.548 & N & 148 & & 2285 & 3983 & -1.396 & -0.289 & -0.074 & 0.082 & 49.419 & 42 \\
				08 02:30:09 & 34.610 & 3.638 & N & 148 & & 2247 & 3876 & 0.152 & -0.008 & 0.130 & 0.082 & 46.173 & -81 \\
				08 22:49:21 & 35.456 & 4.090 & N & 165 & He$\;${\sc i} $D_3$ flare$^{a,b}$ & & & & & & & 21.045 & \\
				09 01:31:57 & 35.569 & 4.151 & N & 163 & He$\;${\sc i} $D_3$ flare$^{a,b}$ & & & & & & & 12.729 & \\
				09 20:41:53 & 36.368 & 4.577 & N & 112 & & 1905 & 2698 & -1.451 & -0.126 & 0.011 & 0.096 & 49.343 & 21 \\
				10 00:43:24 & 36.536 & 4.667 & N & 148 & & 2263 & 3888 & 1.360 & 0.118 & 0.129 & 0.082 & 32.864 & -83 \\
				10 01:29:32 & 36.568 & 4.684 & N & 149 & & 2278 & 3953 & 1.840 & 0.041 & -0.068 & 0.082 & 28.874 & -111 \\
				10 20:34:03 & 37.362 & 5.109 & N & 140 & & 2243 & 3654 & -2.267 & 0.140 & 0.016 & 0.083 & -5.886 & -72 \\
				11 00:34:03 & 37.529 & 5.197 & N & 151 & & 2318 & 4015 & -2.188 & -0.120 & -0.069 & 0.080 & -31.885 & 104 \\
				11 20:33:11 & 38.362 & 5.642 & N & 147 & & 2272 & 3909 & 0.047 & -0.331 & -0.204 & 0.082 & 55.426 & -87 \\
				12 00:53:59 & 38.543 & 5.739 & N & 148 & & 2317 & 3928 & 2.837 & 0.232 & 0.034 & 0.081 & 19.504 & -101 \\
				12 20:28:06 & 39.358 & 6.175 & N & 160 & & 2359 & 4379 & -2.323 & -0.026 & 0.020 & 0.079 & -13.047 & -6 \\
				13 00:27:29 & 39.525 & 6.263 & N & 169 & & 2375 & 4647 & -0.553 & -0.468 & -0.081 & 0.079 & -5.297 & -52 \\
				14 20:21:32 & 41.354 & 7.241 & N & 149 & & 2268 & 3915 & -0.962 & 0.022 & 0.119 & 0.082 & -8.269 & 61 \\
				15 01:21:03 & 41.562 & 7.352 & N & 141 & & 2230 & 3620 & 4.118 & 0.065 & 0.009 & 0.083 & 3.858 & -62 \\
				15 20:34:39 & 42.363 & 7.780 & N & 132 & Moon & 2168 & 3299 & 1.919 & -0.107 & -0.032 & 0.085 & -4.559 & -103 \\
				20 20:38:45 & 47.365 & 10.452 & N & 126 & & 2198 & 3078 & 3.010 & 0.020 & -0.007 & 0.085 & 34.878 & -100 \\
				21 00:57:56 & 47.545 & 10.548 & N & 136 & & 2185 & 3356 & -1.160 & 0.045 & 0.092 & 0.085 & 36.919 & -54 \\
			\end{tabular}}

			\subfloat[Observations of the late 2015 and early 2016 runs.]{\begin{tabular}{cccccccccccccc}
				UTC & BJD & Cycle & Instr. & \sn & Comment & \sn$_I$ & \sn$_V$ & \rvraw\ & \rvfil$_{\rm /ZDI}$ & \rvfil$_{\rm /GP}$ & $\sigma_{\rm RV}$ & EW$_{\hal}$ & B$_{\rm long}$ \\
				2015 Dec & 2457300+ & 1349+ & & & & & & (\kms) & (\kms) & (\kms) & (\kms) & (\kms) & (G) \\
				\hline
				01 22:44:05 & 58.453 & 0.313 & N & 121 & & 1776 & 2974 & 2.464 & 0.058 & 0.050 & 0.102 & -6.540 & -60 \\
				02 02:00:59 & 58.590 & 0.386 & N & 109 & & 1690 & 2561 & 3.011 & 0.461 & 0.122 & 0.107 & 1.045 & -239 \\
				02 22:07:10 & 59.427 & 0.833 & N & 101 & & 1699 & 2360 & -1.825 & -0.080 & 0.008 & 0.107 & 0.196 & -262 \\
				03 01:30:49 & 59.569 & 0.909 & N & 114 & & 1807 & 2794 & -2.520 & 0.226 & 0.066 & 0.101 & -5.859 & -108 \\
				03 21:39:29 & 60.408 & 1.357 & N & 109 & & 1743 & 2543 & 3.040 & 0.203 & -0.115 & 0.104 & -10.697 & -112 \\
				04 02:23:31 & 60.605 & 1.462 & N & 119 & & 1786 & 2804 & 0.070 & -0.168 & -0.009 & 0.102 & 19.800 & -165 \\
				06 22:53:07 & 63.459 & 2.987 & N & 152 & & 1875 & 3764 & -2.324 & -0.086 & -0.002 & 0.097 & 9.991 & -17 \\
				07 01:15:19 & 63.558 & 3.040 & N & 152 & & 1886 & 3853 & -1.401 & 0.091 & 0.004 & 0.097 & -0.787 & -68 \\
				07 23:02:53 & 64.466 & 3.525 & N & 94 & & 1633 & 2122 & -0.690 & 0.173 & 0.093 & 0.110 & 30.595 & -70 \\
				08 01:31:49 & 64.569 & 3.580 & N & 142 & & 1844 & 3566 & -0.195 & -0.186 & 0.008 & 0.099 & 34.071 & -251 \\
				09 21:05:04 & 66.384 & 4.549 & N & 103 & & 1713 & 2409 & -0.775 & -0.116 & -0.083 & 0.105 & 37.568 & -128 \\
				11 22:05:07 & 68.426 & 5.640 & N & 129 & & 2147 & 3344 & 1.638 & -0.028 & 0.028 & 0.086 & 71.742 & -187 \\
				12 00:56:44 & 68.545 & 5.704 & N & 136 & & 2124 & 3424 & 2.077 & 0.065 & -0.159 & 0.087 & 28.515 & -65 \\
				12 22:13:39 & 69.432 & 6.177 & N & 113 & & 2077 & 2927 & -0.153 & 0.050 & 0.015 & 0.089 & -29.140 & -58 \\
				13 01:11:33 & 69.555 & 6.243 & N & 135 & & 2144 & 3458 & 0.521 & -0.283 & -0.032 & 0.086 & -16.825 & -23 \\
				13 22:12:08 & 70.431 & 6.711 & N & 106 & & 1962 & 2553 & 2.337 & 0.433 & 0.186 & 0.093 & 9.573 & -93 \\
				14 01:29:54 & 70.568 & 6.784 & N & 134 & & 2095 & 3343 & -0.445 & -0.204 & -0.035 & 0.088 & 9.052 & -145 \\
				18 01:04:45 & 74.550 & 8.912 & N & 124 & & 2036 & 3124 & -2.902 & -0.225 & -0.046 & 0.090 & 13.281 & -144 \\
				18 21:36:31 & 75.406 & 9.369 & N & 138 & & 2087 & 3452 & 2.918 & 0.224 & -0.000 & 0.089 & -21.444 & -171 \\
				19 01:09:10 & 75.553 & 9.447 & N & 124 & & 1972 & 3016 & 0.091 & -0.426 & -0.039 & 0.093 & 17.394 & -189 \\
				19 20:09:16 & 76.345 & 9.870 & N & 106 & & 1918 & 2497 & -2.441 & -0.020 & -0.023 & 0.095 & 21.741 & -16 \\
				\hline
				2016 Jan & 2457400+ & 1376+ & & & & & & & (\kms) & (\kms) & (\kms) \\
				\hline
				20 22:42:30 & 8.450 & 0.021 & N & 153 & He$\;${\sc i} $D_3$ flare$^{a,b}$ & & & & & & & 23.847 & \\
				20 23:35:42 & 8.487 & 0.040 & N & 154 & He$\;${\sc i} $D_3$ flare$^{a,b}$ & & & & & & & 13.077 & \\
				21 18:28:09 & 9.273 & 0.460 & N & 127 & & 2817 & 3105 & -0.896 & -0.014 & 0.003 & 0.068 & 31.341 & -154 \\
				21 23:28:57 & 9.482 & 0.572 & N & 121 & & 2765 & 2956 & -0.092 & -0.064 & -0.003 & 0.069 & 37.161 & -280 \\
				24 19:12:18 & 12.303 & 2.079 & N & 105 & & 2702 & 2587 & -1.095 & 0.136 & -0.002 & 0.071 & -10.058 & 54 \\
				24 21:38:26 & 12.405 & 2.134 & N & 121 & & 2831 & 3058 & -0.414 & -0.037 & 0.014 & 0.068 & -23.494 & 45 \\
				26 18:19:49 & 14.267 & 3.128 & N & 135 & & 2930 & 3444 & -0.506 & -0.034 & -0.016 & 0.066 & -31.556 & 0 \\
				26 23:23:18 & 14.478 & 3.241 & N & 140 & & 2886 & 3485 & 1.534 & -0.047 & 0.010 & 0.067 & -39.330 & -34 \\
				27 18:13:34 & 15.262 & 3.660 & N & 124 & & 2853 & 3091 & 2.589 & 0.058 & -0.003 & 0.068 & 28.597 & -202 \\
				27 20:34:49 & 15.360 & 3.712 & N & 122 & & 2898 & 3126 & 2.359 & 0.047 & 0.005 & 0.067 & 31.543 & -94 \\
				29 21:36:35 & 17.403 & 4.803 & N & 100 & & 2792 & 2378 & -0.924 & -0.214 & 0.011 & 0.069 & 41.991 & -166 \\
			\end{tabular}}
		\end{table*}

		The telescopes used for photometry at CrAO are AZT-11, a 1.25{~}m telescope with a five-channel photometer-polarimeter, and T60Sim, a 0.60{~}m telescope with a four-channel photometer.
		\begin{landscape}
			\begin{table}
				\centering
				\caption{Information on the \vt\ photometric data. For each table, the first and second columns indicate the time at which the observations were taken, in UTC and BJD format respectively. The third column contains the measured visible magnitude, then columns 4 to 6 list color indexes B-V, V-R$_{\rm J}$ and V-I$_{\rm J}$ provided in the Johnson UBVRI system, and column 7 indicates the name of the telescope used for the observation.}
				\label{tab:pob}
				\subfloat[Photometric measurements of the set 08b+09a (left) and of the first half of the set 09b+10a (right).]{\begin{tabular}{ccccccccccccccc}
					Date & HJD (2454000+) & V (mag) & B-V & V-R$_{\rm J}$ & V-I$_{\rm J}$ & Telescope & | & Date & HJD (2455000+) & V (mag) & B-V & V-R$_{\rm J}$ & V-I$_{\rm J}$ & Telescope \\
					\hline
					03-Aug-2008 & 682.5282 & 10.878 & 1.182 & 1.075 & - & AZT-11 & | & 15-Aug-2009 & 59.5212 & 10.894 & 1.147 & 1.133 & - & AZT-11 \\
					04-Aug-2008 & 683.5300 & 10.896 & 1.192 & - & - & AZT-11 & | & 15-Aug-2009 & 59.5320 & 10.904 & 1.149 & 1.131 & - & AZT-11 \\
					09-Aug-2008 & 688.5258 & 10.857 & 1.169 & 1.046 & - & AZT-11 & | & 15-Aug-2009 & 59.5418 & 10.886 & 1.154 & 1.095 & - & AZT-11 \\
					10-Aug-2008 & 689.5304 & 10.806 & 1.185 & 1.047 & - & AZT-11 & | & 16-Aug-2009 & 60.5209 & 10.812 & 1.164 & 1.063 & - & AZT-11 \\
					12-Aug-2008 & 691.5306 & 10.831 & 1.168 & 1.062 & - & AZT-11 & | & 16-Aug-2009 & 60.5289 & 10.807 & 1.167 & 1.045 & - & AZT-11 \\
					13-Aug-2008 & 692.5308 & 10.918 & 1.203 & 1.059 & - & AZT-11 & | & 16-Aug-2009 & 60.5369 & 10.812 & 1.169 & 1.072 & - & AZT-11 \\
					14-Aug-2008 & 693.5253 & 10.874 & 1.166 & 1.054 & - & AZT-11 & | & 19-Aug-2009 & 63.5237 & 10.857 & 1.152 & 1.126 & - & AZT-11 \\
					14-Aug-2008 & 693.5471 & 10.887 & 1.183 & 1.051 & - & AZT-11 & | & 19-Aug-2009 & 63.5314 & 10.855 & 1.153 & 1.077 & - & AZT-11 \\
					26-Aug-2008 & 705.5112 & 10.842 & 1.188 & 1.025 & - & AZT-11 & | & 19-Aug-2009 & 63.5394 & 10.829 & 1.148 & 1.069 & - & AZT-11 \\
					01-Sep-2008 & 711.5369 & 10.891 & 1.189 & 1.047 & - & AZT-11 & | & 19-Aug-2009 & 63.5474 & 10.838 & 1.157 & 1.084 & - & AZT-11 \\
					01-Sep-2008 & 711.5527 & 10.888 & 1.190 & 1.041 & - & AZT-11 & | & 21-Aug-2009 & 65.5183 & 10.798 & 1.146 & 1.058 & - & AZT-11 \\
					02-Sep-2008 & 712.5160 & 10.855 & 1.182 & 1.049 & - & AZT-11 & | & 21-Aug-2009 & 65.5267 & 10.794 & 1.156 & 1.058 & - & AZT-11 \\
					02-Sep-2008 & 712.5308 & 10.859 & 1.173 & 1.035 & - & AZT-11 & | & 21-Aug-2009 & 65.5348 & 10.796 & 1.143 & 1.054 & - & AZT-11 \\
					02-Sep-2008 & 712.5763 & 10.857 & 1.159 & 1.055 & - & AZT-11 & | & 24-Aug-2009 & 68.4813 & 10.830 & 1.166 & 1.053 & - & AZT-11 \\
					03-Sep-2008 & 713.5146 & 10.868 & 1.184 & 1.047 & - & AZT-11 & | & 24-Aug-2009 & 68.4892 & 10.847 & 1.165 & 1.071 & - & AZT-11 \\
					03-Sep-2008 & 713.5295 & 10.861 & 1.181 & 1.035 & - & AZT-11 & | & 24-Aug-2009 & 68.4982 & 10.841 & 1.159 & 1.066 & - & AZT-11 \\
					04-Sep-2008 & 714.5299 & 10.850 & 1.151 & 1.051 & - & AZT-11 & | & 25-Aug-2009 & 69.5189 & 10.781 & 1.140 & 1.103 & - & AZT-11 \\
					05-Sep-2008 & 715.4831 & 10.861 & 1.148 & 1.051 & - & AZT-11 & | & 25-Aug-2009 & 69.5258 & 10.769 & 1.153 & 1.098 & - & AZT-11 \\
					05-Sep-2008 & 715.5357 & 10.862 & 1.150 & 1.095 & - & T60Sim & | & 25-Aug-2009 & 69.5329 & 10.771 & 1.151 & 1.080 & - & AZT-11 \\
					05-Sep-2008 & 715.5636 & 10.842 & 1.151 & 1.039 & - & AZT-11 & | & 27-Aug-2009 & 71.5171 & 10.788 & 1.141 & 1.042 & - & AZT-11 \\
					06-Sep-2008 & 716.5612 & 10.835 & 1.170 & 1.070 & - & T60Sim & | & 27-Aug-2009 & 71.5242 & 10.783 & 1.159 & 1.036 & - & AZT-11 \\
					07-Sep-2008 & 717.5431 & 10.839 & 1.210 & 1.052 & - & T60Sim & | & 29-Aug-2009 & 73.5157 & 10.810 & 1.157 & 1.072 & - & AZT-11 \\
					08-Sep-2008 & 718.4556 & 10.824 & 1.177 & 1.038 & - & T60Sim & | & 29-Aug-2009 & 73.5229 & 10.808 & 1.162 & 1.071 & - & AZT-11 \\
					08-Sep-2008 & 718.5607 & 10.843 & 1.185 & 1.036 & - & T60Sim & | & 29-Aug-2009 & 73.5297 & 10.810 & 1.162 & 1.073 & - & AZT-11 \\
					11-Sep-2008 & 721.5327 & 10.858 & 1.189 & 1.002 & - & T60Sim & | & 16-Sep-2009 & 91.4903 & 10.892 & 1.155 & 1.079 & - & AZT-11 \\
					30-Sep-2008 & 740.5057 & 10.843 & 1.202 & 1.045 & - & T60Sim & | & 16-Sep-2009 & 91.4972 & 10.887 & 1.147 & 1.083 & - & AZT-11 \\
					30-Sep-2008 & 740.5724 & 10.835 & 1.186 & 1.044 & - & AZT-11 & | & 16-Sep-2009 & 91.5040 & 10.891 & 1.140 & 1.076 & - & AZT-11 \\
					30-Sep-2008 & 740.6480 & 10.826 & 1.175 & 1.044 & - & AZT-11 & | & 20-Sep-2009 & 95.4203 & 10.792 & 1.155 & 1.042 & - & AZT-11 \\
					01-Oct-2008 & 741.5784 & 10.878 & 1.176 & 1.057 & - & AZT-11 & | & 20-Sep-2009 & 95.4286 & 10.787 & 1.149 & 1.045 & - & AZT-11 \\
					04-Oct-2008 & 744.5202 & 10.843 & 1.171 & 1.056 & - & T60Sim & | & 20-Sep-2009 & 95.4360 & 10.773 & 1.158 & 1.037 & - & AZT-11 \\
					04-Oct-2008 & 744.5202 & 10.822 & 1.162 & 1.042 & - & T60Sim & | & 22-Sep-2009 & 97.5267 & 10.745 & 1.154 & 1.022 & - & AZT-11 \\
					23-Oct-2008 & 763.5826 & 10.872 & 1.167 & 1.048 & - & AZT-11 & | & 22-Sep-2009 & 97.5338 & 10.747 & 1.150 & 1.049 & - & AZT-11 \\
					30-Oct-2008 & 770.4631 & 10.822 & 1.173 & 1.056 & - & AZT-11 & | & 22-Sep-2009 & 97.5410 & 10.747 & 1.147 & 1.028 & - & AZT-11 \\
					30-Oct-2008 & 770.6079 & 10.805 & 1.174 & 1.027 & - & AZT-11 & | & 24-Sep-2009 & 99.5012 & 10.743 & 1.166 & 1.017 & - & AZT-11 \\
					07-Nov-2008 & 778.5546 & 10.879 & 1.185 & 1.050 & - & AZT-11 & | & 24-Sep-2009 & 99.5087 & 10.748 & 1.171 & 1.032 & - & AZT-11 \\
					08-Nov-2008 & 779.6318 & 10.859 & 1.169 & 1.052 & - & AZT-11 & | & 24-Sep-2009 & 99.5159 & 10.751 & 1.158 & 1.026 & - & AZT-11 \\
					01-Dec-2008 & 802.5990 & 10.811 & 1.157 & 1.053 & - & AZT-11 & | & 26-Sep-2009 & 101.5720 & 10.770 & 1.162 & 1.014 & - & AZT-11 \\
					01-Jan-2009 & 833.1937 & 10.871 & 1.168 & 1.042 & - & AZT-11 & | & 26-Sep-2009 & 101.5777 & 10.777 & 1.172 & 1.043 & - & AZT-11 \\
					21-Jan-2009 & 853.2120 & 10.863 & 1.181 & 1.047 & - & AZT-11 & | & 26-Sep-2009 & 101.5828 & 10.791 & 1.167 & 1.050 & - & AZT-11 \\
					26-Jan-2009 & 858.2434 & 10.859 & 1.160 & 1.055 & - & AZT-11 & | & 27-Sep-2009 & 102.4691 & 10.896 & 1.180 & 1.030 & - & AZT-11 \\
					& & & & & & & | & 28-Sep-2009 & 103.5756 & 10.774 & 1.181 & 1.042 & - & AZT-11 \\
					& & & & & & & | & 30-Sep-2009 & 105.4924 & 10.788 & 1.187 & 1.053 & - & AZT-11 \\
					& & & & & & & | & 30-Sep-2009 & 105.5064 & 10.792 & 1.176 & 1.042 & - & AZT-11 \\
					& & & & & & & | & 30-Sep-2009 & 105.5868 & 10.778 & 1.186 & 1.051 & - & AZT-11 \\
					& & & & & & & | & 30-Sep-2009 & 105.5923 & 10.777 & 1.190 & 1.016 & - & AZT-11 \\
				\end{tabular}}
			\end{table}
		\end{landscape}
	\begin{landscape}
			\begin{table}
				\ContinuedFloat
				\caption{(Continued from previous page).}
				\subfloat[Photometric measurements of the second half of the set 09b+10a (left) and of the set 10b (right).]{\begin{tabular}{ccccccccccccccc}
					Date & HJD (2455000+) & V (mag) & B-V & V-R$_{\rm J}$ & V-I$_{\rm J}$ & Telescope & | & Date & HJD (2455000+) & V (mag) & B-V & V-R$_{\rm J}$ & V-I$_{\rm J}$ & Telescope \\
					\hline
					09-Oct-2009 & 114.4547 & 10.734 & 1.136 & 1.025 & - & AZT-11 & | & 03-Sep-2010 & 443.5388 & 10.782 & 1.149 & 1.032 & - & AZT-11 \\
					10-Oct-2009 & 115.3930 & 10.850 & 1.178 & 1.043 & - & AZT-11 & | & 03-Sep-2010 & 443.5463 & 10.807 & 1.150 & 1.037 & - & AZT-11 \\
					10-Oct-2009 & 115.4597 & 10.873 & 1.162 & 1.072 & - & AZT-11 & | & 03-Sep-2010 & 443.5538 & 10.801 & 1.163 & 1.049 & - & AZT-11 \\
					11-Oct-2009 & 116.4768 & 10.744 & 1.166 & 1.031 & - & AZT-11 & | & 04-Sep-2010 & 444.5159 & 10.776 & 1.177 & 1.042 & - & AZT-11 \\
					11-Oct-2009 & 116.5391 & 10.763 & 1.167 & 1.032 & - & AZT-11 & | & 04-Sep-2010 & 444.5222 & 10.773 & 1.190 & 1.039 & - & AZT-11 \\
					11-Oct-2009 & 116.6024 & 10.776 & 1.175 & 1.042 & - & AZT-11 & | & 04-Sep-2010 & 444.5284 & 10.776 & 1.167 & 1.037 & - & AZT-11 \\
					19-Oct-2009 & 124.5660 & 10.773 & 1.163 & 1.052 & - & AZT-11 & | & 05-Sep-2010 & 445.5842 & 10.742 & 1.140 & 1.029 & - & AZT-11 \\
					19-Oct-2009 & 124.5733 & 10.808 & 1.176 & 1.034 & - & AZT-11 & | & 10-Sep-2010 & 450.5110 & 10.838 & 1.166 & 1.038 & - & AZT-11 \\
					23-Oct-2009 & 128.3930 & 10.823 & 1.187 & 1.029 & - & AZT-11 & | & 10-Sep-2010 & 450.5175 & 10.839 & 1.171 & 1.025 & - & AZT-11 \\
					23-Oct-2009 & 128.4038 & 10.819 & 1.181 & 1.043 & - & AZT-11 & | & 10-Sep-2010 & 450.5238 & 10.838 & 1.157 & 1.040 & - & AZT-11 \\
					31-Oct-2009 & 136.4814 & 10.846 & 1.138 & 1.069 & - & AZT-11 & | & 11-Sep-2010 & 451.5063 & 10.805 & 1.179 & 1.067 & - & AZT-11 \\
					31-Oct-2009 & 136.4886 & 10.850 & 1.133 & 1.067 & - & AZT-11 & | & 11-Sep-2010 & 451.5126 & 10.804 & 1.177 & 1.051 & - & AZT-11 \\
					31-Oct-2009 & 136.4958 & 10.839 & 1.140 & 1.060 & - & AZT-11 & | & 11-Sep-2010 & 451.5187 & 10.810 & 1.178 & 1.071 & - & AZT-11 \\
					07-Nov-2009 & 143.5181 & 10.870 & 1.180 & 1.046 & - & AZT-11 & | & 14-Sep-2010 & 454.5298 & 10.853 & 1.181 & 1.050 & - & AZT-11 \\
					07-Nov-2009 & 143.6312 & 10.895 & 1.175 & 1.048 & - & AZT-11 & | & 14-Sep-2010 & 454.5382 & 10.849 & 1.180 & 1.048 & - & AZT-11 \\
					08-Nov-2009 & 144.3301 & 10.686 & 1.146 & 1.012 & - & AZT-11 & | & 14-Sep-2010 & 454.5455 & 10.863 & 1.161 & 1.053 & - & AZT-11 \\
					08-Nov-2009 & 144.3365 & 10.699 & 1.148 & 1.007 & - & AZT-11 & | & 15-Sep-2010 & 455.5121 & 10.815 & 1.186 & 1.062 & - & AZT-11 \\
					08-Nov-2009 & 144.5302 & 10.741 & 1.164 & 1.041 & - & AZT-11 & | & 15-Sep-2010 & 455.5190 & 10.815 & 1.182 & 1.062 & - & AZT-11 \\
					08-Nov-2009 & 144.5420 & 10.732 & 1.162 & 1.031 & - & AZT-11 & | & 15-Sep-2010 & 455.5258 & 10.811 & 1.179 & 1.083 & - & AZT-11 \\
					08-Nov-2009 & 144.6125 & 10.765 & 1.158 & 1.045 & - & AZT-11 & | & 16-Sep-2010 & 456.5575 & 10.802 & 1.153 & 1.041 & - & AZT-11 \\
					09-Nov-2009 & 145.3064 & 10.847 & 1.192 & 1.040 & - & AZT-11 & | & 17-Sep-2010 & 457.5121 & 10.787 & 1.174 & 1.055 & - & AZT-11 \\
					09-Nov-2009 & 145.5874 & 10.915 & 1.180 & 1.051 & - & AZT-11 & | & 17-Sep-2010 & 457.5186 & 10.779 & 1.178 & 1.030 & - & AZT-11 \\
					09-Nov-2009 & 145.6245 & 10.862 & 1.178 & 1.053 & - & AZT-11 & | & 17-Sep-2010 & 457.5251 & 10.779 & 1.187 & 1.063 & - & AZT-11 \\
					10-Nov-2009 & 146.5472 & 10.759 & 1.176 & 1.021 & - & AZT-11 & | & 18-Sep-2010 & 458.5115 & 10.770 & 1.160 & 1.020 & - & AZT-11 \\
					10-Nov-2009 & 146.5560 & 10.757 & 1.174 & 1.017 & - & AZT-11 & | & 18-Sep-2010 & 458.5181 & 10.773 & 1.145 & 1.033 & - & AZT-11 \\
					23-Nov-2009 & 159.4806 & 10.717 & 1.160 & 1.019 & - & AZT-11 & | & 18-Sep-2010 & 458.5246 & 10.772 & 1.151 & 1.049 & - & AZT-11 \\
					23-Nov-2009 & 159.4856 & 10.719 & 1.160 & 1.024 & - & AZT-11 & | & 12-Oct-2010 & 482.5118 & 10.842 & 1.172 & 1.034 & - & AZT-11 \\
					21-Dec-2009 & 187.2062 & 10.727 & 1.157 & 1.025 & - & AZT-11 & | & 12-Oct-2010 & 482.5180 & 10.844 & 1.178 & 1.039 & - & AZT-11 \\
					21-Dec-2009 & 187.2141 & 10.729 & 1.147 & 1.030 & - & AZT-11 & | & 12-Oct-2010 & 482.5245 & 10.848 & 1.158 & 1.057 & - & AZT-11 \\
					21-Dec-2009 & 187.2227 & 10.735 & 1.136 & 1.030 & - & AZT-11 & | & 29-Oct-2010 & 499.5370 & 10.819 & 1.164 & 1.056 & - & AZT-11 \\
					21-Dec-2009 & 187.3067 & 10.712 & 1.144 & 1.021 & - & AZT-11 & | & 30-Oct-2010 & 500.5347 & 10.806 & 1.177 & 1.042 & - & AZT-11 \\
					21-Dec-2009 & 187.3151 & 10.715 & 1.150 & 1.012 & - & AZT-11 & | & 01-Nov-2010 & 502.5591 & 10.778 & 1.167 & 1.042 & - & AZT-11 \\
					21-Dec-2009 & 187.3225 & 10.708 & 1.148 & 1.018 & - & AZT-11 & | & 05-Nov-2010 & 506.5768 & 10.776 & 1.172 & 1.051 & - & AZT-11 \\
					24-Dec-2009 & 190.2749 & 10.882 & 1.192 & 1.055 & - & AZT-11 & | & 06-Nov-2010 & 507.5761 & 10.778 & 1.172 & 1.029 & - & AZT-11 \\
					24-Dec-2009 & 190.2824 & 10.892 & 1.191 & 1.076 & - & AZT-11 & | & 13-Nov-2010 & 514.5477 & 10.811 & 1.163 & 1.045 & - & AZT-11 \\
					24-Dec-2009 & 190.3358 & 10.908 & 1.189 & 1.058 & - & AZT-11 & | & 14-Nov-2010 & 515.5455 & 10.805 & 1.172 & 1.018 & - & AZT-11 \\
					26-Jan-2010 & 223.2694 & 10.772 & 1.186 & 1.041 & - & AZT-11 & | & 17-Nov-2010 & 518.5531 & 10.772 & 1.152 & 1.014 & - & AZT-11 \\
					26-Jan-2010 & 223.2732 & 10.766 & 1.171 & 1.039 & - & AZT-11 & | & 18-Nov-2010 & 520.4961 & 10.747 & 1.157 & 1.039 & - & AZT-11 \\
					26-Jan-2010 & 223.2771 & 10.770 & 1.178 & 1.041 & - & AZT-11 & | & 18-Nov-2010 & 520.4939 & 10.752 & 1.155 & 1.024 & - & AZT-11 \\
					26-Jan-2010 & 223.2814 & 10.775 & 1.164 & 1.047 & - & AZT-11 & | & 08-Dec-2010 & 539.4696 & 10.815 & 1.164 & 1.052 & - & AZT-11 \\
					22-Feb-2010 & 250.2121 & 10.884 & 1.187 & 1.048 & - & AZT-11 & | & 08-Dec-2010 & 539.4811 & 10.818 & 1.173 & 1.053 & - & AZT-11 \\
					22-Feb-2010 & 250.2159 & 10.908 & 1.155 & 1.027 & - & AZT-11 & | & 08-Dec-2010 & 539.4859 & 10.813 & 1.179 & 1.052 & - & AZT-11 \\
					22-Feb-2010 & 250.2197 & 10.890 & 1.142 & 1.078 & - & AZT-11 & | & & & & & & & \\
					22-Feb-2010 & 250.2236 & 10.913 & 1.169 & 1.064 & - & AZT-11 & | & & & & & & & \\
				\end{tabular}}
			\end{table}
		\end{landscape}
	\begin{landscape}
			\begin{table}
				\ContinuedFloat
   				\caption{(Continued from previous page).}
				\subfloat[Photometric measurements of the set 11b+12a.]{\begin{tabular}{ccccccccccccccc}
					Date & HJD (2455000+) & V (mag) & B-V & V-R$_{\rm J}$ & V-I$_{\rm J}$ & Telescope & | & Date & HJD (2455000+) & V (mag) & B-V & V-R$_{\rm J}$ & V-I$_{\rm J}$ & Telescope \\
					\hline
					29-Jul-2011 & 772.5187 & 10.801 & 1.154 & 1.049 & 1.767 & AZT-11 & | & 23-Sep-2011 & 828.5423 & 10.848 & 1.193 & 1.059 & 1.778 & AZT-11 \\
					02-Aug-2011 & 776.5394 & 10.816 & 1.176 & 1.030 & 1.762 & AZT-11 & | & 23-Sep-2011 & 828.5512 & 10.857 & 1.184 & 1.048 & 1.787 & AZT-11 \\
					03-Aug-2011 & 777.5088 & 10.847 & 1.183 & 1.059 & 1.787 & AZT-11 & | & 26-Sep-2011 & 831.5396 & 10.883 & 1.180 & 1.057 & 1.801 & AZT-11 \\
					03-Aug-2011 & 777.5158 & 10.834 & 1.203 & 1.050 & 1.780 & AZT-11 & | & 26-Sep-2011 & 831.5461 & 10.878 & 1.185 & 1.056 & 1.801 & AZT-11 \\
					03-Aug-2011 & 777.5221 & 10.844 & 1.191 & 1.069 & 1.794 & AZT-11 & | & 26-Sep-2011 & 831.5527 & 10.883 & 1.187 & 1.062 & 1.803 & AZT-11 \\
					05-Aug-2011 & 779.5239 & 10.823 & 1.185 & 1.042 & 1.783 & AZT-11 & | & 28-Sep-2011 & 833.5811 & 10.877 & 1.198 & 1.059 & 1.803 & AZT-11 \\
					05-Aug-2011 & 779.5312 & 10.826 & 1.184 & 1.035 & 1.789 & AZT-11 & | & 28-Sep-2011 & 833.5880 & 10.878 & 1.185 & 1.072 & 1.799 & AZT-11 \\
					05-Aug-2011 & 779.5383 & 10.831 & 1.186 & 1.054 & 1.790 & AZT-11 & | & 28-Sep-2011 & 833.5944 & 10.893 & 1.187 & 1.068 & 1.815 & AZT-11 \\
					06-Aug-2011 & 780.5127 & 10.742 & 1.163 & 1.012 & 1.715 & AZT-11 & | & 02-Oct-2011 & 837.5672 & 10.844 & 1.194 & 1.059 & 1.783 & AZT-11 \\
					06-Aug-2011 & 780.5191 & 10.742 & 1.148 & 1.021 & 1.720 & AZT-11 & | & 02-Oct-2011 & 837.5745 & 10.832 & 1.188 & 1.055 & 1.782 & AZT-11 \\
					06-Aug-2011 & 780.5259 & 10.738 & 1.148 & 1.025 & 1.718 & AZT-11 & | & 02-Oct-2011 & 837.5813 & 10.848 & 1.182 & 1.059 & 1.795 & AZT-11 \\
					07-Aug-2011 & 781.5278 & 10.815 & 1.199 & 1.035 & 1.773 & AZT-11 & | & 05-Oct-2011 & 840.5208 & 10.757 & 1.154 & 1.041 & 1.730 & AZT-11 \\
					23-Aug-2011 & 797.5141 & 10.775 & 1.158 & 1.029 & 1.734 & AZT-11 & | & 05-Oct-2011 & 840.5275 & 10.758 & 1.156 & 1.028 & 1.737 & AZT-11 \\
					23-Aug-2011 & 797.5221 & 10.755 & 1.150 & 1.027 & 1.729 & AZT-11 & | & 05-Oct-2011 & 840.5341 & 10.766 & 1.150 & 1.027 & 1.741 & AZT-11 \\
					23-Aug-2011 & 797.5289 & 10.758 & 1.152 & 1.031 & 1.731 & AZT-11 & | & 20-Oct-2011 & 855.6103 & 10.801 & 1.179 & 1.071 & 1.766 & AZT-11 \\
					23-Aug-2011 & 797.5168 & 10.765 & 1.190 & 1.019 & 1.725 & AZT-11 & | & 20-Oct-2011 & 855.6176 & 10.821 & 1.166 & 1.045 & 1.757 & AZT-11 \\
					23-Aug-2011 & 797.5241 & 10.755 & 1.158 & 1.033 & 1.721 & AZT-11 & | & 20-Oct-2011 & 855.6247 & 10.808 & 1.197 & 1.045 & 1.761 & AZT-11 \\
					23-Aug-2011 & 797.5308 & 10.756 & 1.160 & 1.035 & 1.721 & AZT-11 & | & 22-Oct-2011 & 857.5245 & 10.834 & 1.151 & 1.048 & 1.770 & AZT-11 \\
					24-Aug-2011 & 798.5194 & 10.831 & 1.173 & 1.054 & 1.786 & AZT-11 & | & 04-Nov-2011 & 870.6130 & 10.816 & 1.160 & 1.062 & 1.780 & AZT-11 \\
					24-Aug-2011 & 798.5278 & 10.828 & 1.182 & 1.050 & 1.783 & AZT-11 & | & 06-Nov-2011 & 872.6375 & 10.882 & 1.173 & 1.066 & 1.799 & AZT-11 \\
					24-Aug-2011 & 798.5347 & 10.828 & 1.177 & 1.054 & 1.781 & AZT-11 & | & 07-Nov-2011 & 873.5742 & 10.854 & 1.178 & 1.062 & 1.781 & AZT-11 \\
					25-Aug-2011 & 799.5576 & 10.815 & 1.166 & 1.049 & 1.773 & AZT-11 & | & 19-Nov-2011 & 885.5749 & 10.793 & 1.149 & 1.046 & 1.761 & AZT-11 \\
					25-Aug-2011 & 799.5638 & 10.821 & 1.163 & 1.054 & 1.771 & AZT-11 & | & 23-Nov-2011 & 889.6055 & 10.908 & 1.205 & 1.051 & 1.807 & AZT-11 \\
					27-Aug-2011 & 801.5312 & 10.864 & 1.181 & 1.060 & 1.797 & AZT-11 & | & 25-Nov-2011 & 891.5719 & 10.878 & 1.183 & 1.080 & 1.803 & AZT-11 \\
					27-Aug-2011 & 801.5385 & 10.869 & 1.175 & 1.051 & 1.794 & AZT-11 & | & 25-Nov-2011 & 891.5787 & 10.893 & 1.196 & 1.080 & 1.801 & AZT-11 \\
					27-Aug-2011 & 801.5457 & 10.861 & 1.180 & 1.057 & 1.793 & AZT-11 & | & 25-Nov-2011 & 891.5855 & 10.867 & 1.196 & 1.075 & 1.799 & AZT-11 \\
					29-Aug-2011 & 803.5316 & 10.877 & 1.191 & 1.056 & 1.797 & AZT-11 & | & 28-Nov-2011 & 894.5180 & 10.741 & 1.154 & 1.020 & 1.719 & AZT-11 \\
					29-Aug-2011 & 803.5380 & 10.876 & 1.198 & 1.053 & 1.803 & AZT-11 & | & 02-Dec-2011 & 898.2574 & 10.732 & 1.144 & 1.030 & 1.714 & AZT-11 \\
					29-Aug-2011 & 803.5445 & 10.872 & 1.190 & 1.050 & 1.797 & AZT-11 & | & 02-Dec-2011 & 898.3005 & 10.727 & 1.127 & 1.017 & 1.726 & AZT-11 \\
					01-Sep-2011 & 806.4727 & 10.818 & 1.142 & 1.036 & 1.761 & AZT-11 & | & 03-Dec-2011 & 899.5691 & 10.841 & 1.193 & 1.067 & 1.797 & AZT-11 \\
					01-Sep-2011 & 806.4800 & 10.809 & 1.159 & 1.026 & 1.753 & AZT-11 & | & 04-Dec-2011 & 900.5568 & 10.787 & 1.174 & 1.022 & 1.757 & AZT-11 \\
					01-Sep-2011 & 806.4873 & 10.798 & 1.159 & 1.034 & 1.746 & AZT-11 & | & 28-Dec-2011 & 924.5089 & 10.722 & 1.157 & 1.006 & 1.718 & AZT-11 \\
					03-Sep-2011 & 808.5388 & 10.739 & 1.142 & 1.009 & 1.719 & AZT-11 & | & 28-Dec-2011 & 924.5157 & 10.717 & 1.159 & 1.010 & 1.716 & AZT-11 \\
					03-Sep-2011 & 808.5456 & 10.743 & 1.141 & 1.003 & 1.723 & AZT-11 & | & 19-Jan-2012 & 946.2393 & 10.817 & 1.183 & 1.058 & 1.780 & AZT-11 \\
					03-Sep-2011 & 808.5525 & 10.752 & 1.142 & 1.026 & 1.722 & AZT-11 & | & 19-Jan-2012 & 946.2463 & 10.819 & 1.177 & 1.056 & 1.781 & AZT-11 \\
					06-Sep-2011 & 811.5790 & 10.832 & 1.183 & 1.059 & 1.786 & AZT-11 & | & 28-Jan-2012 & 955.2345 & 10.869 & 1.175 & 1.058 & 1.798 & AZT-11 \\
					23-Sep-2011 & 828.5314 & 10.848 & 1.189 & 1.059 & 1.787 & AZT-11 & | & & & & & & & \\
				\end{tabular}}
			\end{table}
		\end{landscape}
	\begin{landscape}
			\begin{table}
				\ContinuedFloat
				\caption{(Continued from previous page).}
				\subfloat[Photometric measurements of the set 12b+13a.]{\begin{tabular}{ccccccccccccccc}
					Date & HJD (2456000+) & V (mag) & B-V & V-R$_{\rm J}$ & V-I$_{\rm J}$ & Telescope & | & Date & HJD (2456000+) & V (mag) & B-V & V-R$_{\rm J}$ & V-I$_{\rm J}$ & Telescope \\
					\hline
					13-Aug-2012 & 153.4989 & 10.801 & 1.166 & 1.036 & 1.762 & AZT-11 & | & 16-Oct-2012 & 217.5183 & 10.802 & 1.190 & 1.048 & 1.789 & AZT-11 \\
					13-Aug-2012 & 153.5059 & 10.836 & 1.195 & 1.078 & 1.768 & AZT-11 & | & 17-Oct-2012 & 218.6062 & 10.732 & 1.164 & 1.021 & 1.735 & AZT-11 \\
					13-Aug-2012 & 153.5129 & 10.820 & 1.187 & 1.063 & 1.762 & AZT-11 & | & 17-Oct-2012 & 218.6137 & 10.740 & 1.156 & 1.020 & 1.742 & AZT-11 \\
					18-Aug-2012 & 158.4723 & 10.635 & 1.121 & 1.013 & 1.674 & AZT-11 & | & 17-Oct-2012 & 218.6220 & 10.752 & 1.155 & 1.037 & 1.745 & AZT-11 \\
					18-Aug-2012 & 158.4829 & 10.695 & 1.139 & 1.028 & 1.720 & AZT-11 & | & 20-Oct-2012 & 221.6156 & 10.860 & 1.193 & 1.062 & 1.803 & AZT-11 \\
					18-Aug-2012 & 158.4913 & 10.670 & 1.121 & 1.022 & 1.705 & AZT-11 & | & 21-Oct-2012 & 222.5558 & 10.799 & 1.181 & 1.049 & 1.774 & AZT-11 \\
					18-Aug-2012 & 158.4988 & 10.642 & 1.125 & 1.008 & 1.692 & AZT-11 & | & 21-Oct-2012 & 222.5627 & 10.807 & 1.165 & 1.038 & 1.782 & AZT-11 \\
					21-Aug-2012 & 161.5144 & 10.829 & 1.189 & 1.072 & 1.790 & AZT-11 & | & 21-Oct-2012 & 222.5702 & 10.791 & 1.184 & 1.050 & 1.774 & AZT-11 \\
					21-Aug-2012 & 161.5214 & 10.832 & 1.197 & 1.061 & 1.799 & AZT-11 & | & 23-Oct-2012 & 224.5439 & 10.804 & 1.188 & 1.060 & 1.781 & AZT-11 \\
					21-Aug-2012 & 161.5306 & 10.824 & 1.177 & 1.056 & 1.802 & AZT-11 & | & 23-Oct-2012 & 224.5509 & 10.809 & 1.183 & 1.042 & 1.777 & AZT-11 \\
					22-Aug-2012 & 162.5606 & 10.768 & 1.171 & 1.012 & 1.751 & AZT-11 & | & 23-Oct-2012 & 224.5577 & 10.817 & 1.186 & 1.050 & 1.780 & AZT-11 \\
					22-Aug-2012 & 162.5672 & 10.778 & 1.167 & 1.041 & 1.756 & AZT-11 & | & 25-Oct-2012 & 226.5496 & 10.793 & 1.177 & 1.040 & 1.766 & AZT-11 \\
					23-Aug-2012 & 163.5296 & 10.780 & 1.164 & 1.035 & 1.765 & AZT-11 & | & 25-Oct-2012 & 226.5576 & 10.788 & 1.177 & 1.034 & 1.764 & AZT-11 \\
					23-Aug-2012 & 163.5560 & 10.815 & 1.193 & 1.064 & 1.774 & AZT-11 & | & 25-Oct-2012 & 226.5645 & 10.800 & 1.176 & 1.047 & 1.775 & AZT-11 \\
					24-Aug-2012 & 164.5151 & 10.790 & 1.162 & 1.042 & 1.765 & AZT-11 & | & 26-Oct-2012 & 227.5960 & 10.756 & 1.165 & 1.035 & 1.751 & AZT-11 \\
					24-Aug-2012 & 164.5221 & 10.789 & 1.168 & 1.033 & 1.764 & AZT-11 & | & 09-Nov-2012 & 241.3498 & 10.818 & 1.176 & 1.052 & 1.776 & AZT-11 \\
					24-Aug-2012 & 164.5288 & 10.794 & 1.169 & 1.047 & 1.769 & AZT-11 & | & 09-Nov-2012 & 241.3572 & 10.784 & 1.176 & 1.034 & 1.776 & AZT-11 \\
					25-Aug-2012 & 165.4764 & 10.816 & 1.173 & 1.043 & 1.776 & AZT-11 & | & 11-Nov-2012 & 243.4725 & 10.782 & 1.159 & 1.045 & 1.759 & AZT-11 \\
					25-Aug-2012 & 165.4836 & 10.801 & 1.175 & 1.046 & 1.767 & AZT-11 & | & 11-Nov-2012 & 243.4829 & 10.780 & 1.168 & 1.052 & 1.757 & AZT-11 \\
					25-Aug-2012 & 165.4907 & 10.808 & 1.165 & 1.053 & 1.776 & AZT-11 & | & 11-Nov-2012 & 243.4986 & 10.783 & 1.164 & 1.050 & 1.759 & AZT-11 \\
					26-Aug-2012 & 166.5257 & 10.809 & 1.178 & 1.042 & 1.777 & AZT-11 & | & 12-Nov-2012 & 244.4816 & 10.730 & 1.168 & 1.027 & 1.746 & AZT-11 \\
					26-Aug-2012 & 166.5327 & 10.809 & 1.173 & 1.040 & 1.778 & AZT-11 & | & 12-Nov-2012 & 244.4910 & 10.732 & 1.167 & 1.024 & 1.746 & AZT-11 \\
					26-Aug-2012 & 166.5402 & 10.812 & 1.190 & 1.047 & 1.774 & AZT-11 & | & 12-Nov-2012 & 244.5133 & 10.734 & 1.161 & 1.040 & 1.744 & AZT-11 \\
					31-Aug-2012 & 171.5560 & 10.716 & 1.154 & 1.017 & 1.737 & AZT-11 & | & 17-Nov-2012 & 249.4379 & 10.843 & 1.200 & 1.067 & 1.792 & AZT-11 \\
					31-Aug-2012 & 171.5628 & 10.719 & 1.155 & 1.015 & 1.739 & AZT-11 & | & 17-Nov-2012 & 249.4459 & 10.847 & 1.193 & 1.059 & 1.798 & AZT-11 \\
					31-Aug-2012 & 171.5692 & 10.716 & 1.169 & 1.039 & 1.739 & AZT-11 & | & 17-Nov-2012 & 249.4524 & 10.849 & 1.195 & 1.069 & 1.797 & AZT-11 \\
					11-Sep-2012 & 182.5588 & 10.775 & 1.167 & 1.043 & 1.761 & AZT-11 & | & 08-Dec-2012 & 270.4771 & 10.813 & 1.176 & 1.047 & 1.778 & AZT-11 \\
					11-Sep-2012 & 182.5663 & 10.791 & 1.161 & 1.042 & 1.772 & AZT-11 & | & 08-Dec-2012 & 270.4845 & 10.808 & 1.173 & 1.052 & 1.778 & AZT-11 \\
					13-Sep-2012 & 184.5151 & 10.771 & 1.168 & 1.047 & 1.760 & AZT-11 & | & 08-Dec-2012 & 270.4924 & 10.796 & 1.165 & 1.057 & 1.778 & AZT-11 \\
					13-Sep-2012 & 184.5232 & 10.748 & 1.160 & 1.031 & 1.753 & AZT-11 & | & 31-Dec-2012 & 293.2226 & 10.681 & 1.150 & 1.025 & 1.726 & AZT-11 \\
					13-Sep-2012 & 184.5295 & 10.758 & 1.162 & 1.041 & 1.763 & AZT-11 & | & 31-Dec-2012 & 293.2294 & 10.681 & 1.162 & 1.020 & 1.716 & AZT-11 \\
					17-Sep-2012 & 188.5117 & 10.712 & 1.147 & 1.005 & 1.750 & AZT-11 & | & 01-Jan-2013 & 294.1756 & 10.799 & 1.183 & 1.036 & 1.759 & AZT-11 \\
					17-Sep-2012 & 188.5201 & 10.702 & 1.162 & 1.012 & 1.727 & AZT-11 & | & 01-Jan-2013 & 294.1848 & 10.809 & 1.175 & 1.050 & 1.773 & AZT-11 \\
					17-Sep-2012 & 188.5269 & 10.721 & 1.145 & 1.009 & 1.739 & AZT-11 & | & 01-Jan-2013 & 294.1998 & 10.804 & 1.176 & 1.039 & 1.765 & AZT-11 \\
					23-Sep-2012 & 194.5221 & 10.809 & 1.177 & 1.042 & 1.782 & AZT-11 & | & 14-Jan-2013 & 307.1871 & 10.784 & 1.166 & 1.040 & 1.752 & AZT-11 \\
					23-Sep-2012 & 194.5300 & 10.810 & 1.180 & 1.034 & 1.783 & AZT-11 & | & 14-Jan-2013 & 307.1984 & 10.786 & 1.161 & 1.032 & 1.751 & AZT-11 \\
					23-Sep-2012 & 194.5375 & 10.809 & 1.161 & 1.039 & 1.785 & AZT-11 & | & 14-Jan-2013 & 307.2176 & 10.782 & 1.162 & 1.038 & 1.754 & AZT-11 \\
					25-Sep-2012 & 196.5659 & 10.807 & 1.187 & 1.044 & 1.776 & AZT-11 & | & 19-Jan-2013 & 312.1867 & 10.727 & 1.144 & 1.031 & 1.737 & AZT-11 \\
					25-Sep-2012 & 196.5764 & 10.810 & 1.180 & 1.062 & 1.781 & AZT-11 & | & 19-Jan-2013 & 312.1936 & 10.728 & 1.150 & 1.032 & 1.732 & AZT-11 \\
					25-Sep-2012 & 196.5840 & 10.813 & 1.177 & 1.062 & 1.783 & AZT-11 & | & 19-Jan-2013 & 312.2004 & 10.728 & 1.150 & 1.022 & 1.736 & AZT-11 \\
					26-Sep-2012 & 197.5275 & 10.726 & 1.110 & 1.016 & 1.729 & AZT-11 & | & 05-Feb-2013 & 329.3020 & 10.843 & 1.178 & 1.055 & 1.784 & AZT-11 \\
					26-Sep-2012 & 197.5354 & 10.767 & 1.119 & 1.047 & 1.761 & AZT-11 & | & 09-Feb-2013 & 333.3268 & 10.792 & 1.151 & 1.051 & 1.771 & AZT-11 \\
					26-Sep-2012 & 197.5426 & 10.767 & 1.152 & 1.034 & 1.761 & AZT-11 & | & 09-Feb-2013 & 333.3352 & 10.774 & 1.155 & 1.050 & 1.773 & AZT-11 \\
				\end{tabular}}
			\end{table}
		\end{landscape}
	\begin{landscape}
			\begin{table}
				\ContinuedFloat
				\caption{(Continued from previous page).}
				\subfloat[Photometric measurements of the set 13b+14a (left) and of the set 14b (right).]{\begin{tabular}{ccccccccccccccc}
					Date & HJD (2456000+) & V (mag) & B-V & V-R$_{\rm J}$ & V-I$_{\rm J}$ & Telescope & | & Date & HJD (2456000+) & V (mag) & B-V & V-R$_{\rm J}$ & V-I$_{\rm J}$ & Telescope \\
					\hline
					15-Aug-2013 & 520.5078 & 10.786 & 1.155 & 1.041 & 1.785 & AZT-11 & | & 24-Aug-2014 & 894.5694 & 10.773 & 1.172 & 1.077 & 1.766 & AZT-11 \\
					16-Aug-2013 & 521.5032 & 10.599 & 1.139 & 0.994 & 1.688 & AZT-11 & | & 26-Aug-2014 & 896.5491 & 10.737 & 1.157 & 1.036 & 1.748 & AZT-11 \\
					18-Aug-2013 & 523.5349 & 10.630 & 1.122 & 1.015 & 1.706 & AZT-11 & | & 27-Aug-2014 & 897.5327 & 10.586 & 1.138 & 0.994 & 1.701 & AZT-11 \\
					18-Aug-2013 & 523.5419 & 10.626 & 1.118 & 1.005 & 1.703 & AZT-11 & | & 29-Aug-2014 & 899.5342 & 10.563 & 1.137 & 0.918 & 1.687 & AZT-11 \\
					18-Aug-2013 & 523.5478 & 10.641 & 1.130 & 1.027 & 1.716 & AZT-11 & | & 30-Aug-2014 & 900.5322 & 10.783 & 1.171 & 1.060 & 1.763 & AZT-11 \\
					01-Sep-2013 & 537.4939 & 10.852 & 1.183 & 1.068 & 1.803 & AZT-11 & | & 31-Aug-2014 & 901.5369 & 10.589 & 1.145 & 0.990 & 1.691 & AZT-11 \\
					01-Sep-2013 & 537.5010 & 10.848 & 1.200 & 1.045 & 1.798 & AZT-11 & | & 01-Sep-2014 & 902.5422 & 10.847 & 1.192 & 1.067 & 1.793 & AZT-11 \\
					01-Sep-2013 & 537.5084 & 10.848 & 1.172 & 1.069 & 1.793 & AZT-11 & | & 02-Sep-2014 & 903.5363 & 10.646 & 1.150 & 1.030 & 1.715 & AZT-11 \\
					11-Sep-2013 & 547.4908 & 10.706 & 1.162 & 1.028 & 1.736 & AZT-11 & | & 04-Sep-2014 & 905.4926 & 10.661 & 1.145 & 1.044 & 1.731 & AZT-11 \\
					09-Oct-2013 & 575.5352 & 10.711 & 1.164 & 1.026 & 1.737 & AZT-11 & | & 05-Sep-2014 & 906.5293 & 10.816 & 1.200 & 1.067 & 1.757 & AZT-11 \\
					09-Oct-2013 & 575.5534 & 10.706 & 1.160 & 1.021 & 1.739 & AZT-11 & | & 16-Sep-2014 & 917.5786 & 10.865 & 1.217 & 1.094 & 1.801 & AZT-11 \\
					11-Oct-2013 & 577.5338 & 10.640 & 1.169 & 1.002 & 1.709 & AZT-11 & | & 20-Sep-2014 & 921.5653 & 10.786 & 1.181 & 1.074 & 1.782 & AZT-11 \\
					13-Oct-2013 & 579.5838 & 10.587 & 1.146 & 0.990 & 1.682 & AZT-11 & | & 20-Sep-2014 & 921.5653 & 10.804 & 1.188 & 1.069 & 1.779 & AZT-11 \\
					13-Oct-2013 & 579.5907 & 10.594 & 1.141 & 1.003 & 1.689 & AZT-11 & | & 25-Sep-2014 & 926.4602 & 10.764 & 1.156 & 1.056 & 1.768 & AZT-11 \\
					13-Oct-2013 & 579.5976 & 10.602 & 1.143 & 1.005 & 1.694 & AZT-11 & | & 01-Oct-2014 & 932.6078 & 10.893 & 1.192 & 1.099 & 1.821 & AZT-11 \\
					25-Oct-2013 & 591.5467 & 10.756 & - & 1.020 & 1.751 & AZT-11 & | & 05-Oct-2014 & 936.5958 & 10.767 & 1.191 & 1.075 & 1.755 & AZT-11 \\
					25-Oct-2013 & 591.6288 & 10.758 & - & 1.028 & 1.744 & AZT-11 & | & 15-Oct-2014 & 946.5919 & 10.637 & 1.165 & 1.021 & 1.707 & AZT-11 \\
					26-Oct-2013 & 592.3649 & 10.791 & 1.168 & 1.009 & 1.730 & AZT-11 & | & 19-Oct-2014 & 950.6129 & 10.676 & 1.161 & 1.053 & 1.764 & AZT-11 \\
					27-Oct-2013 & 593.3578 & 10.805 & 1.166 & 1.067 & 1.788 & AZT-11 & | & 26-Oct-2014 & 957.4846 & 10.607 & 1.153 & - & 1.700 & AZT-11 \\
					27-Oct-2013 & 593.5660 & 10.822 & 1.182 & 1.062 & 1.796 & AZT-11 & | & 28-Oct-2014 & 959.5475 & 10.620 & 1.151 & 0.989 & 1.715 & AZT-11 \\
					29-Oct-2013 & 595.5280 & 10.857 & 1.196 & 1.070 & 1.812 & AZT-11 & | & 02-Nov-2014 & 964.4461 & 10.894 & 1.192 & 1.089 & 1.819 & AZT-11 \\
					30-Oct-2013 & 596.5614 & 10.622 & 1.137 & 1.000 & 1.694 & AZT-11 & | & 05-Nov-2014 & 967.5922 & 10.778 & 1.176 & - & 1.776 & AZT-11 \\
					01-Nov-2013 & 598.6273 & 10.779 & 1.163 & 1.000 & 1.726 & AZT-11 & | & 05-Nov-2014 & 967.5959 & 10.792 & 1.150 & - & 1.793 & AZT-11 \\
					08-Nov-2013 & 605.4124 & 10.750 & 1.165 & 1.056 & 1.758 & AZT-11 & | & 13-Nov-2014 & 975.4212 & 10.777 & 1.189 & 1.071 & 1.768 & AZT-11 \\
					08-Nov-2013 & 605.4189 & 10.738 & 1.162 & 1.022 & 1.750 & AZT-11 & | & 14-Nov-2014 & 976.4189 & 10.626 & 1.146 & 1.017 & 1.714 & AZT-11 \\
					08-Nov-2013 & 605.4253 & 10.738 & 1.167 & 1.028 & 1.750 & AZT-11 & | & 13-Dec-2014 & 1005.4167 & 10.808 & 1.173 & 1.057 & 1.781 & AZT-11 \\
					08-Nov-2013 & 605.5687 & 10.669 & 1.153 & 1.036 & 1.726 & AZT-11 & | & 14-Dec-2014 & 1006.5163 & 10.694 & 1.183 & 1.041 & 1.750 & AZT-11 \\
					09-Nov-2013 & 606.5695 & 10.821 & 1.178 & 1.055 & 1.790 & AZT-11 & | & & & & & & & \\
					10-Nov-2013 & 607.4549 & 10.658 & 1.149 & 1.007 & 1.713 & AZT-11 & | & & & & & & & \\
					23-Nov-2013 & 620.5919 & 10.665 & 1.176 & 1.027 & 1.721 & AZT-11 & | & & & & & & & \\
					03-Dec-2013 & 630.4060 & 10.723 & 1.150 & 1.031 & 1.728 & AZT-11 & | & & & & & & & \\
					04-Feb-2014 & 693.2297 & 10.828 & 1.186 & 1.058 & 1.775 & AZT-11 & | & & & & & & & \\
					05-Feb-2014 & 694.2651 & 10.807 & 1.182 & 1.060 & 1.783 & AZT-11 & | & & & & & & & \\
					18-Feb-2014 & 707.2155 & 10.751 & 1.172 & 1.006 & 1.747 & AZT-11 & | & & & & & & & \\
					21-Mar-2014 & 738.2431 & 10.759 & 1.182 & 1.048 & 1.758 & AZT-11 & | & & & & & & & \\
					23-Mar-2014 & 740.2544 & 10.679 & 1.164 & 1.045 & 1.731 & AZT-11 & | & & & & & & & \\
				\end{tabular}}
			\end{table}
		\end{landscape}
	\begin{landscape}
			\begin{table}
				\ContinuedFloat
				\caption{(Continued from previous page).}
				\subfloat[Photometric measurements of the set 15b+16a (left) and of the set 16b+17a (right).]{\begin{tabular}{ccccccccccccccc}
					Date & HJD (2457000+) & V (mag) & B-V & V-R$_{\rm J}$ & V-I$_{\rm J}$ & Telescope & | & Date & HJD (2457000+) & V (mag) & B-V & V-R$_{\rm J}$ & V-I$_{\rm J}$ & Telescope \\
					\hline
					15-Aug-2015 & 250.5211 & 10.864 & 1.166 & 1.109 & 1.779 & AZT-11 & | & 03-Sep-2016 & 635.4773 & 10.735 & 1.156 & - & 1.742 & AZT-11 \\
					16-Aug-2015 & 251.5160 & 10.660 & 1.145 & - & 1.716 & AZT-11 & | & 08-Sep-2016 & 640.5200 & 10.759 & 1.174 & - & - & AZT-11 \\
					17-Aug-2015 & 252.5152 & 10.868 & 1.174 & 1.100 & 1.780 & AZT-11 & | & 10-Sep-2016 & 642.5399 & 10.715 & 1.165 & 0.964 & - & AZT-11 \\
					25-Aug-2015 & 260.5188 & 10.726 & 1.155 & - & 1.745 & AZT-11 & | & 12-Sep-2016 & 644.5383 & 10.714 & 1.166 & - & 1.747 & AZT-11 \\
					27-Aug-2015 & 262.5817 & 10.699 & 1.153 & - & 1.743 & AZT-11 & | & 14-Sep-2016 & 646.5414 & 10.715 & 1.166 & - & 1.741 & AZT-11 \\
					09-Sep-2015 & 275.5796 & 10.731 & 1.162 & - & - & AZT-11 & | & 19-Nov-2016 & 712.5682 & 10.849 & 1.184 & - & 1.782 & AZT-11 \\
					11-Sep-2015 & 277.5817 & 10.696 & 1.161 & - & - & AZT-11 & | & 20-Jan-2017 & 774.2898 & 10.812 & 1.170 & - & 1.768 & AZT-11 \\
					16-Sep-2015 & 282.5832 & 10.890 & 1.185 & - & - & AZT-11 & | & 23-Jan-2017 & 777.1821 & 10.806 & 1.171 & - & - & AZT-11 \\
					18-Sep-2015 & 284.5447 & 10.888 & 1.172 & - & - & AZT-11 & | & 31-Jan-2017 & 785.2764 & 10.736 & 1.149 & - & - & AZT-11 \\
					25-Sep-2015 & 291.5730 & 10.830 & 1.159 & - & 1.768 & AZT-11 & | & 12-Feb-2017 & 797.2023 & 10.827 & 1.183 & - & - & AZT-11 \\
					04-Oct-2015 & 300.4300 & 10.775 & 1.171 & - & 1.756 & AZT-11 & | & 27-Feb-2017 & 812.3250 & 10.873 & 1.167 & 0.984 & 1.795 & AZT-11 \\
					03-Nov-2015 & 330.5253 & 10.831 & 1.181 & - & 1.776 & AZT-11 & | & 09-Mar-2017 & 822.3079 & 10.738 & 1.147 & - & 1.759 & AZT-11 \\
					04-Nov-2015 & 331.5427 & 10.834 & 1.182 & - & - & AZT-11 & | & & & & & & & \\
					05-Nov-2015 & 332.5222 & 10.834 & 1.174 & - & - & AZT-11 & | & & & & & & & \\
					30-Jan-2016 & 418.3730 & 10.785 & 1.182 & - & - & AZT-11 & | & & & & & & & \\
				\end{tabular}}
			\end{table}
		\end{landscape}

	\section{Photometry analysis}
	\label{anx:pha}
	From our photometric data, we retrieved the stellar rotation period at each epoch and derived the photosphere contrast.

	To retrieve the stellar rotation period, we applied two types of models to our V magnitude curves: a periodic fit involving the fundamental frequency and the first two harmonics to each of the 9 datasets individually (as well as a periodic fit involving the fundamental frequency and the first four harmonics to the whole data set), and GPR (see Section{~}\ref{sec:rv}). Since the data sets 15b+16a and the 16b+17a are particularly small (15 and 13 points respectively) and consecutive, we grouped them together for the GPR.

\begin{figure*}
		\centering
		\subfloat[08b+09a]{\includegraphics[width=0.33\linewidth]{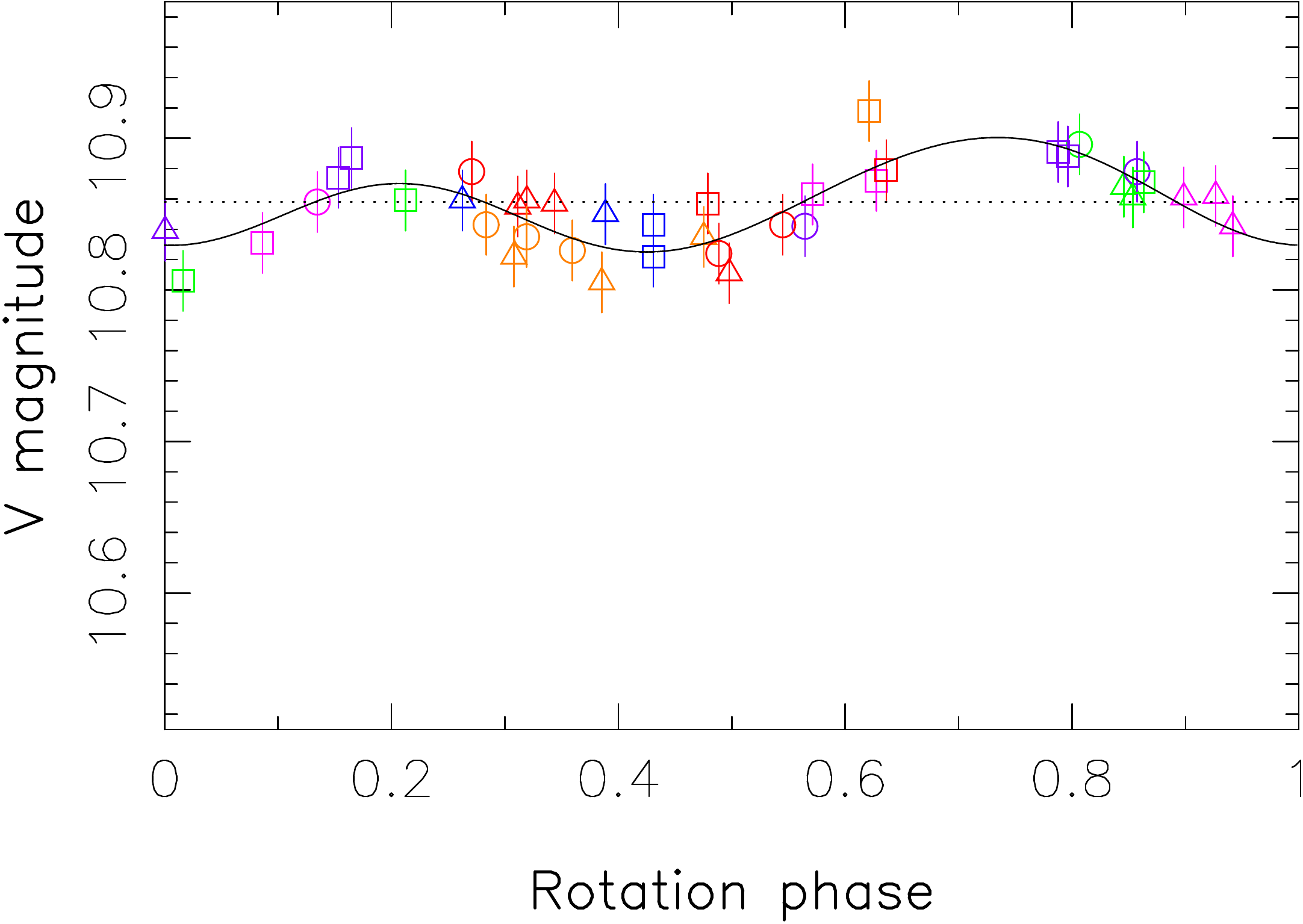}}
		\subfloat[09b+10a]{\includegraphics[width=0.33\linewidth]{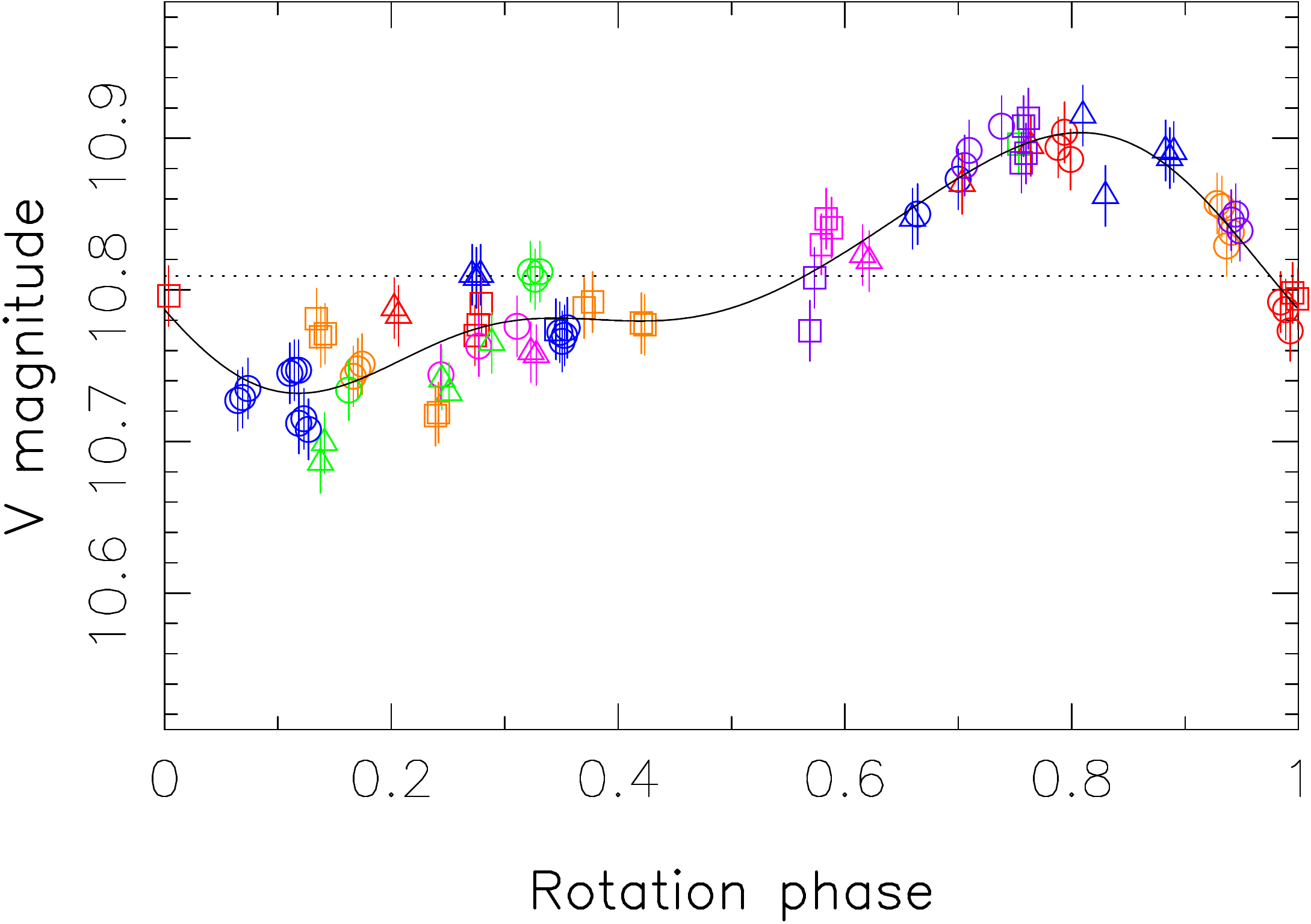}}
		\subfloat[10b]{\includegraphics[width=0.33\linewidth]{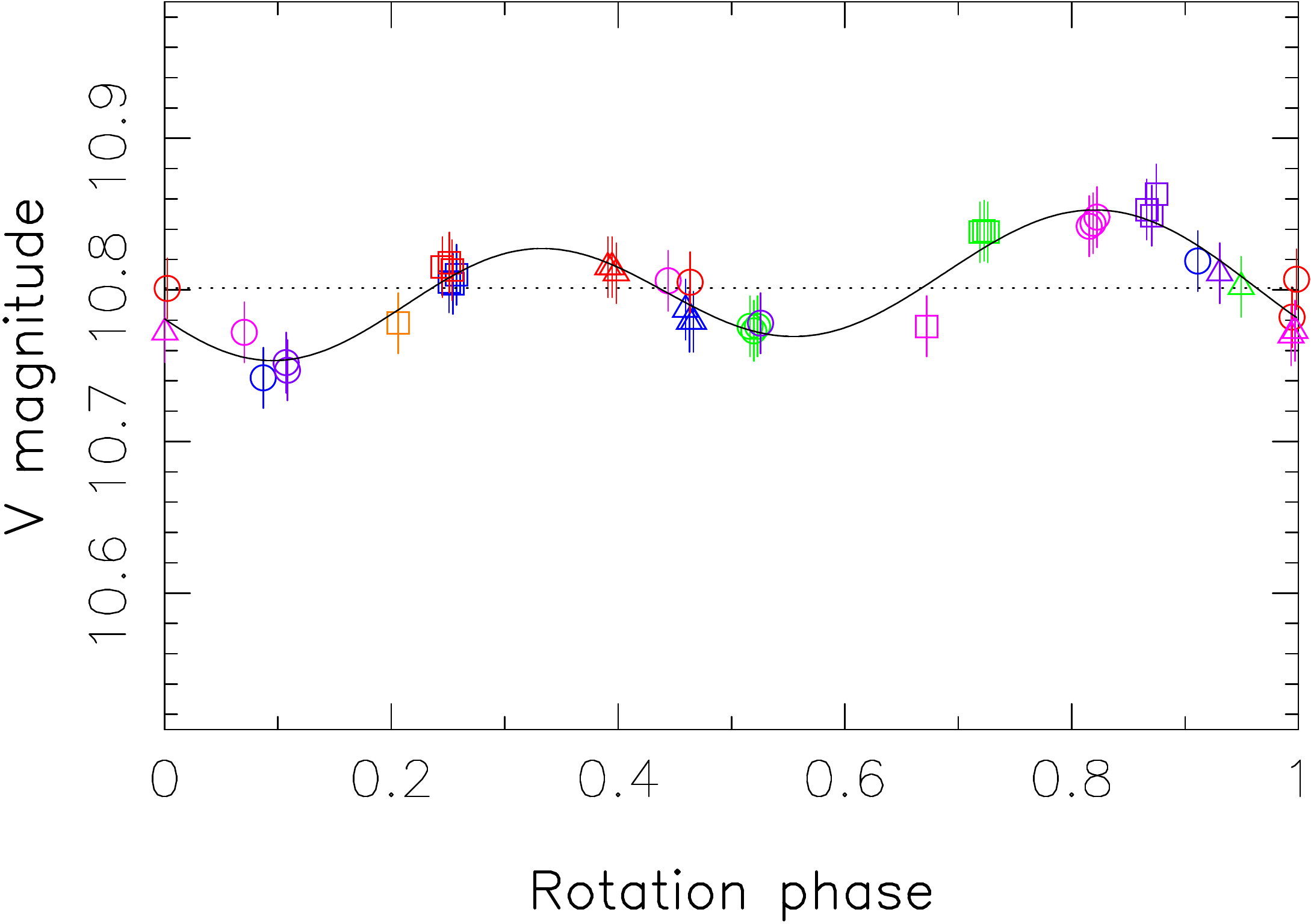}}

		\subfloat[11b+12a]{\includegraphics[width=0.33\linewidth]{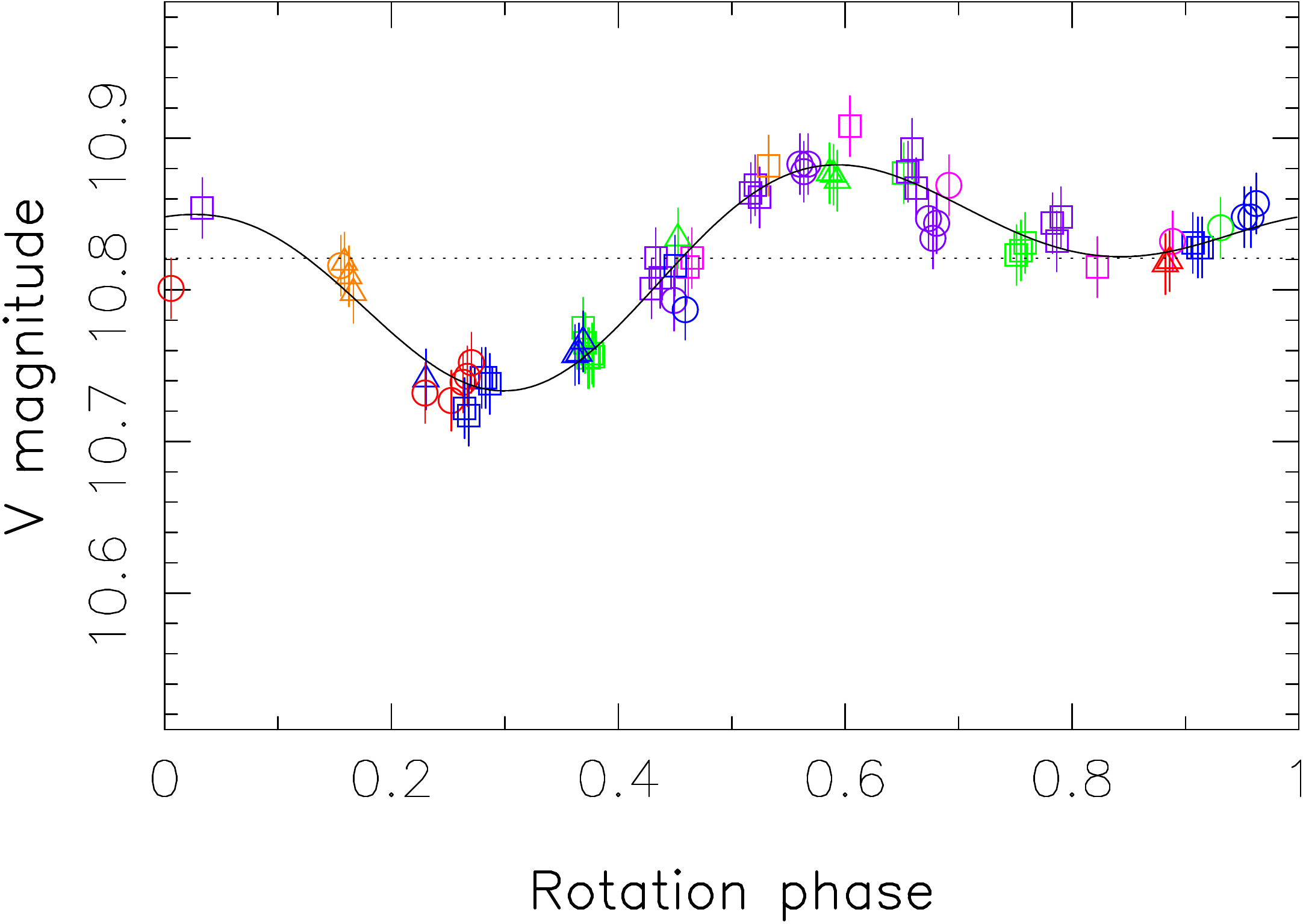}}
		\subfloat[12b+13a]{\includegraphics[width=0.33\linewidth]{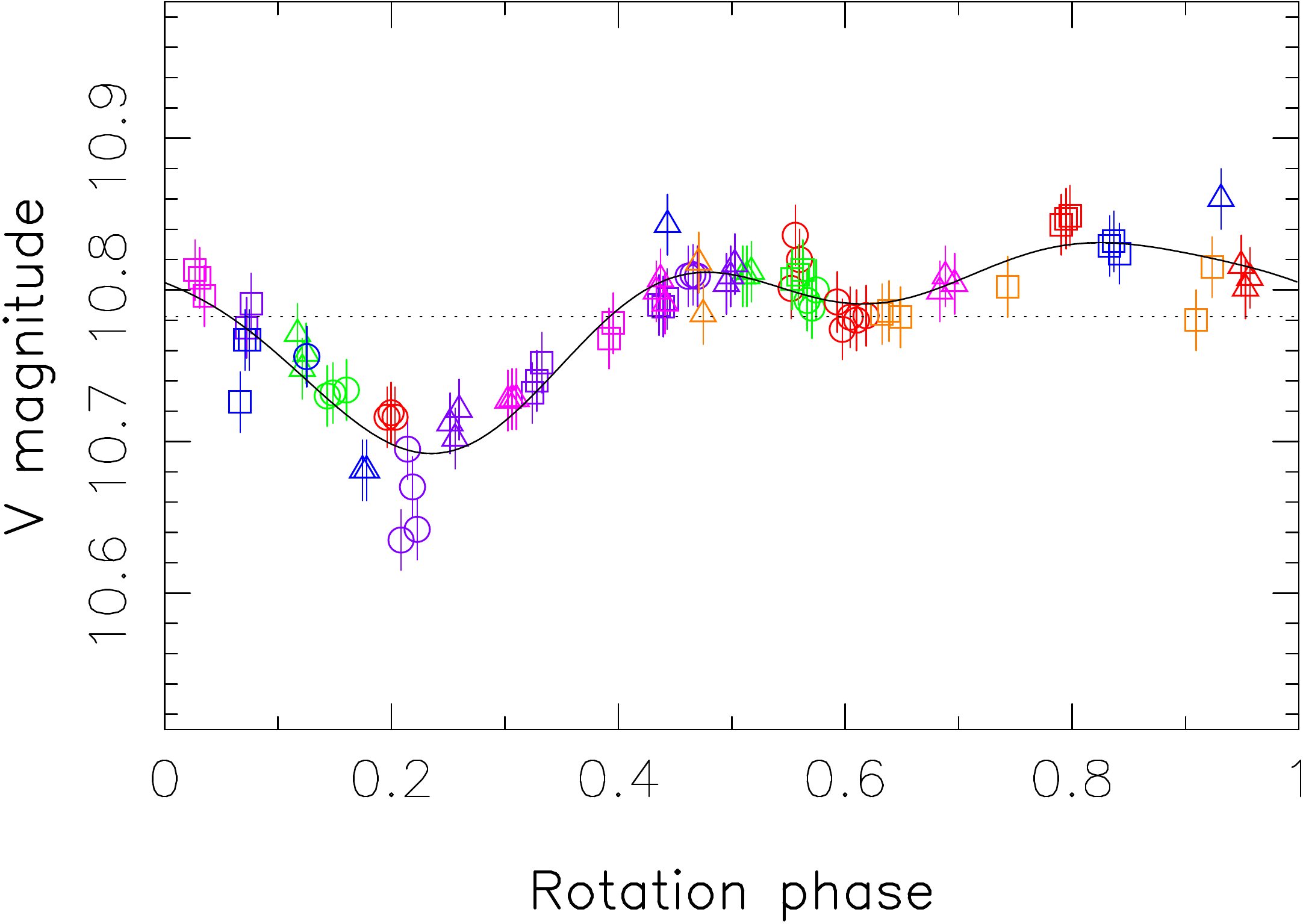}}
		\subfloat[13b+14a]{\includegraphics[width=0.33\linewidth]{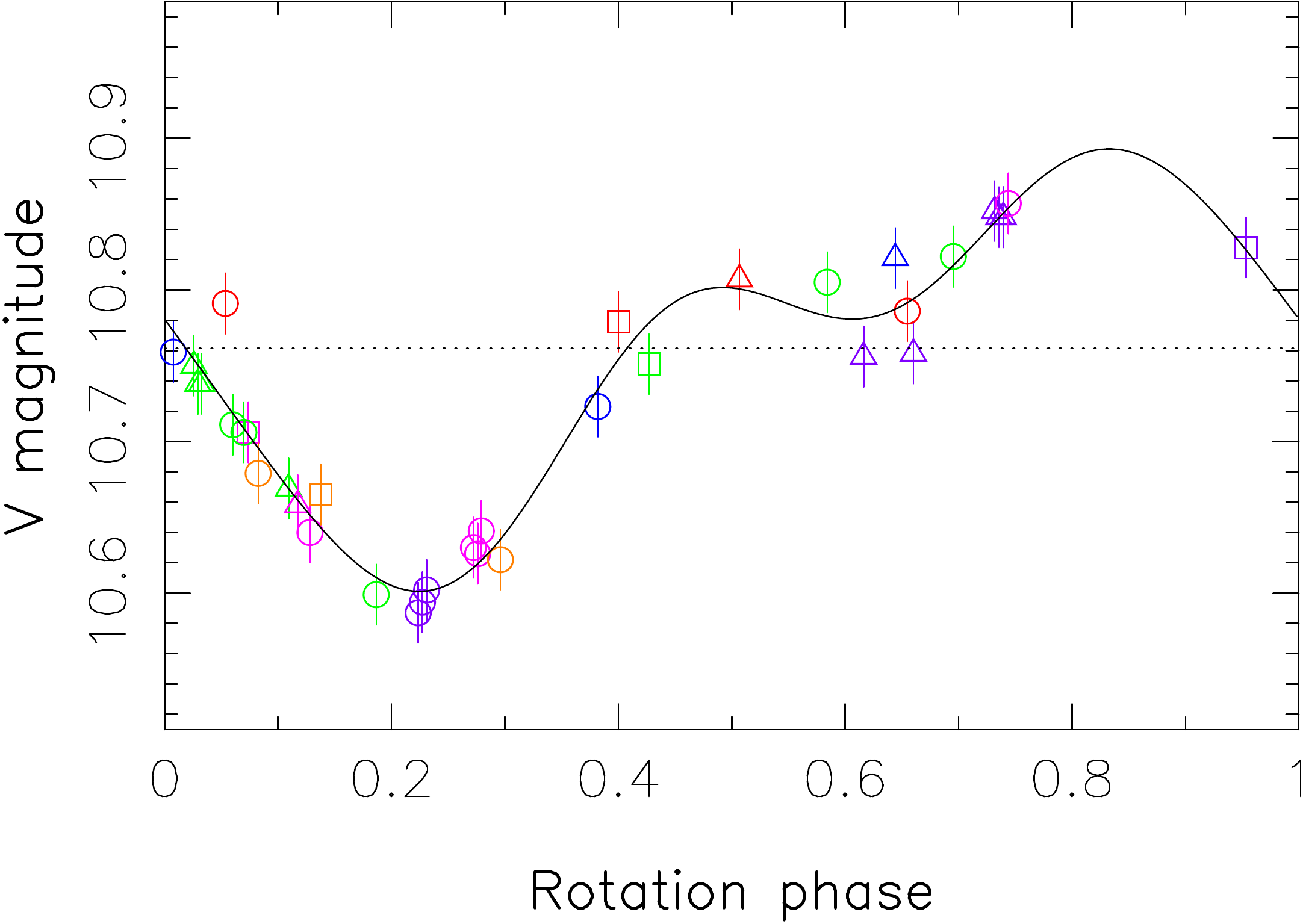}}

		\subfloat[14b]{\includegraphics[width=0.33\linewidth]{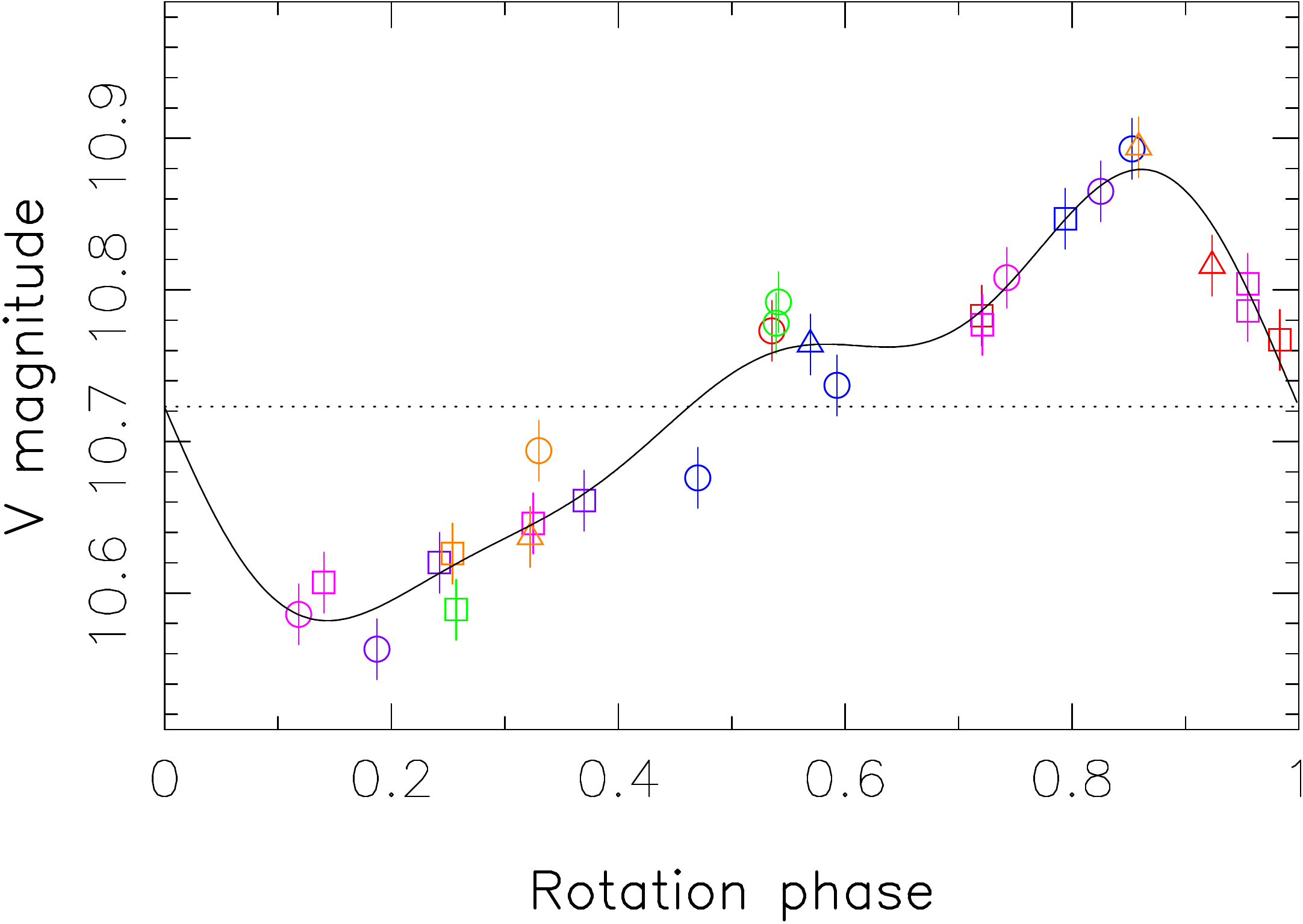}}
		\subfloat[15b+16a]{\includegraphics[width=0.33\linewidth]{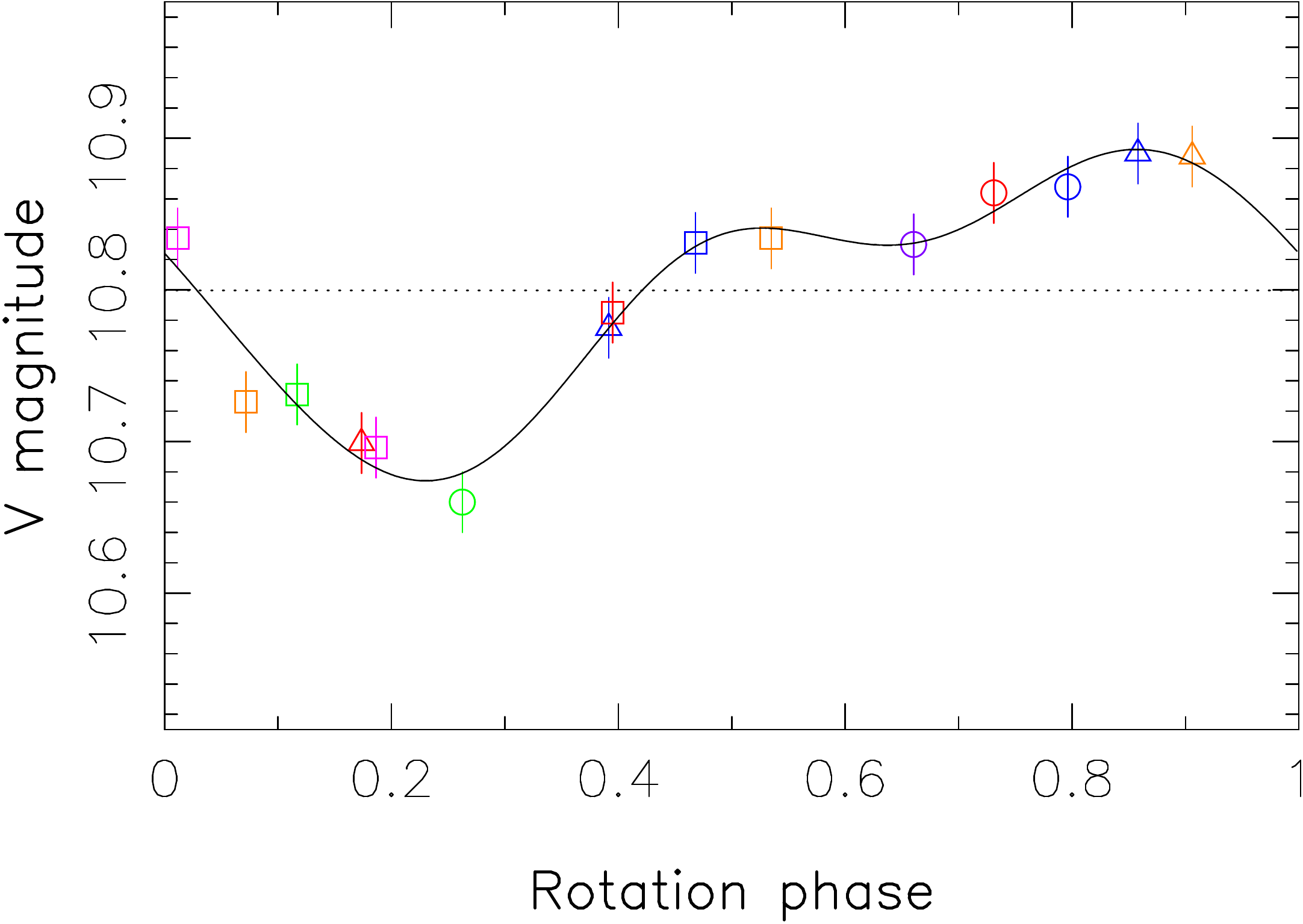}}
		\subfloat[16b+17a]{\includegraphics[width=0.33\linewidth]{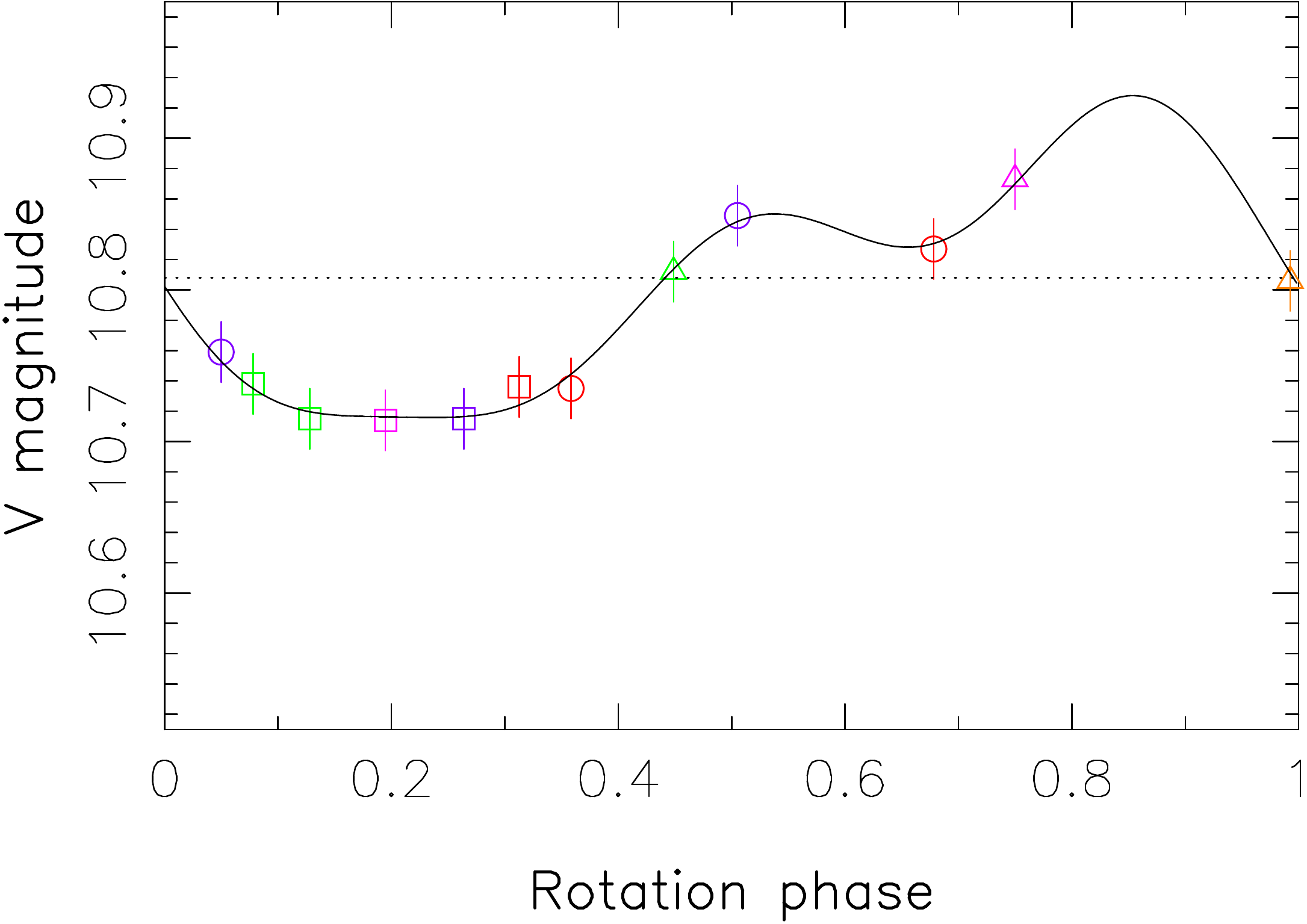}}
		\caption{Fits of the V magnitude measurements at each epoch by a sine curve and two harmonics. Each data set was folded according to the corresponding period listed in Table{~}\ref{tab:sip}. As a consequence, the rotation phase used in these plots does not correspond to the rotation phase in Table{~}\ref{tab:sob}, but rather to the phase in the model sine curves.}
		\label{fig:sp1}
	\end{figure*}

\begin{figure}
		\centering
		\includegraphics[totalheight=0.17\textheight]{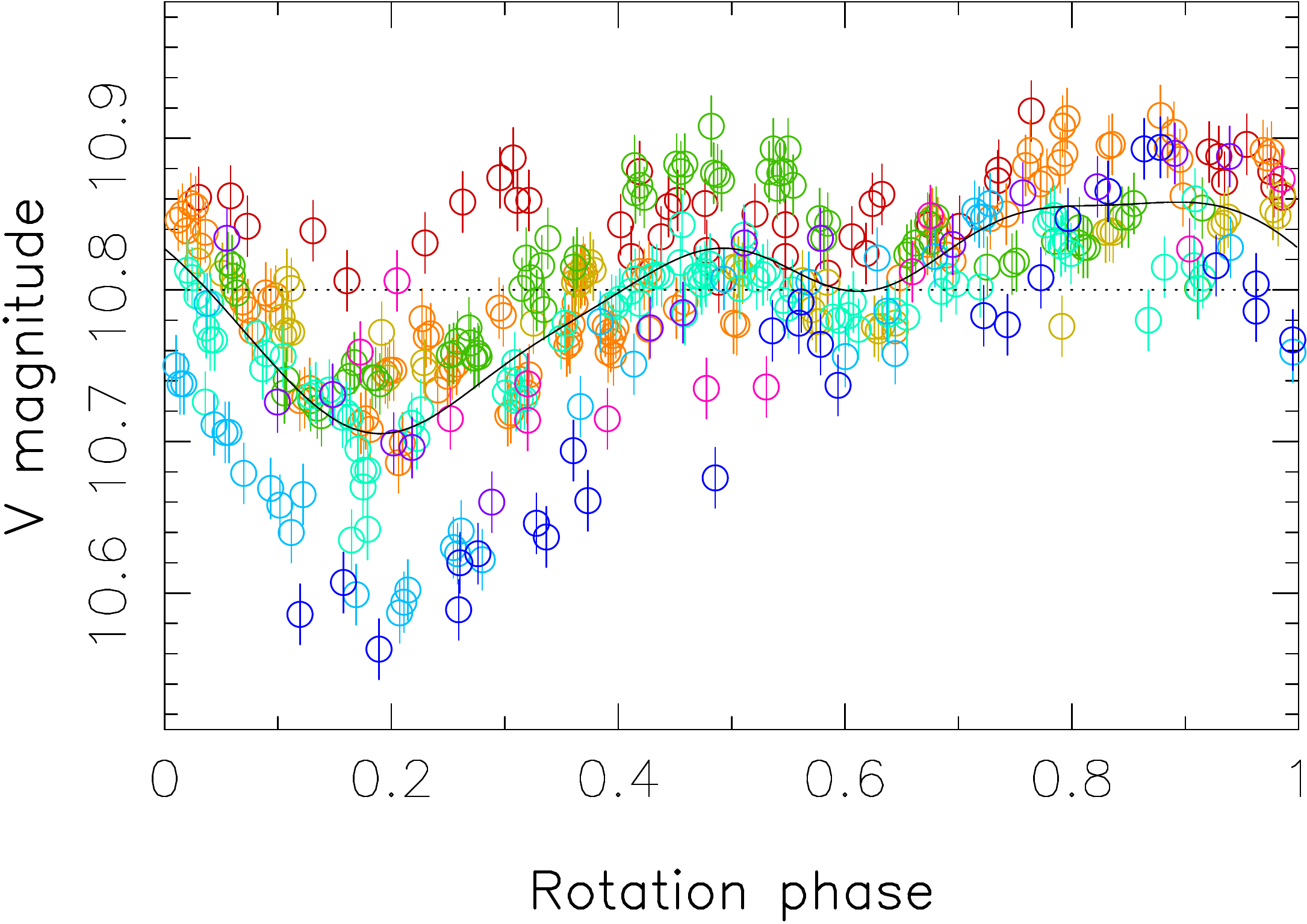}
		\caption{Fit of the V magnitude in the whole data set by a sine curve and four harmonics. The color-to-dataset correspondency is as follows: red=08b+09a, orange=09b+10a, yellow=10b, green=11b+12a, turquoise=12b+13a, cyan=13b+14a, blue=14b, purple=15b+16a, pink=16b+17a.}
		\label{fig:spa}
	\end{figure}

	The results of the sine fits are listed in Table{~}\ref{tab:sip}, and plotted in Figures{~}\ref{fig:sp1} and \ref{fig:spa}. All observation epochs yield a modulation period within 1$\sigma$ of the value we use throughout this paper for the stellar rotation period. We note that the error bar recovered on the whole data set is underestimated since it was measured on the curvature of the \chisqr(\Prot) curve around the minimum, curve which presents many aliased local minima due to the observation sampling.

\begin{figure}
		\centering
		\includegraphics[angle=-90,width=\linewidth]{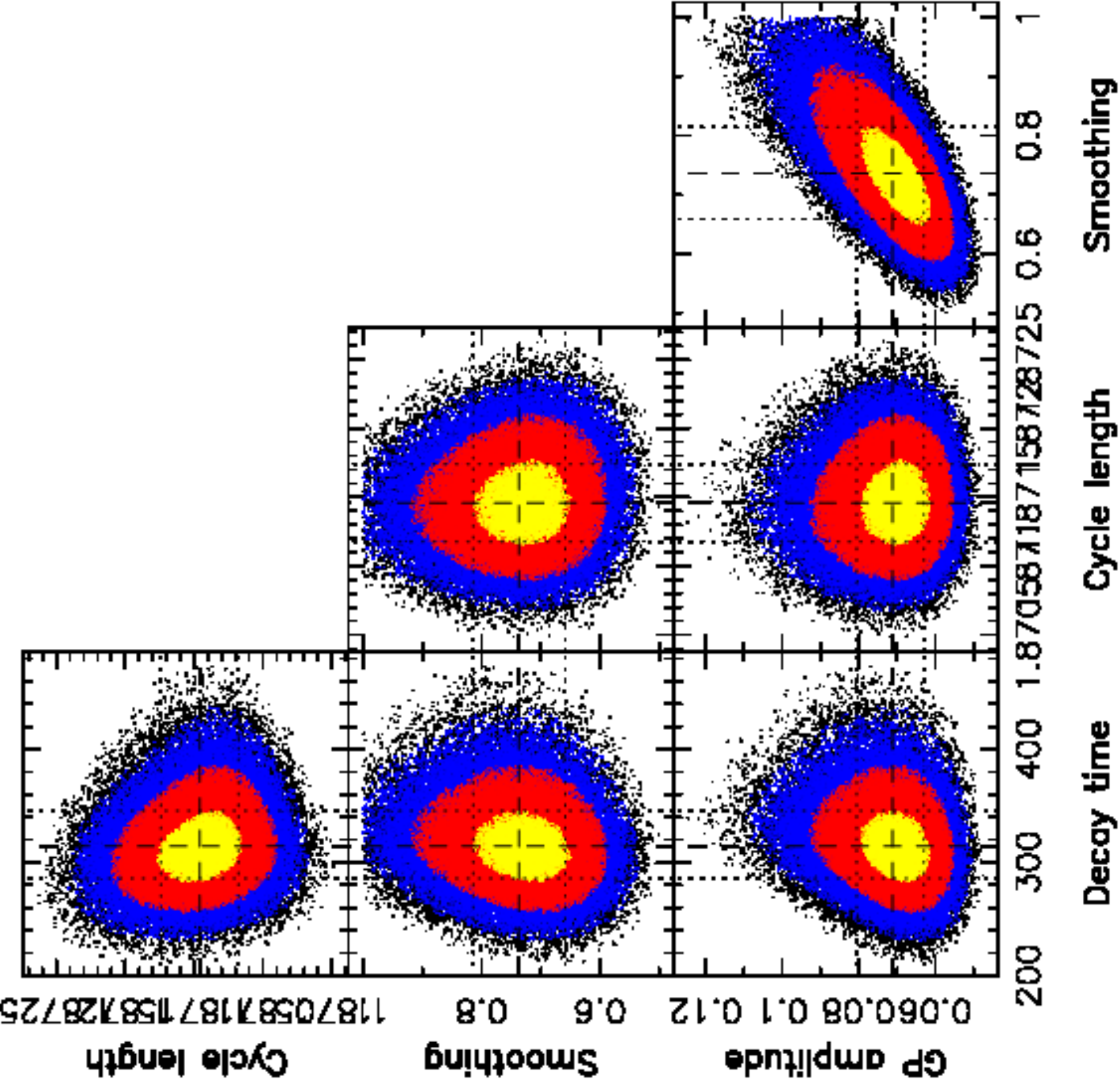}
		\caption{GPR-MCMC phase plot for the entire data set of V magnitudes. GP amplitude ${\theta_1 = 0.0712^{+0.0093}_{-0.0082}}${~}mag, cycle length ${\theta_2 =  1.8715\pm 0.0003}${~}d, decay time ${\theta_3 = 314^{+31}_{-29}}${~}d, smoothing parameter ${\theta_4 = 0.74\pm 0.08}${~}d.}
		\label{fig:pp0}
	\end{figure}

\begin{figure*}
		\centering
		\subfloat[08b+09a]{\includegraphics[angle=-90,width=0.25\linewidth]{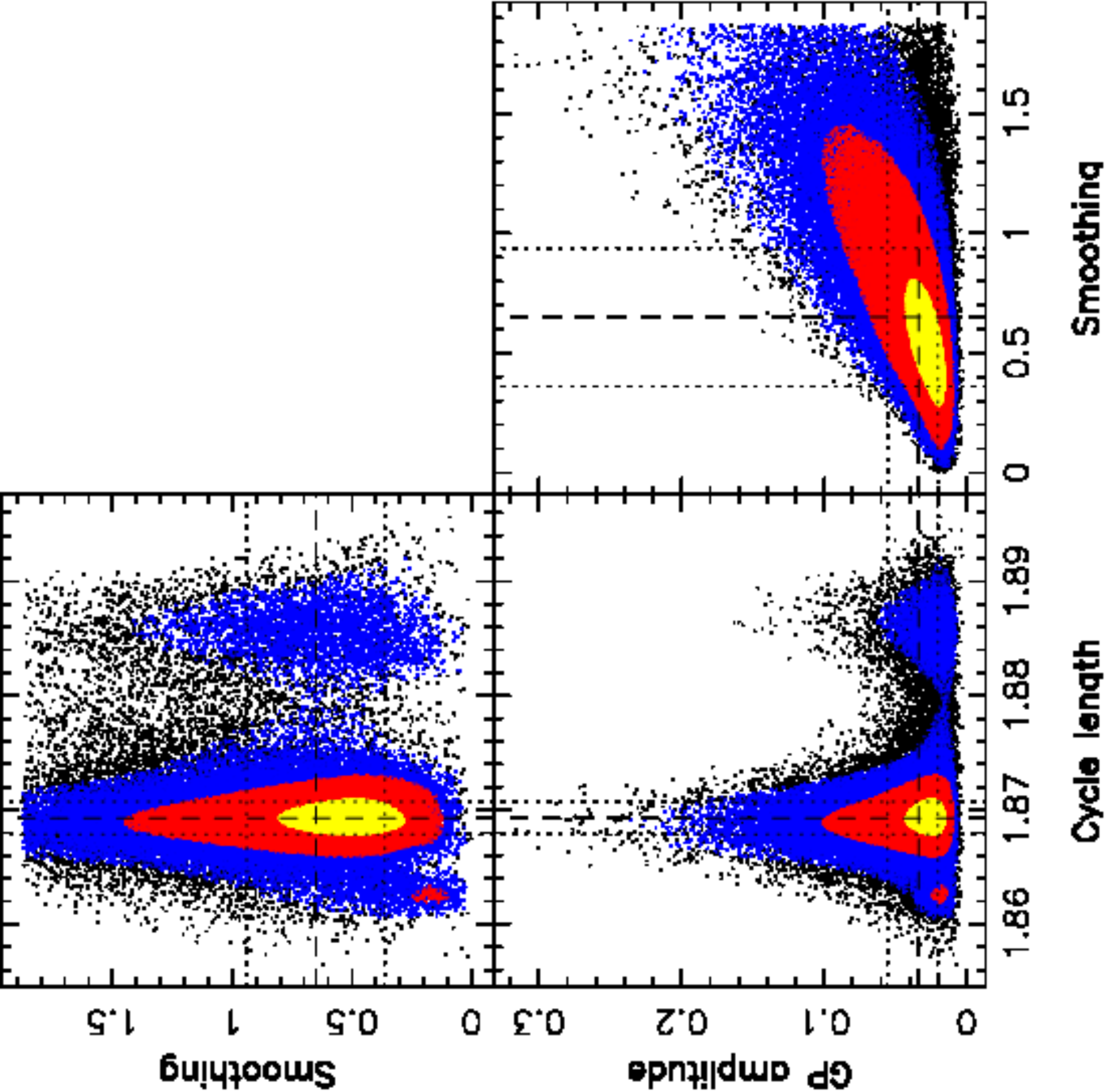}}
		\subfloat[09b+10a]{\includegraphics[angle=-90,width=0.25\linewidth]{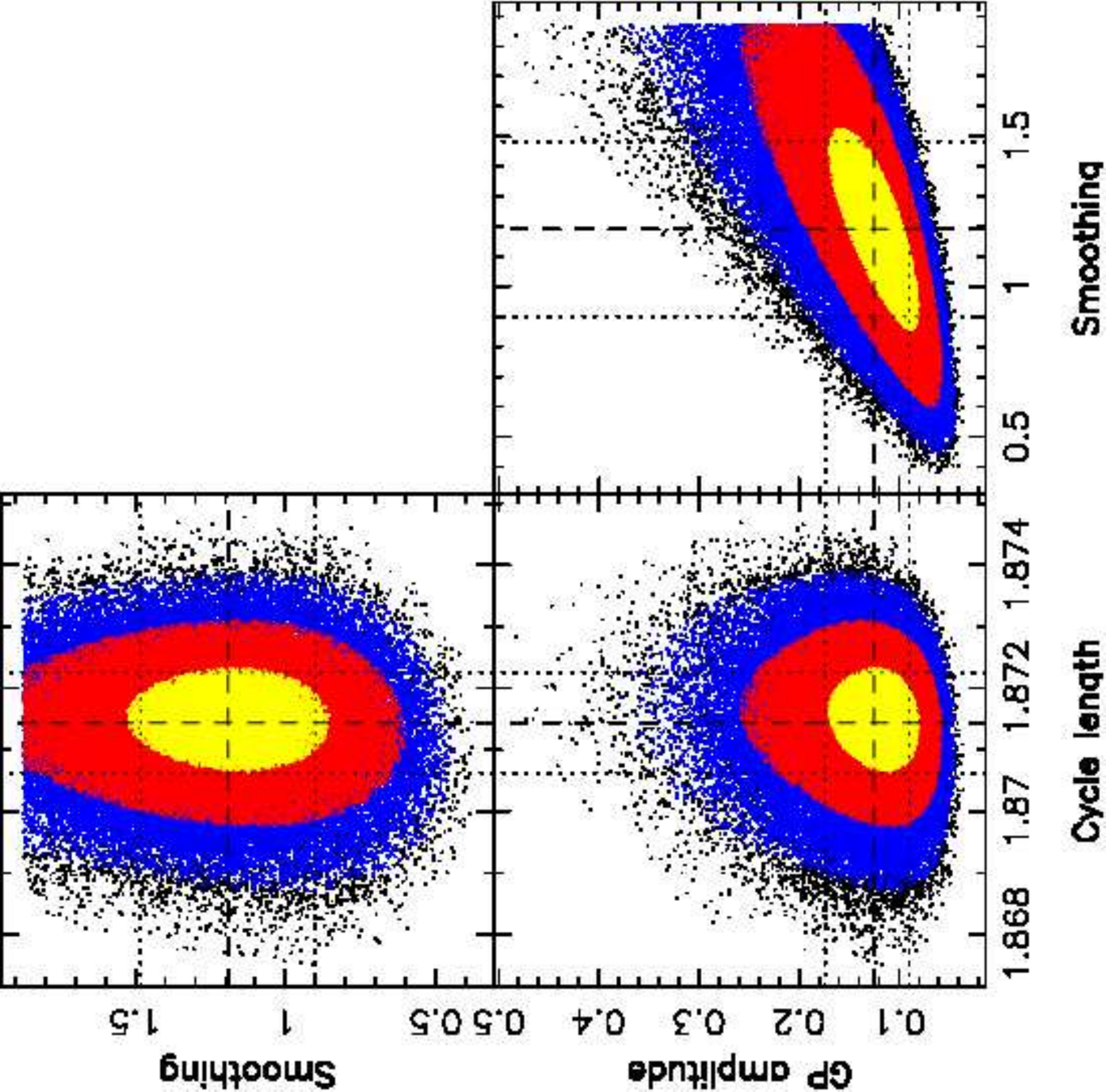}}
		\subfloat[10b]{\includegraphics[angle=-90,width=0.25\linewidth]{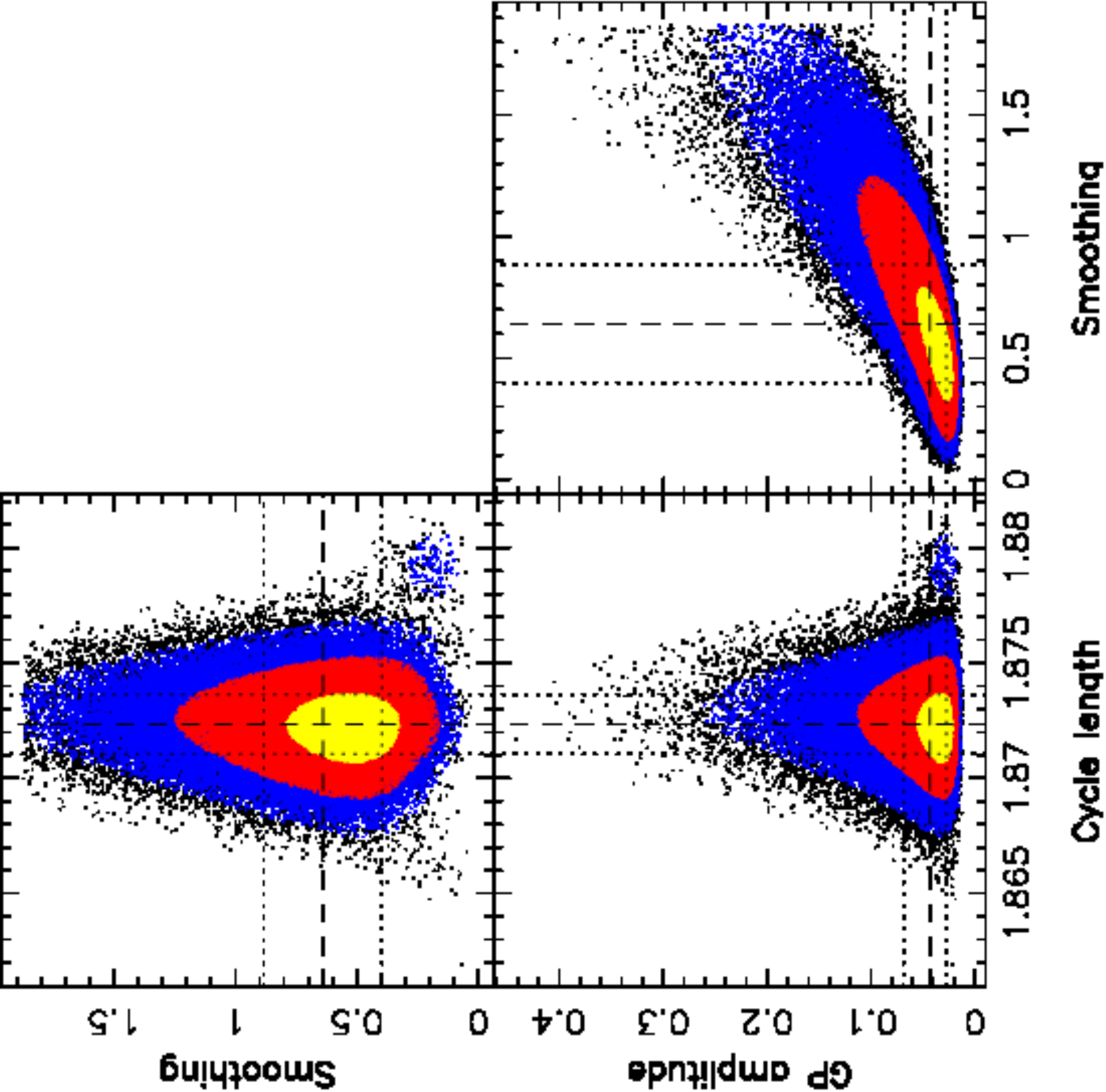}}
		\subfloat[11b+12a]{\includegraphics[angle=-90,width=0.25\linewidth]{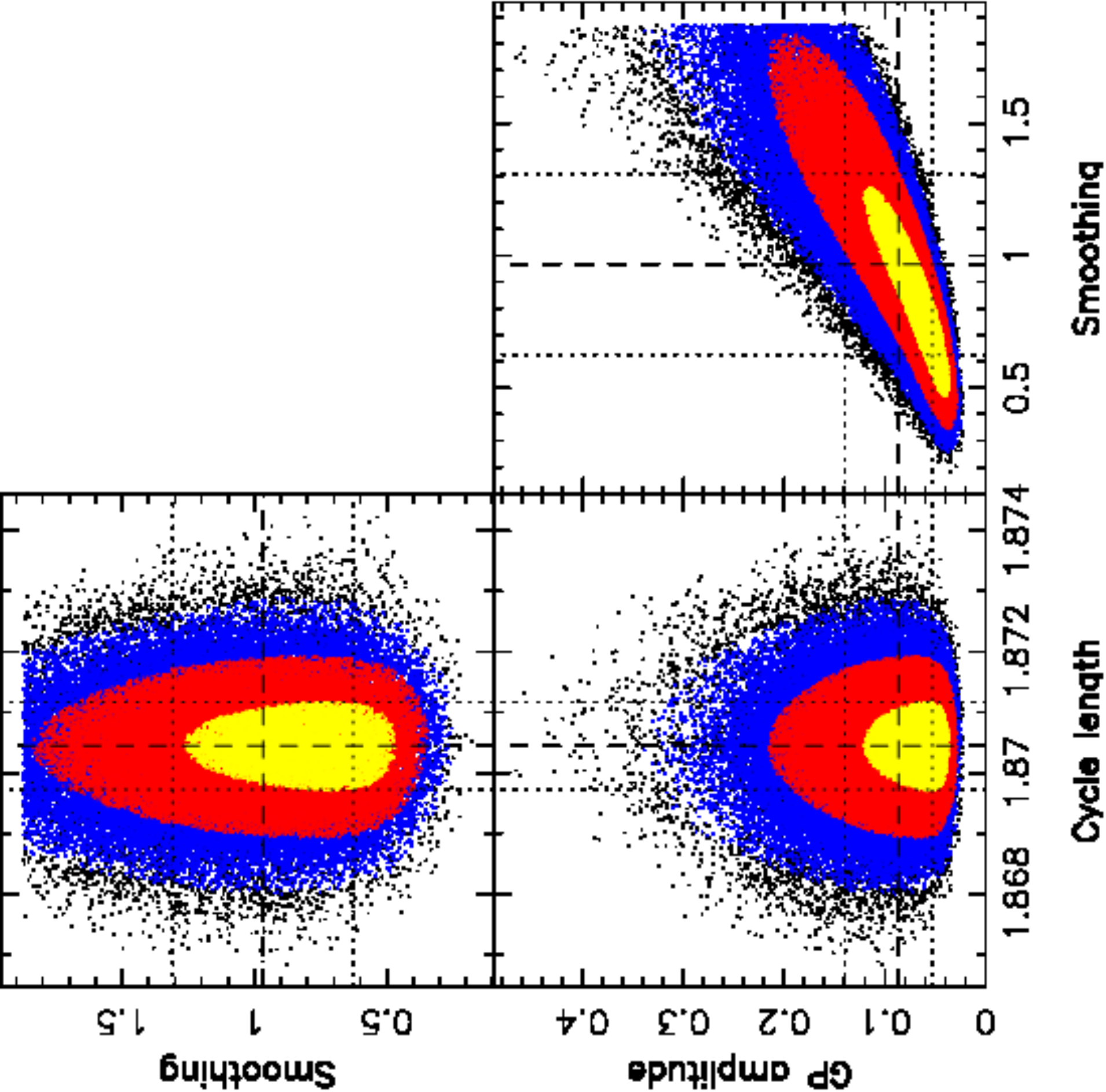}}

		\subfloat[12b+13a]{\includegraphics[angle=-90,width=0.25\linewidth]{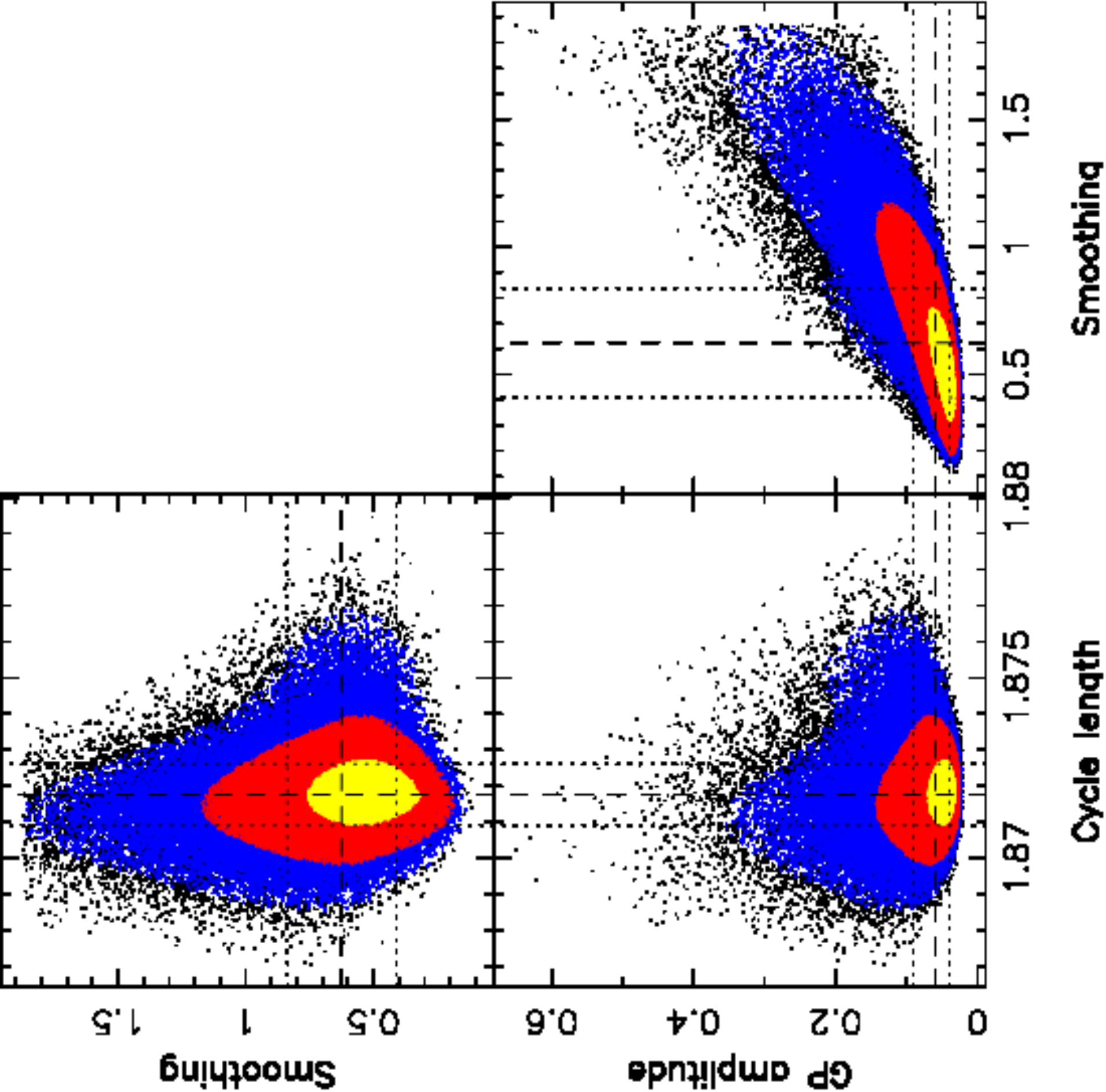}}
		\subfloat[13b+14a]{\includegraphics[angle=-90,width=0.25\linewidth]{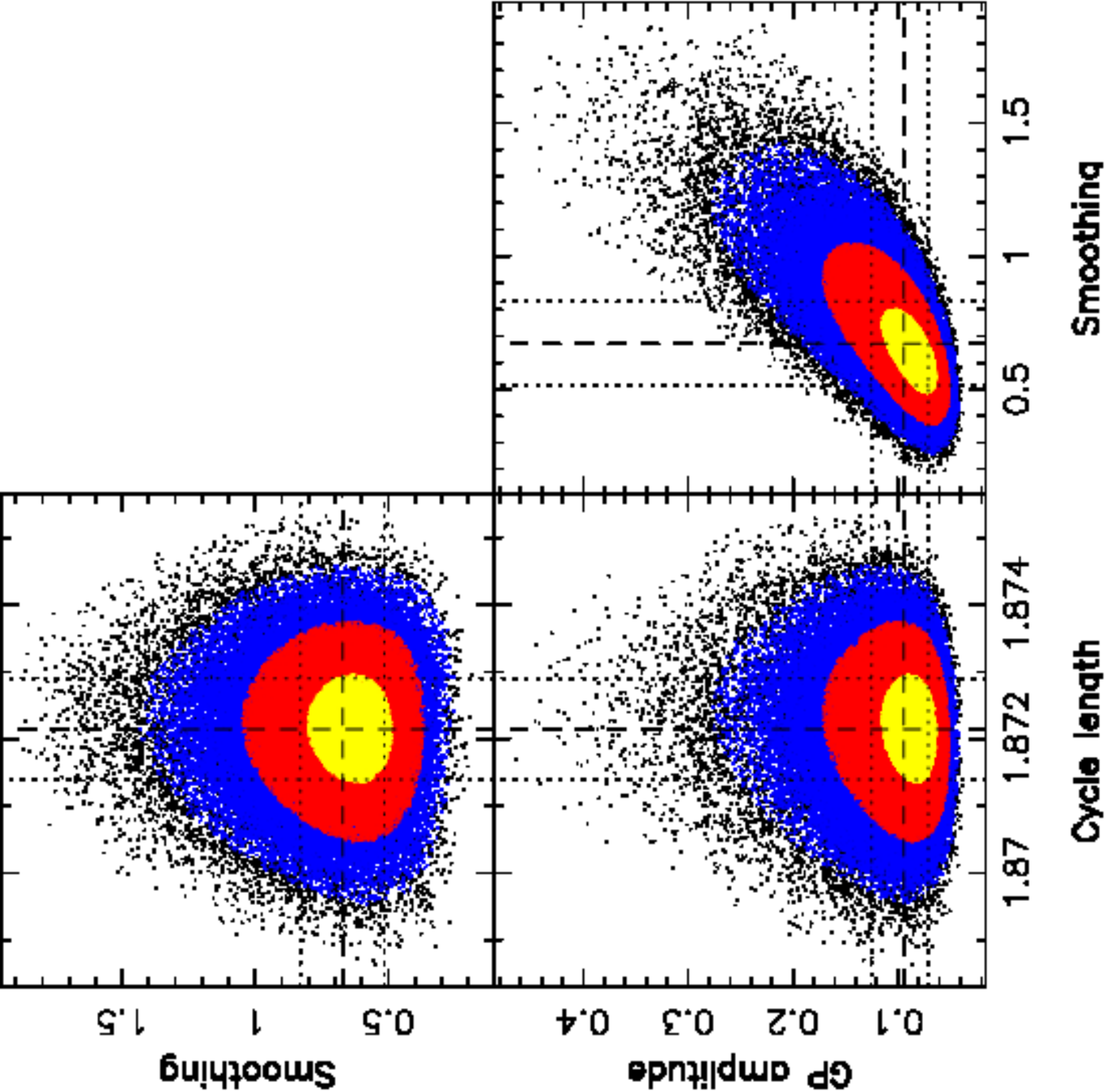}}
		\subfloat[14b]{\includegraphics[angle=-90,width=0.25\linewidth]{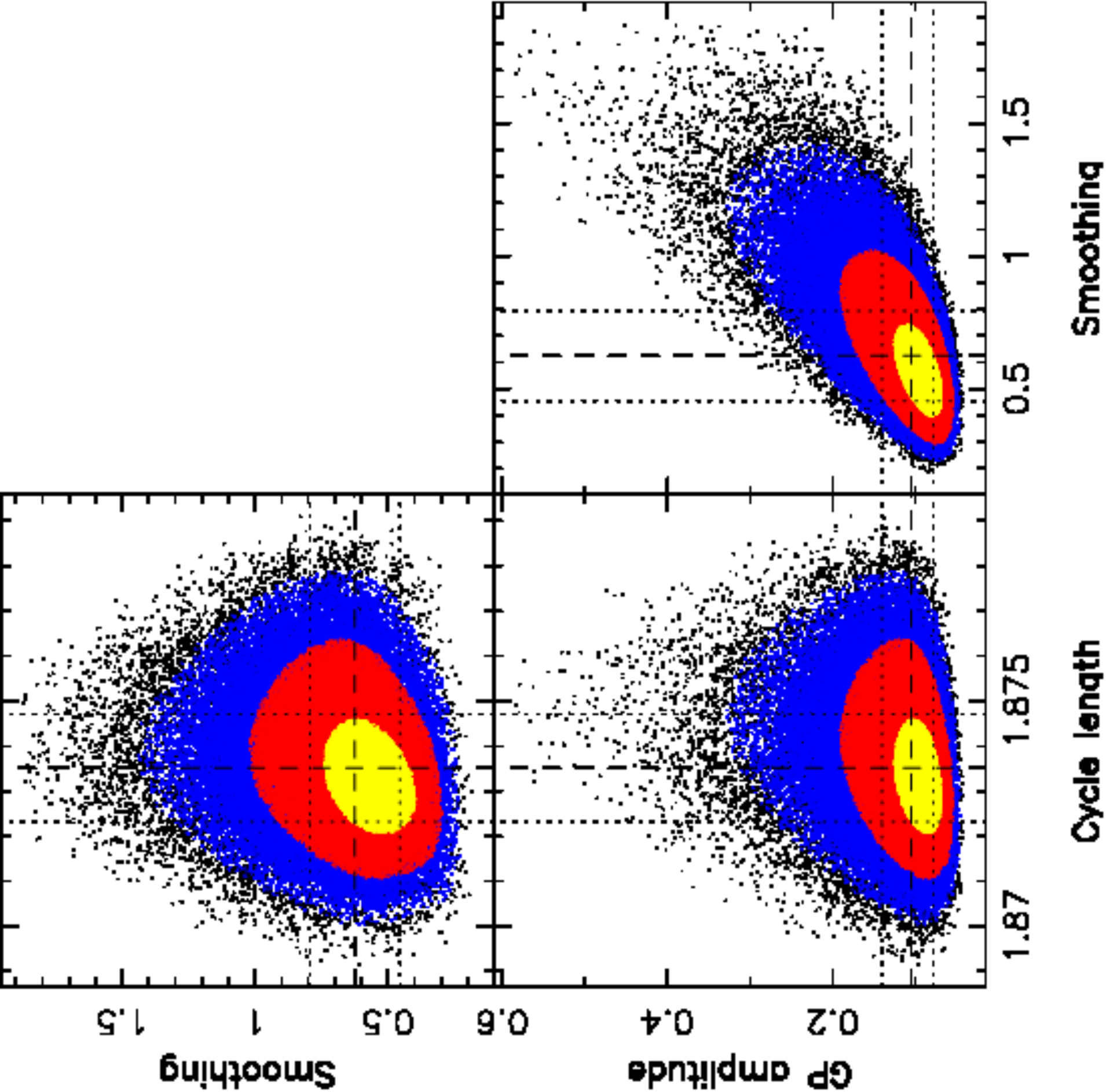}}
		\subfloat[15b+16a+16b+17a]{\includegraphics[angle=-90,width=0.25\linewidth]{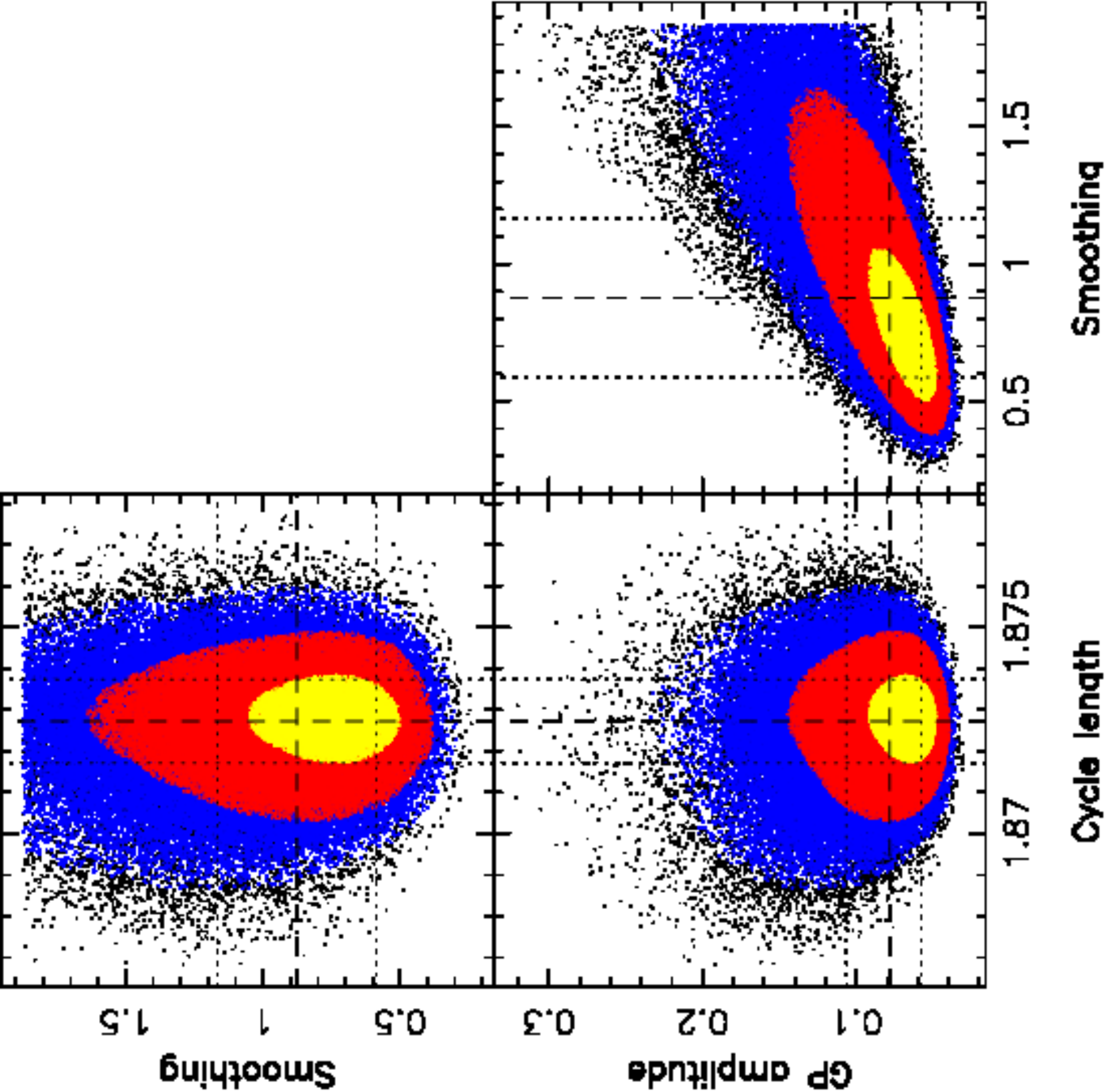}}
		\caption{MCMC phase plots for GPR applied to each of our V magnitude data sets.}
		\label{fig:pp1}
	\end{figure*}

	For the GPR, we made a first run on the global data set (phase plot in Fig.{~}\ref{fig:pp0}) and used its result to freeze the decay time for the modelling of the individual data sets, to avoid degeneracy. The retrieved hyperparameters are given in Table{~}\ref{tab:phg}. The phase plots of the individual data sets are displayed in Figure{~}\ref{fig:pp1}. Again, a neat period around 1.87 is outlined for each data set. The periods found with GPR and with sine fits are generally consistent, but the error bar for the rotation period on the whole data set is more trustworthy when computed statistically from GPR-MCMC than from the local curvature of the sine fit aliased \chisqr\ curve.

\begin{figure}
		\includegraphics[totalheight=0.25\textheight]{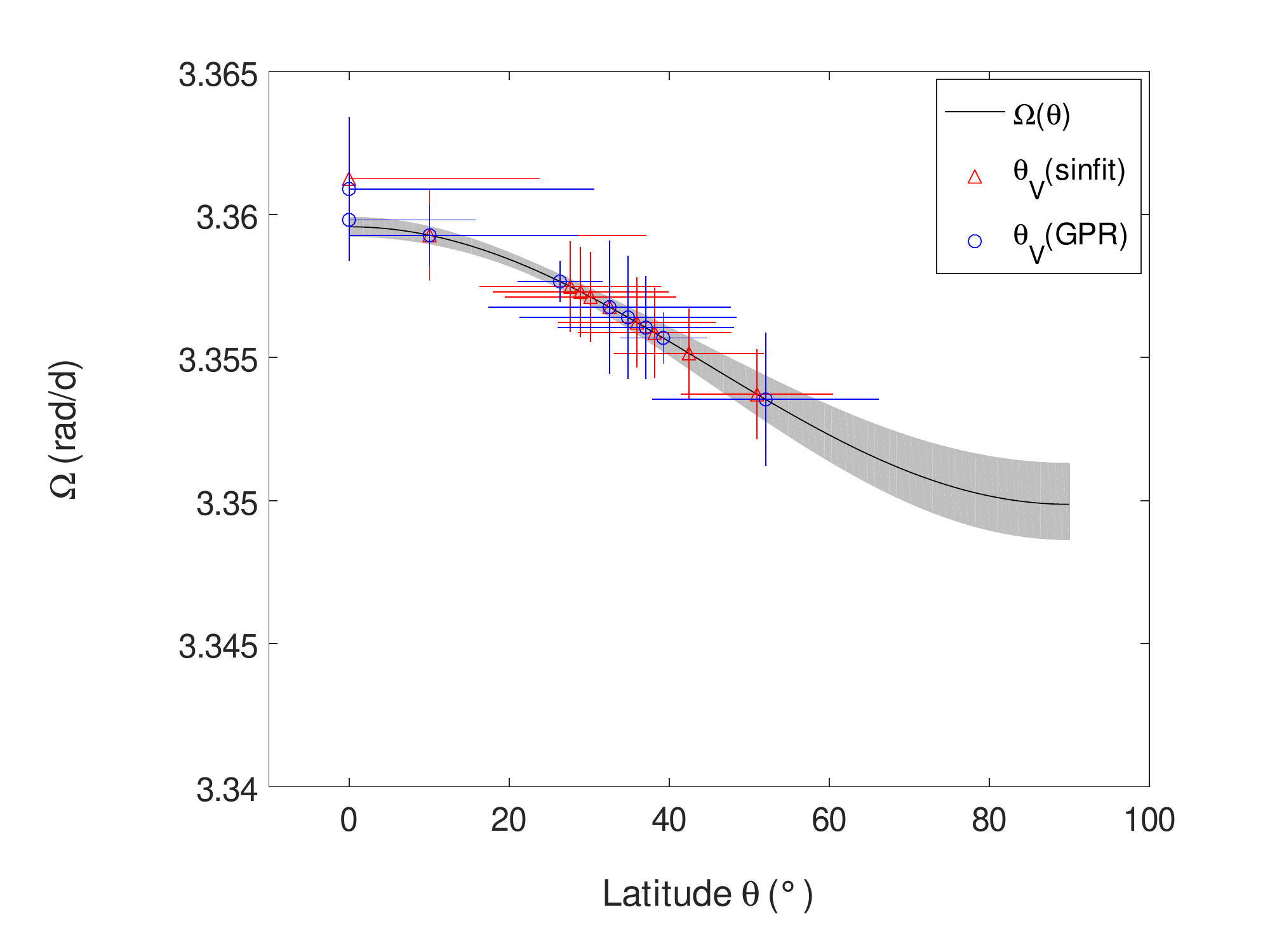}
		\caption{Differential rotation curve in blue, with parameters \omeq\ and \dom\ as defined in the introduction of Section{~}\ref{sec:mod}. Red: \hal\ rotation rates, green: \Bl\ rotation rates, circles: derived from 2013 Dec data set, triangles: derived from 2015 Dec data set, x symbols: derived from the whole data set (143 points for \hal\ and 135 for \Bl). Photometry rotation rates are displayed, those derived with sinfit (Table{~}\ref{tab:sip}) in green and those derived with GPR (Table{~}\ref{tab:phg}) in magenta.}
		\label{fig:drp}
	\end{figure}

\begin{figure}
		\includegraphics[totalheight=0.25\textheight]{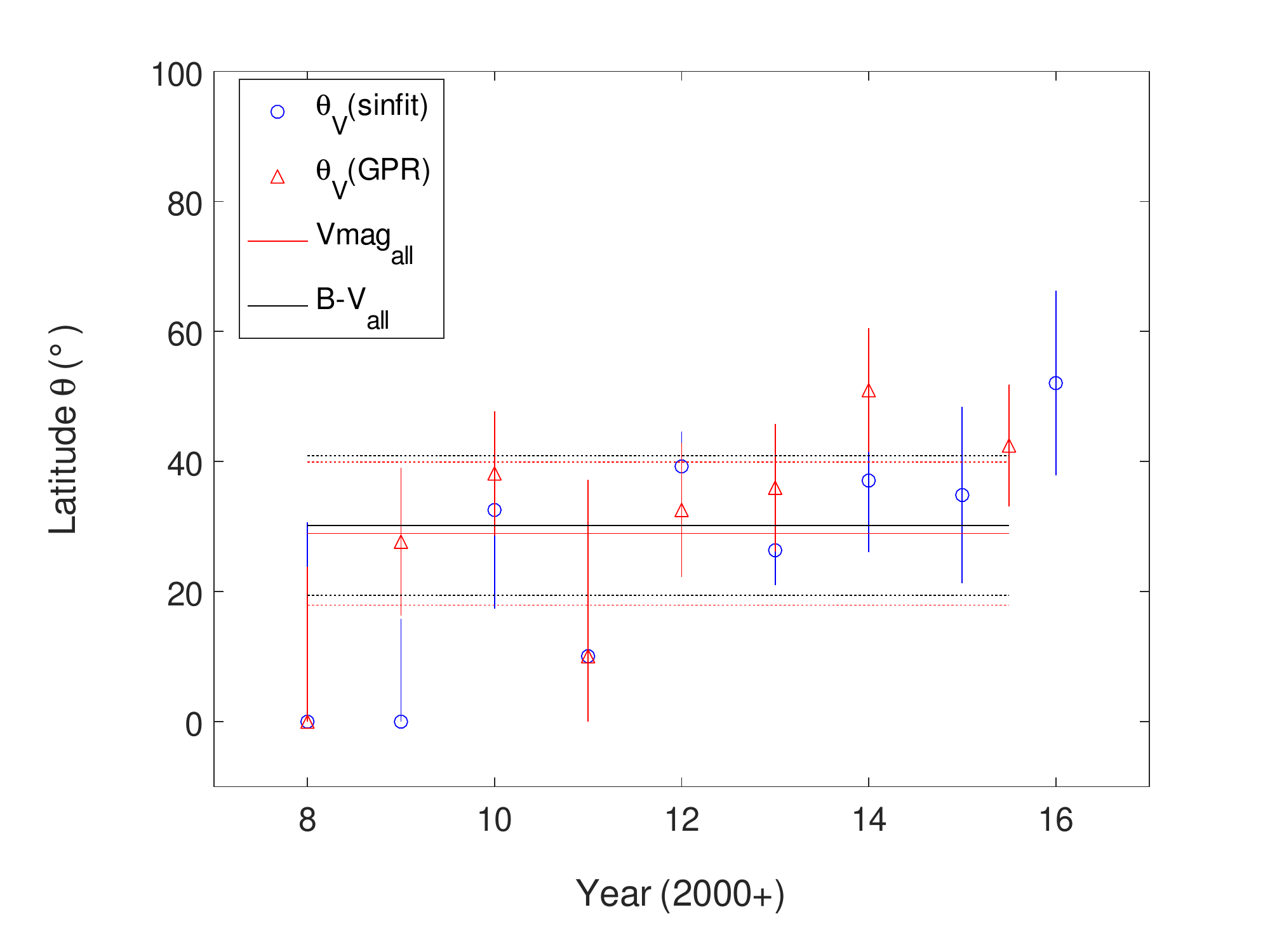}
		\caption{Colatitude found for the V magnitude, for each epoch and for the whole data set with sinfit (x-coordinate: 20) as well as for B-V with sinfit (x-coordinate: 21).}
		\label{fig:cop}
	\end{figure}

	All derived rotation periods, from sine fits and GPR, are plotted against their corresponding latitude using the ZDI-retrieved differential rotation in Figure{~}\ref{fig:drp}), and the thus-derived latitudes are plotted against time in Figure{~}\ref{fig:cop}, showing a global increasing trend of that latitude, regardless of the period retreval method.

\begin{figure}
		\includegraphics[totalheight=0.25\textheight]{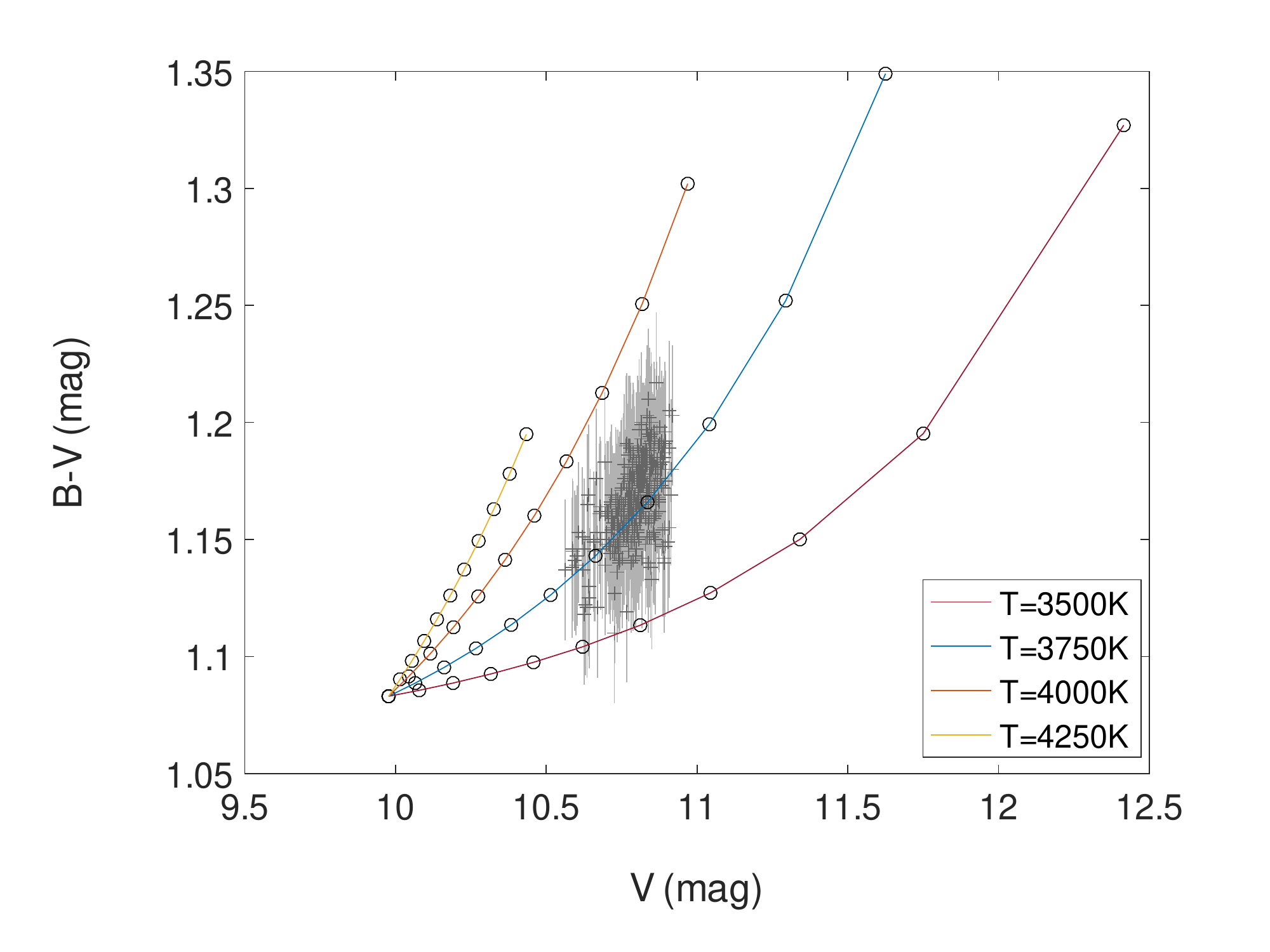}
		\caption{Fit of the B-V(V) curve with Kurucz models, with a photosphere temperature of 4500{~}K, \logg\ of 3.5, E(B-V) of 0.10. Each full line corresponds to a particular value of the spot temperature, and dots mark the spot coverage with steps of 10\% (the dot at V=10.0 and B-V=1.08 corresponding to a 0\% spot coverage). The extension of our data correspond to a spot coverage constantly between 50\% and 75\%.}
		\label{fig:vbk}
	\end{figure}

	We computed B-V(V) models from the Kurucz models for colors of main sequence stars with log(g)=3.5, T$_{\rm eff}$=4500{~}K and E(B-V)=0.10{~}mag \citep{Kurucz93}: we fit a two-temperature model with a photospheric temperature of 4500 K and different values for the spot temperature. Then, for each tested spot temperature, for all values of spot coverage from 0 too 100 \%, we computed the resulting B and the resulting V using the following formulas, from which we derived the B-V. The resulting models are plotted in Figure{~}\ref{fig:vbk}. We find that a spot temperature of 3750{~}K fits our B-V measurements well, from which we deduce that the extension of our data imply a spot coverage on \vt\ between 50 and 75\%, in agreement with the assumption in Section{~}\ref{sec:evo}.
	\begin{eqnarray*}
		V(r) & = -2.5 \log_{10} (r10^{-\frac{V_{\rm spot}}{2.5}}+(1-r)10^{-\frac{V_{\rm star}}{2.5}}) \\
		B(r) & = -2.5 \log_{10} (r10^{-\frac{B_{\rm spot}}{2.5}}+(1-r)10^{-\frac{B_{\rm star}}{2.5}})
	\end{eqnarray*}

	\begin{table}
		\centering
		\caption{Sinfit results on photometric data.}
		\tiny
		\begin{tabular}{cccc}
			Data & Period (d) & Amplitude (mag) & Dispersion (mag) \\
			\hline
			08b+09a & 1.8695{~}$\pm${~}0.0014 & 0.019{~}$\pm${~}0.005 & 0.015 \\
			09b+10a & 1.8701{~}$\pm${~}0.0004 & 0.069{~}$\pm${~}0.004 & 0.020 \\
			10b & 1.8718{~}$\pm${~}0.0013 & 0.016{~}$\pm${~}0.005 & 0.011 \\
			11b+12a & 1.8704{~}$\pm${~}0.0006 & 0.045{~}$\pm${~}0.004 & 0.014 \\
			12b+13a & 1.8724{~}$\pm${~}0.0005 & 0.051{~}$\pm${~}0.003 & 0.018 \\
			13b+14a & 1.8713{~}$\pm${~}0.0004 & 0.114{~}$\pm${~}0.006 & 0.018 \\
			14b & 1.8722{~}$\pm${~}0.0010 & 0.117{~}$\pm${~}0.006 & 0.021 \\
			15b+16a & 1.8720{~}$\pm${~}0.0012 & 0.089{~}$\pm${~}0.007 & 0.014 \\
			16b+17a & 1.8736{~}$\pm${~}0.0013 & 0.088{~}$\pm${~}0.013 & 0.006 \\
			\hline
			All V mag & 1.871254{~}$\pm${~}0.000030 & 0.0568{~}$\pm${~}0.0032 & 0.047 \\
		\end{tabular}
		\label{tab:sip}
	\end{table}

	\begin{table*}
		\caption{Results of the GPR-MCMC runs on our V magnitude measurements.}
		\begin{tabular}{ccccc}
			\hline
			Data set & GP Period (d) & Decay time (d) & Smoothing (d) & Amplitude (mag) \\
			& (best) & (best) & (best) & (best) \\
			\hline
			08b+09a & 1.8693{~}$\pm${~}0.0014 (1.8694) & 311.1333 & 0.7{~}$\pm${~}0.3 (0.5) & 0.03$^{+0.02}_{-0.01}$ (0.02) \\
			& & & & \\
			09b+10a & 1.8714{~}$\pm${~}0.0008 (1.8715) & 311.1333 & 1.2{~}$\pm${~}0.3 (1.2) & 0.12$^{+0.05}_{-0.04}$ (0.11) \\
			& & & & \\
			10b & 1.8723{~}$\pm${~}0.0013 (1.8722) & 311.1333 & 0.6{~}$\pm${~}0.2 (0.5) & 0.04$^{+0.03}_{-0.02}$ (0.03) \\
			& & & & \\
			11b+12a & 1.8704{~}$\pm${~}0.0007 (1.8705) & 311.1333 & 1.0{~}$\pm${~}0.3 (0.7) & 0.09$^{+0.05}_{-0.03}$ (0.05) \\
			& & & & \\
			12b+13a & 1.8718{~}$\pm${~}0.0009 (1.8718) & 311.1333 & 0.6{~}$\pm${~}0.2 (0.5) & 0.06$^{+0.03}_{-0.02}$ (0.04) \\
			& & & & \\
			13b+14a & 1.8721{~}$\pm${~}0.0008 (1.8722) & 311.1333 & 0.7{~}$\pm${~}0.2 (0.6) & 0.09$^{+0.03}_{-0.02}$ (0.08) \\
			& & & & \\
			14b & 1.8735{~}$\pm${~}0.0012 (1.8732) & 311.1333 & 0.6{~}$\pm${~}0.2 (0.5) & 0.11$^{+0.04}_{-0.03}$ (0.09) \\
			& & & & \\
			15b+16a & 1.8727{~}$\pm${~}0.0010 (1.8729) & 311.1333 & 0.9{~}$\pm${~}0.3 (0.7) & 0.08$^{+0.03}_{-0.02}$ (0.06) \\
			+16b+17a & & & & \\
			\hline
			All V mag & 1.8715{~}$\pm${~}0.0003 (1.8714) & 314$^{+31}_{-29}$ (311.1333) & 0.74{~}$\pm${~}0.08 (0.73) & 0.071$^{+0.009}_{-0.008}$ (0.070) \\
			& & & & \\
			\hline

		\end{tabular}
		\label{tab:phg}
	\end{table*}

	\section{Activity proxies}
	\label{anx:act}
	This section shows the line profiles of \hal, \hei\ and \caii, as well as some results on \Bl.
	\subsection{H$\alpha$}
\begin{figure*}
		\subfloat[2008 Dec]{\includegraphics[totalheight=0.25\textheight]{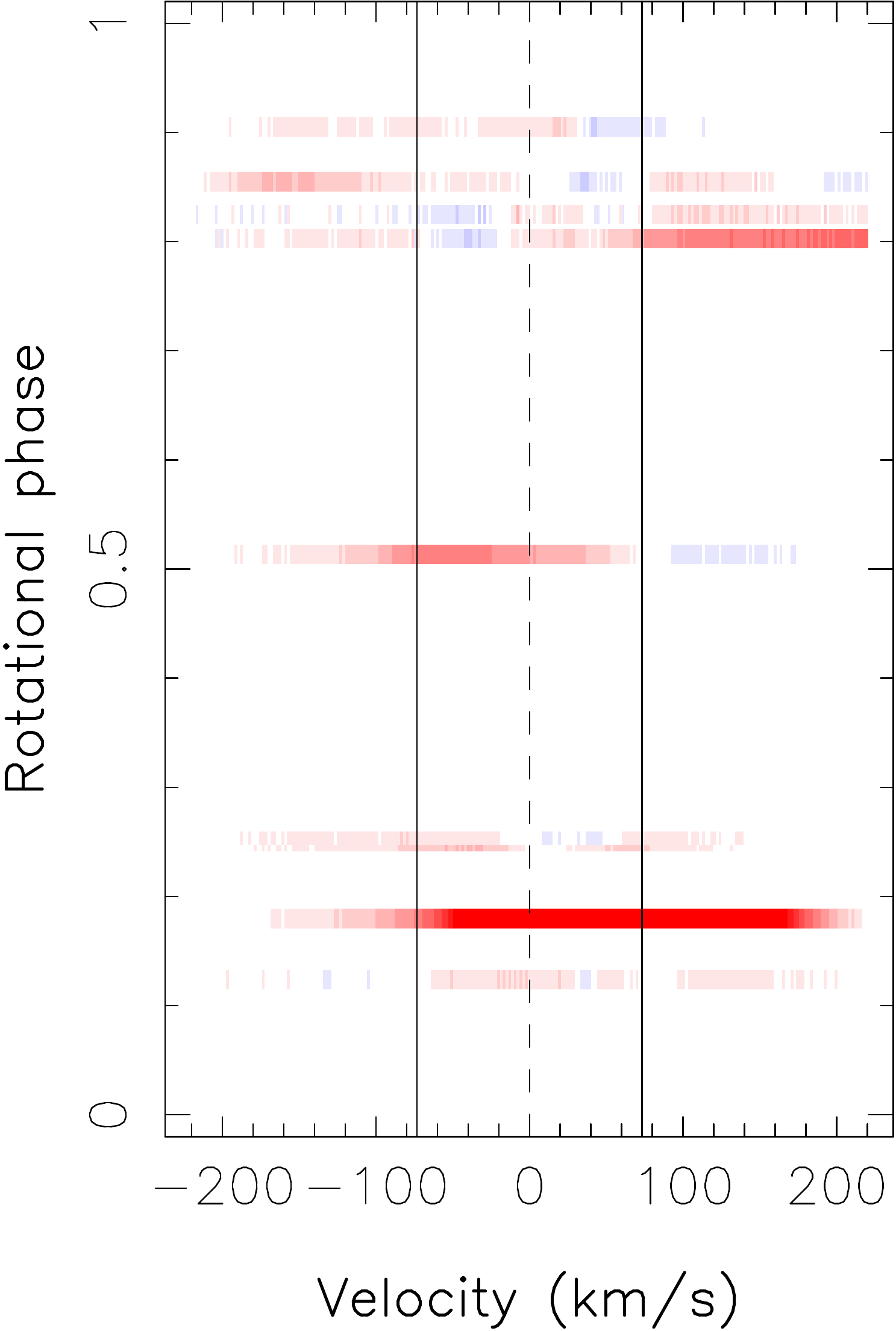}}
		\hfill
		\subfloat[2009 Jan cycles 0-4]{\includegraphics[totalheight=0.25\textheight]{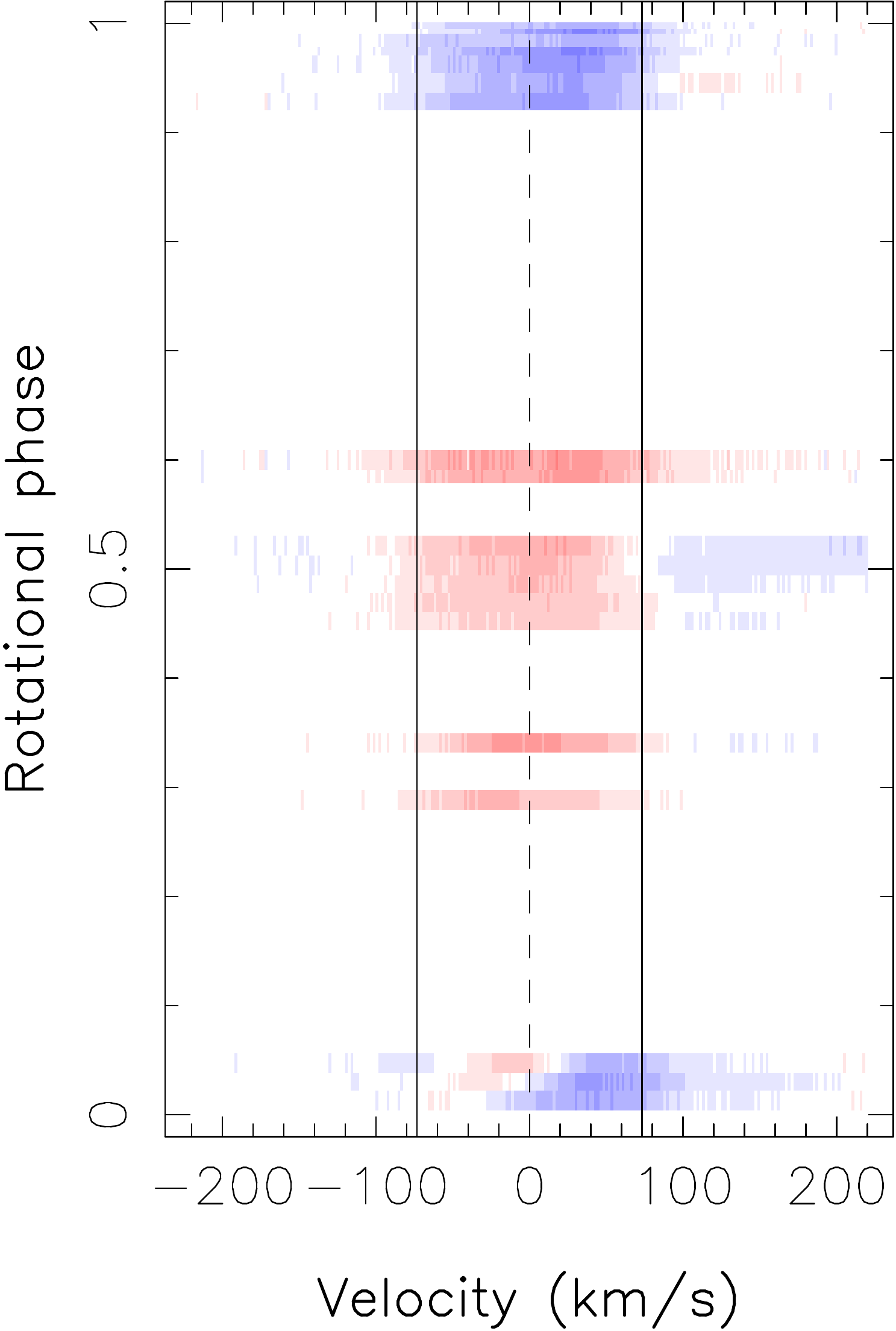}}
		\hfill
		\subfloat[2009 Jan cycles 5-8]{\includegraphics[totalheight=0.25\textheight]{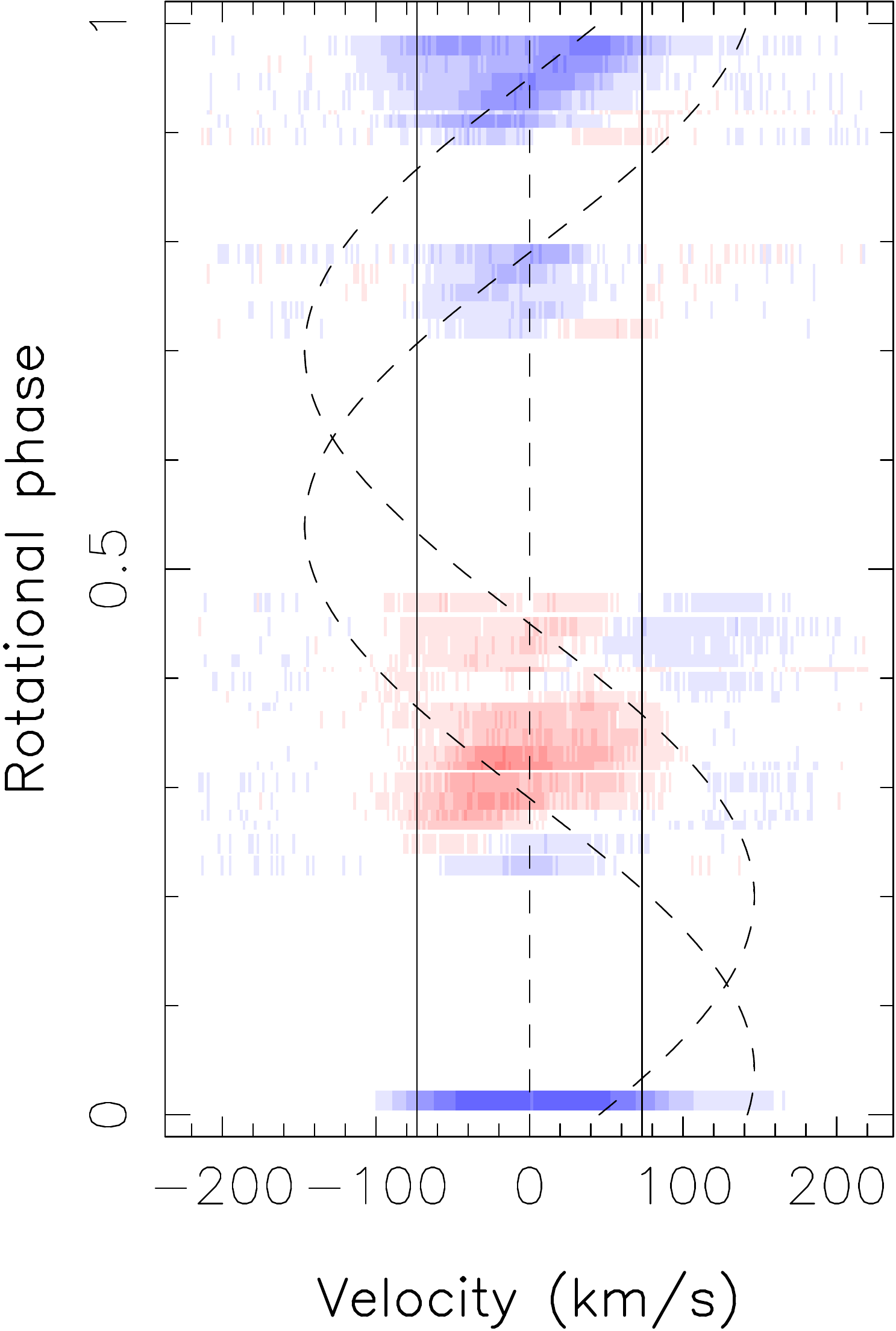}}
		\hfill
		\subfloat[2011 Jan]{\includegraphics[totalheight=0.25\textheight]{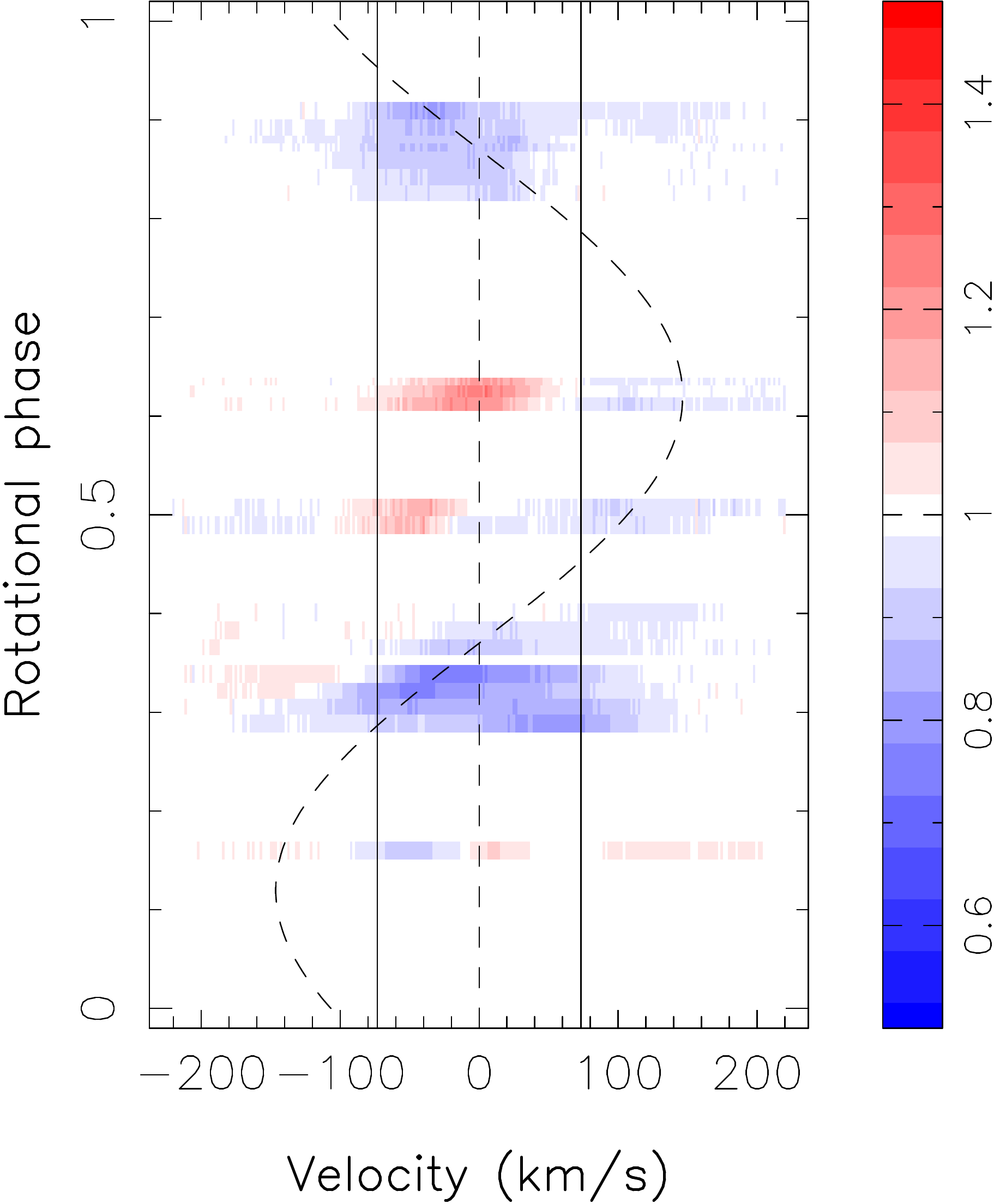}}

		\subfloat[2013 Dec]{\includegraphics[totalheight=0.25\textheight]{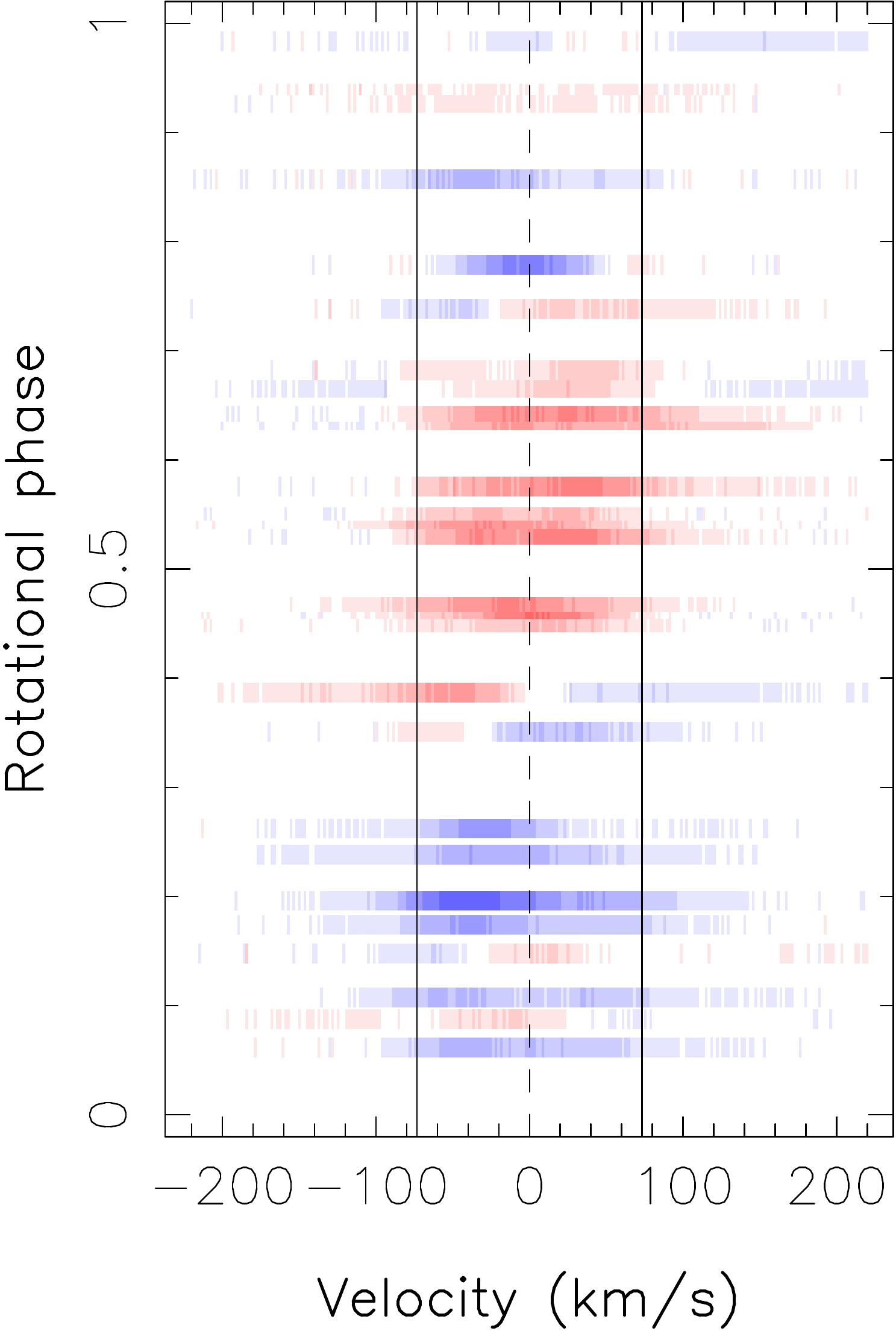}}
		\hfill
		\subfloat[2015 Dec]{\includegraphics[totalheight=0.25\textheight]{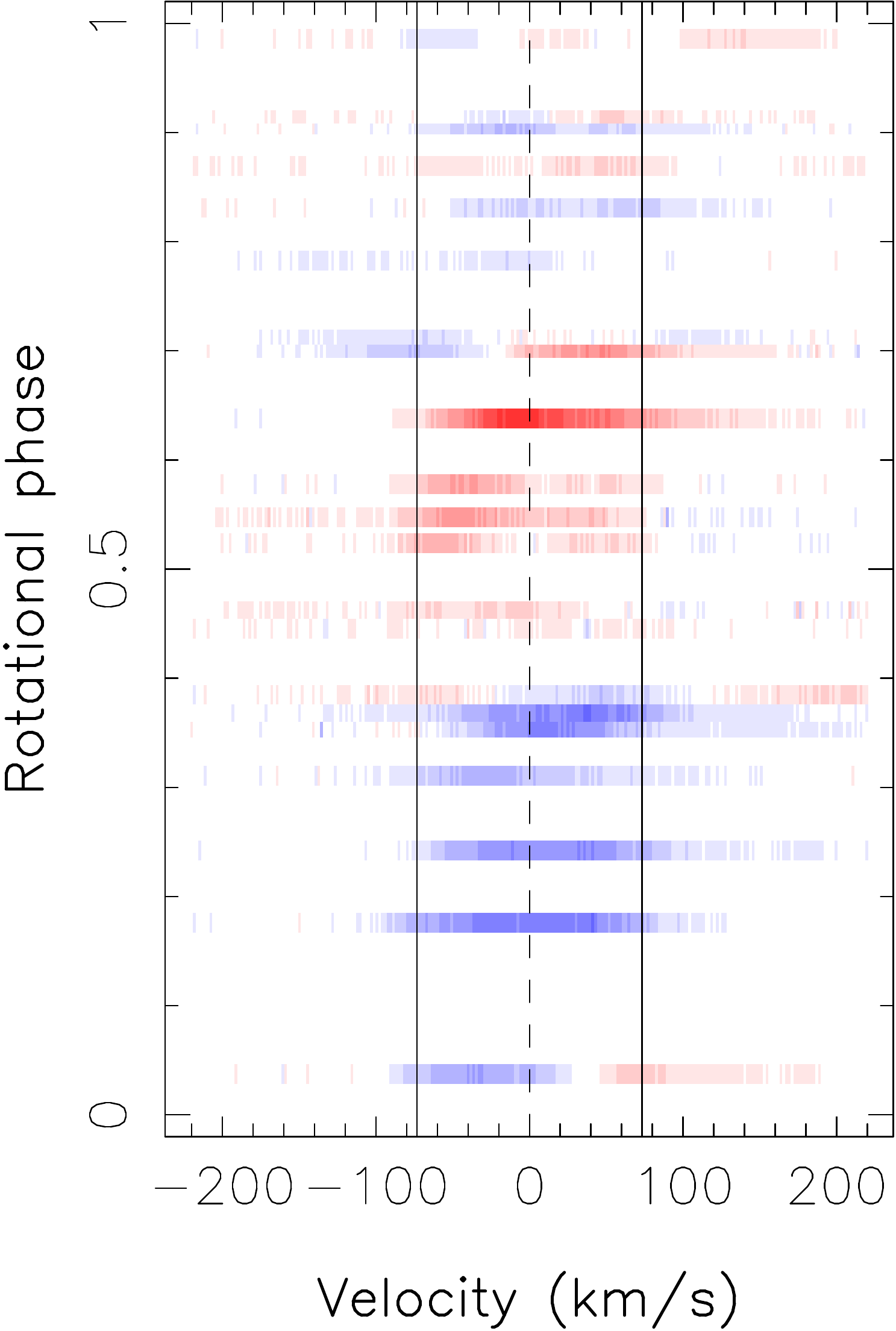}}
		\hfill
		\subfloat[2016 Jan]{\includegraphics[totalheight=0.25\textheight]{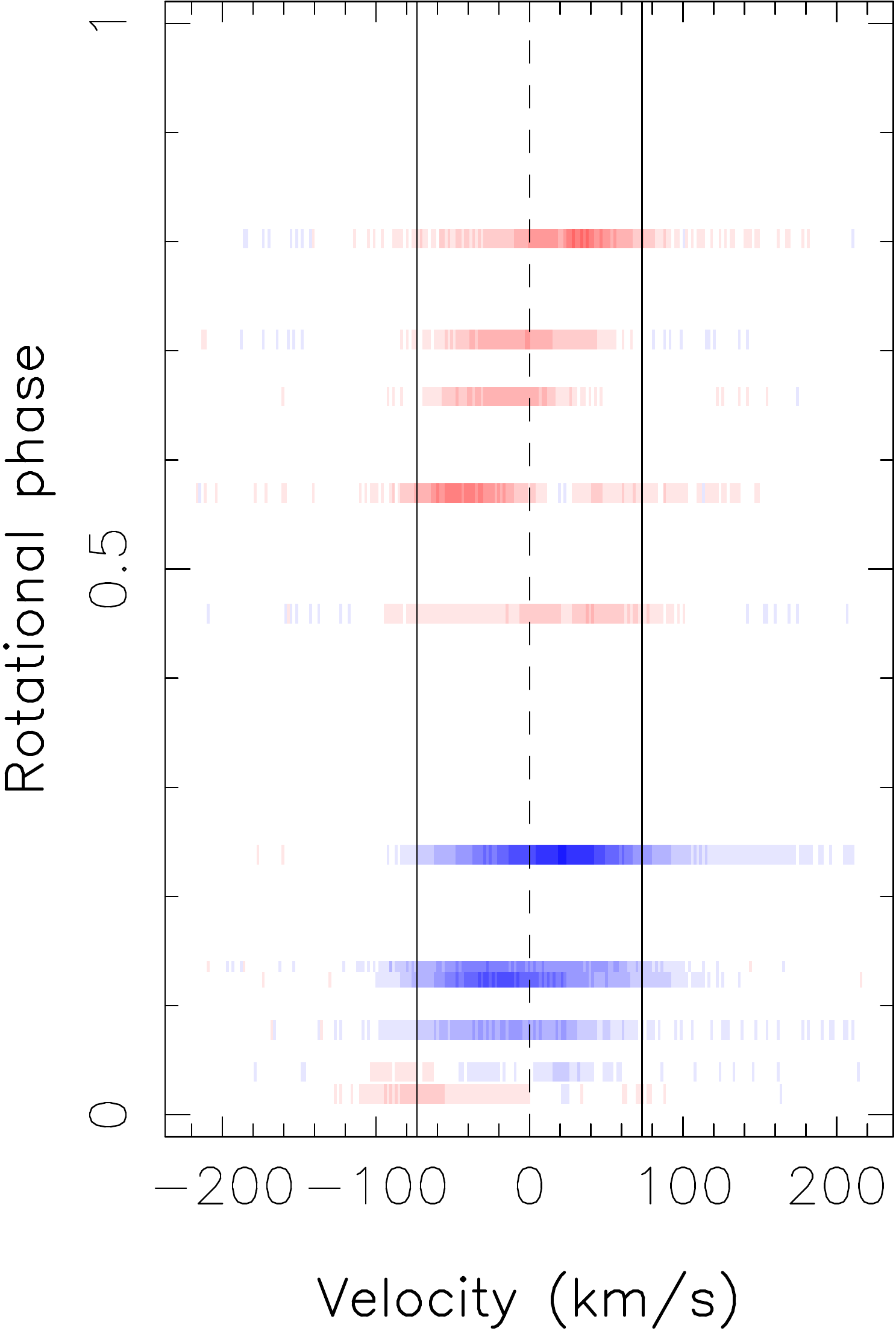}}
		\caption{\hal\ dynamical spectra for epochs 2008 Dec (a), 2009 Jan (b,c), 2011 Jan (d), 2013 Dec (e), 2015 Dec (f) and 2016 Jan (g).}
		\label{fig:hal}
	\end{figure*}

	H$\alpha$ dynamic spectra are plotted in Figure{~}\ref{fig:hal}, with the 2009 Jan data set being split in half to better see the absorption feature around phase 0.95. We also see two other absorption features in 2009 Jan around phase 0.80 and in 2011 Jan around phase 0.35, and we fit a sine curve in each to determine the potential altitude of a prominence or cloud that could be the origin of these absorption features. We find a sine semi-amplitude of $\sim$2{~}\vsini\ for each of them.

	Lomb-Scargle periodograms for individual epochs are plotted in Figure \ref{fig:hap}, and the periodogram for the whole data set is shown in Figure{~}\ref{fig:haa}, showing a neat peak at the rotation period.

	\begin{figure*}
		\centering
		\subfloat[2009 Jan]{\includegraphics[totalheight=0.25\textheight]{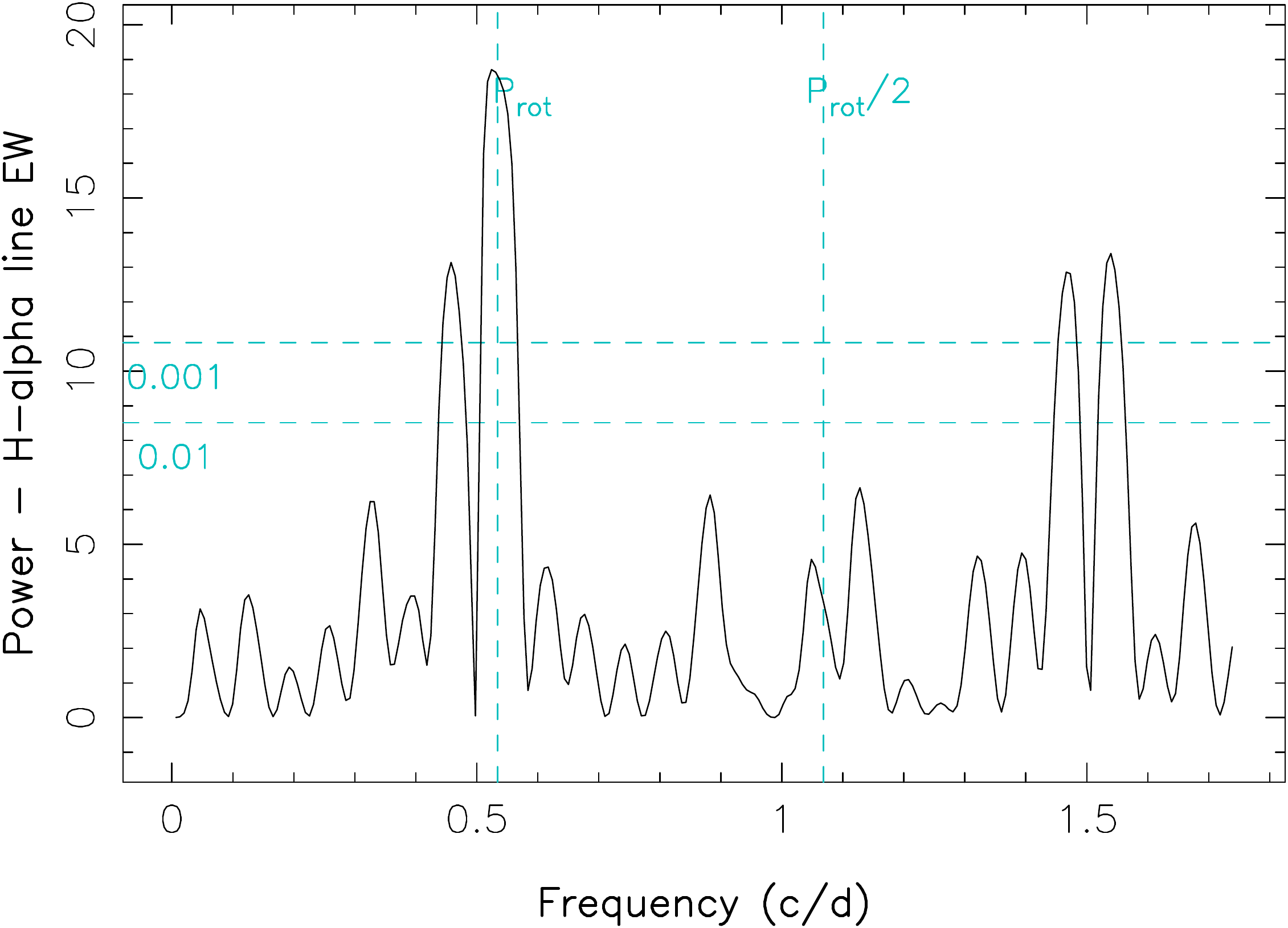}}
		\hfill
		\subfloat[2011 Jan]{\includegraphics[totalheight=0.25\textheight]{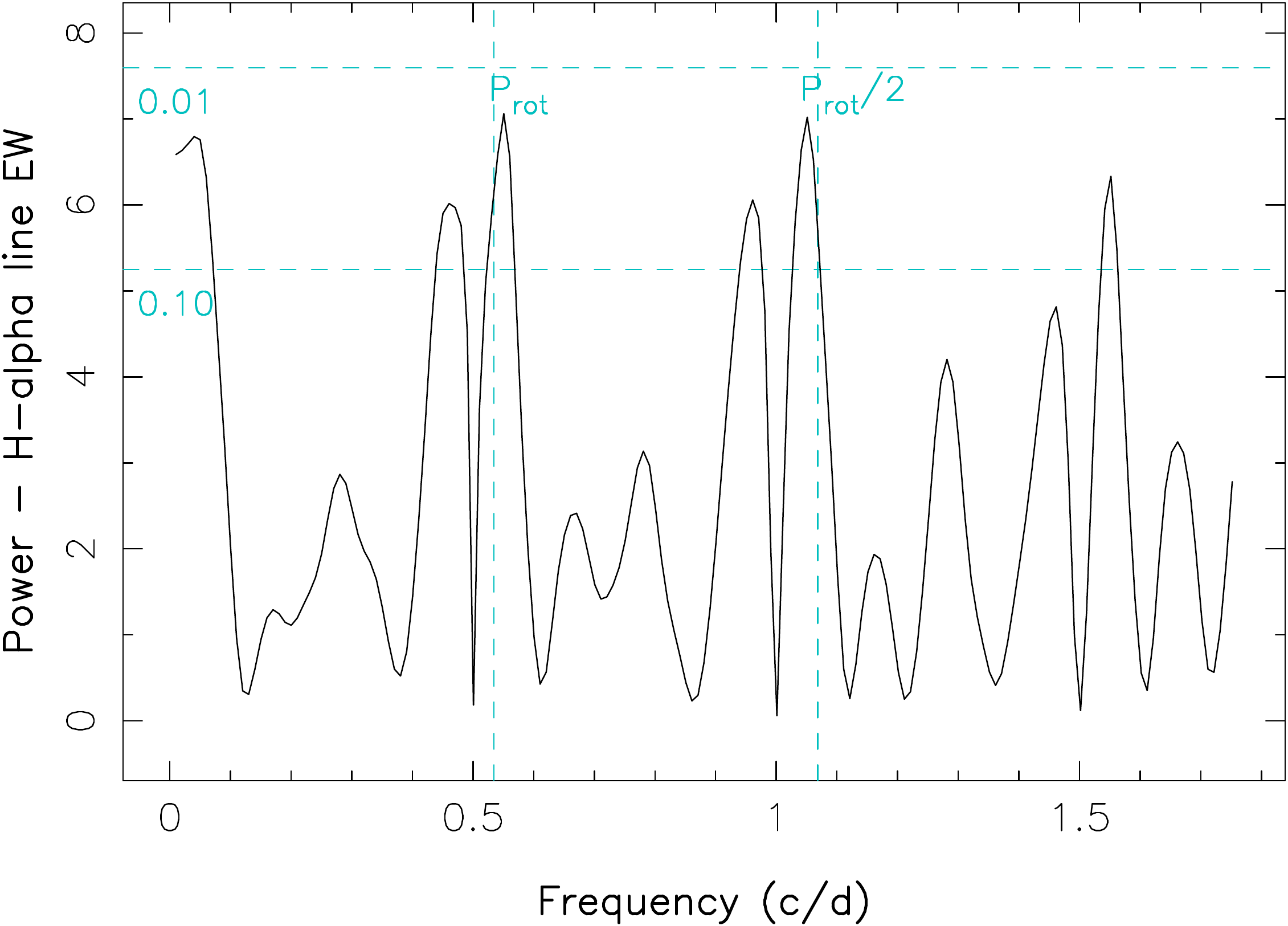}}
		\caption{Periodograms for the \hal\ line EW, for observation epochs 2009 Jan (a), 2011 Jan (b), 2013 Dec (c) and 2015 Dec (d). False-alarm probability levels of 1\% and 0.1\% are represented as horizontal cyan dashed lines, and \Prot\ and its first harmonic as vertical cyan dashed lines.}
		\label{fig:hap}
	\end{figure*}
	\begin{figure*}
		\centering
		\ContinuedFloat
		\subfloat[2013 Dec]{\includegraphics[totalheight=0.25\textheight]{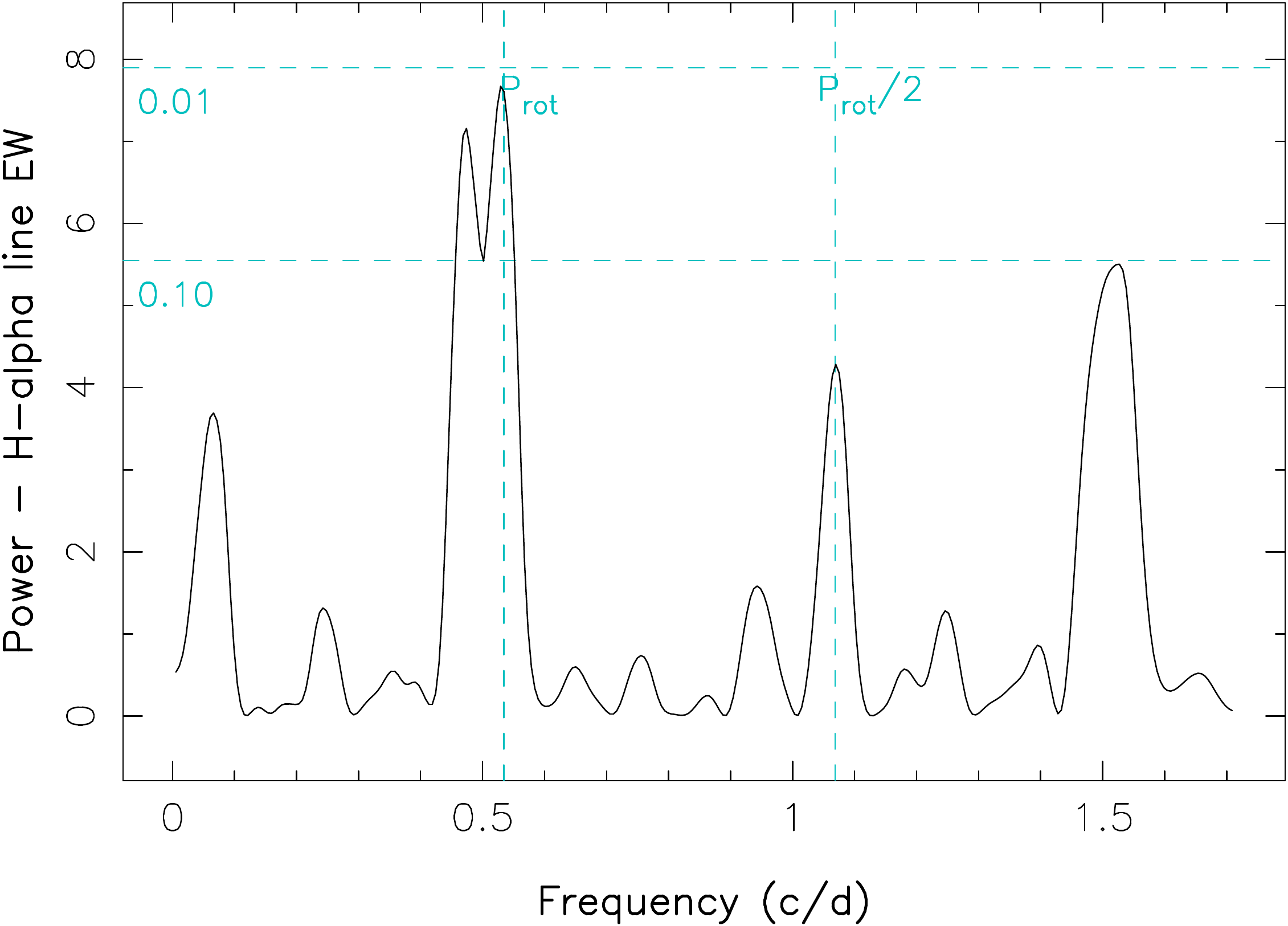}}
		\hfill
		\subfloat[2015 Dec]{\includegraphics[totalheight=0.25\textheight]{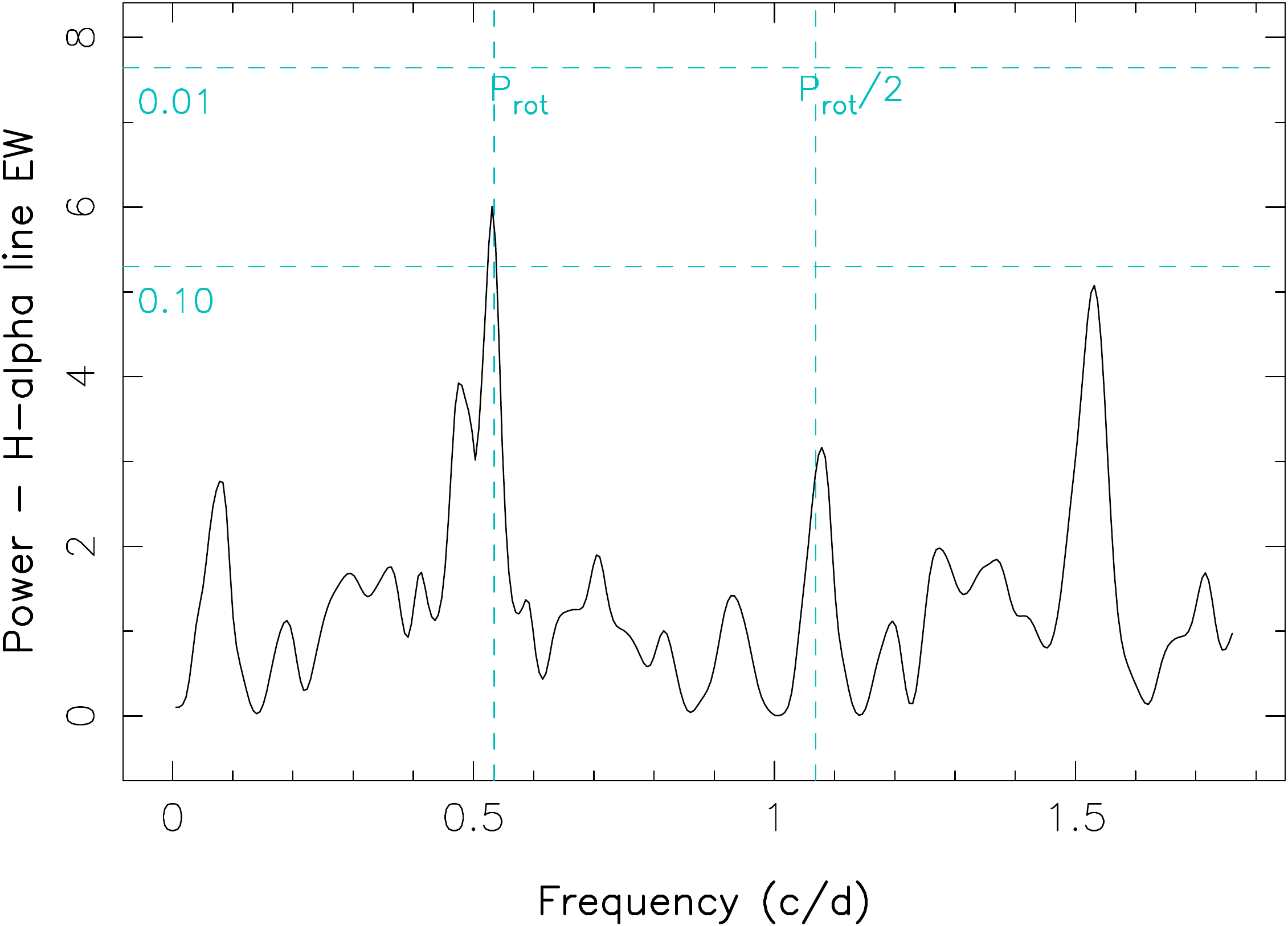}}
		\caption{(Continued from the previous page).}
	\end{figure*}
	\begin{figure*}
		\includegraphics[width=\linewidth]{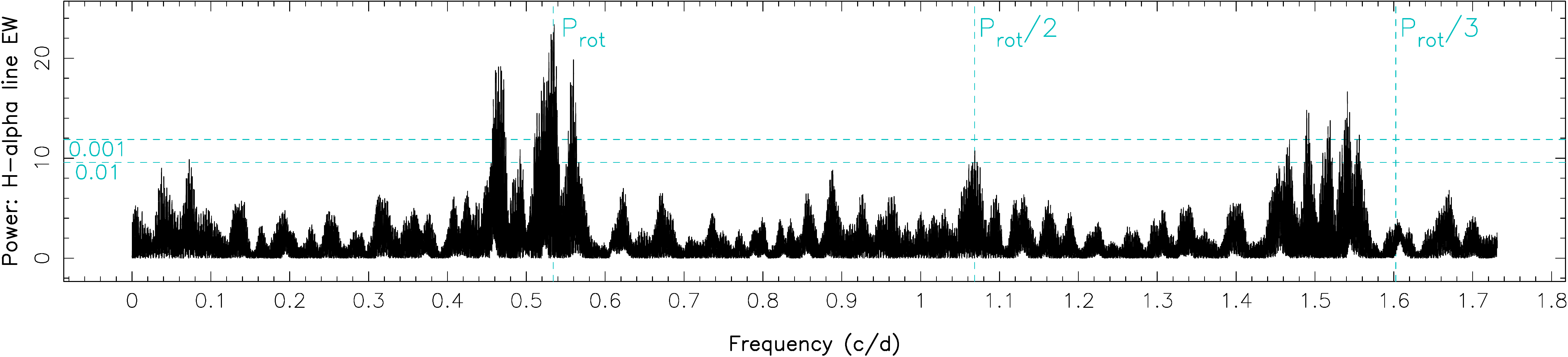}
		\caption{Periodogram of the equivalent width of the \hal\ line. The maximum power is found at 3.3636 rad/d, or 1.8680{~}d.}
		\label{fig:haa}
	\end{figure*}
	\begin{figure}
		\includegraphics[angle=-90,width=\linewidth]{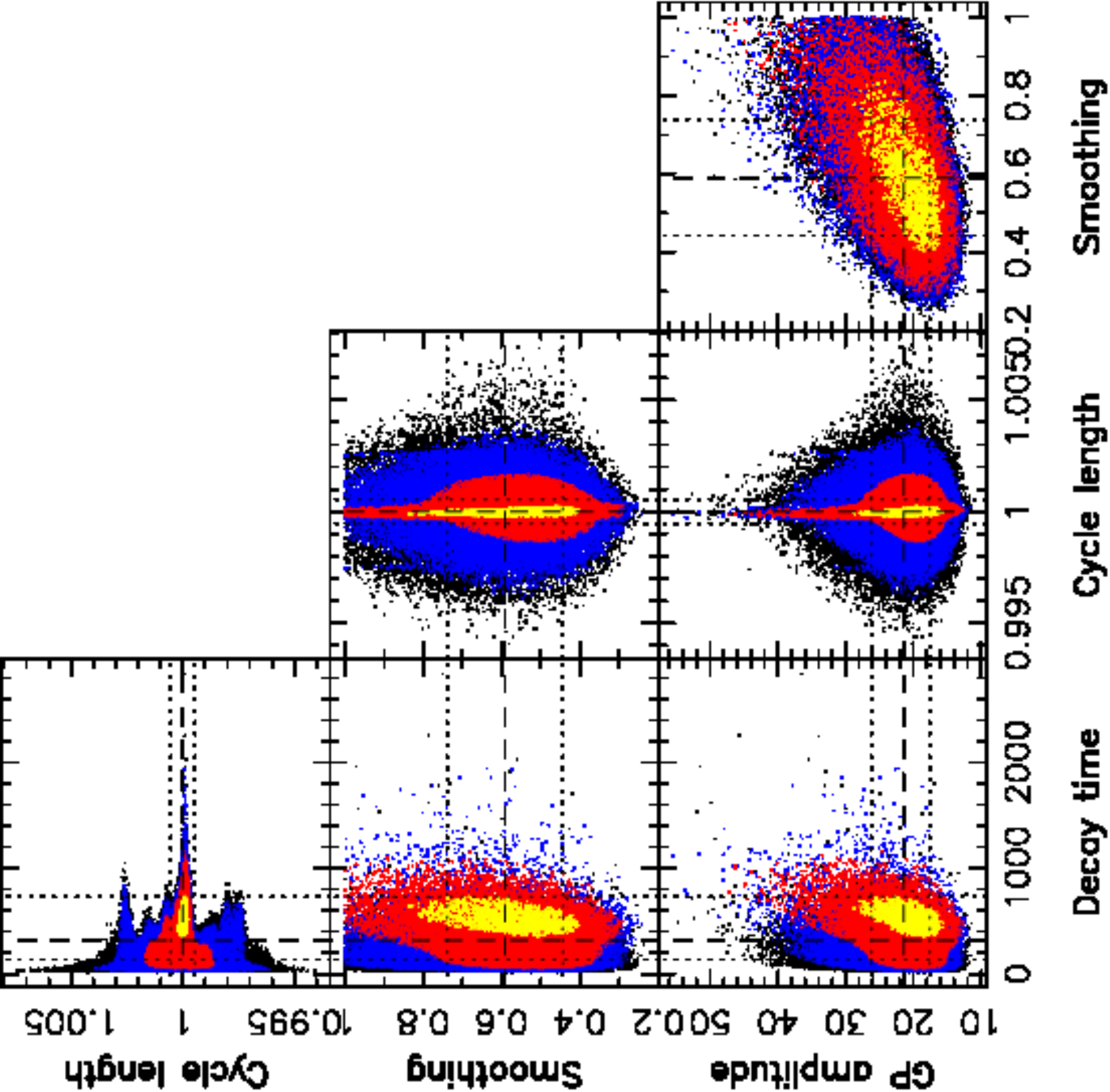}
		\caption{GPR-MCMC phase plot for our \hal\ equivalent width data. Amplitude ${\theta_1 = 21.4^{+4.7}_{-3.9}}${~}\kms, decay time ${\theta_3 = 315^{+414}_{-179}}${~}\Prot, Cycle length ${\theta_2 = 1.0000\pm 0.0005}${~}\Prot, Smoothing ${\theta_4 = 0.59\pm 0.15}${~}\Prot.}
		\label{fig:hag}
	\end{figure}

	\subsection{He$\;${\sc i} $D_3$}
\begin{figure*}
		\centering
		\subfloat[2008 Oct]{\includegraphics[totalheight=0.25\textheight]{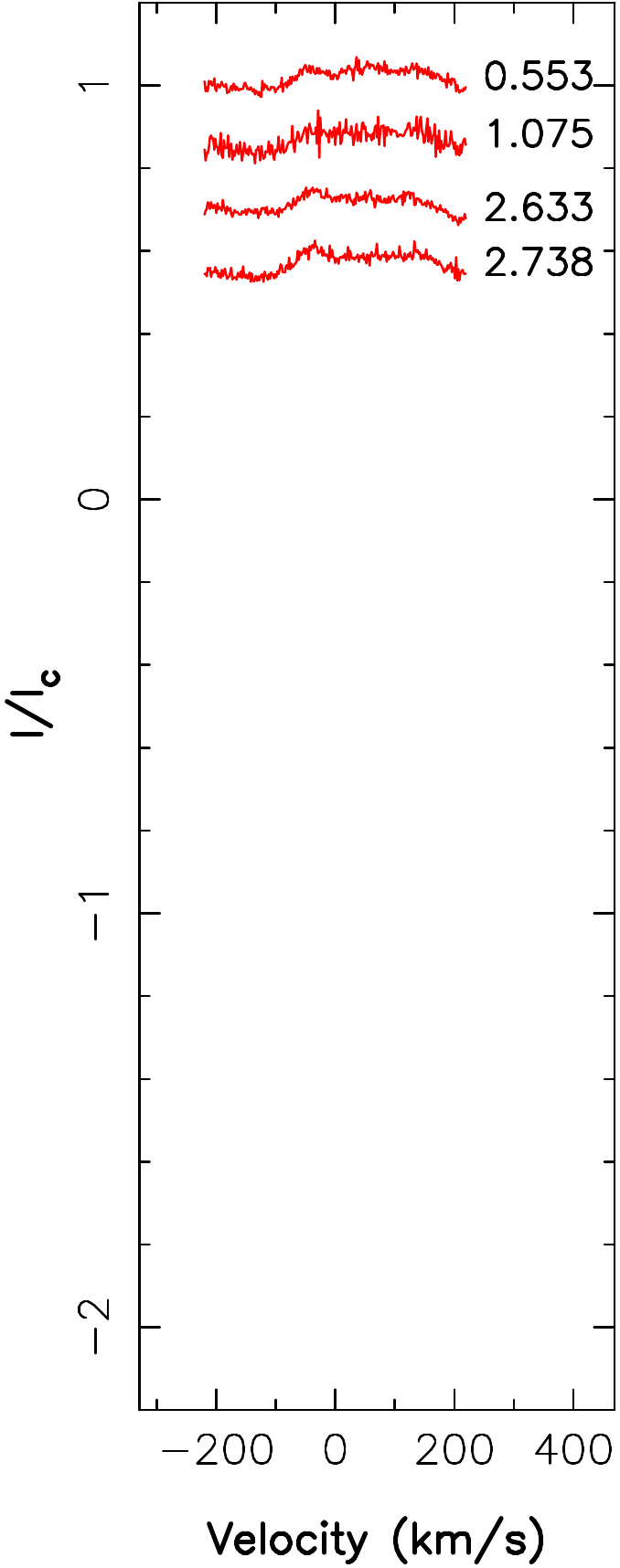}}
		\subfloat[2008 Dec]{\includegraphics[totalheight=0.25\textheight]{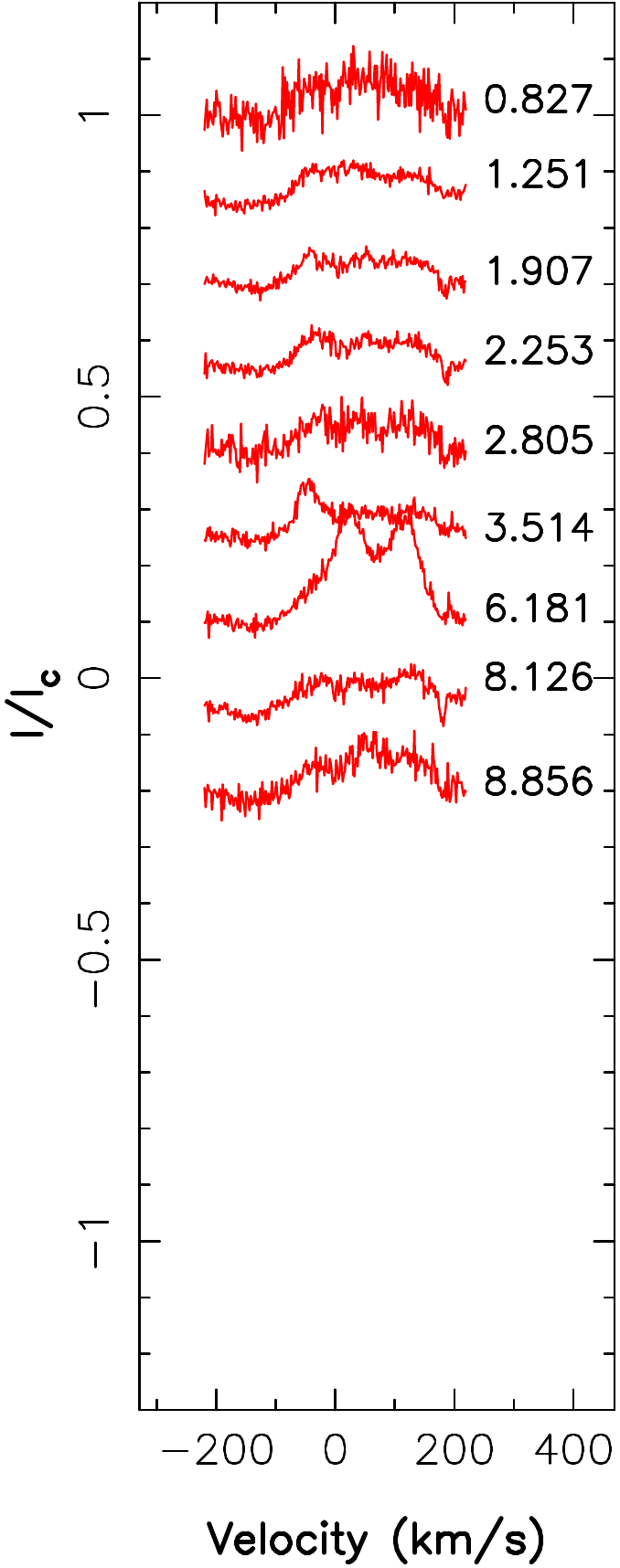}}
		\subfloat[2009 Jan]{\includegraphics[totalheight=0.25\textheight]{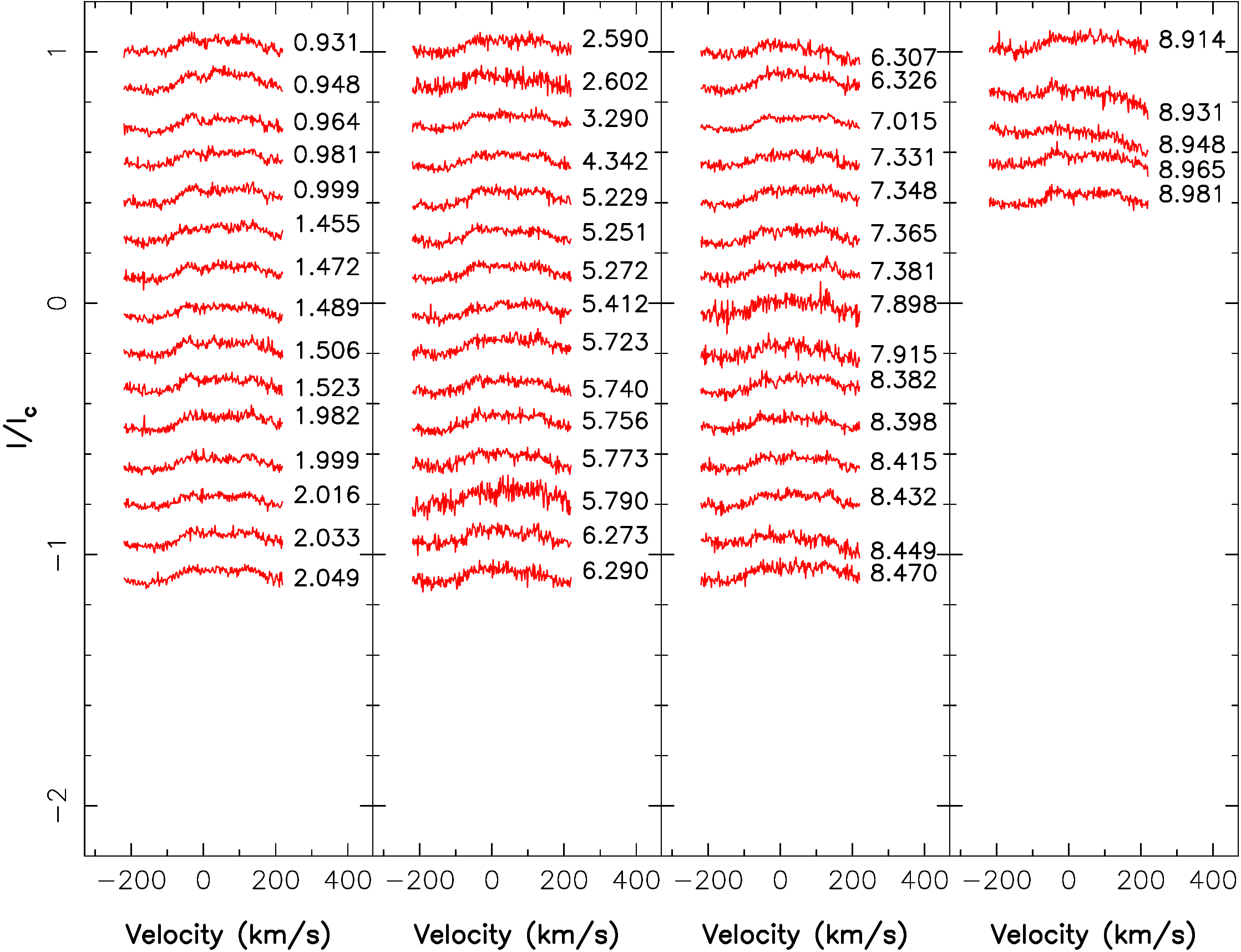}}
		\subfloat[2011 Jan]{\includegraphics[totalheight=0.25\textheight]{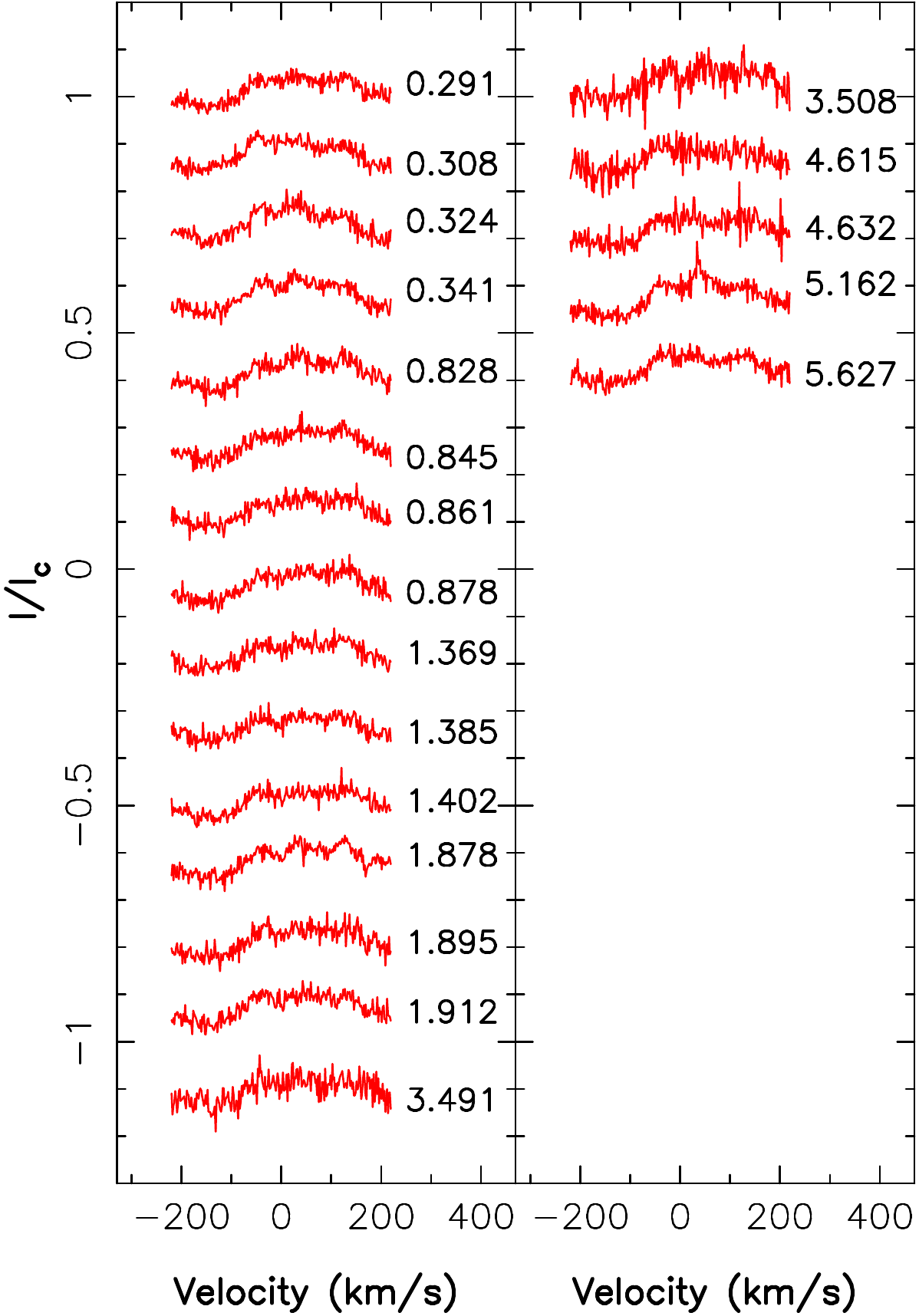}}

		\subfloat[2013 Nov]{\includegraphics[totalheight=0.25\textheight]{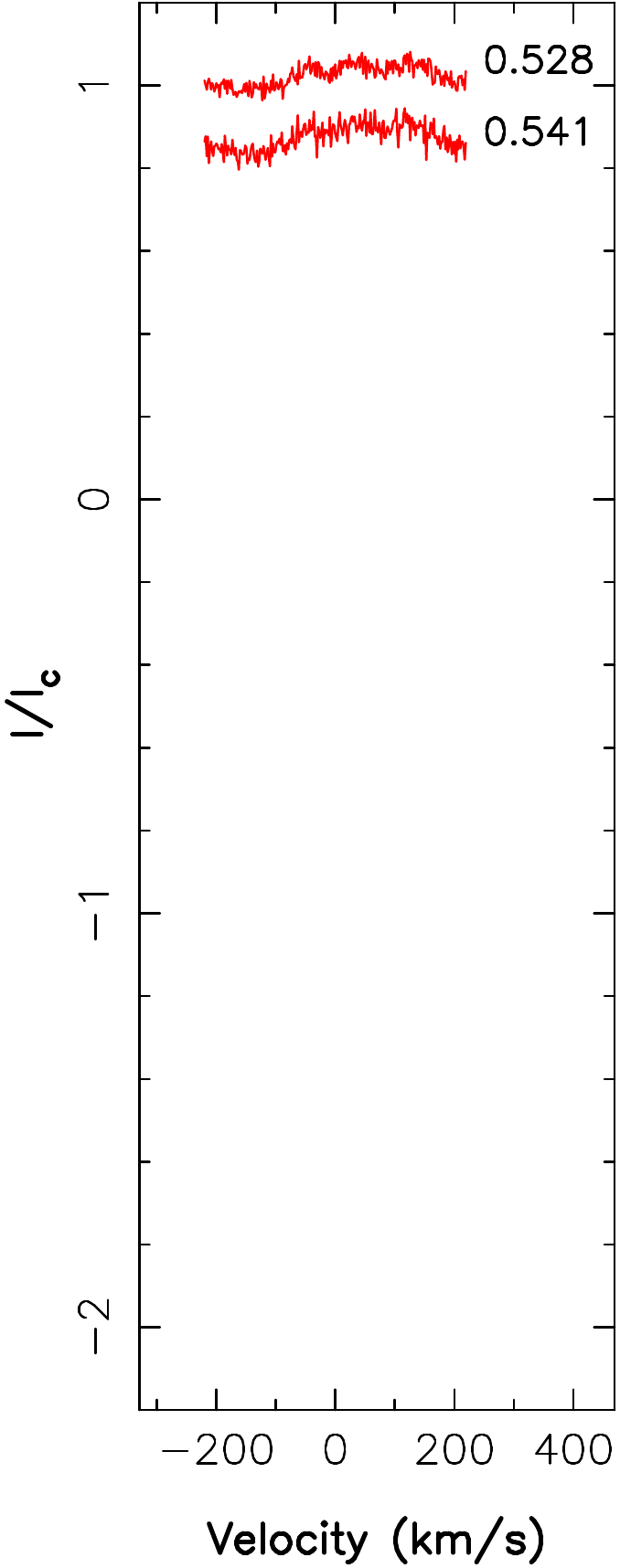}}
		\subfloat[2013 Dec]{\includegraphics[totalheight=0.25\textheight]{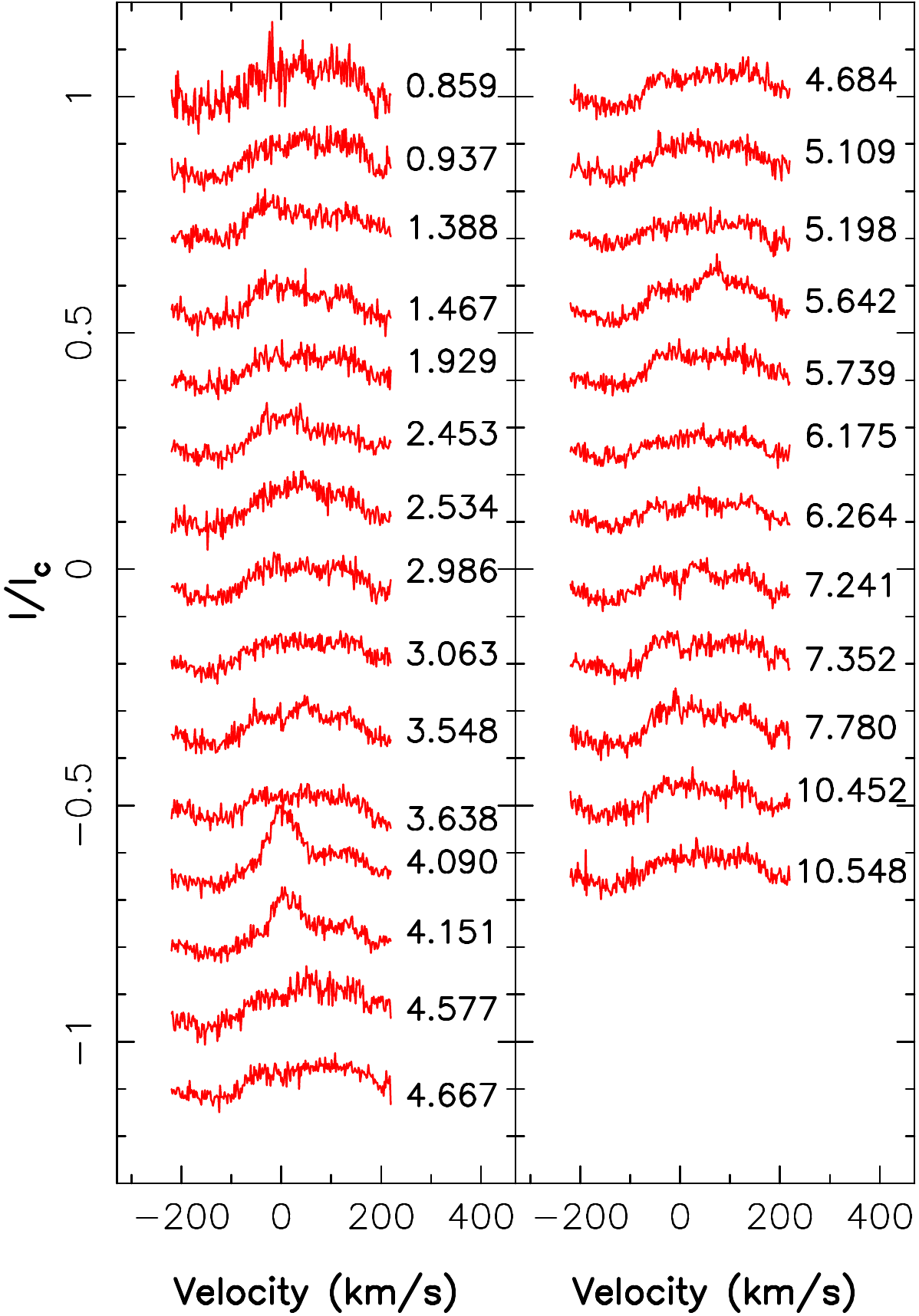}}
		\subfloat[2015 Dec]{\includegraphics[totalheight=0.25\textheight]{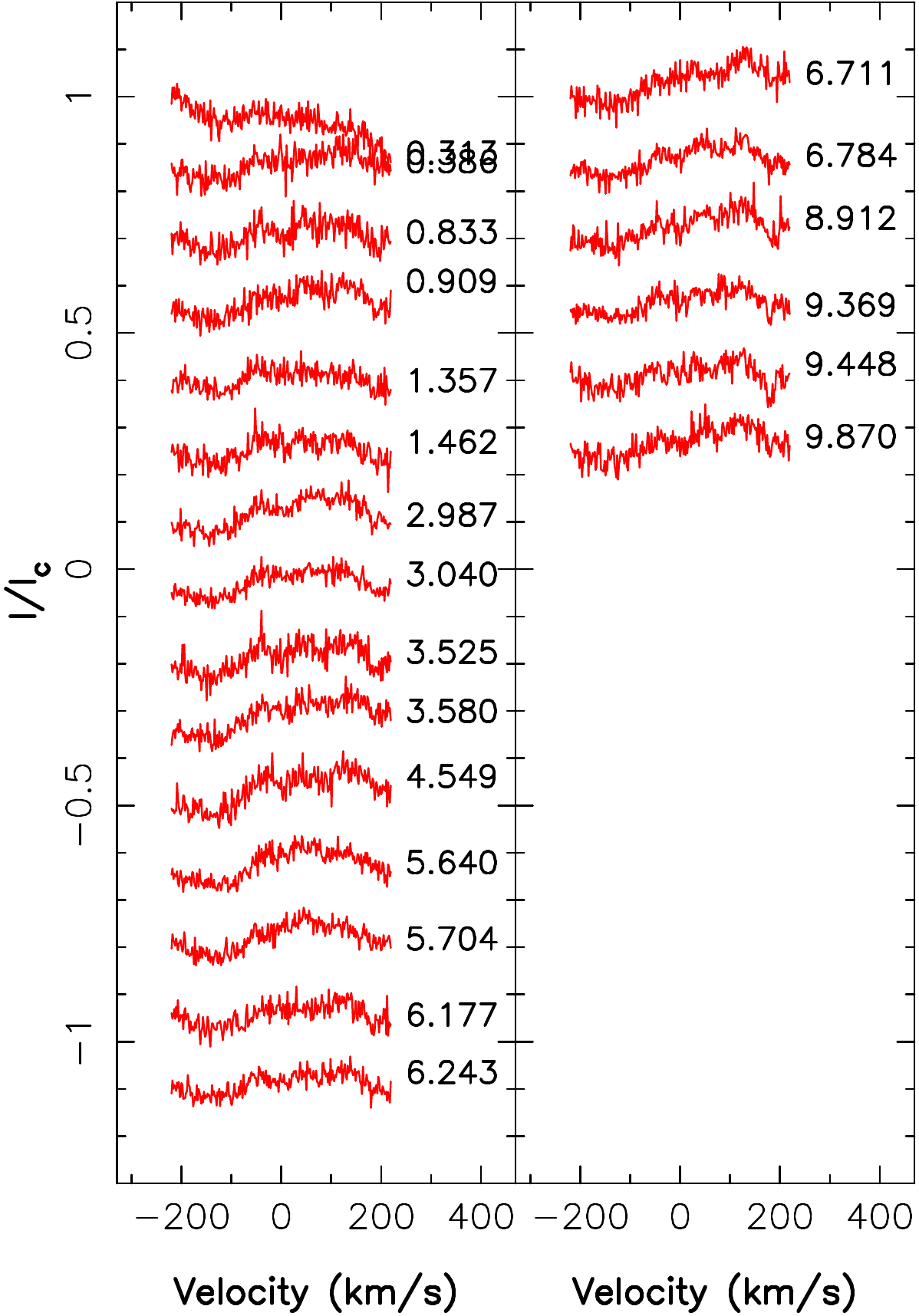}}
		\subfloat[2016 Jan]{\includegraphics[totalheight=0.25\textheight]{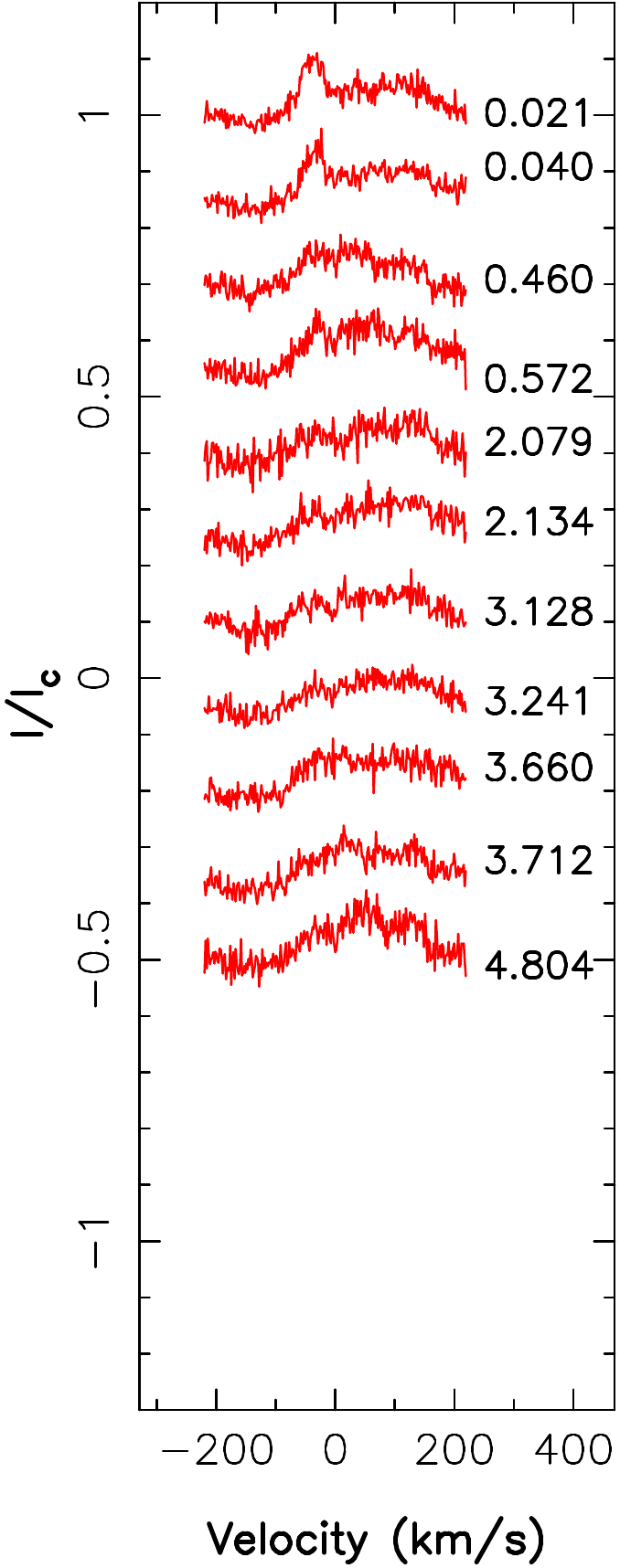}}
		\caption{. He$\;${\sc i} $D_3$ Oct 2008 (ref cycle: -42), Dec 2008 (ref cycle: -15), Jan 2009 (ref cycle: 0), Jan 2011 (ref cycle: 397) Nov 2013 (ref cycle: 946), Dec 2013 (ref cycle: 959), Dec 2015 (ref cycle: 1349) and Jan 2016 (ref cycle: 1376)}
		\label{fig:he1}
	\end{figure*}

	The \hei\ line profiles are shown in Fig.{~}\ref{fig:he1}. We can clearly see the flares at the dates marked in Table{~}\ref{tab:pob}.

	\subsection{Ca$\;${\sc ii}}
\begin{figure*}
		\centering
		\subfloat[2008 Oct]{\includegraphics[totalheight=0.25\textheight]{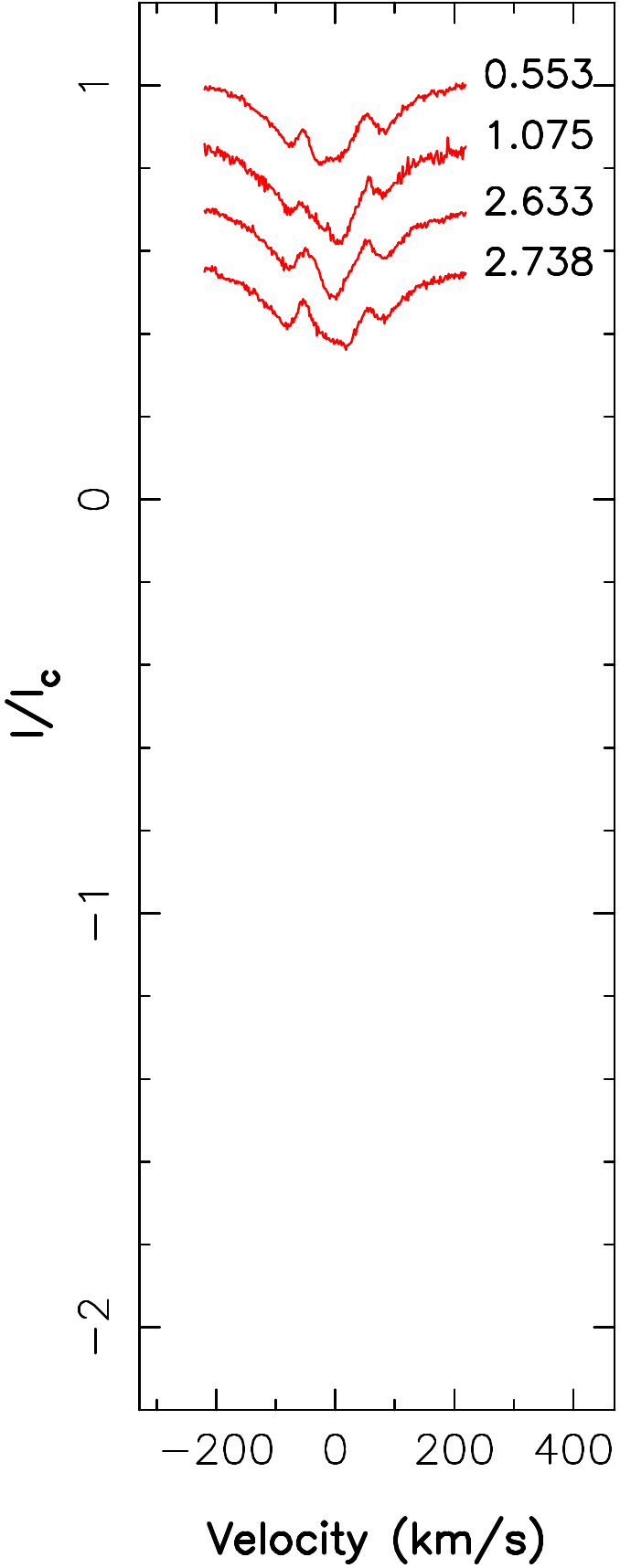}}
		\subfloat[2008 Dec]{\includegraphics[totalheight=0.25\textheight]{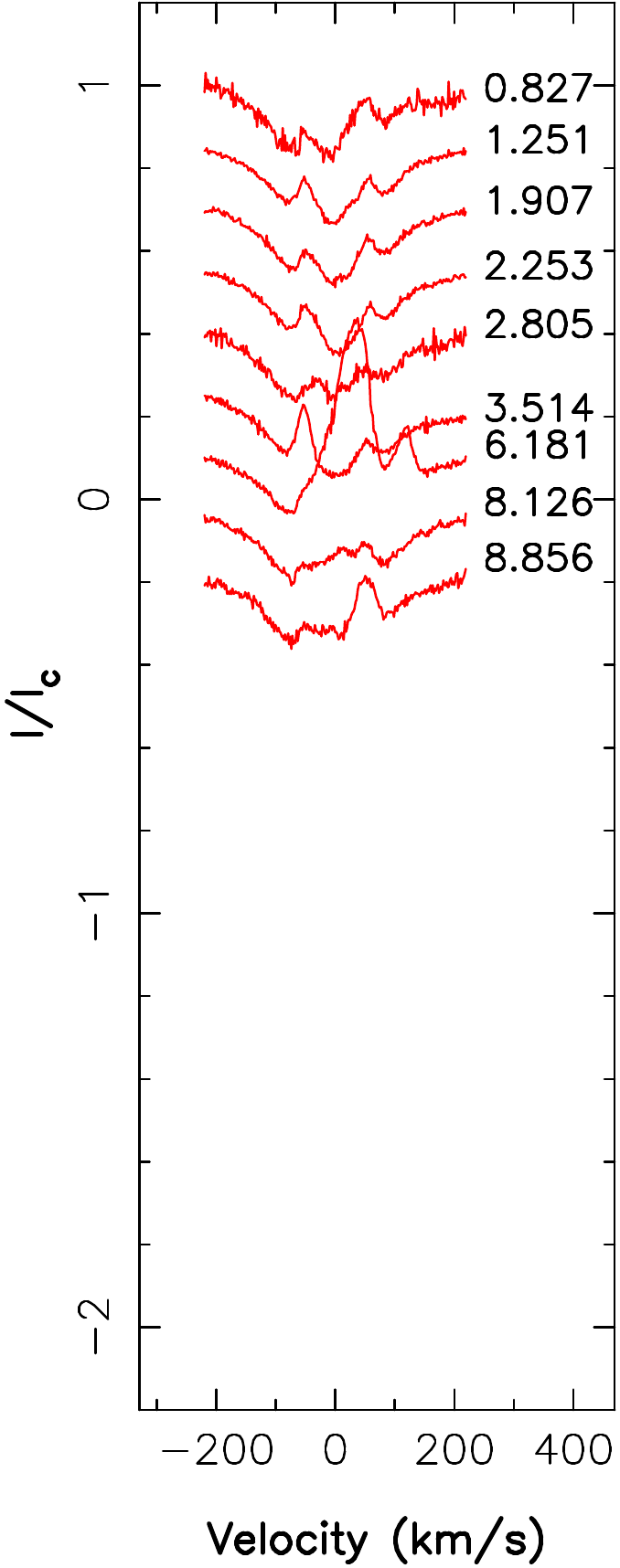}}
		\subfloat[2009 Jan]{\includegraphics[totalheight=0.25\textheight]{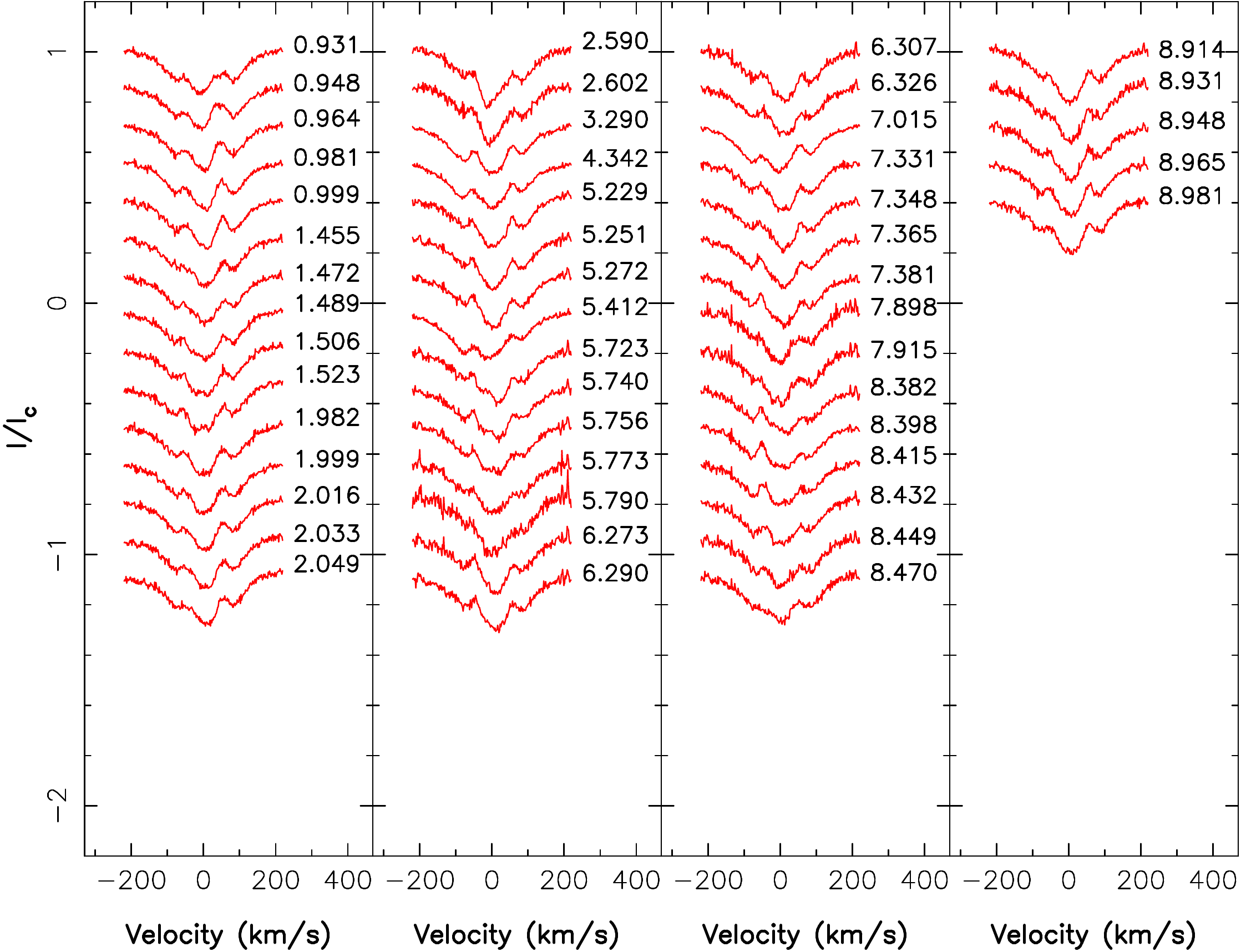}}
		\subfloat[2011 Jan]{\includegraphics[totalheight=0.25\textheight]{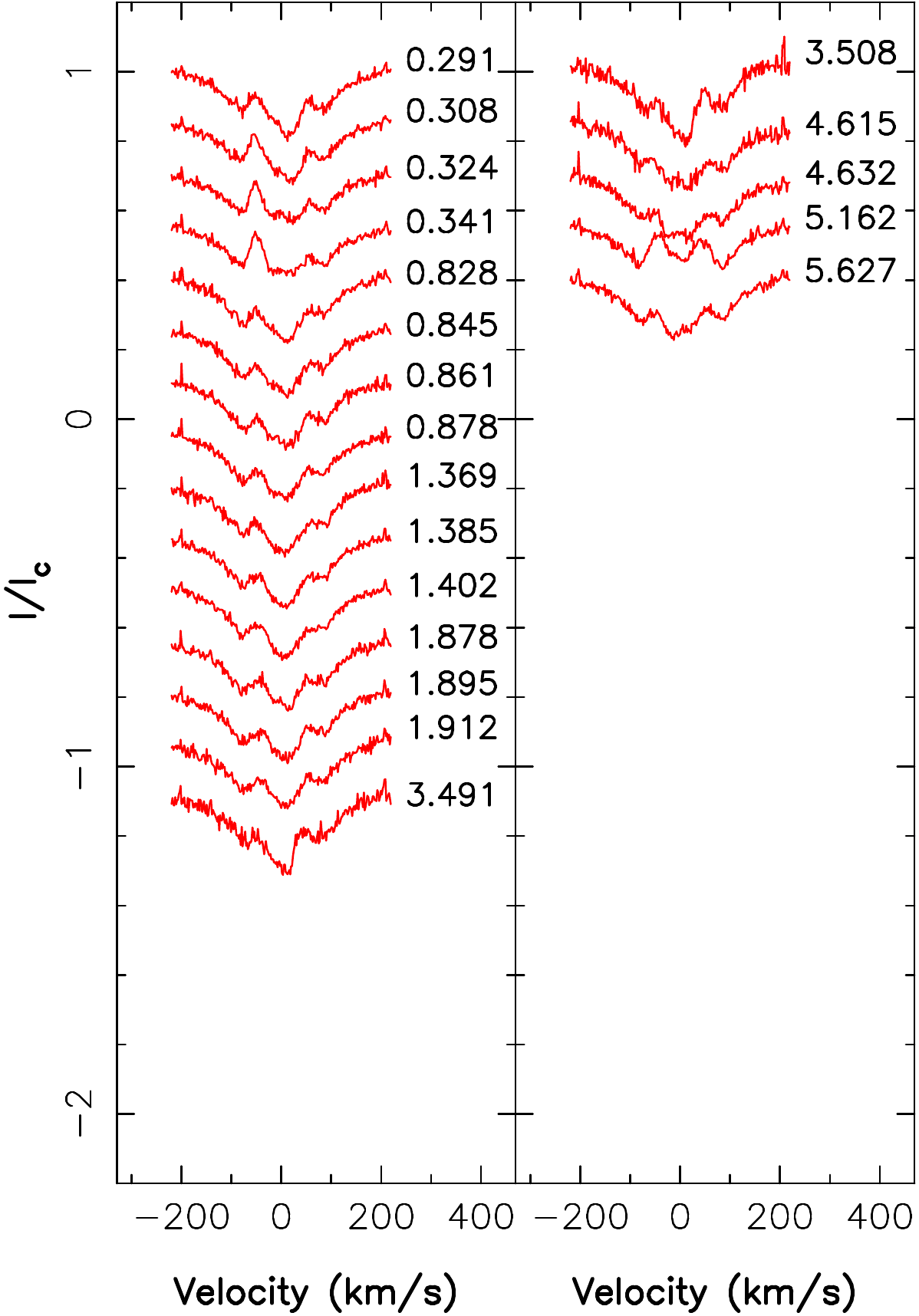}}

		\subfloat[2013 Nov]{\includegraphics[totalheight=0.25\textheight]{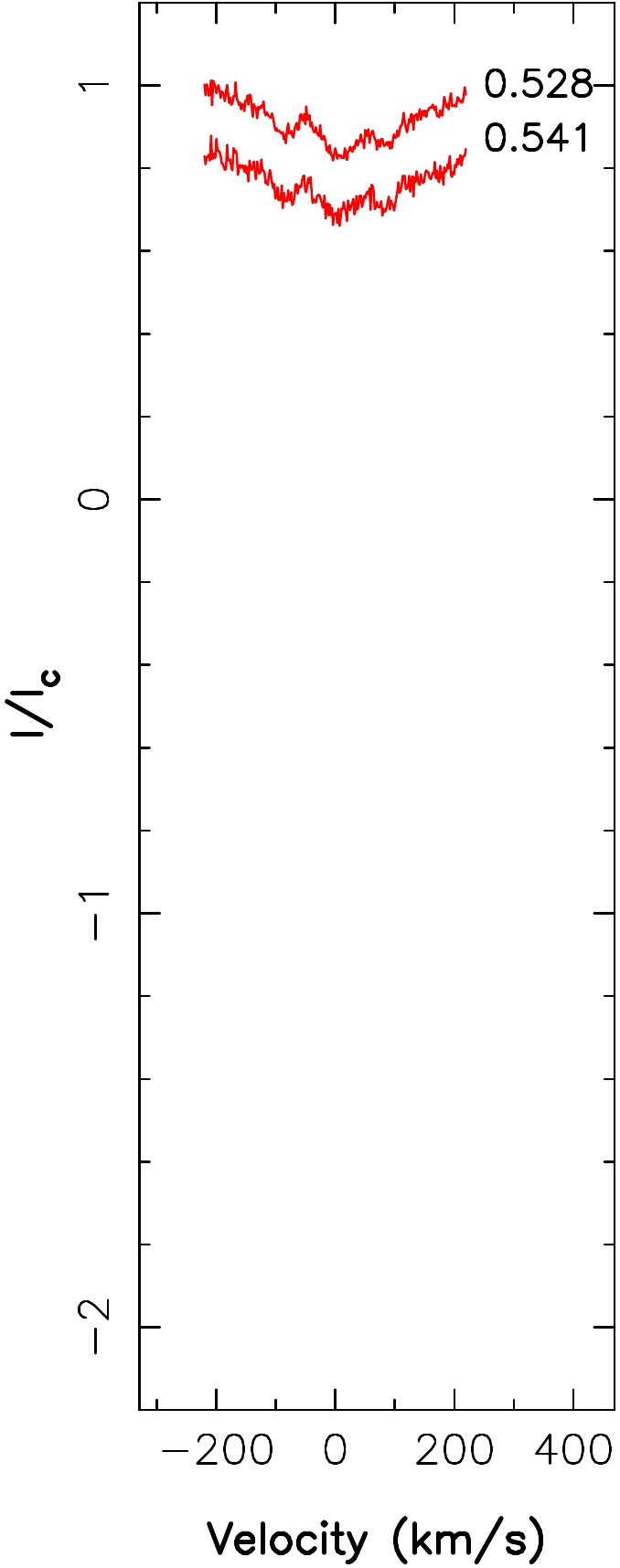}}
		\subfloat[2013 Dec]{\includegraphics[totalheight=0.25\textheight]{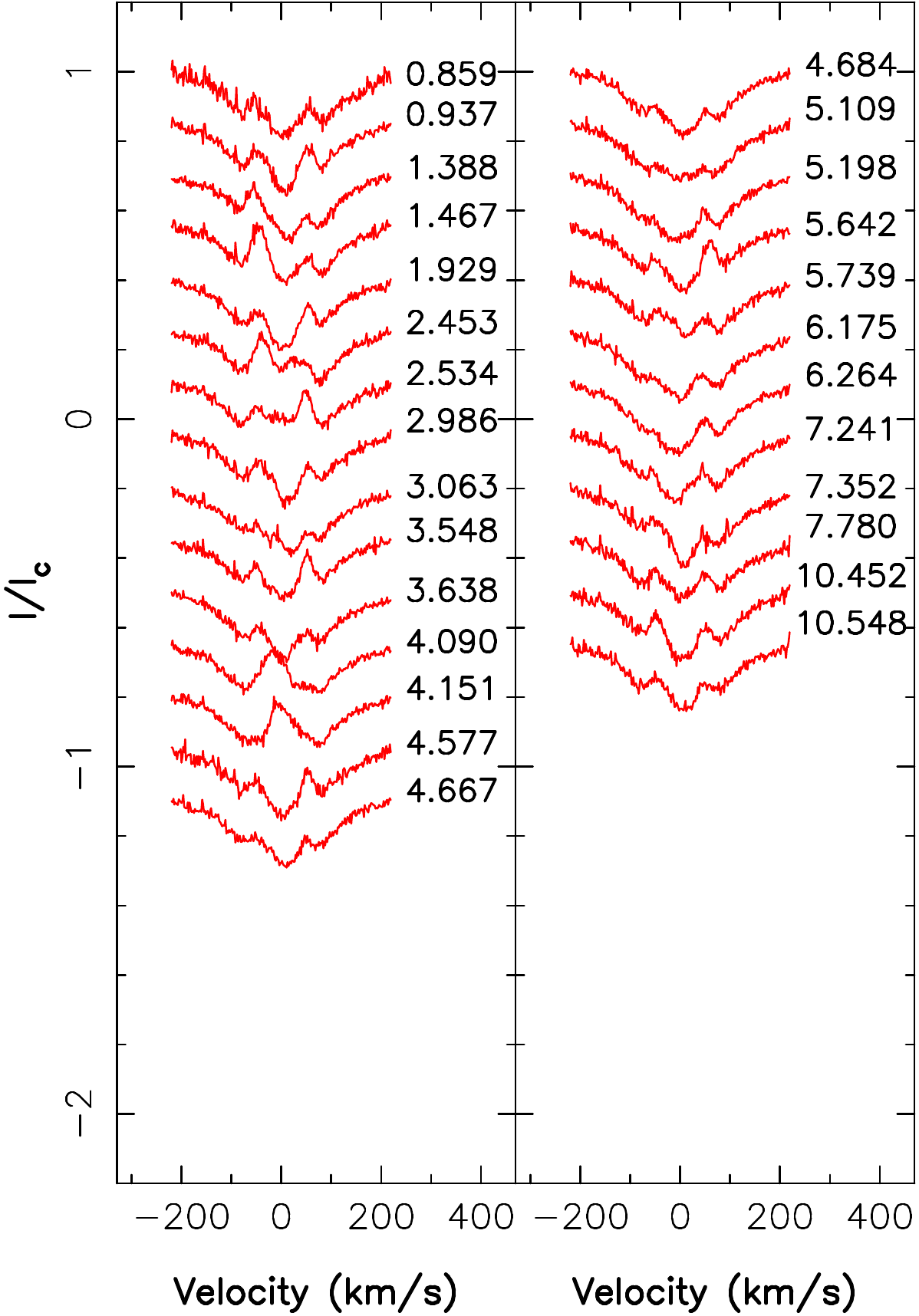}}
		\subfloat[2015 Dec]{\includegraphics[totalheight=0.25\textheight]{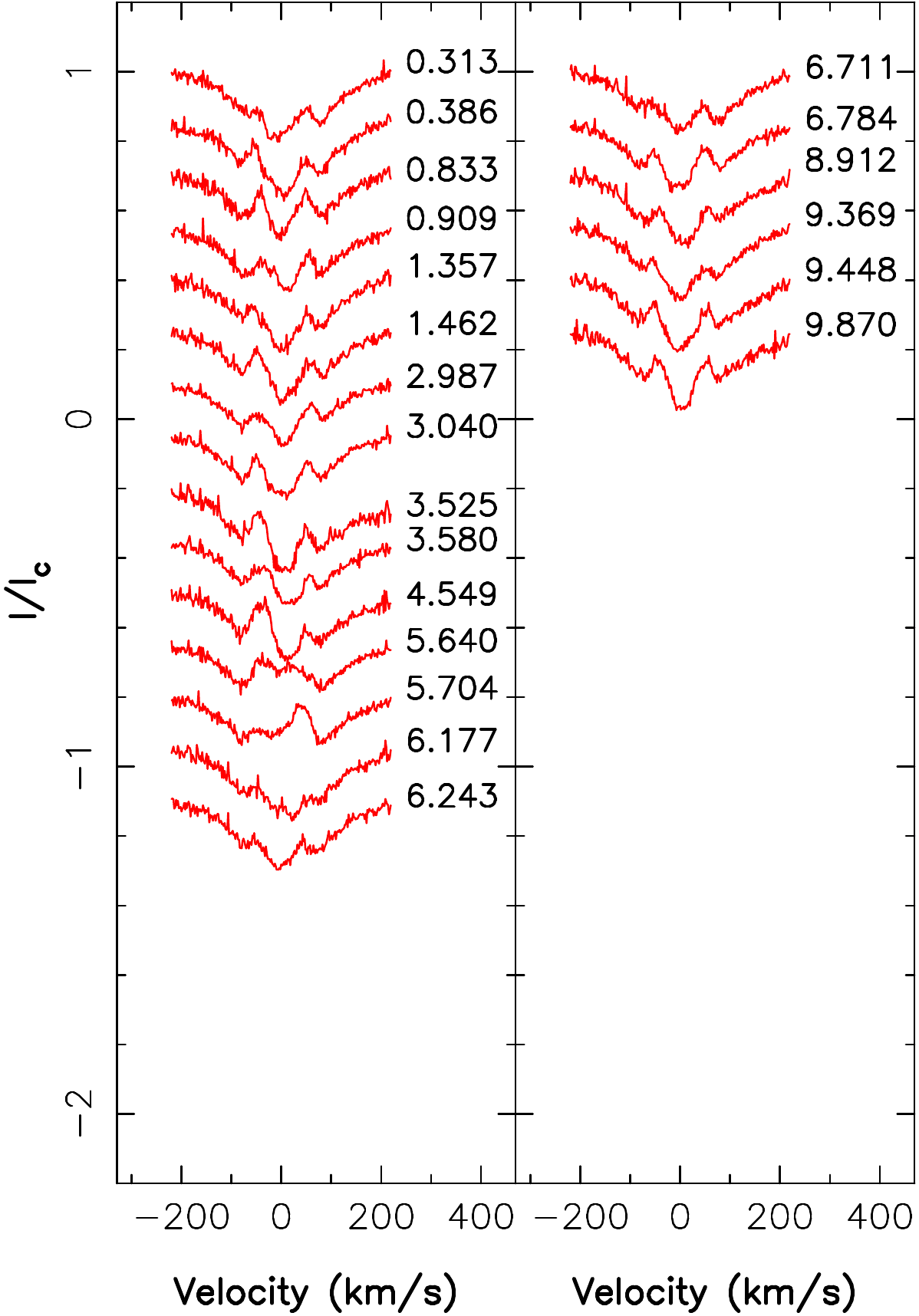}}
		\subfloat[2016 Jan]{\includegraphics[totalheight=0.25\textheight]{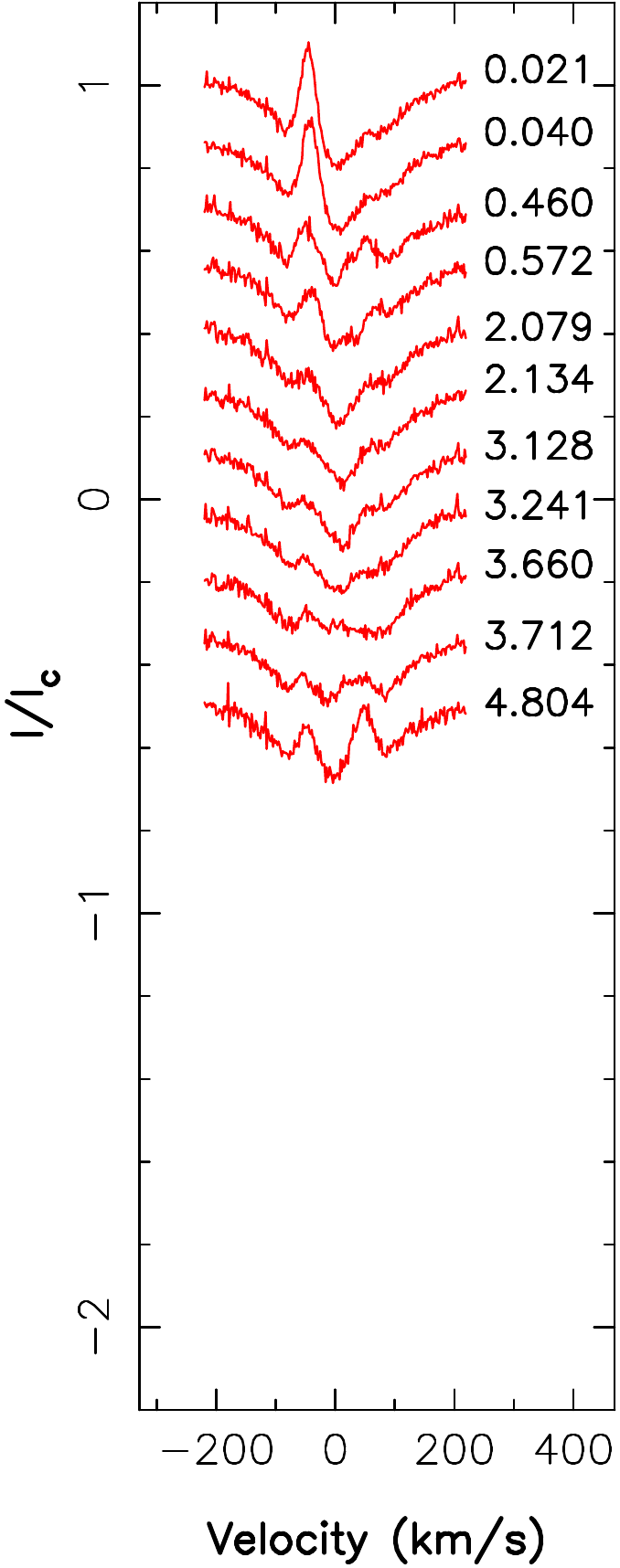}}
		\caption{Ca$\;${\sc ii} $D_3$ Oct 2008 (ref cycle: -42), Dec 2008 (ref cycle: -15), Jan 2009 (ref cycle: 0), Jan 2011 (ref cycle: 397) Nov 2013 (ref cycle: 946), Dec 2013 (ref cycle: 959), Dec 2015 (ref cycle: 1349) and Jan 2016 (ref cycle: 1376)}
		\label{fig:ca1}
	\end{figure*}

	The \caii\ line profiles are shown in Fig.{~}\ref{fig:ca1}.

\subsection{\Bl}
\begin{figure}
		\includegraphics[angle=-90,width=\linewidth]{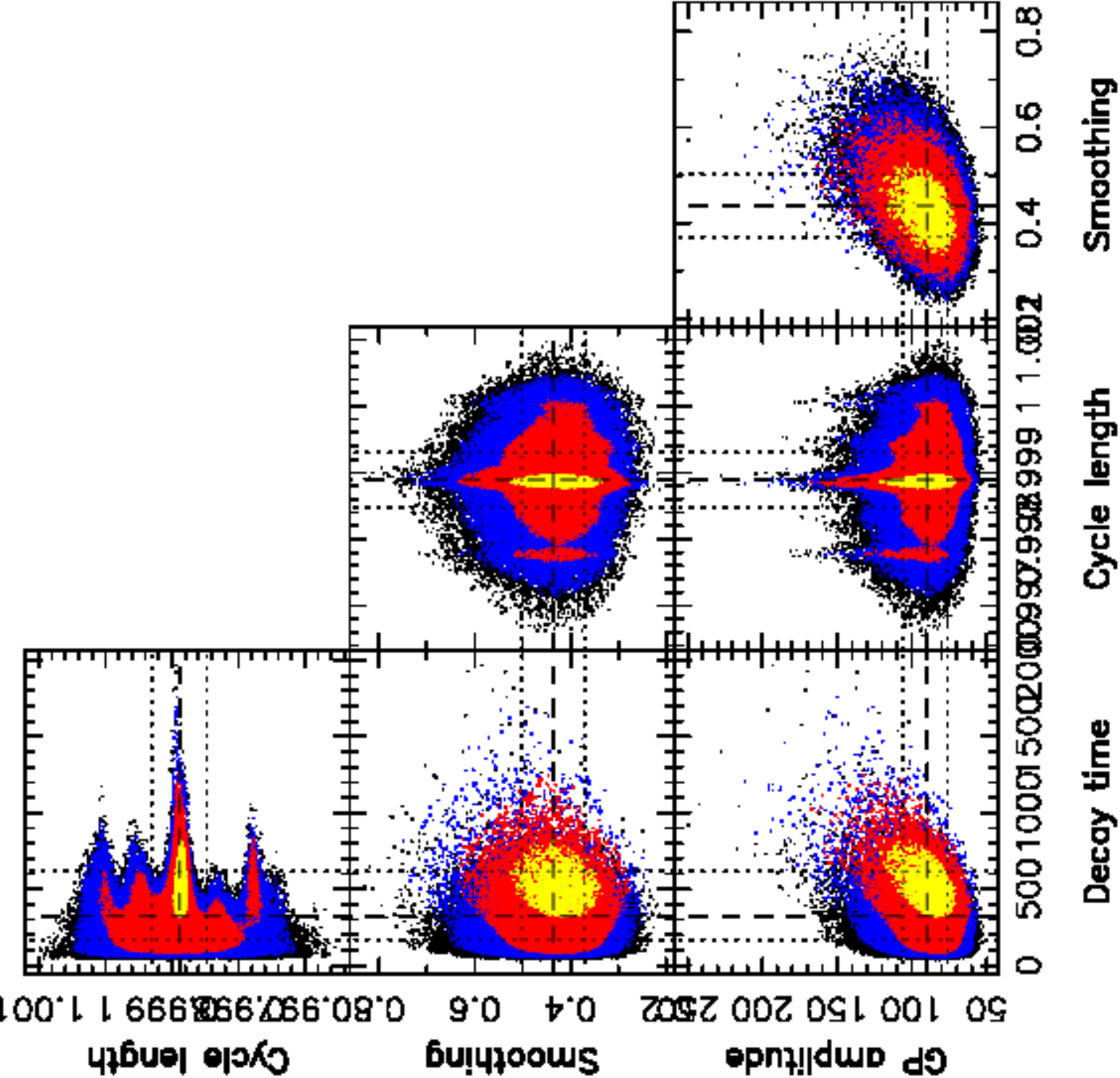}
		\caption{GPR-MCMC phase plot for \Bl. GP amplitude ${\theta_1 = 90^{+16}_{-14}}${~}G, cycle length ${\theta_2 = 0.9989\pm 0.0004}${~}\Prot, decay time ${\theta_3 = 322^{+295}_{-154}}${~}\Prot, smoothing ${\theta_4 = 0.436\pm 0.066}${~}\Prot.}
		\label{fig:blg}
	\end{figure}

	We derived longitudinal magnetic field values as first-order moments of our \stv\ LSD profiles, and applied a GPR-MCMC run on them. The phase plot is shown in Figure{~}\ref{fig:blg}.

	\bsp	
	\label{lastpage}
\end{document}